\documentclass[letterpaper,12pt,peerreviewca,draftcls]{IEEEtran}
\usepackage{csmMBFeb07,graphicx,url}

\usepackage{setspace}
\usepackage{multicol}
\usepackage{amsmath}
\usepackage{amssymb}
\usepackage[usenames, dvipsnames]{color}
\usepackage{bm}
\usepackage{cite}
\usepackage{color,soul}
\usepackage{xcolor}
\usepackage{soul}

\usepackage{caption}
\usepackage{subcaption}
\usepackage{float}
\usepackage{textcomp}

\usepackage[T1]{fontenc}
\usepackage{amsfonts}
\usepackage{xspace}
\usepackage{latexsym}
\usepackage{graphicx}
\usepackage{amssymb}
\usepackage{boxedminipage}
\usepackage{url}
\usepackage{color}

\title{Nonlinear System Identification\\
{\Large{A User-Oriented Roadmap}\\Preprint submitted to IEEE Control Systems Magazine}
}
\author{Johan Schoukens$^1$ and Lennart Ljung$^2$ \\  \\$^1$ Dep. INDI, Faculty of Engineering, Vrije Universiteit Brussel, Belgium\\$^1$ Dep. of Electrical Engineering, Eindhoven University of Technology,  The Netherlands \\ $^2$ Reglerteknik, ISY, Link\"{o}pings Universitet, Sweden}

\begin{document}
\maketitle
\CSMsetup

\section{Summary}
The goal of this article is twofold.  Firstly, nonlinear system identification is introduced to a wide audience, guiding practicing engineers and newcomers in the field to a sound solution of their data driven modeling problems for nonlinear dynamic systems. In addition, the article also provides a broad perspective on the topic to researchers that are already familiar with the linear system identification theory, showing the similarities and differences between the linear and nonlinear problem. The reader will be referred to the existing literature for detailed mathematical explanations and formal proofs. Here the focus is on the basic philosophy, giving an intuitive understanding of the problems and the solutions, by making a guided tour along the wide range of user choices in nonlinear system identification.  Guidelines will be given in addition to many examples, to reach that goal. 

\section{Introduction}

Nonlinear system identification is a very wide topic, every system that is not linear is nonlinear. That makes it impossible to give a full overview of all aspects of the field. For that reason, the selection of the topics and the organization of the discussion is strongly colored by the personal journey of the authors in this nonlinear universe. 

Identification of linear dynamic systems started in the late 1950's. Zadeh \cite{Zadeh1962} put the need for a well developed system identification framework on the top of the agenda, followed by early overviews of the field \cite{Astrom1971}. Eventually a series of books established the field \cite{BoxandJenkins1970,Astrom1970,Eykhoff1974,Ljung1987,Soderstrom1989}. Linear system identification booked many successes, and data driven modeling became an enabling factor in modern design methods. 

\begin{figure}[h] 
\centering
\includegraphics[scale=0.5]{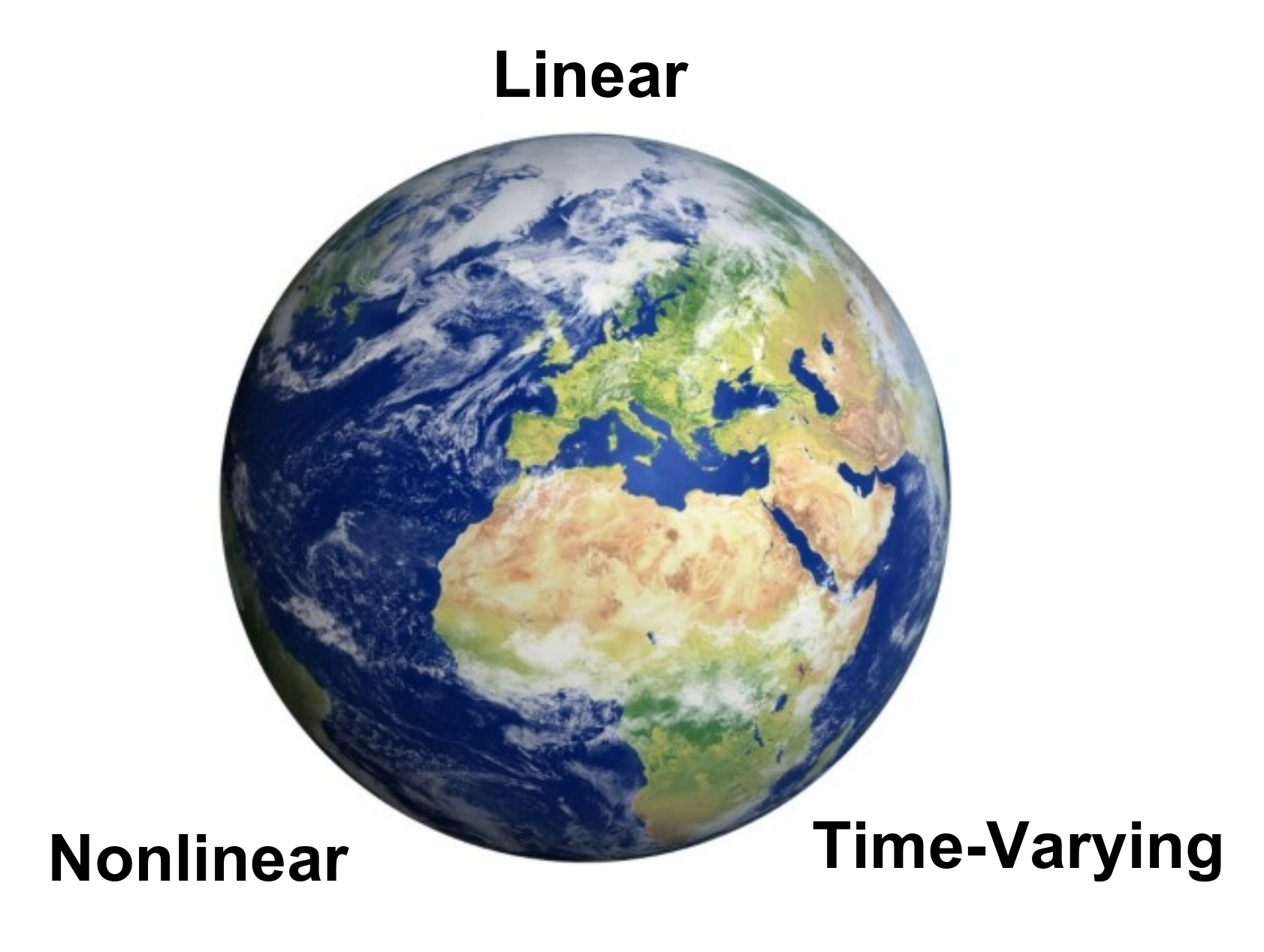}
\caption{Linear system identification is successfully applied to a wide variety of problems coming from many different fields. However,  the ever increasing demand for higher performance and efficiency pushes the systems in a nonlinear operation mode so that nonlinear models are required for their design and control. The real world is nonlinear and time-varying, and these aspects cannot be ignored. Data driven nonlinear model building has applications in traditional industrial and in emerging new high technological applications coming, amongst others, from the mechanical, electrical, electronic, telecommunication, and automotive field. Also biomechanical and biomedical applications can take full advantage of a nonlinear modeling framework. Good nonlinear models provide designers with (intuitive) insight that can guide them towards better solutions for tomorrow's products.}
\label{fig:RealWorld}
\end{figure}

Nonlinear system identification comes into the picture where linear system identification \cite{Soderstrom1989, Ljung1987, Pintelon2012} fails to address the users questions. The real world is nonlinear and time-varying, and in some applications these aspects can be no longer ignored (see Figure \ref{fig:RealWorld}) so that linear models become imprecise, or do not reproduce essential aspects of the behavior of the system under test. This article is focused on nonlinear system identification. An overview of time-varying system identification is given in \cite{Pintelon2015} and the references therein. 

\begin{figure}[h] 
\centering
\includegraphics[scale=0.5]{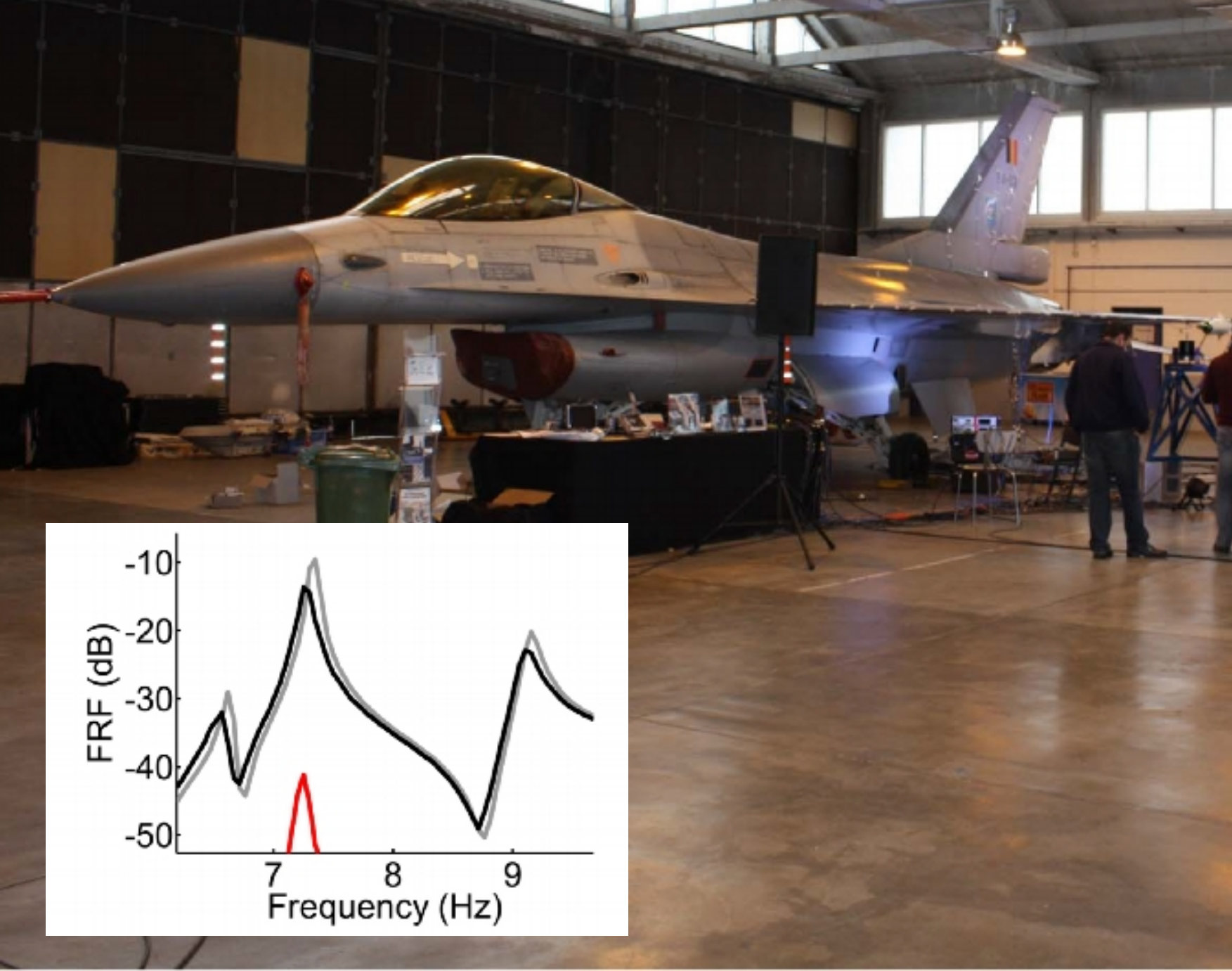}
\caption{Variation of the resonance frequency and damping of a torsion mode of the wing of an air-fighter for varying excitation levels. The right wing is excited at the tip with a shaker. The FRF from the input force to the tip-acceleration is measured at small (grey line) and medium (black line) excitation levels. The red line gives the level of the nonlinear distortions. The nonlinearity is due to friction and gaps in the bolted connection between the wing and the missile. In the inset, the FRF of the best linear approximation $G_{BLA}(f)$ is shown in the frequency band of interest  (see ``Linear Models of Nonlinear Systems'') \cite{Vaes2015,Dossogne2015}.}
\label{fig:F16}
\end{figure}

Nonlinear behaviour appears in many engineering problems. In mechanical engineering, nonlinear stiffness and damping, nonlinear interconnections, etc. are troubling ground vibration tests of airplanes and satellites, resulting amongst others in resonance frequencies and dampings that vary with the excitation level  (see Figure \ref{fig:F16}). In telecommunication, power amplifiers are pushed in a nonlinear operation regime to improve the power efficiency. Distillation columns exhibit nonlinear dynamic behavior. Many biological systems (eye, ear, sense of touch) apply first a nonlinear compression, known as the Weber-Fechner law, in order to cover the very large dynamic range of the inputs. The human brain is governed by nonlinear relations between the neurons. 

The need for nonlinear system identification goes far beyond the control application field. Nonlinear models are instrumental to get a basic understanding in very different problems like brain activity modeling, chemical reactions, ... where researchers still struggle with the question: How does it work? In these applications, (high) accuracy is not always needed, qualitative results can be very helpful to isolate the dominant terms. So,  structural model errors (that is, deficiencies in the chosen model structure) become more important than noise disturbances, and the system identification tools should be properly tuned to deal with dominating structural model errors. This shows that there are many different reasons to move from linear towards nonlinear models.  It is very important to keep in mind the motivation of the researchers who developed nonlinear identification tools to relate the different methods. Without understanding the drive of the researcher/scientist it is often very difficult to understand and appreciate the choices they made.

The outline of the article is organized along a number of sections that cover the main flow of the nonlinear system identification process, supported by sidebars to provide additional information and background information on some topics. The article starts with a short discussion on the lead actors in (nonlinear) system identification. These are still the same as those identified in the early 1960's by Zadeh \cite{Zadeh1962}: the data, the model, and the matching criterion. These are discussed from a nonlinear identification perspective in the Section "Lead Actors in Nonlinear System Identification".  Next, in the Section "Why is Nonlinear System Identification so Involved", it is clarified why the complexity of the modeling process grows very fast when moving from linear to nonlinear system identification. The goal of the nonlinear modeling process strongly affects the required user effort. This is discussed in the Section "Goal of the Nonlinear System Identification Process". Nonlinear system identification is much more involved than that linear identification is. For that reason it should be a well informed decision to move towards nonlinear methods. This is discussed in the Section "Linear or Nonlinear System Identification: A Users Decision".  There are many more nonlinear model structures than there are for linear systems. Making a proper choice along this wide range of possibilities is one of the major difficulties for newcomers in the field. Guidance to make a proper choice is given from a systems behavior perspective (for example: hysteresis, chaos, fading memory, etc.), and from a users point of view (physical or black box model, model that is linear-in-the-parameters, etc.). This is discussed respectively in the Section "The Palette of Nonlinear Models" and the Sidebar ``External or internal nonlinear dynamics''. Black box nonlinear models are more difficult to access and to understand than physical models (``Black Box Models Complexity: Keeping the Exploding Number of Parameters Under Control; Increased Structural Insight; Model Reduction''). Often it is hard to get intuitive insight in the modeled behavior because the number of terms in the model becomes very large. Eventually, some attention is paid to experiment design. A number of (extensive) sidebars provide more detailed background information and highlight some important aspects of the nonlinear identification process so that the main flow of the article remains very clear. For example, in the Sidebar "Retrieving Structural Information", methods are discussed to increase the insight and to reduce the number of terms by searching for hidden structural information in the raw nonlinear black box models. The Sidebar ``Impact of Structural Model Errors'' makes the reader familiar with the huge impact that structural model errors have on the best identification practices.

Many aspects are illustrated on examples, and supporting software is made publicly available. Guidelines for the user, to pinpoint the main lessons and conclusions are given at the end of most sections. 
\section{The Lead Actors in (Nonlinear) System Identification}
Any identification procedure to build models using observed input-output data is characterized by three main components, the \emph{data}, the set of candidate \emph{models}, and the estimation \emph{method}. To the lead actors should also be counted the process to gain confidence of the estimated model, the \emph{validation} procedure. These four actors will be briefly presented in this section, while more comprehensive treatments will follow in forthcoming sections.
\subsection{The Data}
The input-output data that is used to select a model is the fundamental information source. To select the signals to be measured, to decide how the input should be configured, and to collect the data with appropriate sampling procedures will have a major impact on the quality of the resulting model. This is the task of \emph{Experiment Design}. It is important to realize that no model can be a perfect description of the true system  under investigation. Any model will be an approximation of the truth, and it will be affected by the aspects of the system that are excited during the experiment. 

For that reason it is important to design the experiment to cover the intended use of the model. The power spectrum (e.g. white or colored noise) and the amplitude distribution (e.g. uniform, Gaussian, or binary) should be properly set. Of course the classical linear identification rules remain also valid (persistency of excitation, maximum Fisher information), but these should be balanced against the other requirements to identify a well behaving approximating model (see also "Impact of Structural Model Errors").

Experiment design is further detailed in the section with that title.

\emph{User guideline}: Make sure that the experiment covers the domain of interest and brings out all essential system features of interest.

\subsection{The Model Structure: the Set of Candidate Models}
For linear identification the choice of model sets is quite easy to grasp: Settle  for a state-space structure of certain dimension or transfer functions of certain orders. In contrast, the choice of a model set for nonlinear identification is a major problem and offers a very rich range of possibilities. It is driven by the users preferences and directed by the system behavior. In fact, aspects of model properties and considerations of model choices dominate this article. An overview of the users choices along possible and useful nonlinear model structures is given in the section ``The Palette of Nonlinear Models''  arranged by the amount of prior physical knowledge about the system that is incorporated in the model. In addition, the systems behavior imposes whether the nonlinearity should be captured in a dynamic closed loop or not. This may have essential impact on the behavioral patterns of the model, and is further discussed in the Sidebar ``External of Internal Nonlinear Dynamics''.

A further important distinction is whether disturbance sources enter before the nonlinearity or not. In the former case, proper stochastic treatment of the model becomes more cumbersome and may require advanced tools. This is discussed in the Side Bar ``Process Noise in Nonlinear System Identification.''

In any case, a model should be capable of producing a \emph{model output} $\hat y(t)$ for the output at time $t$ based on previous input-output measurements. This could be computed as a formal prediction of the output, or it can be based on other considerations.
The set of candidate models is typically parameterized by a parameter vector $\theta$, and the notation 
\begin{equation}
\hat y(t|\theta)
\end{equation}
 will be used for the model output corresponding to the model parameter $\theta$.

\subsubsection{User guideline}
The choice of the model structure is directed by behavior and structural aspects.
\begin{itemize}
\item \emph{Behavior aspects} are imposed by the system: Can the selected model reproduce the observed macroscopic behavior like shifting resonances, hysteresis, etc? 
\item \emph{Structural aspects} are a user choice that is set by the level of physical insight that the user desires to inject in the model, ranging from white box physical models to black box models.
\item  \emph{Structural model errors} in the model result if the chosen model structure is not rich enough to contain a true description of the system
\end{itemize}

\subsection{The Estimation Method}
With a given data set and model set, the identification task is to \emph{select that model ($\theta$) that best describes the observed data}. Most such estimation methods are based on a \emph{criterion of fit} between the observed output $y(t)$ and the model output $\hat y(t|\theta)$,  which can conceptually be written as
\begin{align}
  \label{eq:critfit}
  \hat \theta_N = \text{argmin}_\theta\sum_{t=1}^N \|y(t)-\hat y(t|\theta)\|^2
\end{align}
If indeed the model output is computed as a one-step ahead prediction based on the model set and data available at time $t-1$, and the prediction error is Gaussian, then the model estimate $\hat \theta_N$ will be the \emph{Maximum Likelihood Estimate, MLE}. But the conceptual method (\ref{eq:critfit}) can be interpreted also in more pragmatic terms without making a statistical motivation, see ``Impact of Structural Model Errors''.

The cost function \eqref{eq:critfit} can be extended with a regularization term to include prior knowledge, or to impose a desired behavior like smoothness or exponential decay to the solution (see also \eqref{eq:CostNLTDReg} in ``Black Box Models Complexity: Keeping the Exploding Number of Parameters under Control; Increased Structural Insight; Model Reduction''). 

\emph{User guidelines}:  The criterion of fit can be based on a statistically grounded choice if the noise disturbances dominate over the structural model errors. If the latter dominate, a weighting function can be selected that reduces the impact of structural model errors in the domain of interest (for example using a user selected frequency weighting), at a cost of getting larger errors outside the domain of interest.

\subsection{Model Validation}
When a model has been estimated, the question to ask is ``Does it solve our problem?'' and/or is it in conflict with either the data or prior knowledge? This is the essential procedure of model validation, which is further discussed in the section with this title. 

Often the decision is that the model is not ``good enough'', so some choices have to be  revised. Typically other model sets have to be tested or the conclusion might be that the data was not informative enough, so the experiment design must be reworked. This is the reason why identification  often is seen as an iterative problem with an ``identification loop''. See, e.g. Figure 17.1 in \cite{Ljung1987}.

Model validation is further detailed in the section with that title.

\emph{User guidelines:}
Validating of a model is a rather subjective and pragmatic problem. Check on a rich validation data set that covers the intended use of the model if the estimated model meets the user expectations.

\section{Why is Nonlinear System Identification so Involved?}
Nonlinear system identification is experienced to be much more involved than linear identification. Three aspects contribute to this observation: i) Nonlinear models live on a complex manifold in a high dimensional space, while linear models live on a simple hyperplane that is much easier to characterize. ii) Structural model errors are often unavoidable in nonlinear system identification, and this affects the three major choices: the experiment design, the model selection, and the selection of the cost function. iii) Process noise entering before the nonlinearity requires new numerical tools to solve the optimization problem.

\subsection{From Hyperplane to Manifold}
\begin{figure*}[h]
    \centering
    \begin{subfigure}[t]{0.5\textwidth}
        \centering
        \includegraphics[scale=0.43]{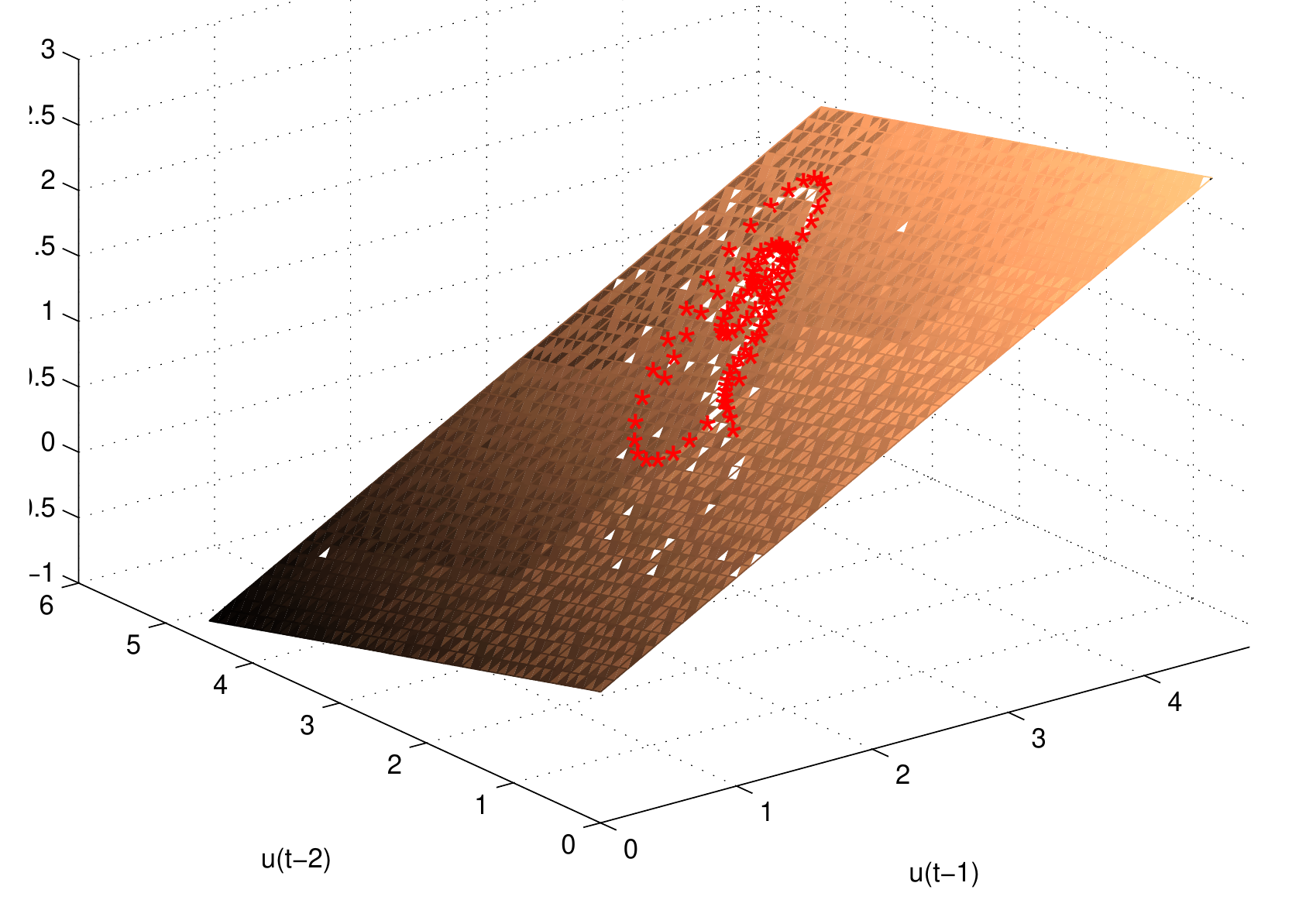}
        \caption{}
    \end{subfigure}%
    ~ 
    \begin{subfigure}[t]{0.5\textwidth}
        \centering
        \includegraphics[scale=0.43]{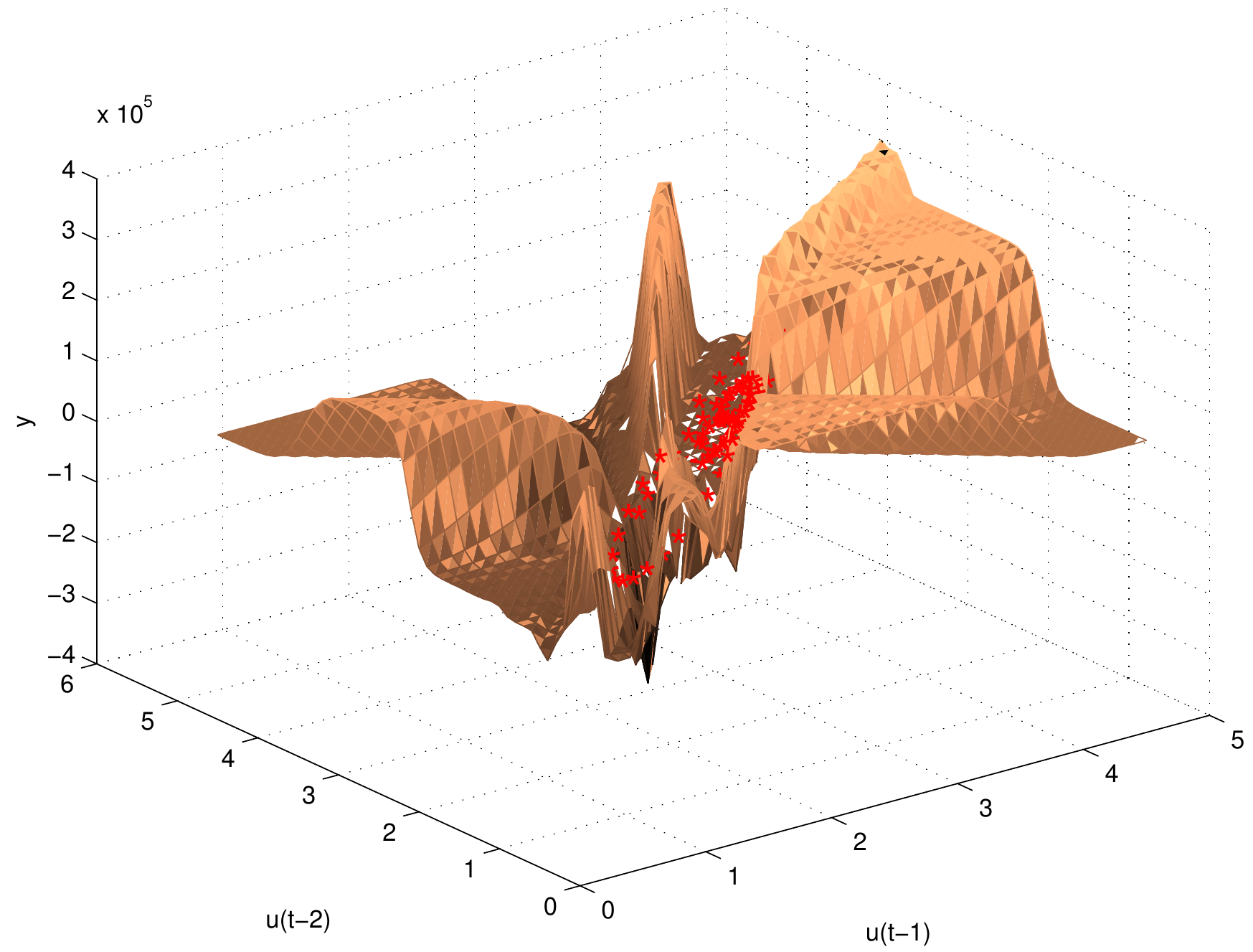}
        \caption{}
    \end{subfigure}
   \caption{Nonlinear models live on a manifold $\hat{y}=h(u(t-1),u(t-2))$ (b) in the input-output space that is much more complex than the hyperplane $\hat{y}=a_1u(t-1)+a_2u(t-2)$ (a) that characterizes a linear system. This affects the whole identification process from experiment design, choice of the model structure, choice and minimization of the cost function, generating initial values, etc.  The complexity of the problems to be solved grows fast with an increasing dimensionality, for example the number of delayed input and outputs.}
\label{fig:LennartNLcomplexity}
\end{figure*}

A linear dynamic system is described by a linear relation between the lagged inputs and outputs, for example, for a simple two tabs FIR (finite impulse response) model,
\begin{equation}
\hat{y}(t)=a_{1}u(t-1)+a_{2}u(t-2),
\end{equation}
that is shown in Figure \ref{fig:LennartNLcomplexity} (a). The output is confined to a hyperplane in the three-dimensional space. For a NFIR (nonlinear finite impulse response) model, the relation can become arbitrary complex, 
\begin{equation}
\hat{y}(t)=h(u(t-1),u(t-2))
\end{equation}
as shown in Figure \ref{fig:LennartNLcomplexity} (b). The estimation task is to estimate this surface. It is clear that the linear hyperplane can be characterized with only a few points, while it is impossible to tell anything about the complex manifold outside the domain where the function is sampled.

This reveals immediately a number of issues in nonlinear system identification that are less pronounced or even not present at all in linear system identification. 1) Experiment design will be extremely important because it should be guaranteed that the full domain of interest is covered. Extrapolation of the model should be avoided at all cost, unless there is physical insight that provides a natural description of the manifold. 2) Finding the parameters that describe the manifold results often in a highly nonlinear optimization problem. Good initial values are needed to make sure that the global minimum is reached. In practice this is often impossible, and the user has to be satisfied with a good local minimum. 3) Because the manifold can be very complex, it is often not possible to propose a model structure that is flexible enough to reproduce it exactly. This leads to the presence of structural model errors. These affect the whole identification process.

\emph{Summary}: Nonlinear systems are intrinsically more involved and complex than linear systems. This affects the experiment design and the model selection of the system identification process. 

\subsection{System identification in the presence of structural model errors}
 As explained in the previous section, it is very hard to avoid structural model errors in nonlinear system identification.  In "Impact Of Structural Model Errors", a formal definition is given of structural and random model errors, followed by a discussion how to deal with structural model errors that dominate the noise disturbances. Special attention is paid to the user choice how to shape the structural model errors, and on the impact of structural model errors on the variance estimate of the model.

\emph{Summary}: Be aware that there is a high risk for dominating structural model errors (over the noise disturbances) in nonlinear system identification which makes some classical choices and results of the (linear) identification theory invalid. Proper actions are needed to deal with this new situation.

\subsection{Impact of process noise on the system identification problem}

The output of a (nonlinear) system depends not only on the known or measured inputs, the system can also be affected by signals that are not known to the user. These unknown inputs $w(t)$ are called \emph{process noise}. While the measurement noise $v(t)$ does not affect the evolution of the system, process noise does as is clearly seen in the state space representation of a nonlinear system:

\begin{equation}
\begin{aligned}
 x(t+1)&=f(x(t),u(t),w(t))\\
y(t)&=h(x(t),u(t))+v(t)
\end{aligned}
\label{eq:ProcNoiseMeasNoise}
\end{equation}

Process noise can have a structural impact on the behavior of a system because it affects also the system's internal signals and not only the measurements.. To set the ideas, we can consider without loss of generality the simple static nonlinear system
\begin{equation}
\begin{aligned}
y(t)&=(u(t)+w(t))^{2}+v(t)\\
&=u(t)^{2}+(w(t)^2+2w(t)u(t))+v(t).
\end{aligned}
\label{eq:PN1}
\end{equation}
At the output $y_{p}(t)=w(t)^2+2w(t)u(t)$ now depends on the input and does not have typical noise properties. Its mean value is different from zero. Moreover, it is shown in \eqref{eq:BLAStatic1} that also the apparent gain of the system can change for odd nonlinearities. 

This simple example illustrates that the presence of process noise increases the complexity of the system identification problem significantly. If the process noise enters before the nonlinearity, its effect on the output  $y_{p}(t)$ is affected by the nonlinear operations so that it is no longer realistic to use the Gaussian framework to formulate the cost function. Unknown distributions that depend on the input and on the model parameters are faced. For that reason, the output error framework, that assumes that all noise enters at the output of the system has to be abandoned, and a generalized multivariate probabilistic framework is needed. The complexity of the methods to solve these problems goes far beyond the linear system identification methods and a completely new set of tools is needed. For that reason it is important to detect the presence of process noise that passes through the nonlinearity, and to select the proper tools when needed. This is discussed in detail in "Process Noise In Nonlinear System Identification" and ``Identifying Nonlinear Dynamical Systems in the Presence of Process Noise''.

\emph{User guidelines}: Check if process noise passes through the nonlinearity, and select the proper tools as needed.  

\section{Goal of the (Nonlinear) System Identification Process}
The goal of the modeling effort strongly affects the complexity of the system identification process. Issues that need to be addressed are: 1) Simulation or prediction models, 2) Physical models or black box models, 3) Application driven models. All these aspects are briefly discussed below.

\subsection{Models for simulation or models for control?}

\emph{Prediction model}:  In layman's terms, a prediction model estimates the output of the system one-step-ahead  at time $t+1$, using the measured input up to time $t+1$, and the measured outputs up to time $t$. Prediction models are central for modern control applications, where essentially the predicted output is controlled. 

\emph{Simulation model}: The alternative is that the measured outputs are not used at all, but the output is calculated from inputs only. This is called a simulation model, and it can be used to simulate the behavior of the system for new inputs. These models are useful to test what happens in new situations, to design systems and controllers, to mimic physical systems, etc. 

It is much harder to get a good nonlinear simulation than prediction model. Simulation models can become unstable, and it is harder to get small structural model errors than it is for one-step-ahead prediction models. Even a very simple linear model can often provide a good one-step-ahead prediction if the sample frequency is high enough. 

A detailed discussion is given in "Simulation Errors And Prediction Errors". Figure \ref{fig:SB2} shows results of linear and nonlinear simulation and prediction models for the forced Duffing oscillator. 

\emph{User guideline}: Do not spent effort to get a complex simulation model if a simple prediction model can do the job. However, keep in mind that a good prediction can fail completely to generate a reliable simulation.

\subsection{Physical or black box models?}
\emph{A physical model} is built on a deep insight in the internal behavior of the system. Detailed physical descriptions are made at the level of the (microscopic) subsystems, and next they are linked together in a macroscopic model that is built on (thousands of) nonlinear (partial) differential equations. Alternatively, simple use of physical insight can be used to guide the model process. Eventually, in both cases, the model depends upon a number of parameters that can be obtained from dedicated measurements (for example, the friction coefficient of a wheel on the road). These models are highly preferred in the industry, but they can be very expensive to construct and difficult to use in real time control. For that reason, such models are often not affordable.

\emph{Black box models} become very attractive when a physical model is too expensive to develop. Black box models describe the input-output behavior of a system, and are tuned directly from experimental data. Black box models are simple to use and can be applied in real time computations.

Of course there are many possibilities in between both extremes. These are discussed in "The Palette Of Nonlinear Models".

 \emph{User guideline}: Select the model level that best balances the need for physical insight - behavior insight and the expenses to build and use the model.

\subsection{Models constrained for particular applications}
The ideal model should cover all possible applications, providing good output simulations for all possible excitations. Of course this is an unattainable ideal, and a more restricted goal should be defined. This is done keeping the application in mind, the model should be able to cover those situations and signals that are important for the actual application, and not more than that \cite{Hjalmarsson2009}. If a system will be mainly driven by low frequent sine excitations, no effort should be spent to develop the model also for wide band random noise excitations. This can again significantly reduce the modeling effort.

\emph{User guideline}: Select carefully the domain and application of interest and focus the modeling effort on it. 

\section{Linear or Nonlinear System Identification: a Users' Decision}

As discussed before, nonlinear system identification is much more involved than the identification of a linear system. The experiment design is more tedious, the model selection is much more involved, and the parameter estimation is more difficult. For that reason, moving from the well established linear identification tools towards the more advanced nonlinear identification methods is an important decision that significantly affects the cost of the identification effort (time, money, experimental resources) that should be well informed. Is a nonlinear model needed to reach the required model quality? Is the quality of the data good enough to improve the results of a linear identification approach? How much can be gained if a linear model is replaced by a nonlinear one? Often, additional information is needed to address these questions. 

If it is possible to apply periodic excitation signals, a full nonparametric analysis can be made that requires no user interaction, while the experimental cost with respect to a linear study remains almost the same (see ``Nonparametric Noise and Distortion Analysis Using Periodic Excitations''). On the basis of the results, the user can detect the presence of nonlinearities, quantify their level, and find out if it are even or odd nonlinearities. With this information the user can make a well informed decision on what approach to use, and how much can be gained by switching from linear to nonlinear modeling.

\subsection{Detection, separation, and characterization of the nonlinear distortions and the disturbing noise}
In this article, only a basic introduction to nonlinear distortion analysis is given. A detailed theoretical analysis is given in \cite{Pintelon2012}, and illustrations on practical examples (fighter jet, diesel engine \cite{Criens2016}, industrial robot \cite{Wernholt2008}) are discussed in \cite{SchoukensJ2016}. In this article, the ideas are illustrated on the forced Duffing oscillator that is discussed in full detail in ``Extensive Case Study: the Forced Duffing Oscilator''. A detailed introduction to the nonlinear distortions analysis is given in ``Nonparametric Noise and Distortion Analysis Using Periodic Excitations''. 

\begin{figure}[h] 
\centering
\includegraphics[scale=0.7]{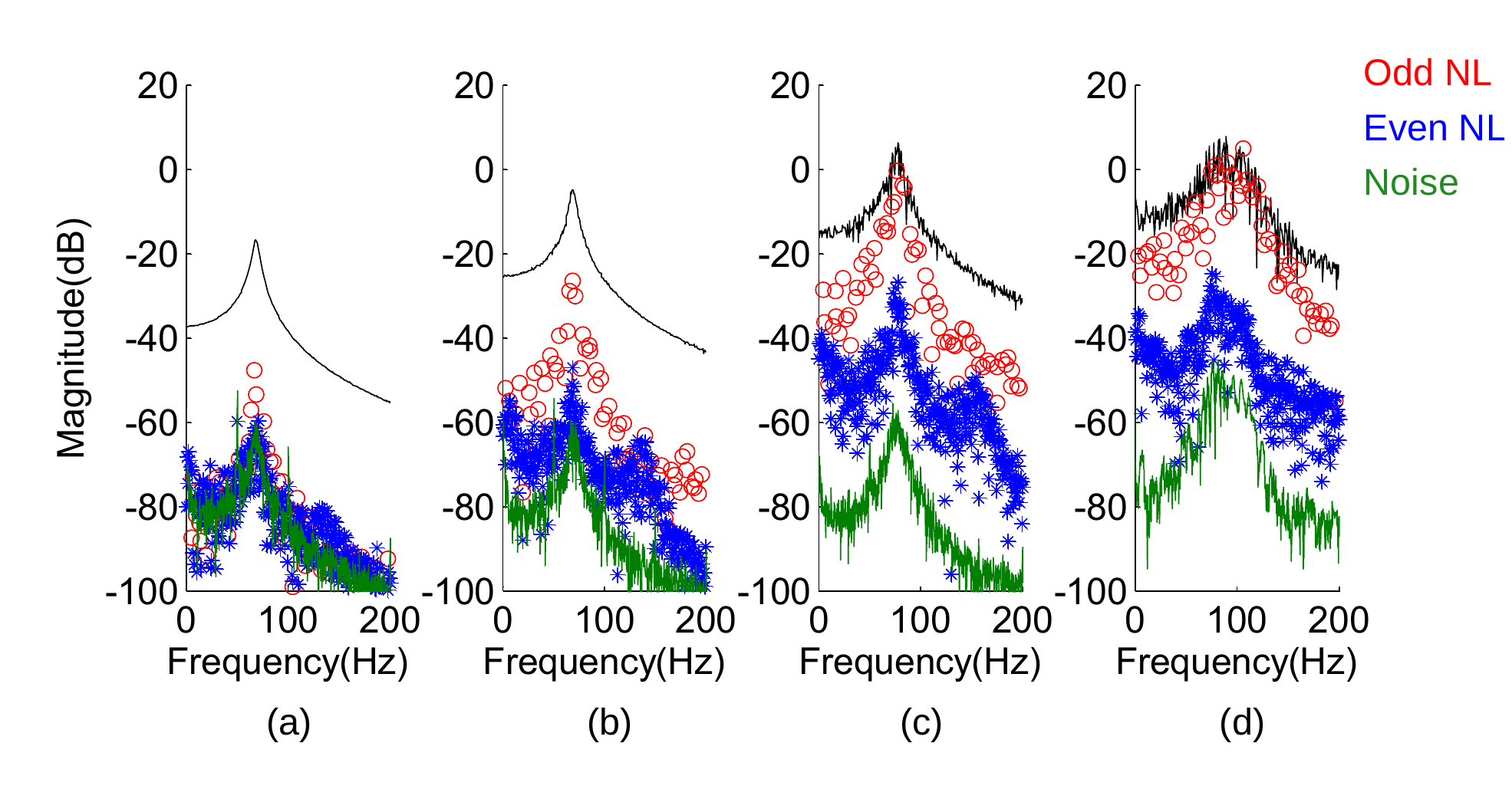}
\caption{Nonparametric analysis of the nonlinear distortions on the forced Duffing oscillator (see Figure \ref{fig:SB1}). The system is excited at a well-selected set of frequencies \cite{SchoukensJ2016}. The nonlinearities become visible at the unexcited frequencies. Black dots: output at the excited frequencies; Red bullets: odd nonlinearities; Blue stars: even nonlinearities; Green line: disturbing noise level. The excitation level is growing from (a) to (d). Observe that the level of the nonlinear distortions grows with the excitation level, the disturbing noise level remains almost constant.}
\label{fig:SilverboxNLdistortions}
\end{figure}

\emph{Nonlinear distortion analysis}: The  basic  idea  is  very simple  and  starts  from  a  periodic input signal with period $T=1/f_{0}$. Only a well-selected set of odd frequencies (odd means that $f$ is an odd multiple of $f_{0}$) is excited, all the other frequencies have zero amplitude (see Figure \ref{fig:Multisine NonLin detection}). This excitation signal is applied to the nonlinear system under test. Effects from even nonlinearities (simplest even nonlinearity is $y=u^2$) show up at the even frequencies, while odd nonlinearities (like $y=u^3$) are present only at the odd frequencies, see \cite{SchoukensJ2016} and references therein. At the odd frequencies that are not excited at the input, the odd 
nonlinear distortions become visible at the output because the  linear  part  of  the  model  does  not  contribute  to  the output at these frequencies. By using a different color for each of these contributions, it becomes easy to recognize these in an amplitude spectrum plot of the output signal. This is illustrated in Figure \ref{fig:SilverboxNLdistortions}. The forced Duffing oscillator (see Figure \ref{fig:SB1}) is excited at different excitation levels, and the output is plotted for the excited frequencies, the even and odd nonlinearities, and the disturbing noise level. For small excitation levels, the nonlinear distortions are at the 10\% level (-20 dB below the output), while for the high excitation levels the nonlinear distortions dominate the output. 

\emph{Linear or nonlinear model?} On the basis of this information the user can make a well informed decision. For example, a linear model can be used if it is known that only small excitations will be applied, and if 10\% errors can be tolerated. However for the large excitation levels, this is no longer an option because the nonlinearities are too large. In that case a nonlinear model is needed. 

\emph{Noise floor}: In Figure \ref{fig:SilverboxNLdistortions}, it is seen that he nonlinear distortions are more than 40 dB or a factor 100 above the noise level (see ``Nonparametric Noise and Distortion Analysis Using Periodic Excitations''), also called the noise floor of the measurements \cite{IEEEstandard1057}. The noise floor is the maximum power level over a given frequency band in the frequency domain of components that are not due to the applied signal, harmonics, or spurious signals (this are signals that are out of the control of the user, for example a disturbance picked up from the mains). The noise floor is estimated by analyzing the variations over the periodic repetitions of the output. This shows that the quality of the data is very high. The distance between the nonlinear distortion levels and the noise floor is a measure for the potential improvement that can be obtained by using a nonlinear model.  In this case a gain of a factor 100 is possible for the high excitation levels if a good nonlinear model can be obtained.

\emph{User guideline}: Make a nonparametric distortion analysis whenever it is possible to apply periodic excitations. Use this information to decide if a linear or nonlinear approach is needed, and to check how much can be gained by turning towards nonlinear system identification.

\subsection{Linear modeling in the presence of nonlinear  distortions}
Deciding to go on with the linear identification approach implies also the presence of structural model errors in the results. This may be an acceptable solution, but the user should fully understand the impact of the structural model errors on the validity of the results. This is discussed in "Impact of Structural Model Errors". The major conclusions are that i)  the experiment should be tuned to the application (same class of excitation signals), and ii) no reliable theoretical uncertainty bounds can be provided. These should be obtained from repeated experiments with different excitation signals generated from the relevant class of input signals \cite{Hjalmarsson1992}. 

The problem of linear modeling in the presence of nonlinear distortions is studied and discussed in "Linear Models of Nonlinear Systems". The classical linear identification methods will lead to consistent estimates of the best linear approximation.

For random excitations, the nonlinear effects at the output will look very similar to noise, and it is very difficult for an unexperienced user to recognize their presence.  Their power spectrum can be estimated using a parametric noise model that is simultaneously identified with the plant model (for example using a Box-Jenkins model, see e.g. \cite{Ljung1987}). The noise model can be used  in a control design to make the disturbance analysis, but uncertainty bounds calculated from it are no longer valid because the errors are not independent of the input. Again there is no theoretical framework available today to solve that problem. The variance of the estimates should be obtained by repeating the experiment multiple times with a random varying excitation as is explained in "Impact of Structural Model Errors".

\emph{Summary}: Linear models can be very useful, even in the presence of (strong) nonlinearities, because it is much easier to deal with it. Make sure to understand the impact of the nonlinear distortions on the best linear approximation.

\subsection{Nonlinear system identification}
A nonlinear system identification approach is justified if the preprocessing step indicates too high nonlinear distortion levels that are well above the noise floor of the data. It is this problem that will be further addressed in this article. What identification methods to use? How to select a model class? How to design the experiments? This are questions that will be addressed in this article. Again the user will have many options, starting from simple nonlinear models that are good enough to solve the problem, to complex models that include also the fine details that are deeply hidden in the data. 

\emph{User guideline}: Select nonlinear system identification only if there is enough evidence that linear models will not solve the problem.

\section{The Palette of Nonlinear Models} 
\subsection{The Multitude of Nonlinear Models}
A major challenge in dealing with nonlinear system identification is that very many nonlinear model structures have been suggested. A user easily can be quite confused getting familiar with the different choices that are available and how to choose a structure that suits his or her particular situation. A clear perspective on this wide variety of choices is needed to make a well informed choice. Nonlinear model structures can be ranked along two axis that are directed respectively by the user's preference and by the systems behavior. 

\begin{itemize}
\item \emph{Users preference}: A first classification of the models is made in terms of how much prior knowledge about the system is used, \cite{Ljung2010}. Based on the familiar concept of \emph{black-box} models for general flexible structures with no physical insights the user can then select from a whole palette of models of different shades of gray to delineate many approaches to common nonlinear models. 
\item \emph{System behavior}: An alternative to use varying degrees of \emph{structural physical insights}, is to use \emph{behavioral aspects} of the system: Does it, for example, show  behaviors like chaos, shifting resonance frequencies and varying damping, or hysteresis, as discussed by Pearson,\cite{Pearson2003}. Then, the main selection is to include the nonlinearity in a feedback loop or not.  Observe that this selection is not a free user choice, it is imposed by the system behavior. A detailed ranking along this line is discussed in ``Static Nonlinearities'' and  "External Or Internal Nonlinear Dynamics".
\item \emph{Remark}: There are many other dimensions in which the different approaches can be classified like ``universality'', ``computational effort'', or ``suitability'' to deal with unstable systems. These aspects are not further elaborated in this article.
\end{itemize}

The user should combine both aspects in the final selection of the model.

\subsection{General Structure of Nonlinear Models}
In general the measured system input and output at time $t$ will be denoted by $u(t)$ and $y(t)$.
All measured data up to time $t$ will be denoted by
\begin{align}
  \label{eq:data}
Z^t={u(s),y(s); s\le t}
\end{align}
The models can be expressed in \emph{discrete time} or \emph{continuous time}.  Most real life systems evolve in continuous time, but often discrete time models are preferred to simplify the numerical simulations. Linear systems can be perfectly represented by discrete time models for zero-order-hold (ZOH) excitations \cite{Ljung1987,Soderstrom1989,Astrom1970}, but this approach cannot be generalized to nonlinear systems because the ZOH nature of the signals is lost inside the nonlinear system. Alternatively, the discrete time approximation can be done within the bandlimited setup \cite{SchoukensJ1994} (see'' Approximating a Continuous Time NLSS with a Discrete Time NLSS Model''). The discretization error can be made arbitrarily small by increasing the sampling frequency and/or the model complexity as shown in "Approximating a Continuous Time NLSS with a Discrete Time NLSS Model". Both type of models will be considered in this section. In the discrete time case, the time variable will simply be enumerated by time instances $t=1,2,3, \ldots$ (``constant sampling interval of one time unit'').

The general structure of all the nonlinear models can be put in the form
\begin{align}
  \label{eq:genstr}
q(t)=F(p(t))
\end{align}
with $p,q$ vectors that are built on the signals turning around in the model, like inputs, outputs, states, internal variables. The function $F$ is a static nonlinearity. The properties and parameterization of static nonlinearities are discussed in more detail in ``External or Internal Nonlinear Dynamics'', which also illustrates how it can be used in various nonlinear dynamic models.
 
The bottom line is that a model is an expression that allows the computation of the next output $y(t)$ based on previous observations
\begin{align}
  \label{eq:pred}
\hat y(t|\theta,Z^{t-})
\end{align}
This model output will depend on a parameter vector $\theta$ that is used to parameterize the model class. The notation $Z^{t-}$ denotes that $y(t)$ is excluded. In discrete time it means $Z^{t-1}$. Remark: if a direct term is needed in the model, the regressors $Z^{t-1}$ can be extended with $u(t)$.

Following the discussion in ``Simulation Errors and Prediction Errors'', the model output $\hat y(t|\theta)$ will be a \emph{simulation output} if it only depends on past inputs, and the corresponding model is  a \emph{simulation} or \emph{output error} model.
If the model output also depends on past outputs, it is a \emph{predicted output} and the corresponding model a \emph{prediction model}. This term does not necessarily imply that it is based on correct probabilistic treatment of the stochastic signals involved in the model.

In the list of models that follows, it will be indicated how they comply with the general structures (\ref{eq:genstr}) and (\ref{eq:pred}).

\emph{Summary}
The user selected model set should reflect both the observed behavior aspects (for example, the presence of chaos or shifting resonances requires a closed loop around the nonlinearity) which leads to a selection imposed by the system, and the available physical insight, leading from snow-white to pit-black models, which is a user choice.

\subsection{Snow-white Models}
In the end, a model of a dynamical system is a collection of mathematical expressions that relate signals and variables that characterise the system behaviour. Important players are
\begin{itemize}
\item $y(t)$ the output signal(s) of the system. These are the signals that are primary interest to be modeled.
\item $u(t)$: the input signal(s) to the system. These are measurable signals that affect the outputs. They may or may not be manipulated by user.
\item $w(t)$ disturbance signals. These are unmeasurable signals that affect the outputs. They are typically described by random processes.
\item $x(t)$ auxiliary signals that are used in the model description
\end{itemize}

So, a model is a collection of equations involving $y,u,w$ and $x$. To be a useful  model it must to be possible  to infer something about $y$ from the other measured variables.

\paragraph{Deterministic Models} If no disturbances $w$ are present, Then it should be possible to compute $y(t)$ from previous values of $u(s); s\le t$.  That means that the model is \emph{deterministic} or a \emph{simulation model} or  \emph{output-error model}.

\paragraph{Stochastic Models} If there are stochastic disturbances $w$ present, the outputs also become stochastic variables. Then  a specific value to $y(t) $ cannot be assigned based on the other variables, but  a stochastic characterization of it must be used instead, like its probability density, or mean value $\hat y(t)$. Since $w$ are not observed,  their values up to time $t$ have to be inferred from the values of the observations $y(s)$ and $u(s)$, $s < t$ The model values of $y(t)$ will then be conditioned on these variables and the (conditional) mean $\hat y(t)$ will really be a \emph{prediction} based on past values.

To deal  with these computability questions, it is necessary to be more specific about the structure of the collections of equations involving the model variables.
\subsubsection{Deterministic DAE Models}
Introduce for simpler notation, the vector $z(t)=[y(t),x(t)]^T$ for the outputs, and auxiliary variables. Assume that there are no disturbances.
When writing down the equations that correspond to the physical knowledge of the system, typically differential equations are used, in addition to algebraic relations among the variables. That means that the equations can be written as
\begin{align}
  \label{eq:dae}
  F(z(t),\dot z(t), u(t))=0
\end{align}
where $\dot z$ is the time derivative of $z$ (It is sufficient to consider first order derivatives, since higher order ones can be rewritten with the aid of extra $x$-variables. Also if $\dot u$ were to appear in (\ref{eq:dae}), $u$ can be included in $z$ and the basic form can still be applied.). This a \emph{Differential Algebraic Equation, DAE} model. There is a clear conceptual relation between (\ref{eq:dae}) and the general structure (\ref{eq:genstr}).

It is a normal case that (\ref{eq:dae}) can be solved for $z$ for given $u$, and most software packages, e.g. \cite{Tiller:01} for modeling systems contain DAE solvers. There is an extensive literature, e.g. \cite{brenan:89}, \cite{LjungG:16} that addresses this problem of solvability of (\ref{eq:dae}).

\subsubsection{Deterministic Statespace Models}
If $\dot z$ can be solved explicitly from (\ref{eq:dae}), it follows that
\begin{subequations}
\label{eq:dss}
\begin{align}
  \dot x(t) &= f(x(t),u(t))\\
y(t) &= h(x(t),u(t))
\end{align}
\end{subequations}
which is  a standard, nonlinear statespace description for the relationship from $u$ to $y$. If an initial state $x(0)$ is given, this equation has as unique solution $y$ under general and mild conditions. $y(t)$ will be a function of past $u(s)$,
\begin{align}
  \label{eq:dsssol}
  y(t) = y(t|u^t); \quad u^t=\{u(s), s < t\}
\end{align}

\subsubsection{Stochastic Statespace Models}
With disturbances present it is customary (and necessary) to represent these as white noise $\nu$ filtered through certain filters and assume that
\begin{subequations}
  \label{eq:sss}
\begin{align}
  x(t&+1) = f(x(t),u(t),w(t))\\
&w(t) \,\text{is a sequence of independent} \nonumber\\ &\text{random variables with  pdf $g_w(\cdot)$}\\
y(t) &= h(x(t),u(t))+v(t)\\
&v(t) \,\text{is a sequence of independent} \nonumber \\ & \text{random variables with  pdf $g_v(\cdot)$}
\end{align}
\end{subequations}

 To avoid intricate issues with continuous time white noises only the discrete time case is considered here.

With the state-space forms (\ref{eq:dsssol}), (\ref {eq:sss}) the links with the general predictor model structure (\ref{eq:pred}) become clear.

 The problem to find the conditional distribution or conditional mean $\hat y(t|t-1) $   given past input- output signals is the well known (nonlinear) prediction or filtering problem. In case $f$ and $h$ are linear and $\nu$ and $e$ are Gaussian variables, this is solved by the familiar \emph{Kalman Filter}, \cite{AndersonM:79}. In the general case, the filtering problem has no closed form solution, but there is recent progress for numerical solution  in terms of \emph{particle filters}, e.g. \cite{Schon2011}, \cite{DoucetT:03},

A simplistic (but mostly erroneous) way to deal with the nonlinear filtering problem is to assume that all disturbances can be collected as white noise added to the output. Then the model will correspond to an output error model like (\ref{eq:dss}) (see also ``Process Noise in Nonlinear System Identification'').

\emph{Summary}:
The work to construct a snow-white model, sometimes called \emph{First Principles modeling} or \emph{Mechanistic modeling}  is typically time-consuming and laborious, and does as such not involve any system identification. It is supported by software like the object oriented languages \emph{Modelica}, \cite{fritzson14} or \emph{Simscape}, \cite{simscape}.

\subsection{Off-white Models}
In the snow-white modeling work, it typically happens that the model includes one or several physical constants whose numerical values are not known. If the values cannot be established by separate measurements, they have to be included as a parameter $\theta$ in the model. In the deterministic state space case the model then takes the form
\begin{subequations}
 \label{eq:dssoff}
\begin{align}
  \dot x(t) &= f(x(t),u(t),\theta)\\
\hat y(t|\theta) &= h(x(t),u(t),\theta)
\end{align}
\end{subequations}
where $\hat y(t|\theta)$ signifies that it is the output corresponding to the specific parameter value $\theta$. In the stochastic case, the symbol will mean the predicted output time $t$ based on the model with parameter $\theta$. Since its value is unknown, the model is no longer snow-white but has become an \emph{off-white} model. Such models are also known as \emph{grey-box} models, e.g. \cite{Bohlin:06}, but as follows from the sequel, there are several shades of grey.

To build an off-white model is the same work as to build a snow-white model, so a considerable amount of work might be required. It can also be remarked that it may not be possible to retrieve all the physical parameters from an identification experiment. 

The off-white model (\ref{eq:dssoff}) is a clear cut case of the general predictor structure (\ref{eq:pred}). It also complies with (\ref{eq:genstr}) by taking $F$ to be solver of the underlying nonlinear differential equation.

An application of off-white model identification to a cascaded tanks system is given in  \eqref{eq:Tanks1} in Section ``Examples''.

\emph{Summary}: Off-white model identification requires substantial modeling work. It is important to realize that any deficiencies in the physical model may cause corruption in the physical parameter estimates.

\subsection{Smoke-grey models: Semi-physical Modeling}
By \textsl{semi-physical modeling} is meant doing physical modeling with a more leisurely attitude to the physics.

It could be a matter of \emph{using qualitative reasoning} rather than formal equations. Take for example a voltage controlled DC motor, with input applied voltage $u$ and output motor shaft angle $y$, a known load disturbance torque $L$ also acts on the shaft. Well known physical laws tell us that the applied voltage to the rotor circuit is split between the internal resistance in the rotor winding and the back emf resulting from the motion of the winding in the magnetic field. The latter is proportional to    $\omega$, the rotational speed. The torque $T$ from the magnetic field is proportional to current in the winding and the resulting torque on the shaft is $T$-$L$ minus frictional torque, which typically proportional to $\omega$. The resulting torque will by Newton's law of motion be proportional to the rotor acceleration $\dot\omega$. All this means that the voltage $u$ will be a linear combination of $\omega, \dot\omega$ and $L$. Since $\omega$ is the derivative of $y$ it follws that
\begin{subequations}
  \label{eq:dc}
\begin{align}
  x&=
     \begin{bmatrix}
       y\\ \omega
     \end{bmatrix}\\
\dot x &= 
           \begin{bmatrix}
             0 &1\\
0&\alpha
           \end{bmatrix}
x +
        \begin{bmatrix}
          0\\
\beta
        \end{bmatrix}
u +
  \begin{bmatrix}
    0\\1
  \end{bmatrix}
L
\end{align}
\end{subequations}
This is of course the same model that would have been obtained with careful ``white'' modeling with the difference that $\alpha,\beta$ would have been expressed in physical constants of the motor, like internal resistance, friction coefficient, rotor moment of inertia, and magnetic field characteristics. That is not really essential, since the only physical constants that can be retrieved from input-output data are the combined expressions $\alpha,\beta$.

Another useful form of semi-physical modeling is finding \emph{nonlinear transformations of the measured data}, so that the
transformed data stand a better chance to describe the system in a
linear relationship. To give a trivial example, consider a process
where water is heated by an immersion heater. The input is the voltage
applied to the heater, and the output is the temperature of the
water. Any attempt to build a linear model from voltage to temperature
will fail. A moment's reflection  
tells us that it is the power of the heater that is the
driving stimulus for the temperature: thus let the squared voltage be
the input to a linear model generating water temperature at the output.
Despite the trivial nature of this example, it is good to keep it
as a template for data preprocessing. Many identification attempts
have failed due to lack of adequate semi-physical modeling. See, e.g.,
\cite{Ljung1987}, Examples~5.1 and pages 533 - 536 for more examples of
this kind.

 Also   \emph{nonlinear re-calibration of time scale} can be counted to the family of non-linear transformations of measured data. Several systems have a natural time-maker. It could be a rotational system where the angle of the rotation is a natural time unit or it could be a flow system where the accumulated amount of transported substance gives a natural time flow (see the example ``Buffer Flow System'').

The perhaps most common and important application of semi-physical modeling is to \emph{concatenate known submodels} A simple example is the DC case above, which is a concatenation of a DC motor model from $u$ to $\omega$ and an integrator. But having libraries of basic element models from which more complex models are built up using simple physics and logics is now perhaps the most common way for modeling in many application areas. Modeling languages like \textsc{Modelica} are based on this principle, and extensive \textsc{Modelica} libraries exist for most applications.

\emph{User guideline}: It is always important to think over the physics of the system to be identified, even if a complete off-white model is not constructed. Such semi-physical modeling can give insights into important non-linear transformations than can be essential components in the model.
\subsection{Steel-grey Models: Linearization Based Models}
\subsubsection{Linear Models of Nonlinear Systems in Identification}
It is well known how a nonlinear model like (\ref{eq:dssoff}) can be linearized around a stationary point $x^*$ (see ``Linear Models of Nonlinear Systems''). In fact, using linearized models may be the most common way to deal with nonlinear systems in practice. Identifying a linearized model is normally not done by going through the linearizing differentiations (since the nonlinear models is typically unknown). Instead one simply postulates a linear model structure and applies normal linear system identification. If the excitation keeps the system in a close vicinity of the stationary point $x^*$, the identified model will be close to the ``stationary point-linearization''. In general, the identified model will be a ``stochastic linearization (BLA)''  reflecting the input and output signal spectra of the identification data. In any case, the usefulness of the model may be limited since it only describes the system in the vicinity of $x^*$.
\subsubsection{Local Linear Models}
An obvious remedy to the local nature of the linearized model is to work with several linearized models and connect them in some way to cover the global behavior of the system. The many ways to do this have lead to an extensive literature on \emph{local linear models} \cite{Nelles2001} (several other terms used as well). 

The basic idea  is to divide the ``state space'' into \emph{Regions} in within each of which a linear model is used to describe the system.
To fix ideas, around a common special case,  use a collection of measurable ``\emph{regime points}'' $P=\{p_i,1=1,\ldots, d\}$ which define \emph{centers} of the regions. Each point is like a stationary point $x^*$ of a statespace model (but needs not be formally defined like that). Each of these points is associated with a linear model, generically characterized by its output prediction $\hat y_i(t|\theta,Z^{t-1})$. The type of linear model could be arbitrary, and some examples are given  below. When the system is at the regime point $p_i$ there is a clear linear model for predicting the output, At other points $p(t)$   the prediction can be interpolated from the values in $P$ by
\begin{align}
  \label{eq:llm}
  \hat y (t|\theta,Z^{1-1})=\sum_{i=1}^d w(p(t),p_i) \hat y_i(t|\theta,Z^{t-1})
\end{align}
This is the archetype of a \emph{local linear model}. This is explicitly of the form (\ref{eq:pred}). In terms of (\ref{eq:genstr}) the nonlinearity hides in the weights $w$ and the shifts between linear models that they represent.

For a concrete case the following items need to be specified
\begin{enumerate}
\item What are the measurable \emph{regime points}?
\item How is the collection $P$ determined?
\item What type of linear models $y_i(t|\theta,Z,^{t-1})$ are used?
\item How to choose the interpolation rule $w(p,p_i)$?
\end{enumerate}

Some comments:
\paragraph{1: Regime points} They are often naturally defined by the application, and may typically be part of the state vector. They correspond to operation conditions that are known to give well defined behavior. Typical cases could be fluid levels in flow systems applications and speed and altitude in flight applications. The regime points can also be determined from data, like a tree-based construction in LOLIMOT, \cite{Nelles2001}.
\paragraph{3: Linear Models}. There is a wide range of linear model types available. The simplest one is ARX-models, that predict the output as a linear combination of past inputs and outputs:e.g.  $\hat y (t|\theta) =-a_1y(t-1)+b_1u(t-1)$ in a first order case. Associating  the $r^{th}$ regime point model with parameters with superscript $(r)$ the complete model (\ref{eq:llm}) will be
\begin{equation}
\begin{aligned}
  \label{eq:llmex}
  &\hat y (t|\theta,Z^{1-1})=\\
&\sum_{r=1}^d w(p(t),p_r) [b_1^{(r)}u(t-1)-a_1^{(r)}y(t-1)]
\end{aligned}
\end{equation}
which is a linear regression with parameters $\theta=[a_1^{(r)}, b_1^{(r)}, r=1,\ldots, d]$.
There is a vast literature on multiple and local linear models,  like \cite{Nelles2001}, \cite{MurrayJ:97}, \cite{TakagiS:85}. The latter article uses fuzzy sets for the model interpolations.

\subsubsection{LPV Models} A related concept is that of \emph{Linear Paramater Varying, LPV}, models. In state space form they can be described as
\begin{subequations}
  \label{eq:lpv}
\begin{align}
x(t+1)&=A(p(t))x(t)+B(p(t))u(t)\\
y(t)&=C(p(t))x(t)
\end{align}
\end{subequations}
(correspondingly in continuous time). Disturbances can also be added. $p(t)$ is a measured \emph{regime variable}.  Formally (\ref{eq:lpv}) is not a non-linear system, but rather a linear, time-varying system. But if $p(t)$ in some way depends on the state $x$, this can be a handy way of dealing with a nonlinear system. There is an extensive literature on dealing with and identifying LPV models, e.g. \cite{LeeP:99}, \cite{Toth10}, \cite{Tothetal:12}. If $p(t)$ only assumes a finite number of different values, (\ref{eq:lpv}) is really  a collection of local linear models, and several identification schemes can be based on the ideas in the previous subsection, and/or on handling time-varying linear systems. 

An important difficulty for LPV systems is keeping track of the state space basis when $p(t)$ is changing.

\emph{Summary:} Using linearlzation is  a standard tool to handle nonlinear physical systems. Many possibilities exist to glue together linearized pieces into a good nonlinear model.

\subsection{Slate-grey Models: Block-Oriented Models}
\begin{figure}[h] 
\centering
\includegraphics[scale=0.5]{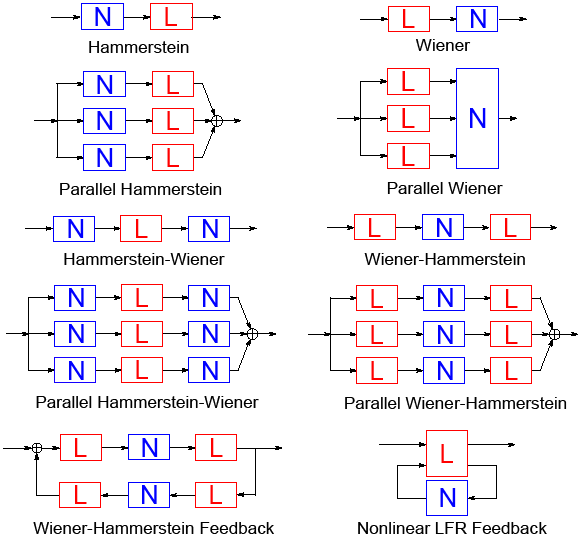}
\caption{Examples of block-oriented models. The linear blocks L capture all dynamics, while the static nonlinear blocks N are used to model the nonlinearity. The Hammerstein model can include the actuator nonlinearities, sensor nonlinearities are covered by the Wiener model. The combined effects are covered by the Hammerstein-Wiener model. The Wiener-Hammerstein model describes a static nonlinearity with an input and output matching network. The parallel Wiener-Hammerstein structures are universal approximators for fading memory systems \cite{Boyd1985,Palm1979}. The feedback structure is a generalized representation of the Duffing oscillator that is studied in "Simulation errors and prediction errors". Remark that in the parallel structures either a MIMO (for example in the parallel Wiener model) or a set of  SISO nonlinear blocks (for example in the parallel Wiener-Hammerstein model) can be used.}
\label{fig:EID1}
\end{figure}

A common and useful family of models is obtained by concatenation blocks of two types:
\begin{itemize}
\item Linear Dynamic Models: $y=G(s)u$
\item Static Nonlinearities: $y(t)=f(z(t))$
\end{itemize}
Many such combinations have direct physical interpretations, like (See Figure  \ref{fig:EID1}) 
\begin{itemize}
\item The Wiener model: A linear model followed by a static nonlinearity $y=f(G(s)u)$, describes linear plant with a nonlinear output sensor.
\item The Hammenstein model. A static nonlinearity followed by a linear system, $y=G(s)f(u)$ depicts a linear model controlled via nonlinear, say saturating, actuator.
\end{itemize}


So, such \emph{block-oriented} models have a flavour of the smoke-gray, semiphysical models, constructed by simple engineering insights of qualitative nature. 


But beyond that, block-oriented model possess interesting approximation properties. It is for example known, \cite{Boyd1985}, that the parallel Wiener model in Figure \ref{fig:EID1} with sufficiently many linear branches, can arbitrarily well approximate fading memory systems  $u\to y$. The convergence rate can be improved by switching to parallel Wiener-Hammerstein models \cite{Palm1979}.  So, block oriented models can be used for many nonlinear systems in appropriate configurations without any physical interpretation and achieve good modeling results. This is much in the spirit of black-box models (next subsection). That motivates giving block-oriented models a darker shade of gray than ``smoke-gray''. See Figure \ref{fig:hydcrane}, for an application that is inspired by such simple reasoning but also benefitting from the extra flexibility by the input nonlinearity.

The block-oriented models clearly fit the general structure (\ref{eq:genstr}) with linearly manipulated model signals in combination with static nonlinearities. To find the predictor (\ref{eq:pred}), it is necessary to follow the signal flow through the linear and nonlinear parts. If there is process noise a correct calculation of the predictor may require the use of advanced statistical techniques, like particle filters, \cite{Wills2013}(see also ``Identifying Nonlinear Dynamical Systems in the Presence of Process Noise''.

The estimation of a block-oriented model basically follows the general minimization of fit between model (predicted) output and measured output. But before that can be done, considerable work may have to be done to come up with the structure and initial parameter values. The concept of the best linear approximation (see ``Linear Models of Nonlinear Systems'')  is useful in this context. A comprehensive account of estimation techniques for block-oriented model is given in the survey\cite{SchoukensTiels2017, Giri2010}. 

The Hammerstein, Wiener, Hammerstein-Wiener and Wiener-Hammerstein, including the parallel structures, are all examples of nonlinear systems with external nonlinear dynamics (see ``External or Internal Nonlinear Dynamics''): the nonlinear blocks are not captured in a dynamic feedback loop. The Wiener-Hammerstein feedback and the LFIR feedback systems are both nonlinear systems with internal nonlinear dynamics because the nonlinear block N is part of a dynamic closed loop. Dealing with multiple branches \cite{Schoukensm2015a} and feedback structures \cite{Schoukens2008} turns out to be very challenging. 

\emph{Summary:} Block-oriented models form a powerful and intuitive tool to handle nonlinear systems. It is often useful to try a simple Wiener or Hammerstein model to see if nonlinear model components show significant improvements over linear models.

\subsection{Black models: universal approximators}
So far, the starting point has been  some kind of physical or behavioral aspect of the system when constructing the model. But in the end, the model -- the Predictor -- is a mapping from past input-output data to the space where the output lives. Lacking insights into the system the focus could be on building general, flexible such mappings that are universal and effective approximators of any reasonable predictor function. That is the idea behind \emph{Black (box) Models.}

\subsubsection{A General Structure of the Mapping}
 A general way to generate very flexible mappings from $Z^{t-1}$ to
  $\hat y$ is to construct a state $x$ from past input output data, and let the predictor be a general function of this state:
\begin{subequations}
\label{eq:bbgen}
  \begin{align}
    x(t+1)&=f(x(t),u(t),y(t),\theta)\\
\hat y(t|\theta)&=h(x(t),\theta)
  \end{align}
  \label{eq:NLSSuy}
\end{subequations}
where $f$ and/or $h$ are flexible static nonlinear functions of their arguments $x(t),u(t),y(t)$.  The possibilities of parameterizations to reach flexibility are discussed in general terms in ``Static Nonlinearities''.

The model (\ref{eq:bbgen}) is directly of the general predictor form (\ref{eq:pred}) and links to the structure (\ref{eq:genstr}) through the nonlinear maps $f$ and $h$.

Working with both $f$ and $h$ may be too general, and  two cases will be further discussed.

\subsubsection{Unknown State Transition Function: nonlinear state space models (NLSS)}
Perhaps the most natural general black box approach is to postulate a general statespace model like (\ref{eq:dssoff}) in discrete time (see also ``Approximating a Continuous Time NLSS with a Discrete Time NLSS Model''):
\begin{subequations}
\label{eq:bb1}
   \begin{align}
    x(t+1)&=f(x(t),u(t),\theta)\\
\hat y(t|\theta)&=h(x(t),\theta)
  \end{align}
   \label{eq:NLSSu}
\end {subequations}
If state transition function $f$  is not known, it could be parameterized with $\theta$ e.g. as a flexible basis function expansion (\ref{eq:basisf}).

Similar expansions can be applied to $h$. In some cases, the state $x$ is measurable and the $h$ would be known. In such cases, \cite{RasmussenW:06} has applied a Gaussian Process model to  $f$ which corresponds to a basis expansion in terms of eigenfunctions associated with the kernel (the covariance function for the Gaussian Process).

\emph{Nonlinear state space models with internal and external dynamics} 
In ``External or Internal Nonlinear Dynamics'' the character of the nonlinear dynamics  is discussed. 
This reflects whether the signals are fed around a nonlinearity or not. For  a NLSS that property depends on the structure of $f$:


If $f$ is lower triangular,
\begin{equation}
x_{j}(t)=f(x_{1,\ldots,j-1},u(t),\theta),
\label{eq:EIDyn4}
\end{equation}
the states can be solved for explicitly from $u$ and $\theta$.
Then it is possible to write (\ref{eq:NLSSu})  in the form $\hat{y}(t|\theta)=h(u(t),\ldots,u(t-n_{b}))$ which is a system with external nonlinear dynamics (actually an NFIR model). Observe that in (\ref{eq:EIDyn4}), the state $x_{1}(t)$ does not depend upon other states, it only depends on $u(t)$.

If the nonlinear state-space cannot be written under the lower triangular form (\ref{eq:EIDyn4}), it is in general not possible to solve the equations explicitly as a function of $u(t)$, and the function \eqref{eq:NLSSu} becomes a system with internal nonlinear dynamics. For these systems, there will be at least one state $x_{i}(t)$ that is a nonlinear function of the past values of some of the other states.  This can be interpreted as the presence of a feedback of an unobserved output of the system.  

\subsubsection{NARX Models}
A special case of (\ref{eq:bbgen}) is the NARX model, where the state is constructed as a finite number of past inputs and outputs
\begin{subequations}
\label{eq:NLARX}
\begin{align}
  x(t)&=\varphi(t) = [y(t-1),\ldots,y(t-na),\\ \nonumber
&u(t-1),\ldots
        u(t-nb)]^T\\
\hat y(t|\theta) &= h(\varphi(t),\theta) 
\end{align}
\end{subequations}
If $h$ is a linear function, this predictor is the familiar simple ARX structure for a linear model. But as indicated, a general nonlinear static function $h$ that can be expressed e.g. in basis function expansion (\ref{eq:basisf}). This structure is therefore known as \emph{NARX} (nonlinear ARX). If $na=0$ it is known as an \emph{NFIR} (nonlinear FIR)

NARX models are a very common class of nonlinear models and can describe a large class of nonlinear systems \cite{Billings2013, Leontaritis1985}. However,  they are not as general as the nonlinear state-space models discussed before. For example the nonlinear system $y(t)=(G(q)u(t))^{2}$, cannot be represented  in an input-output presentation (since the even nonlinearity $x^{2}$ cannot be inverted). 

NARX models come in many different shapes, depending on how $h$ is parameterized.
See ``Sidebar: Static Nonlinearities''. They include Volterra systems, {\eqref{eq:Volterra_u}}, Neural Networks (\ref{eq:nn1}),  Gaussian Processes \cite{Liu2018, RasmussenW:2006}, as well as custom made, semiphysical models, (\ref{eq:cust}).

The case where the nonlinear function $h$ is written as a linear combination of known  basis functions,  (\ref{eq:basisf}), \cite{Sjoberg1995, Juditsky1995}, simplifies the identification problem to a linear regression. Then no iterative optimization procedure is needed \cite{Ljung1987, Pintelon2012}. This is one  reason why NARX models are very popular and have been successfully applied to many industrial problems. 

The number of terms $M$ of a NARX model with basis expansion (\ref{eq:basisf}) may grow very fast with the memory length. Special model pruning methods have been developed to keep only the most dominant terms in the model, e.g. \cite{Billings2013}.

 \subsubsection{User guideline} Lacking physical insights it may be necessary to use black-box model structures. Many flexible and useful such structures exist. But keep in mind that they all have a strong \emph{curve-fitting} flavor and may not pick up any intrinsic system features. They basically reflect the properties of the estimation data which must be chosen with great care in these cases.
 

 
\subsection{Pit-black models: nonparametric smoothing}
\begin{figure}
  \centering
  \includegraphics[scale=0.5]{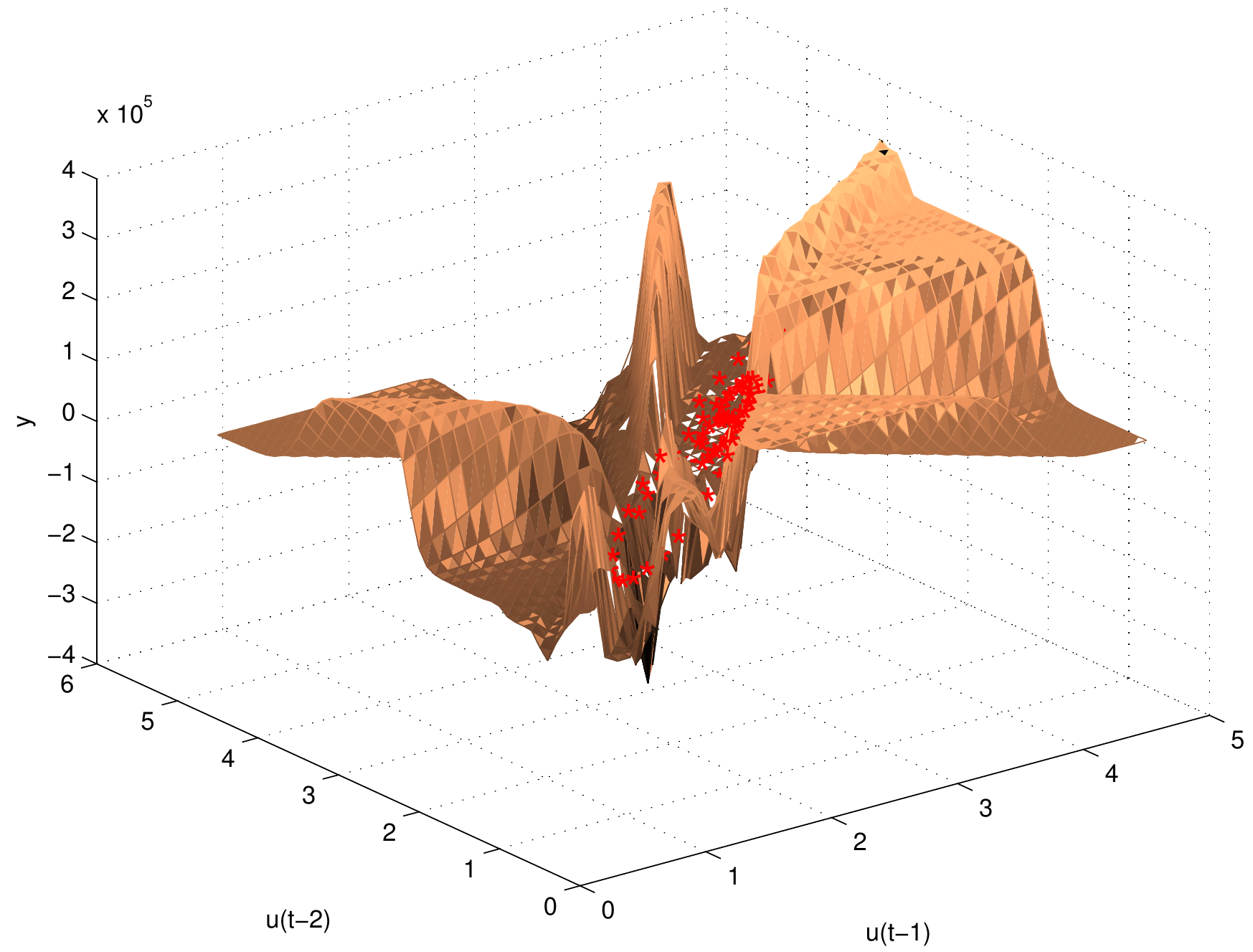}
  \caption{The cloud of observed data (red dots) and the model as a surface in the data space.}
  \label{fig:geom}
\end{figure}

So far models  have been  described   essentially  in parametric and analytic terms. But there is also another possibility: A \emph{pit-black model} takes a ``geometrical'' view on the observed data  set and the model construction. This approach is outside the scope of the current survey, but a    few basic facts can be provided to make the model discussion more complete:

 The model is a relation between the predictor and the output, i.e.  between all past observations and the observed next output. Denote by $\varphi(t)$ an $n$-dimensional vector of relevant representation of past observations (it is like the the state in (\ref{eq:bbgen})) Then the model is a relation
\begin{align}
  \label{eq:pit}
  y(t) = g(\varphi(t)) + e(t)
\end{align}
Here $g$ will act like the predictor in the previous sections, so it corresponds to the general model structure (\ref{eq:pred}). But instead of focusing on estimating $g$ by some parameterization,  the whole data record can be viewed ``geometrically''. Assume that $y(t)$ is scalar. Then each value pair $y(t),\varphi(t)$  is a point in  a $n+1$ -dimensional space $Z$. So the data set is a ``point-cloud'' in this space. In that perspective the modeling task is to find a surface $g$ in $Z$ that as well as possible describes this cloud. See Figure \ref{fig:geom}.
 This can be accomplished by various ways to ``smooth'' the raw data cloud. The basic assumption is that the model surface is ``smooth'', i.e. that $g(\varphi_1)$ and $g(\varphi_2)$ are ``close'' if $\varphi_1$
and $\varphi_2$ are. This should mean that if   the observed points $[y_i,\varphi_i], i=1,\ldots, N$ are available  an estimate $\hat g(\varphi_*)$ could be constructed for any point $\varphi_*$ from observed $y_i$ at neighbouring points:
\begin{align}
  \label{eq:nonpar}
  \hat g(\varphi_*) = \sum_{i=1}^N y_i w(|\varphi_*-\varphi_i|)
\end{align}
where the kernel $w$ weights the value of the observation $y$ by its distance to the sought regressor point $\varphi_*)$.  Such smoothing kernels and nonparametric estimates  $\hat g(\varphi_*)$ are discussed extensively in the statistical literature, e.g. \cite{FanG:96},\cite{Nad:64}. In the control literature, they have been discussed under the name of  ``just-in-time-models'', since the estimate at point $\varphi_*$ is constructed from raw measured data only when it is requested.

A very simple, and common approach is  to let $w$ be zero except  for the $\varphi(t)$ that is closest to $\varphi_*$ in the data base. That makes the estimate $\hat g$ equal to its \emph{nearest neighbour}. Many other kernels have been suggested and studied and  a commonly used one is a parabola bottom, turned upside down, the Epanechikov kernel, \cite{Epanechnikov:69}.
\begin{align}
  \label{eq:epa}
  w(x) = C(1-x^2)_+ \quad x = \|\varphi_*-\varphi(t)\|
\end{align}
where $(\cdot))_+ = \max[\cdot,0]$ and $C$ is a normalization constant.
 Another approach to selecting the weights $w$ in (\ref{eq:nonpar}) is \emph{Direct Weight Optimization, DWO},  e.g. \cite{RollNL:05a}. Here $w$ is selected so that an upper bound of the error in the estimate $\hat g(\varphi_*)-g(\varphi_*)$ is minimized using a quadratic programming technique.

\paragraph{Manifold Learning} Another approach to this ``geometrical'' construction of linear models is to find lower dimensional manifolds in the cloud where has a clear concentration of points.

 A linear model version of this idea can be used for illustration: in the ``data cloud'' setup  a model is  a hyperplane in the data space. Then many techniques are well known to fit the hyperplane to data, including principal component analysis (PCA) for finding essential linear subspace descriptions. That means that the modeling can be concentrated to the selected subspace with lower dimensional models. The nonlinear counterpart is to find a lower dimensional manifold $\Upsilon = r(Z)$ and express a model in terms of the lower dimensional image of the variables $\varphi$ under this mapping. Finding such a manifold is a challenging problem that has been discussed in an extensive literature under the name of \emph{Manifold Learning}, e.g. \cite{Tenenbaumetal:00} (Isomap), and \cite{RoweisS:00} (Local Linear Embedding).

\emph{Summary:} Pit-black models constructing model predictors directly from data, not employing explicitly a parameterized model is an interesting option for nonlinear identification that has not been used that much in the control community.

\section{Experiment Design} 

The experimental data is the fundamental information source for the data driven modeling process. Practical (easy access to a nonparametric noise and nonlinear distortion analysis to guide the model selection process) and theoretical concerns (maximize the information in the data with respect to the selected model structure) should be addressed during the design of the experiment.  This leads directly to the following guidelines:
\begin{itemize}
\item \emph{Practical concerns}: Use periodic excitations whenever it is possible, because these give a direct access to a nonparametric distortion analysis, without any user interaction (see ``Nonparametric Noise and Distortion Analysis Using Periodic Excitations'').
\item \emph{Theoretical concerns}: Design the (amplitude) distribution and power spectrum of the excitation to maximize the information with respect to the parameters that need to be estimated in the selected model structure. Keep in mind that this is still no guarantee that the full domain of interest is covered (see for example Figure \ref{fig:SilverboxPhasePlane}).
\item \emph{Warning}: Because structural model errors often dominate the noise induced errors, it is necessary to select excitations signals that reflect the later use of the model in order to keep the structural model errors `small' in the domain of interest (see Section "System identification in the presence of structural model errors" and ``Impact of Structural Model Errors''). 
\end{itemize}

\subsection{Design of periodic excitation signals}
The most simple periodic excitation is 
\begin{equation}
u(t) = U_{1}cos(2\pi f_{0}t).
\end{equation}
The period of this signal is $T=1/f_{0}$. In most experiments, $u(t)$ is generated and processed in discrete time $t=lT_{s}$, with $l=1,\ldots,N$ and $T_{s}=1/f_{s}$ the inverse sample frequency. The sample frequency and period length are `matched' to each other by chosing $T=NT_{s}$, so that $N$ samples fit exactly in one period of the signal. This relates also the frequency $f_0$ to the sample frequency $f_{s}$ by $f_{0}=f_{s}/N$.

More general periodic signals are represented by their Fourier series as the sum of harmonicaly related sines and cosines with frequencies at integer multiples of $f_{0}$:
\begin{equation}
u(t) = \sum_{k=1}^{F}U_{k}cos\left(2\pi kf_{0}t+\varphi_{k}\right).
\label{eq:Multisine}
\end{equation}
Such a signal is called a multisine \cite{SchoukensJ1988}, and has a period $T=1/f_{0}$ and frequency resolution $f_{0}=f_{s}/N=1/T$. The amplitude spectrum $U_{k}$, the phases $\varphi_{k}$, the number of excited frequencies $F$, and the frequency resolution $f_{0}$ are user choices that define the periodic signal.  These can be set by the following guidelines (see also ``Nonparametric Noise and Distortion Analysis Using Periodic Excitions''):

\emph{User guidelines to design a multisine} \cite{Schoukens2016, Pintelon2012, Schoukens2012}:
\begin{itemize}
\item \emph {Spectral resolution} $f_{0}=f_{s}/N$: should be chosen high enough so that no sharp resonances are missed \cite{Geerardyn2013}.
\item \emph{Period length $N$}: is set by $T=1/f_{0}=N/f_{s}$. A higher frequency resolution requires a longer measurement time. 
\item \emph{Amplitude spectrum} $U_{k}, k=1,\ldots,F$: should be chosen such that the frequency band of interest is covered. 
\item \emph{Phases}: Use random phases, mutually independent for $k\neq l$, and  $E\{e^{j\varphi_{k}}\}=0$. For example, select $\varphi_{k}$ from a uniform distribution on the interval $[0,2\pi)$. For this  choice, it follows from the central limit theorem, that $u(t)$ is asymptotically Gaussian distributed for $F \rightarrow \infty$.
\item \emph{Signal amplitude}: should be scaled such that it also covers the input amplitude range of interest.
\item \emph{Length of the experiment}: At least one, and preferably a few, for example 3, periods should be measured.
\item If possible, repeat the experiment with a new realization of the random phase multisine and average the results over the multiple realizations to get improved estimates of the nonlinear distortion levels. The additional data sets are also very valuable for model validation on different but very similar excitations, and the generation of more reliable uncertainty bounds in the presence of structural model errors (see also ``Impact of Structural Model Errors'').
\end{itemize}

Remark: By optimizing the phases, it is possible to create randomized signals with a user controlled amplitude distribution and power spectrum \cite{Schoukens1988}.  For example, signals with a uniform amplitude distribution that excite a specified frequency band can be generated.

\subsection{Experiment design: most informative experiment}

The goal of the experiment design can be formalized as a procedure to obtain the minimum required information needed to reach the modeling goals at the lowest experimental cost (time, power consumption, disturbance of the process, etc.). 

Originally, optimal experiment design was completely focused on maximizing the information content of the experiment \cite{Mehra1974}, quantified by the 'Fisher Information Matrix' $M$ \cite{Ljung1987, Soderstrom1989, Pintelon2012} that is directly linked to the smallest possible covariance matrix of the parameter estimates $P_{\theta}=M^{-1}$  . A scalar measure, for example, the determinant $det(M)$ or the trace $tr(M)$, is used to quantify the information in the experiment. 

Although an optimal experiment design is not always done for each experiment, it is very useful to make the exercise on a number of typical problems to be studied, because it provides a lot of intuitive insight in what turns an experiment into a good one. On the basis of that experience the quality of the experiments can be significantly boosted, even without designing explicitly an optimal input. 

For models that are linear-in-the-parameters, the information matrix $M$ does not depend on the actual parameter variables. This does not hold true if the output of the model is a nonlinear function of the parameters, in that case $M(\theta_{0})$ depends explicitly on the true but unknown parameters $\theta_{0}$. Often, the true parameters $\theta_{0}$ are replaced by an estimate $\hat{\theta}$ obtained from an initial experiment. 

The more structured a model is, the higher the gain that can be obtained by a specific optimally designed experiment. The optimal experiment for a first order linear system, described by its nonparametric impulse response representation $g(k), k=1,\ldots,n$ is a white noise excitation. If the same system is represented by its transfer function model $G(s)=\frac{1}{1+\tau s}$, the optimal experiment is a single sine excitation at a frequency $f=1/\tau$ \cite{Goodwin1977}. 

\subsubsection{Optimal input design for linear systems}
Optimal input design for linear systems is fully understood today. In the basic problem, an excitation is designed that maximizes the determinant of the information matrix $det(M)$.  The optimal design minimizes the normalized variance of the estimated transfer function and is retrieved by solving a convex optimization problem.  Since the problem depends only on the second order properties of the input signal, its solution is given by an optimal power spectrum of the excitation signal \cite{Mehra1974, Goodwin1977, Pronzato2008}. The actual shape of the signal (amplitude distribution) does not affect the information, but practical constraints can have a strong impact on it, leading for example to signals ranging from filtered white noise (having a Gaussian distribution), to binary excitations with a user imposed power spectrum. 

Soon it became clear that this simple problem statement does not meet the full user needs. The optimal experiment design should be plant friendly (not all excitations are acceptable for the operators) \cite{Rivera2003}. Moreover,  the uncertainty on the estimated model should be tuned to result in a (control) design that meets the global goals of the project. These initiated a search of application oriented input design \cite{Gevers2005,Bombois2012,Forgione2015,Hjalmarsson2009,annegren2017}, resulting in a design that pushes the uncertainty in a direction where it does not hurt the quality of the application: the uncertainty ellipsoid of the model is matched to the contour plots of the application cost.

\subsubsection{Optimal input design for nonlinear systems}
While the optimal input design for linear systems is well understood, it is much harder to provide general guidance of the optimal input design for nonlinear systems. Although it is possible to retrieve (numerically) the Fisher information matrix for nonlinear systems, it is very difficult to interpret these equations and to translate them into an optimal input \cite{Mahata2016}. This is due to the dependence of the Fisher information matrix $M$ on the higher order moments of the input for nonlinear systems. This leads to the design of the multivariate probability distribution of the input (for all moments and all lags) \cite{Gevers2012, Mahata2016} which is a highly non-convex problem. 

In the case of a static nonlinear system that is linear-in-the-parameters, the problem becomes convex again. The solution depends only on the amplitude distribution of the excitation and results in a signal that is concentrated around a discrete set of excitation levels. The order of the samples does not influence the solution so that the power spectrum is completely free in this case.

The optimal input design for nonparametric impulse response estimation (resulting in a white noise excitation) can be generalized to nonlinear systems using the nonparametric Volterra representation. It leads to a design that combines the properties of the linear impulse response design (white noise excitation) with that of optimal input design for static nonlinear systems (discrete set of amplitude levels)  \cite{Birpoutsoukis2018}. This idea was also the starting point for a numerical design, leading to numerical procedures that make a brute force search for discrete level signals \cite{Forgione2014,DeCock2016}.

Solving the full optimal experiment design problem for nonlinear systems is tackled today using brute force numerical optimization methods. The solutions can be generalized and simplified using a proper normalization of the problem. The choice of the signal constraint (power constraint at the input or output, amplitude constraint at the input, output, or at an intermediate signal) turns out to be most important. A natural choice seems to restrict the amplitude range at the input of the nonlinear sub-system \cite{DeCock2017PhD}.

In some problems, the interest is in the identification of a single parameter in a nonlinear model. Its variance can be reduced using a well designed feedback law that can be applied in real time \cite{Chianeh2011}.

\subsection{Summary}
\begin{itemize}
\item Optimal input design that is solely based on the Fisher information matrix $M$ and its related variance expressions should only be applied if there are no dominant structural model errors present.
\item For linear systems, optimal input design is a well developed field and the nature of the solutions is fully understood, even if numerical procedures are needed to calculate the optimal input. 
\item Optimal input design is closely linked to the intended application.
\item For nonlinear systems, numerical procedures are available (for some nonlinear model structures). A full understanding of the important aspects of a good solution is still lacking.
\end{itemize}

\section{Model Validation}
Model validation addresses the question ``Does the model solve our problem?'', and/or is it in conflict with either the data or prior knowledge? A number of linear and nonlinear validation tools are discussed below.
\subsection{Cross Validation}
One of the most common and pragmatic tools for model validation is \emph{Cross Validation}, that is to  check how well the model is able to reproduce the behaviour of new data sets  ---\emph{Validation Data}  --- that were not used to estimate the model. One way is to use the input of the validation data to simulate the model, to produce a \emph{simulated model output} $\hat y_S(t)$ and compare how well this model output reproduces the output $y(t)$ of the validation data. The comparison could simply be a subjective, ocular inspection of the plots, to see if essential aspects of the system (for the intended application) are adequatly reproduced.

The comparison can also be done by computing numerical measures of the fit between the two signals. These are naturall based on the distance between $y(t)$ and $\hat y_s(t)$. A common numerical measure is the \emph{fit}, used in the system identification toolbox, \cite{Ljungtb:18}.
\begin{align}
  \label{eq:fit}
  fit = 100 \left(1 - \frac{\sqrt{\|\sum y(t)-\hat y_s(t)\|^2}}
  {\sqrt{\sum |y(t)-mean(y(t))\|^2}}\right)
 \;\text{(in \%)}
\end{align}
So the fit tells, in percent, how much of the variation of the output is correctly reproduced by the model.

For models that contain integration or are used for control design it  may be more revealing to evaluate the model's prediction capability. Then the $k$-step ahead predicted output for validation data $\hat y_p(t|t-k)$ is computed using the model ($\hat y_p(t|t-k)$ that depends on all relevant past input, and the output up to time $t-k$).
The prediction can then be compared with the measured validation output by inspecting the plots or by the fit criterion (\ref{eq:fit}).

Apart from these simple simulation and prediction applications, cross validation can be used in several sophisticated ways, discussed in the statistics literature, see e.g. \cite{modval}. 
\subsection{Nonparametric validation for periodic excitations}
Using periodic excitations gives access to a nonparametric noise and distortion analysis (see ``Nonparametric Noise and Distortion Analysis Using Periodic Excitations''). Adding the residues $y(t)-\hat y(t)$ to this plot shows how well the model captured the nonlinear contributions (how far are the validation errors below the nonlinear distortion levels), and if the errors drop to the noise floor.

\emph{Actions}
\begin{itemize}
\item \emph{The error level is above the noise floor}: Structural model errors are detected. The user should decide if these errors are acceptable or not.
\item \emph{Nonparametric analysis of the errors}: If the special periodic excitations of ``Nonparametric Noise and Distortion Analysis Using Periodic Excitations'' are used in the validation test, it is possible to find out the nature of the dominant errors (even or odd nonlinearities are missed in the model), giving indications how the model can be improved.
\end{itemize}

\subsection{Linear validation tools}
Linear validation tools check the whiteness of the residuals (auto-correlation test) and verify if no linear relations between the input and the residuals are left (cross-correlation). Although this are both second order moment tests that reveal only a subset of the possible problems (the higher order moments are not tested), it still provides valuable information.

\emph{Actions}
\begin{itemize}
\item \emph{Cross-correlation detected}: A more flexible linear part of the model can reduce the linear dependency at a low cost.
\item \emph{Auto-correlation detected}: The residuals are still colored. Using a linear noise model it is possible to reduce the prediction errors. This can improve the efficiency of the estimation procedure.
\end{itemize}

\subsection{Nonlinear validation}

\subsubsection{Higher order moment tests}
A full validation of a nonlinear model requires also the higher order moments to be ``white'' (strictly spoken this term applies only to the second order moment), and no higher order cross-correlations should exist. In practice these tests are often not made for the following reasons: i) Moments of order $n$ are $n-1$ dimensional objects. Visual evaluating these at all lags becomes very cumbersome and time consuming;   ii) The required experiment length to estimate the higher order moments with a given precision grows very fast with the order $n$, making such a test often unfeasible. 

For some dedicated problems, higher order tests are proposed to detect the presence of nonlinearities \cite{Enqvist2007}.

\subsubsection{Change the nature of the input}
The behavior of a nonlinear system strongly depends on the nature of the excitation signal, even if the maximum input amplitude and power spectrum remain the same. For that reason, it is a very strong requirement for a model to cover a very wide range of input signals which can be tested during the validation. In Figure \ref{fig:SB3} it is shown on the Duffing oscillator that the nonlinear NARX model that was tuned for the tail data failed to give a good simulation on the sweep data.

\emph{Action}:
Verify the quality of the model on all relevant classes of excitation signals.

\subsubsection{Check of the domain}
\begin{figure}[h] 
\centering
\includegraphics[scale=0.5]{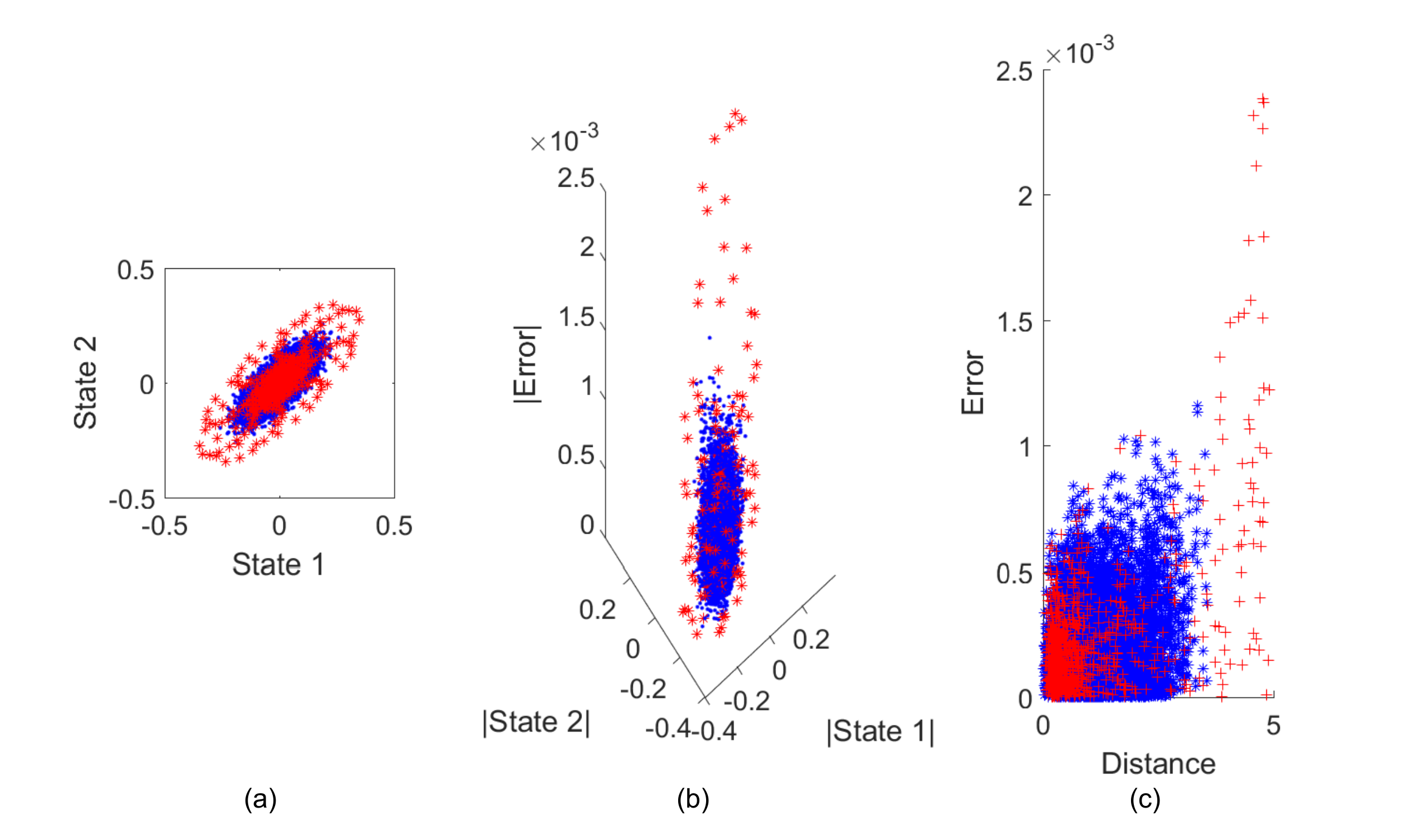}
\caption{Model validation on the forced Duffin oscilator. A NLSS model is identified (see Figure  \ref{fig:NARXresults1}) using a part of the tail data described in ``Extensive Case Study: the Forced Duffing Oscilator''. Next the models are validated on another data set from the tail part (blue) and from the sweeping part (red). The input signals in both data sets have almost the same maximum amplitude and power spectrum, as shown in Figure \ref{fig:SBsignals}. Figure (a) plots the state trajectory for both validation experiments. Observe that the domain covered by the sweeping signal is much larger than that of the tail signal. In (b) the absolute value of the simulation error for both experiments is plotted as a function of the states. The errors on the states that are outside the domain covered by the tail signal are much larger. This is clearly visible in (c) where the absolute error is plotted as a function of the normalized distance $s^{t}C_{s}^{-1}s$ with $s$ the states, and $C_{s}$ the covariance matrix of the tail (blue) states. Observe that the red errors in (b) and (c), for the large values of the distance, are much larger then the blue errors. This shows that a nonlinear model that is only validated on the blue data fails to explain the red data that cover a larger domain, illustrating the risk of extrapolation in nonlinear system identification.} 
\label{fig:SilverboxPhasePlane}
\end{figure}

Changing the nature of the excitation changes also the domain on which the internal nonlinear function \eqref{eq:genstr} that is present in each nonlinear model is evaluated. The best guarantee to get a valid model is given by making sure that the complete domain of interest is covered during the estimation and validation. Although this is not always possible, it might be helpful to check the covered domain by plotting the phase plane trajectory for the estimated model for different excitations. This is illustrated on the Duffing oscillator example in Figure \ref{fig:SilverboxPhasePlane} and Figure \ref{fig:NARXresults1}.

\subsection{Check the uncertainty bounds}
As explained in the section ``Generation of uncertainty bounds'' in ``Impact of Structural Model Errors'', and illustrated in Figure \ref{fig:WindTunnel}, it is very hard to provide reliable uncertainty bounds in the presence of structural model errors that dominate the noise disturbances. The actual observed variability of the model is larger than the theoretically expected one.  For that reason it is indispensable to verify the validity of the theoretically calculated uncertainty bounds. 

\emph{Action}: Estimate the selected model structure with fixed complexity for different realizations of the excitation and verify if the actual observed standard deviation agrees with the theoretical one.

\subsection{User guidelines:}
\begin{itemize}
\item Validating of a model is a rather subjective and pragmatic problem. Check on a rich validation data set that covers the intended use of the model if the estimated model meets the user expectations.
\item The final modeling goal should be kept in mind during the model validation: in many problems, structural model errors that are below a user defined level can be tolerated, even if these errors are clearly detected in the model validation step.
\item Check also the theoretically obtained uncertainty bounds. In the presence of structural model errors these are under estimating the actual variability.
\item Make sure that the validation tests cover the full domain of interest.
\end{itemize}

\section{Examples of Nonlinear System Identification: From White to Black Box Models}
In the next series of examples, the use of different levels of physical insight in the nonlinear system identification process are illustrated.

  \section{Examples I: Off-White Model\\ Tank System}
\subsection{The system}
\begin{figure}[h] 
\centering
\includegraphics[scale=1]{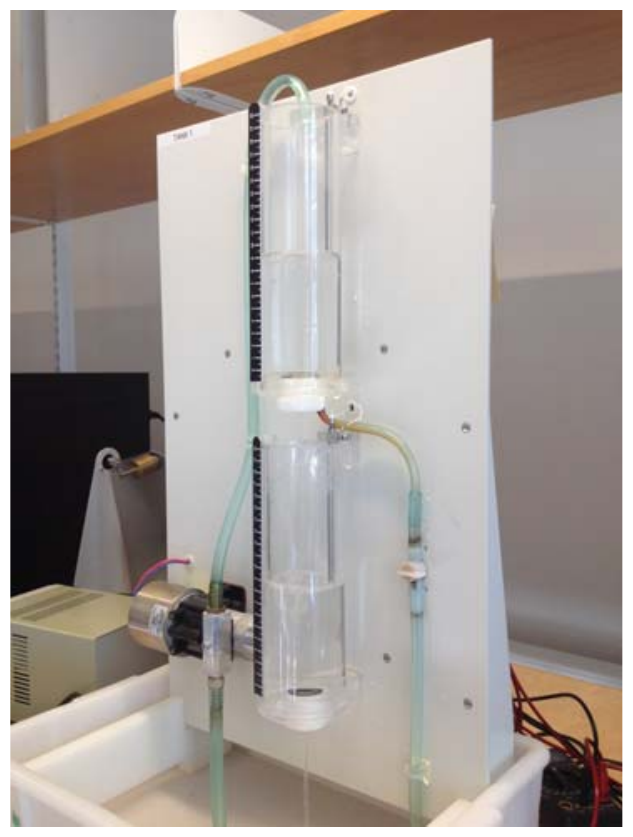}
\caption{ The cascaded tanks system: The input signal controls a water pump that delivers the water from a reservoir into the upper water tank from where it flows through a small opening into the lower tank, and finally through a small opening from the lower tank back into the reservoir. When the amplitude of the input signal is too large, an overflow can happen in the upper tank, and with a delay also in the lower tank. When the upper tank overflows, part of the water goes into the lower tank, the rest flows directly into the reservoir. This effect is partly stochastic, hence it acts as an input-dependent process noise source. The overflow saturation nonlinear behavior of the lower tank is clearly visible in the output signals that saturate at level 10 (see Figure \ref{fig:Tank2}). The input is the pump voltage, the output is the water level of the lower tank \cite{SchoukensM2016b}.}
\label{fig:Tank1}
\end{figure}

The cascaded tanks system is a benchmark system that was setup at the University of Uppsala \cite{SchoukensM2016b}. It is a fluid level control system consisting of two tanks with free outlets fed by a pump that is described in detail in Figure \ref{fig:Tank1}. 

\subsection{A physical model}

When no overflow occurs, a model can be constructed based on Bernoulli's principle and conservation of mass:

\begin{equation} \label{eq:Tanks1}
\begin{aligned}
\dot{x}_{1}(t) &= -k_{1}\sqrt{x_{1}(t)}+k_{4}u(t)+w_{1}(t), \\
\dot{x}_{2}(t)  &= k_{2}\sqrt{x_{1}(t)}-k_{3}\sqrt{x_{2}(t)}+w_{2}(t),  \\
y(t)&=x_{2}(t)+e(t).     
\end{aligned}
\end{equation}
where $u(t)$ is the input signal, $x_{1}(t)$ and  $x_{2}(t)$ are the states of the system,  $w_{1}(t),w_{2}(t)$
and $e(t)$ are additive noise sources, and $k_{1},\ldots,k_{4}$ are constants depending on the system properties.

The relation between the water flowing from the upper tank to the lower tank and the water flowing from the lower tank into the reservoir are weakly nonlinear functions if there is no overflow (\ref{eq:Tanks1}), while in the presence of overflow, hard nonlinearities need to be identified. 

To model the overflow, $x_{1}(t)$ and  $x_{2}(t)$ are constrained to their maximum value, and an additional term $w_{3}(t)$ is added to the second equation in (\ref{eq:Tanks1}) for $x_{1}(t)>x_{1max}$ \cite{Rogers2017}. 

In \cite{Rogers2017} it is also proposed to add adiitonal terms $k_{5}x_{1}(t)$ to $\dot{x}_{1}(t)$ and $k_{6}x_{2}(t)$ to  $\dot{x}_{2}(t)$, to include also the losses in the fluid flow. The losses are proportional to the verlocity of the fluid squared and, therefore, proportional to the height of the fluid in each tank. It turned out \cite{Rogers2017} that this additional flexibility in the model is also used to accomodate other imperfections of the model, leading to an improvement that goes far beyond the expected impact of the loss terms in this system.

\subsection{The data}
\begin{figure}[h] 
\centering
\includegraphics[scale=1]{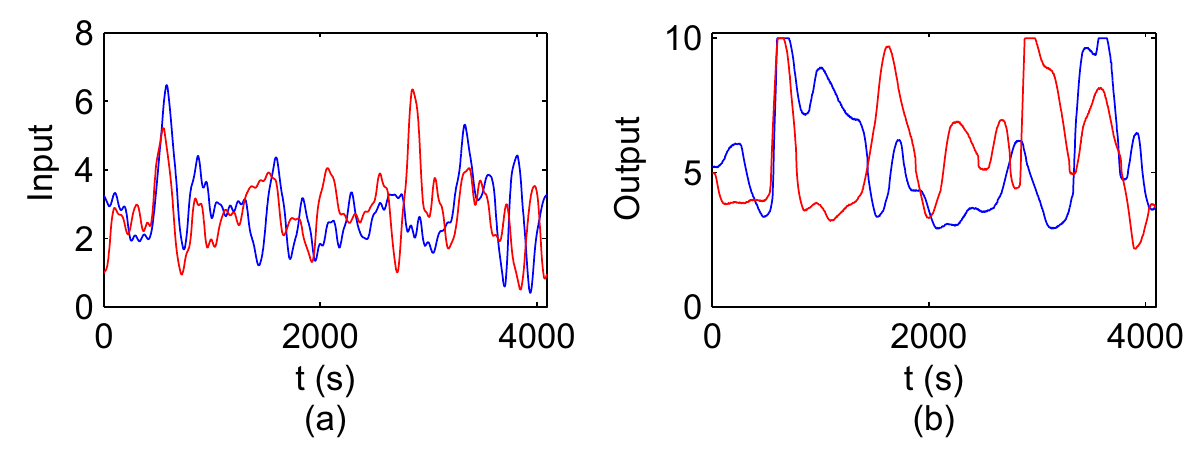}
\caption{Input (a) and output (b) signals of the estimation (blue) and test (red) data \cite{SchoukensM2016b}. Observe the saturation in the output (level second tank) when the output $y(t)=10$.}
\label{fig:Tank2}
\end{figure}

The input signals are ZOH-multisine signals which are 1024 points long, and excite the frequency range from 0 to 0.0144 Hz, both for the estimation and test case  (see Figure \ref{fig:Tank2}). The lowest frequencies have a higher amplitude then the higher frequencies. The sample period $T_{s}$ is equal to 4 s, the period length is 4096 s. Two similar data sets were collected, one for estimation and one for test (validation). The water level is measured using capacitive water level sensors, the measured output signals have a signal-to-noise ratio that is close to 40 dB. The water level sensors are considered to be part of the system,
they are not calibrated and can introduce an extra source of nonlinear behavior.

Note that the system was not in steady state during the measurements. The system states have an unknown initial value at the start of the measurements. This unknown state is the same for both the estimation and the test data record and needs also to be estimated.

\subsection{Cost function}
The cost function used to match the model to the data is
\begin{equation}
V=\frac{1}{N}\sum_{t=1}^{N}(y(t)-\hat y(t,\theta))^{2},
\end{equation}
with $y_{mod}(t,\theta)$ the model output.
Observe that either the simulation or one-step ahead prediction error can be minimized.

\subsection{Results}
In \cite{Rogers2017}, the parameters of the simple and extended physical model are directly estimated using well-selected numerical optimization procedures. The fit error \eqref{eq:fit} of the simulated output equals $fit=4.23\%$ on the estimation data, and $fit=5.93\%$ on the test data. For the extended model that includes also the 'loss' terms, this error drops to $fit=1.02\%$ on the estimation data, and $fit=1.78\%$ on the test data. 

As a reference, also a black box Gaussian Process NARX model (see also Section ``The Palette of Nonlinear Models: Black Models: NARX'') with 15 lags for the input and output was estimated,  resulting in $fit=4.62\%$ for the simulation error, and $fit=0.057\%$ for the one-step-ahead prediction error. This illustrates once more that it is much easier to get a small prediction error than it is to get a small simulation error.

\begin{figure}[h] 
\centering
\includegraphics[scale=0.6]{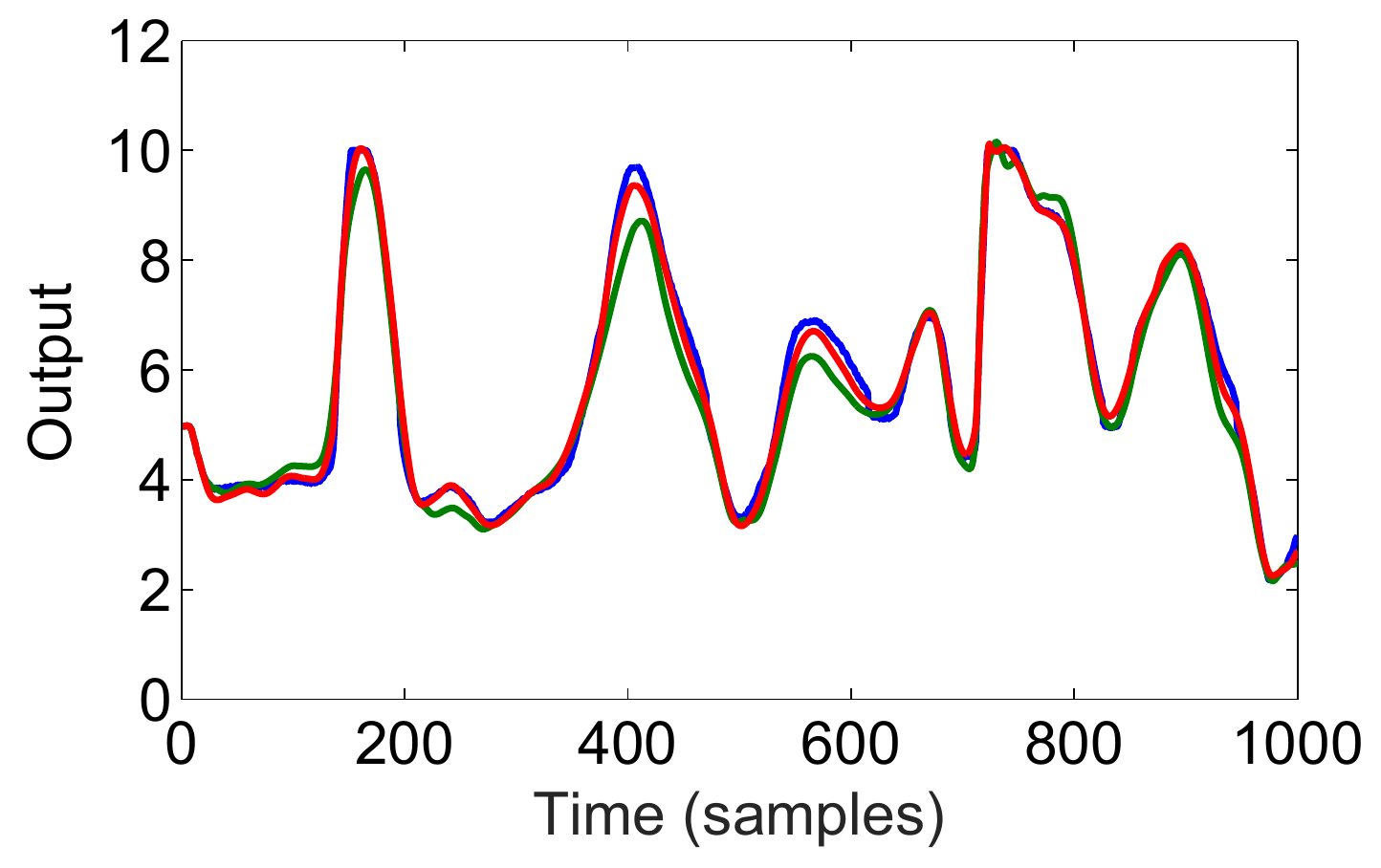}
\caption{The cascaded tanks are modeled using the simple and extended physical model (\ref{eq:Tanks1}) and validated on the test data \cite{Rogers2017}. The extended physical model (red) simulates the measured output (blue) much better ($fit=1.02\%$) than the simple model (green) with $fit=5.93\%$ does. Observe that also the saturations starting at $t=150$ and $t=750$ are well retrieved by the extended model.}
\label{fig:CascadedTanksWhiteBoxWorden}
\end{figure}

\section{Examples II: Smoke-Grey Model (Semiphysical Model): An Industrial Buffer Flow System}
\subsection{The system}
\begin{figure}
  \centering
   \includegraphics[scale=0.3]{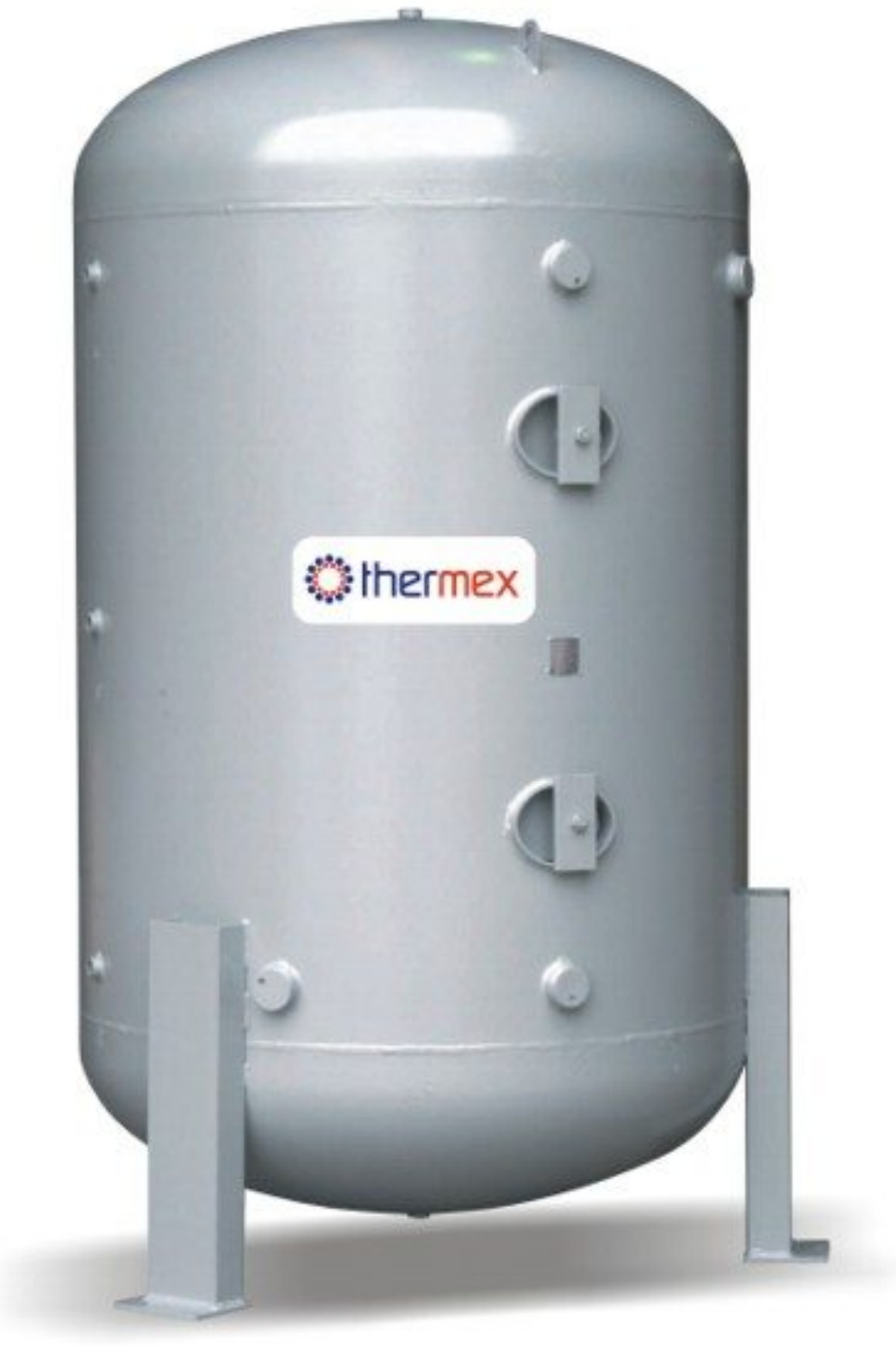}
 \includegraphics[scale=0.27]{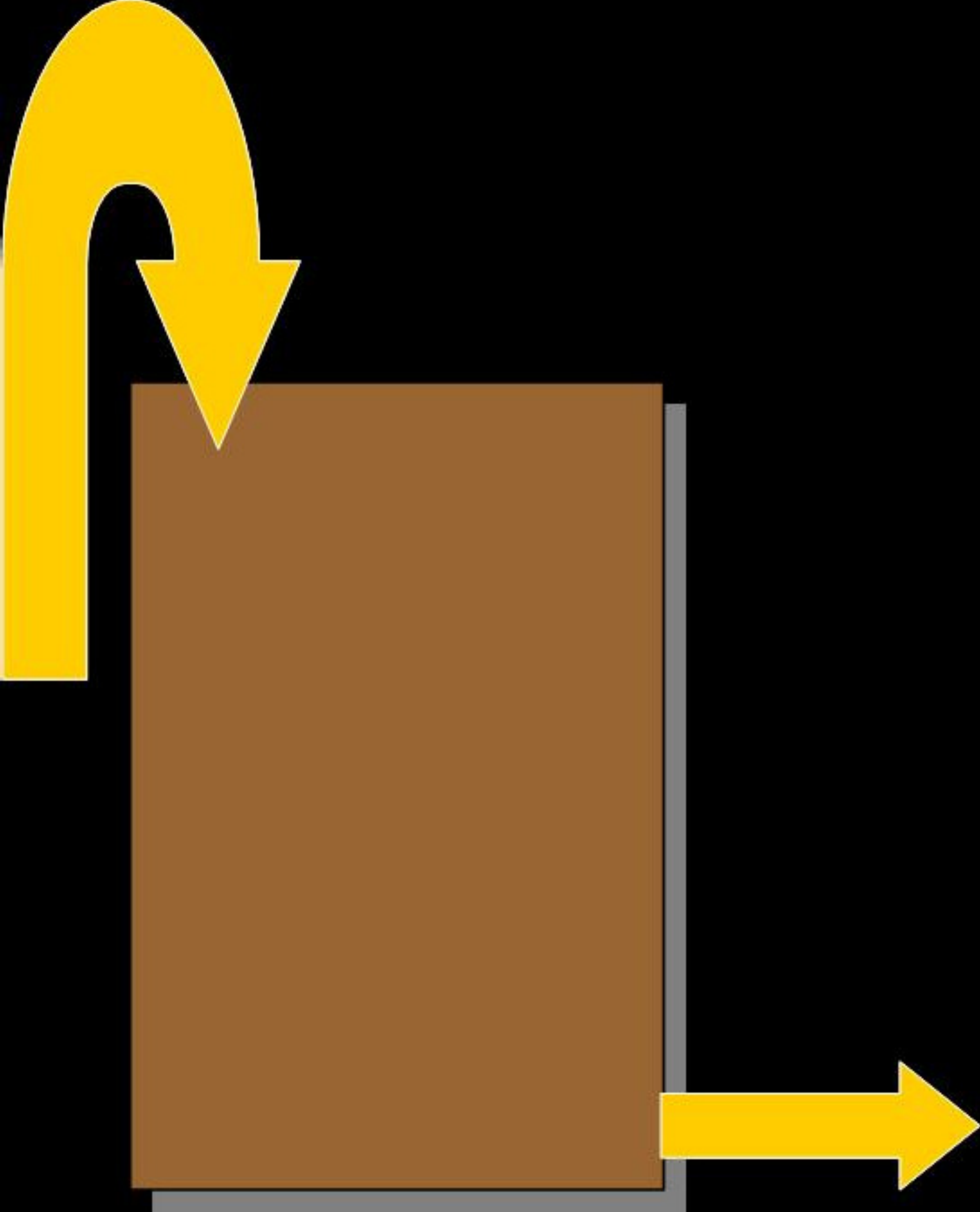}
  \caption{Left: Physical buffer  Right: Schematic picture of a buffer vessel with inflow and outflow}
  \label{fig:buffer}
\end{figure}

This is an example of the usefulness of recalibration of the time scale,  \cite{Andersson&Pucar:95} as explained in Section ``Smoke-grey Models: Semi-physical modeling''. The process is a buffer vessel in a pulp factory, in Skutsk\"ar in Sweden. The pulp spends some 48 hours in the different stages of the process, cooking, washing, bleaching, etc. It passes through several buffers to allow for a smooth continuous treatment. It is important to know the residence time in the buffers for proper bookkeeping. The so called $\kappa$ number is a property of the pulp, measuring its lignin contents. The buffer vessel is schematically depicted in Figure \ref{fig:buffer}.
 So the problem is to find a model for the dynamics of the buffer vessel that allows evaluating the residence time in the vessel and to ``time mark'' the pulp as it passes through the several vessels.
\subsection{The data}
\begin{figure}
\centerline{\includegraphics[scale=0.4]{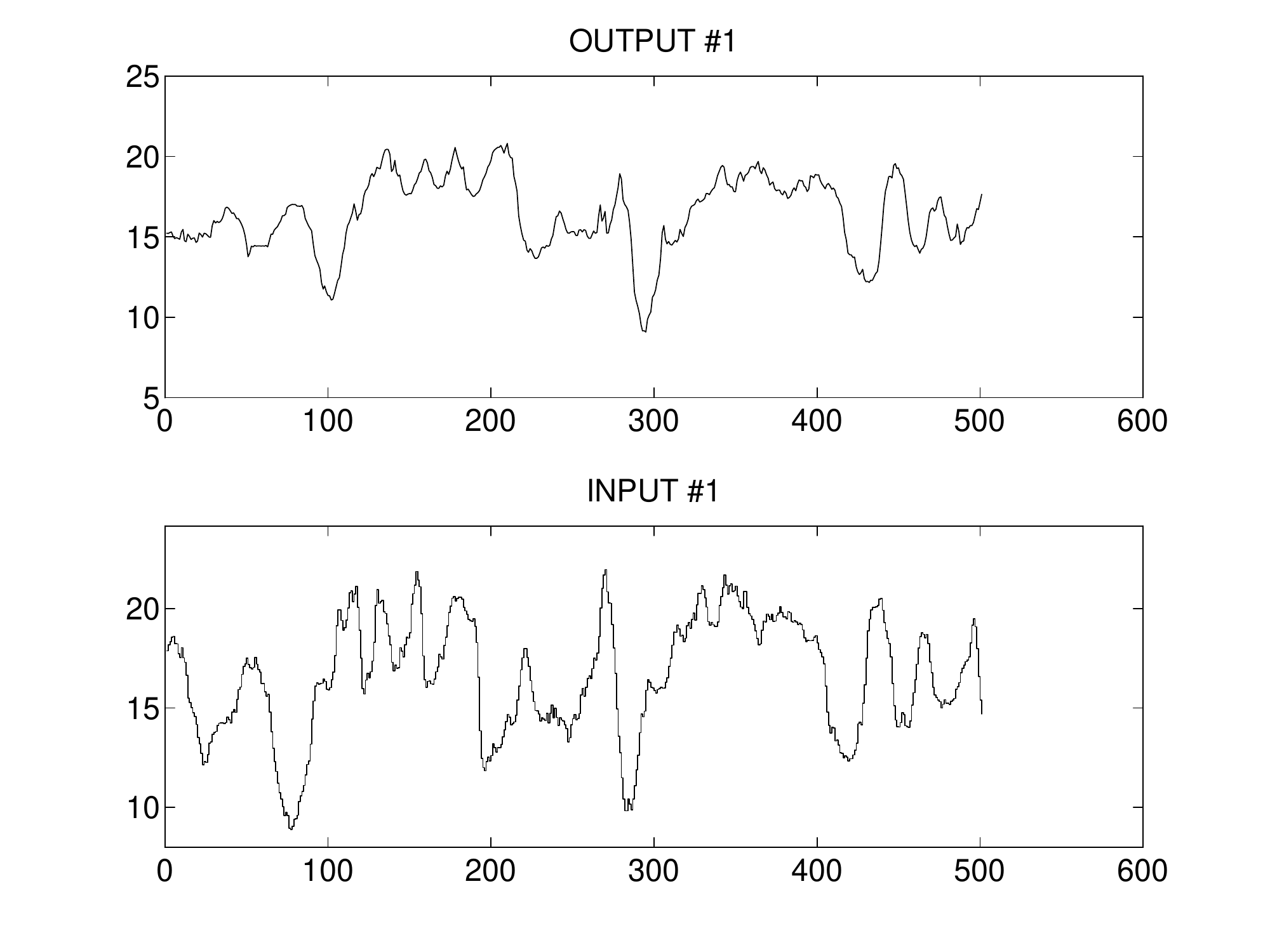}}
\caption{Top: The output $\kappa$-number $y$. Bottom: The input $\kappa$-number $u$}
\label{fig:flows}
\end{figure}

\begin{figure}
  \centering
 \centerline{\includegraphics[scale=0.4]{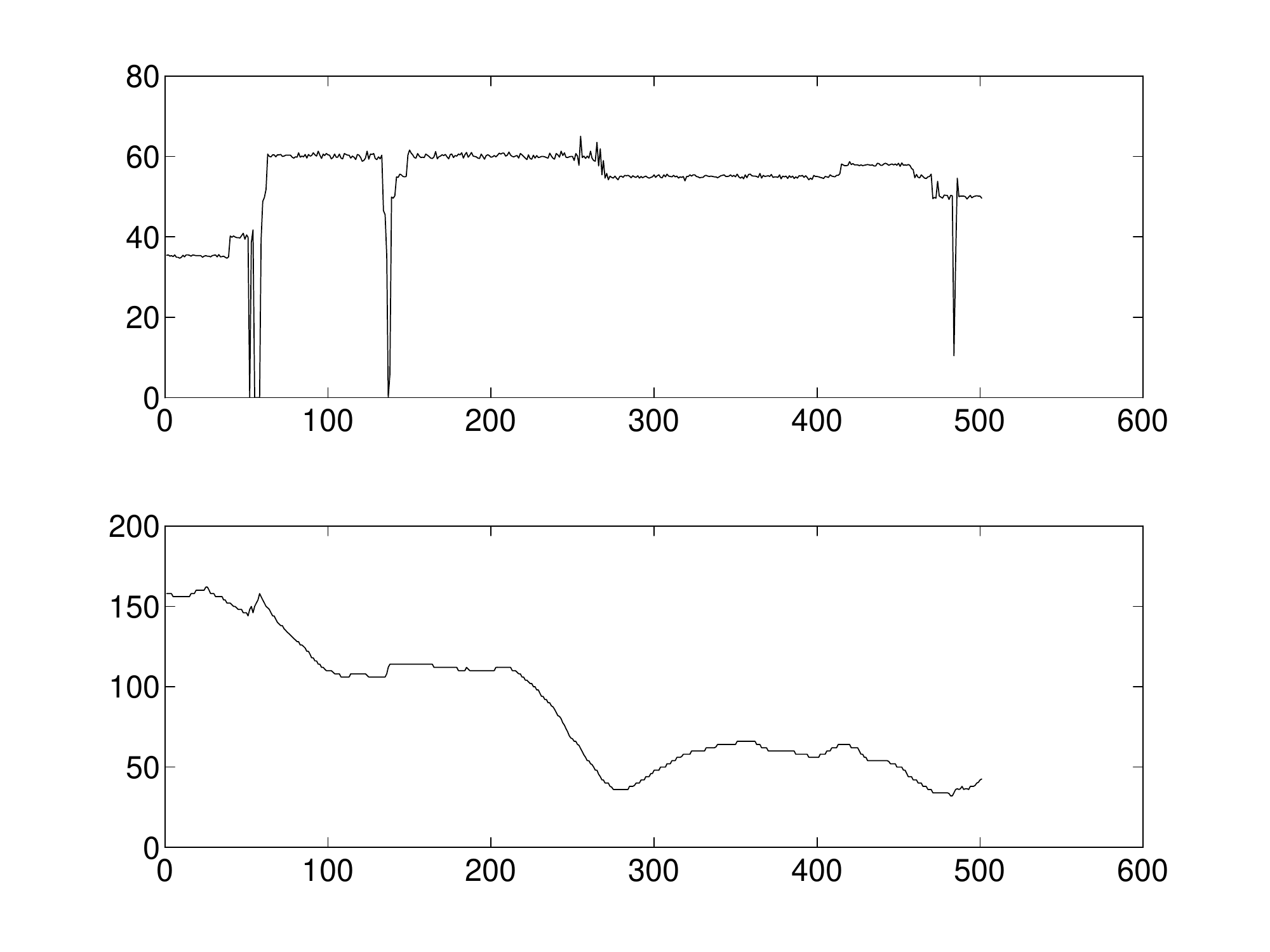}}
  \caption{Top: The buffer flow. Bottom The buffer level.}
  \label{fig:levels}
\end{figure}

In a particular buffer, the $\kappa$-number of the outflow, the output $y$, was meausured along with  $u$,
the $\kappa$-number of the inflow.  See Figure \ref{fig:flows}. The vessel level and flow were also measured, Figure~\ref{fig:levels}.


\subsection{First Model attempt: Linear Model Based on Raw Data }
\begin{figure}
  \centering
  \includegraphics[scale=0.4]{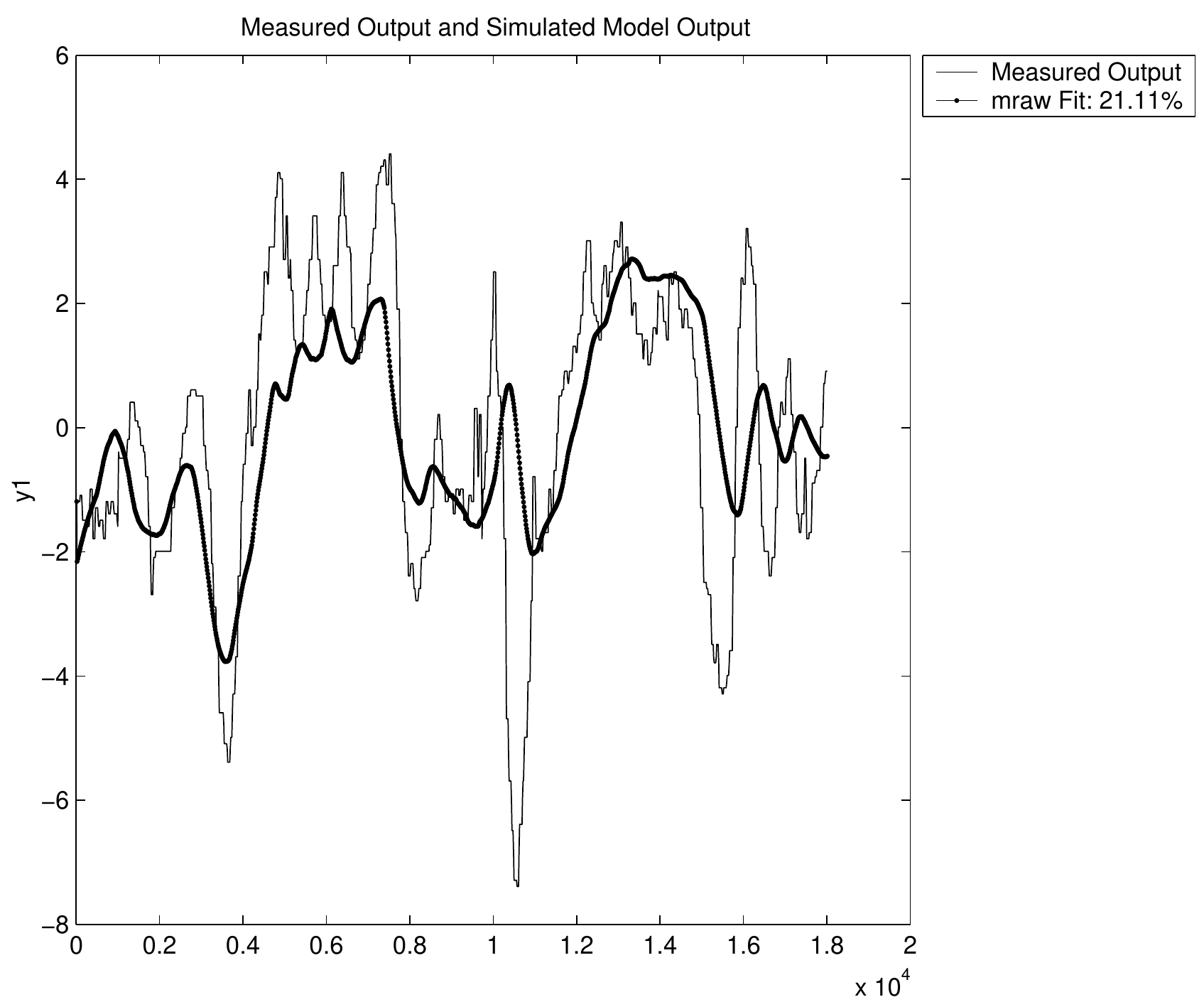}
  \caption{Linear Model: The model output (thick line) compared to the measured output (thin line). The fit (\ref{eq:fit}) for this model is 21.1\%}
  \label{fig:rawmodel}
\end{figure}

First estimate a linear process model using the input-output data $u,y$ using the first half of the data.
That gave the model $G(s) = \frac{0.818}{1+676s}e^{-480s}$. This model was simulated with the output and the model output is compared with the measured output in Figure \ref{fig:rawmodel}.
 

This linear model is quite bad. The simulated output differs quite substantially from the measured output
\subsection{Apply ''semiphysical modeling''}    The physics behind the flow system needs to be taken into account.  How does the flow and buffer level affect the buffer dynamics? 

It there is no mixing in the vessel: (``Plug flow''), the  vessel is just a pure time delay for the pulp flow: Delay time: Vessel Volume/Pulp Flow (dimension time.)


If there is perfect mixing in tank, the system is a text-book first order system with gain=1 and time constant = Volume/Flow


 So if Volume and Flow are changing, the system  is non-linear.

 The natural time variable is really Volume/Flow,  (which has been measured). The observed data can be resampled according to this natural time variable.

\subsection{Recalibrate the time scale}
Apply a nonlinear transformation to the raw data by re-sampling it to the natural time variable: In \textsc{Matlab} this becomes

\noindent
{\tt  z = [y,u]; pf = flow./level;
\newline t = 1:length(z)
\newline newt = interp1([cumsum(pf),t],[pf(1):sum(pf)]');
\newline newz = interp1([t,z], newt);}

\begin{figure}
  \centering
  \includegraphics[scale=0.4]{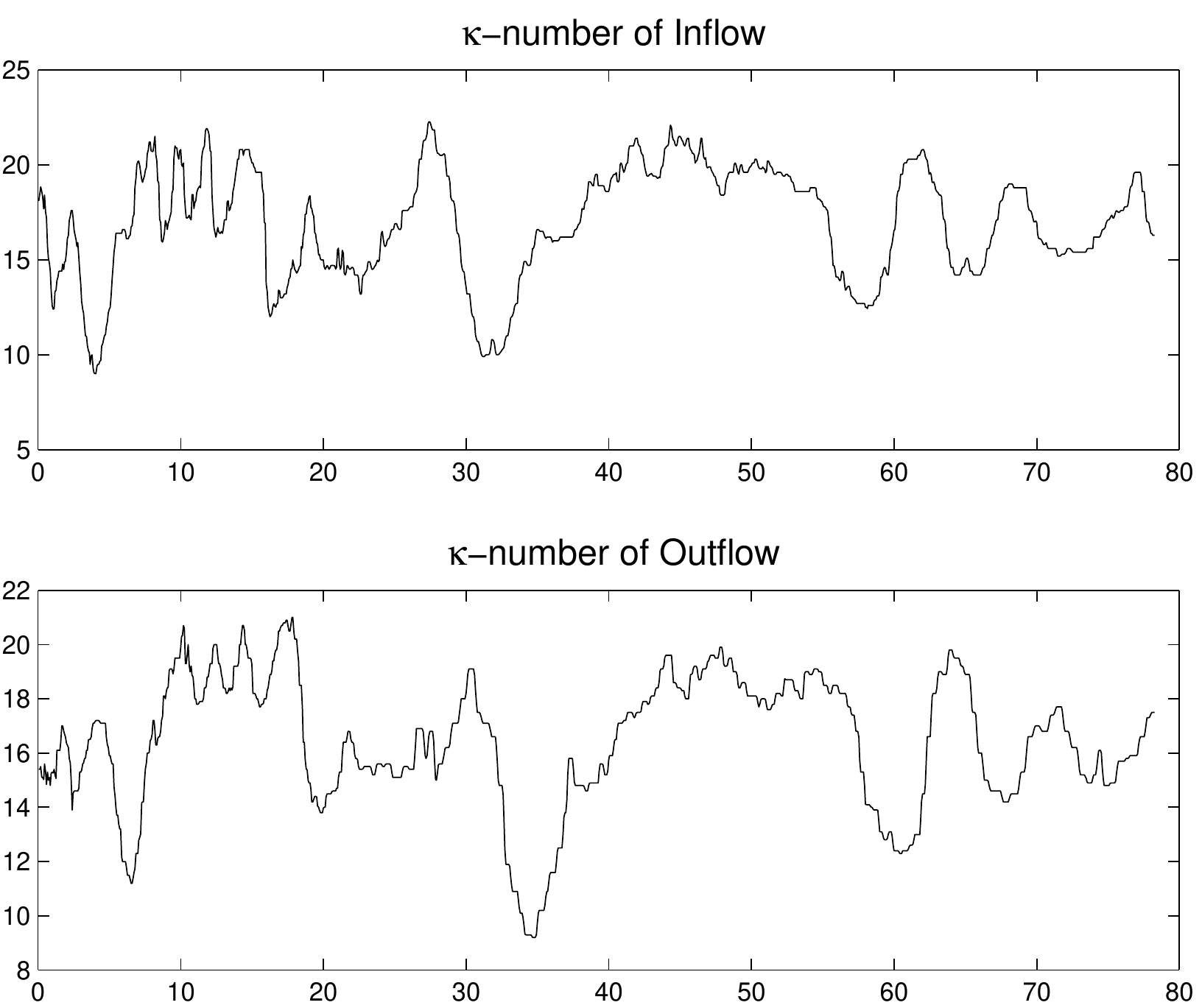}
  \caption{The resampled output (top) and input (bottom).}
  \label{fig:resam}
\end{figure}

The resampled inputs and outputs are shown in Figure \ref{fig:resam}.

 \subsection{Second model attempt: Linear model based on resampled data}


Building a linear process model from the first half of the resampled inputs and outputs gives the model$ G(s) = \frac{0.8116}{1+110.28s}e^{-369.58s}$

\begin{figure}
  \centering
   \includegraphics[scale=0.4]{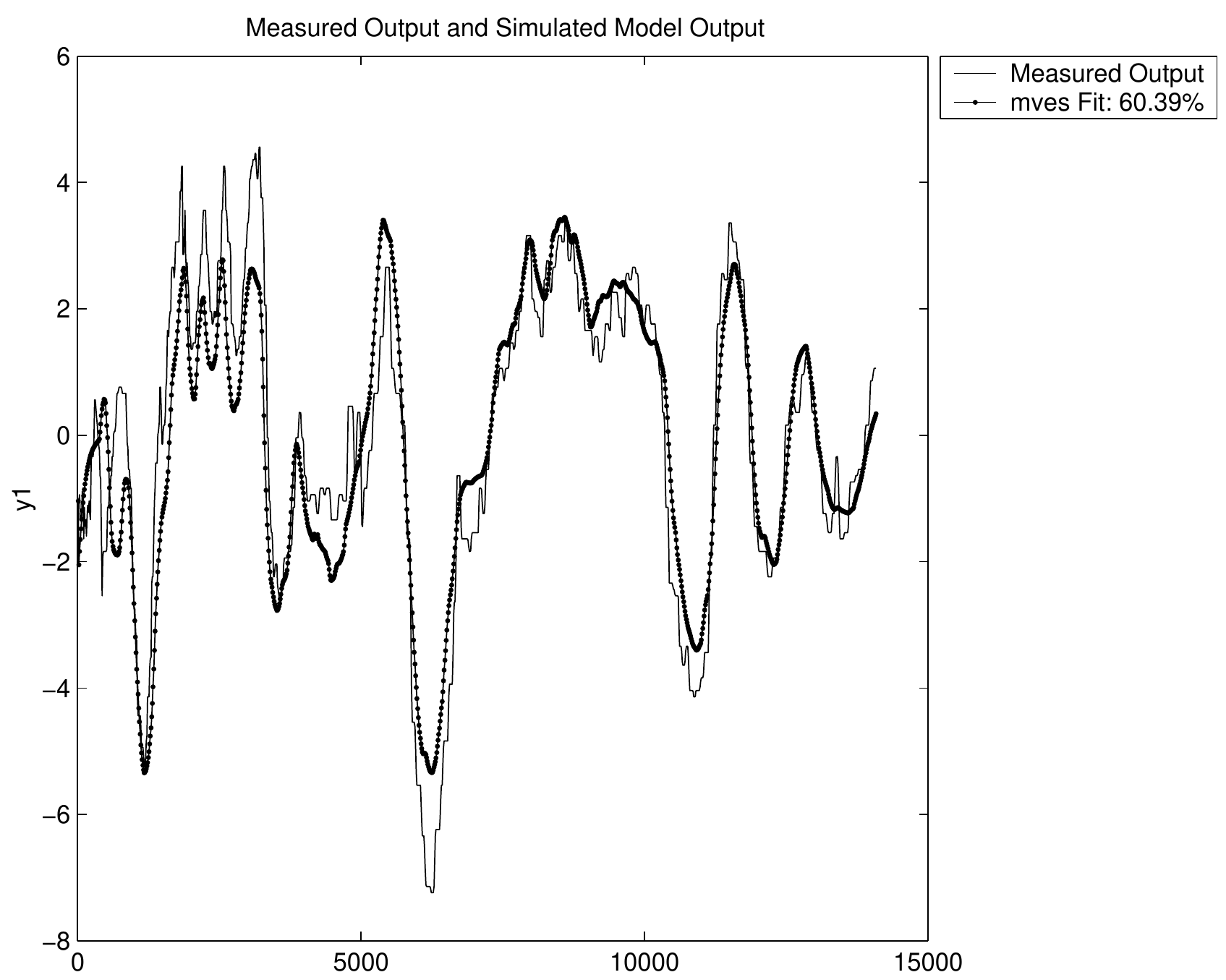}
  \caption{Simulated model output (thick line) and measured output (thin line) for resampled data. The fit (\ref{eq:fit}) for this model is 60.4\%}
  \label{fig:sammodel}
\end{figure}

Simulating that model and comparing with the measured output (for resampled data) gives a much better fit as shown in Figure \ref{fig:sammodel}, analoguous to Figure \ref{fig:rawmodel}.


    The semi-physical  model is linear in the resampled data, but nonlinear in the original raw data in Figures  \ref{fig:flows}-\ref{fig:levels}. It gives a sufficiently good description of the buffer, to allow proper time-marking of the pulp before and after.

\section{Examples III: Steel-Grey Model (Linearization-Based Model): The High Pressure Fuel Supply System}

The team of Oliver Nelles (University of Siegen, Germany) developed a Local Linear Modeling based approach and used it to identify a high pressure fuel supply system (HPFS) that is used in a common rail direct fuel injection for diesel engines. The reader is referred to \cite{Heinz2017, Heinz2018, Nelles2001} for a detailed description of the project and the general methodology. This section is fully based on these references.

\subsection{The system}
\begin{figure}
  \centering
   \includegraphics[scale=0.8]{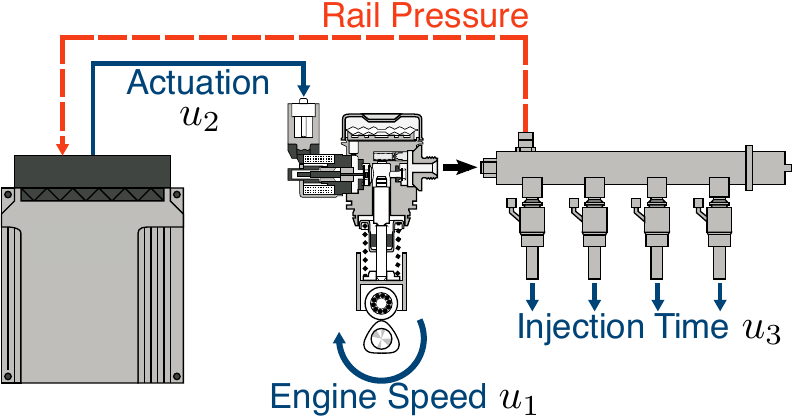}
  \caption{Main components of the HPFS system, inputs (continuous lines), and output (dashed line) \cite{Tietze2014}}
  \label{fig:FuelSystem}
\end{figure}

The main components of a HPFS system are the high pressure rail, the high pressure fuel pump, and the ECU (Engine Control Unit; see Figure \ref{fig:FuelSystem}). The pump is actuated by the crankshaft of the engine. A demand control valve in the pump allows to control the delivered volume per stroke. A pressure-relief valve is also included in the pump, but should never open, if possible. Hence, we want to limit the maximum pressure
during the whole measurement process. The pump transports the fuel to the rail, which contains the pressure sensor. From there, it is injected into the combustion chambers. The system has three inputs and one output. The engine speed $n_{mot}$ affects the number of strokes per minute of the pump and the engine's fuel consumption. The fuel pump actuation $M_{SV}$ gives the fuel volume which is transported with every stroke of the pump. It is applied by opening and closing the demand control valve accordingly during one stroke of the pump. The injection time $t_{inj}$ is a variable calculated by the ECU, which sums up the opening times of the single injectors and is, thus, related to the discharge of fuel from the rail.

During the measurement procedure, the injection time is not varied manually but set by the ECU. The permissible times depend on many factors and a wrong choice could extinguish the combustion or even damage components. The engine load would have a major influence on the HPFS system via the injection time, but is omitted to prevent the necessity of a vehicle test bench.

\subsection{The model}
A NARX model \eqref{eq:NLARX} (see also Section ``The Palette of Nonlinear Models: Black Models: NARX'') with external nonlinear dynamics (see ``External or Internal Nonlinear Dynamics'')  
\begin{equation}
\label{eq:NARXmodel}
y(t)=h(u(t),\ldots,u(t-n_{b}),y(t-1),\ldots,y(t-n_{a}))
\end{equation}
 will be used to model the pressure $y(t)$ of the rail as a function of three inputs $u=[u_1, u_2, u_3]$, with  $u_1(t)=n_{mot}$, $u_2(t)=M_{SV}$, and $u_3(t)=t_{inj}$. The local linear modeling method \eqref{eq:llm} is selected to represent and identify the nonlinear function $h$ \cite{Nelles2001} (see also the section on Steel-grey models in The Palette of Nonlinear Models). These are specified by the choice of i) the regime points, ii) the validity functions, and iii) the local linear models.

\emph{The regime points}  $p(k)$, also called $z$-variables in \cite{Heinz2017,Heinz2018}, that are the entries of the validity functions $w(p(t),p_i)$ (weighting functions) in \eqref{eq:llm} are reduced to one time delayed process inputs and output  in this application: $p(k)=[u_1(k-1),u_2(k-1),u_3(k-1),y(k-1)]$. This choice takes into account, that the actual operating point is defined by the level of the actual process inputs and output. The first (and higher) derivatives of the model inputs and output are assumed to be insignificant to describe the operating point.

\emph{The validity functions} $w(p(t),p_i)$ in this contribution are constructed using the hierarchical local model tree (HILOMOT) algorithm \cite{Nelles2006}. This incremental growing tree construction algortihm divides the input space with axes-oblique splits. The validity functions are generated by sigmoid splitting functions that are linked in a hierarchical, multiplicative way, see \cite{Nelles2006} for more details.

\begin{figure}
  \centering
   \includegraphics[scale=0.8]{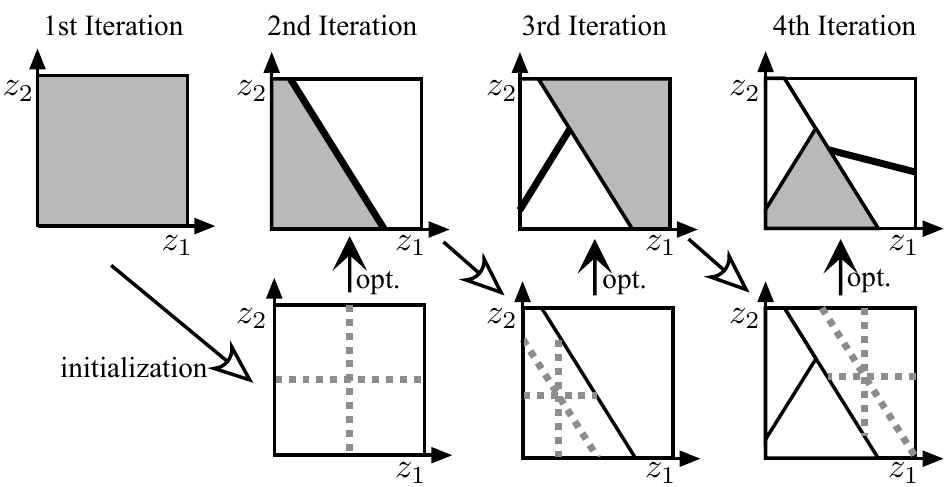}
  \caption{The first Iterations of HILOMOT using a two-dimensional $z$-input space \cite{Nelles2006}.}
  \label{fig:NellesHILOMOT}
\end{figure}

The procedure of the HILOMOT algorithm can be explained with the help of Figure \ref{fig:NellesHILOMOT}. Starting with a global affine model, in each iteration an additional local affine model is generated. The local model with the worst local error measure (gray areas in Figure \ref{fig:NellesHILOMOT}) is split into two submodels, such that the spatial resolution is adjusted in an adaptive way. 

\emph{The local linear models} $\hat y_i(t|\theta,Z^{t-1})$ are here ARX models of order 3, $Z^{t-1}=[u_1(t-1),u_1(t-2),u_1(t-3),\ldots,u_3(t-1),u_3(t-2),u_3(t-3),y(t-1),y(t-2),y(t-3)]$.

\subsection{The data}
\emph{Experiment design}: Two main aspects are driving the experiment design. The experiments should be rich enough to get a good estimate for the local linear models, but even more important, the experiments should be designed such that the regime points cover the full space of interest \cite{Heinz2017, Heinz2018, Nelles2001}. The excitation signal should also comply with the constraints imposed by the process, like restrictions on the amplitude level and signal gradients. In \cite{Heinz2018} a procedure is described that searches for randomized step like sequences that meet all these constraints. The step time is set by the dominant time constant of the system, while the step sequences are designed to assure that the regime points are well distributed over the full operational space of the system. Detailed information, including the design of excitations for multiple input systems is given in \cite{Heinz2018}. 

\emph{The data}: The optimized experiment is called OptiMized Nonlinear InPUt Signal (OMNIPUS), and it will be compared to a second experiment where the excitation is a combination of a ramp and a chirp sequences proposed in \cite{Tietze2015}. The measurement time for each signal is limited to 10 minutes. The sampling frequency $f_s=100Hz$, resulting in a signal length of $N=60 000$ samples.

\subsection{Cost function}
The models are estimated by minimizing the squared errors \eqref{eq:critfit} between the measured $y(t)$ and modeled output $\hat y(t|\theta)$ \eqref{eq:NARXmodel}. No weighting is applied.

\subsection{Results}
\begin{figure}
  \centering
   \includegraphics[scale=0.6]{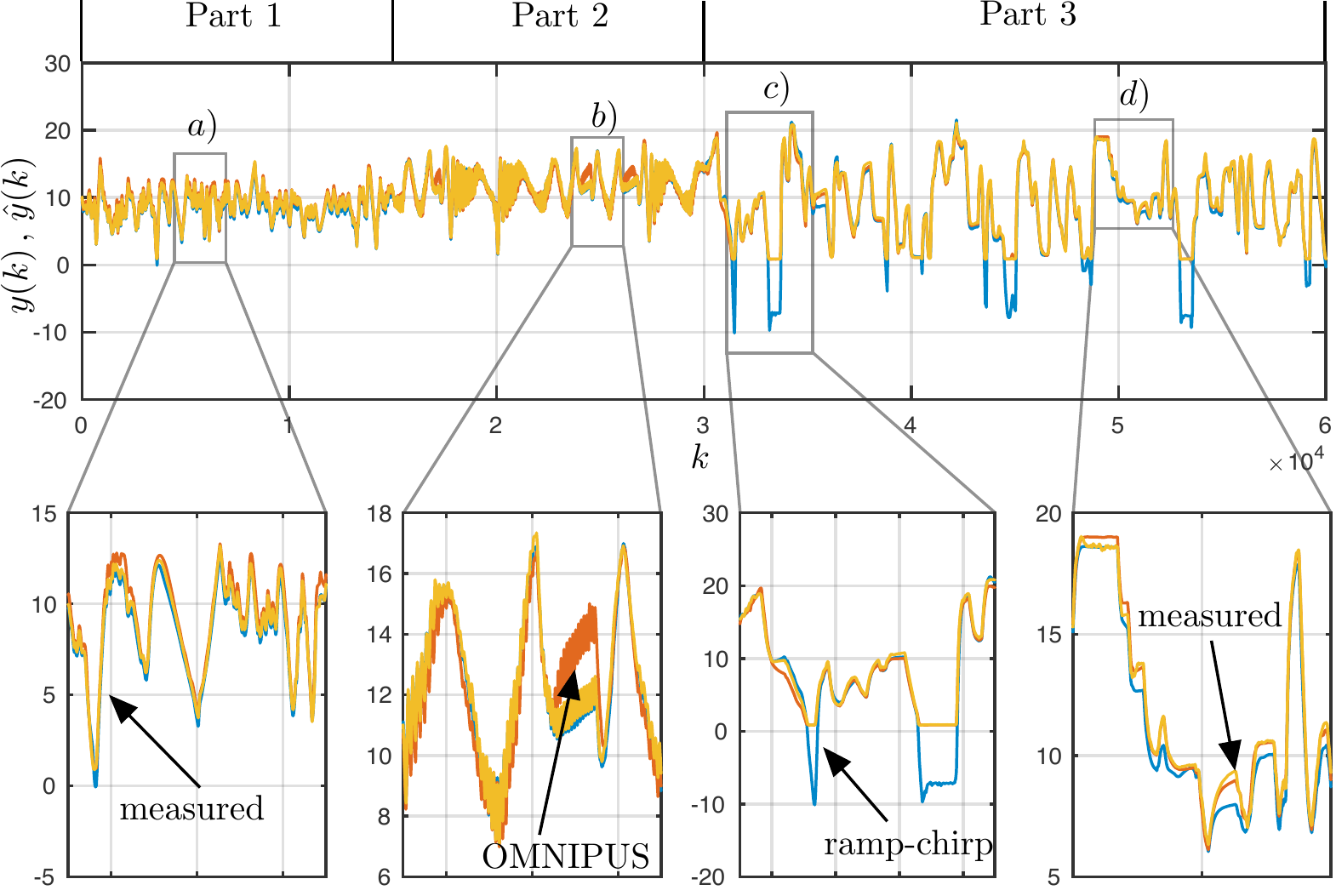}
  \caption{Two local linear model networks are identified, the first being estimated on an OMNIPUS excitation, the second on a ramp-chirp excitation. Both models are used to simulate the measured plant output on a test data set that consists of a new realization of a ramp-chirp and OMNIPUS excitation. The simulated output of the ramp-chirp (blue) and the OMNIPUS HILOMOT (red) model \cite{Heinz2017} on the test data is shown and compared to the measured output (yellow) that is given as reference. The test signal is designed such that the model behavior outside the training domain can be analyzed: Part 1 consist of ramp test sequences, Part 2 consists of chirp test sequences, and Part 3 consists of the OMNIPUS test sequence. The ramp-chirp model was trained on the ramp-chirp training data (similar to the signals in Part 1 and 2). Thus, the second half of the test signal (Part 3) is outside the training domain of the ramp-chirp model. The other way round, the OMNIPUS model was trained on the OMNIPUS training data (Part 3). Thus, the first half of the test signal (Part 1 and 2) is outside the training domain. This effect can be seen on the two subfigures b) and c). In a) good model fit for both models on the ramp-chirp sequence is obtained. In b) a plant model mismatch of the OMNIPUS model on the ramp-chirp sequence is visible, and in c) the plant model mismatch of the ramp-chirp model on the OMNIPUS sequence can be observed. In d) a good model fit for both models on the OMNIPUS sequence is obtained.}
  \label{fig:NellesResults}
\end{figure}

Two local linear model networks (LMN) are identified, the first being estimated on the OMNIPUS data, the second on the ramp-chirp data. Both models are used to simulate the measured plant output on a test data set that consists of a new realization of a ramp-chirp and OMNIPUS excitation. The simulated output of both models on the test data are shown and discussed in Figure \ref{fig:NellesResults}.  The error for both local linear models is in most cases small (e.g. Figure \ref{fig:NellesResults} (a) and (b)).  But there are also some significant mismatches for example in Figure \ref{fig:NellesResults} (c). The nonlinear behavior of the process is not well identified by the ramp-chirp LMN. This major mismatch between process and model indicates a poor modeling most likely because informative data in this area of operation are missing. Figure \ref{fig:NellesResults} (b) shows a minor mismatch between process and the OMNIPUS LMN.

\begin{figure}
  \centering
   \includegraphics[scale=0.8]{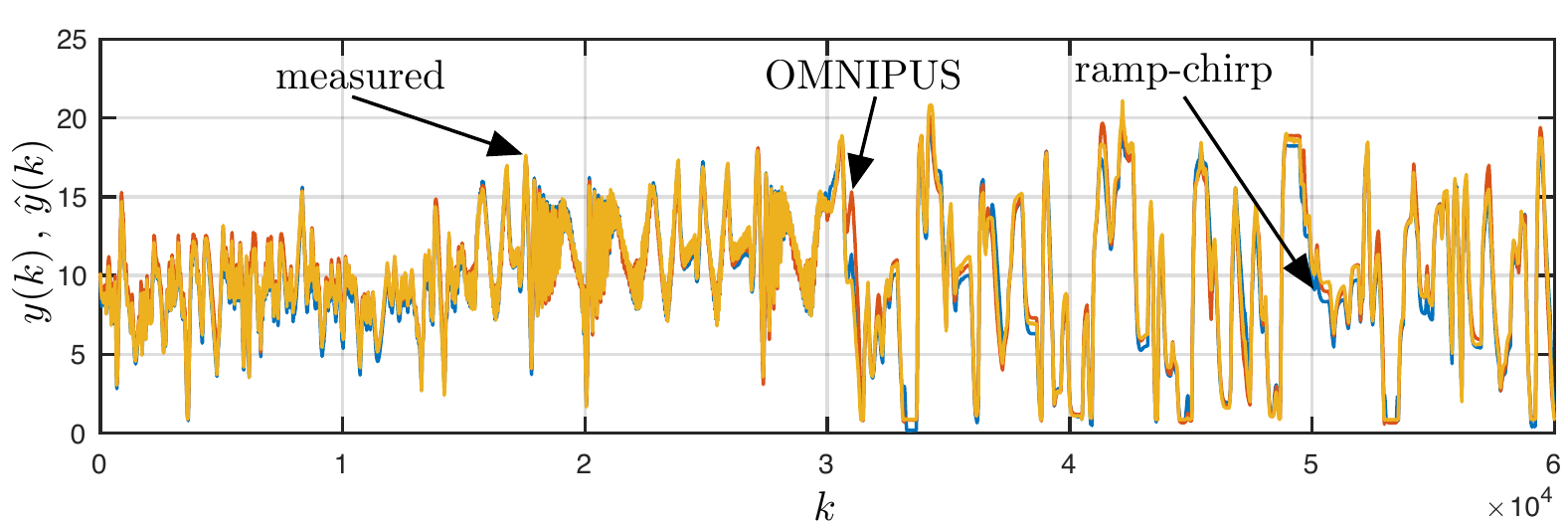}
  \caption{Output signal of the ramp-chirp GPM (blue) and the OMNIPUS GPM (red) on test data \cite{Heinz2017}. The measured output (yellow) is given as reference .}
  \label{fig:NellesResultsGPM}
\end{figure}

As a reference, the same data were also processed using Gaussian process models (GPM) for the $h$ function in \eqref{eq:NARXmodel} \cite{Heinz2017}, and shown in Figure \ref{fig:NellesResultsGPM}. The GPMs do not show dramatical mismatches between process and the identified models. In the last half of the signal the ramp-chirp model seems to be slightly worse because of some slightly discrepancies. But overall, the GPMs seem to be valid in the range of operation. A quantitative comparison shows that the GPM and LMN models behave quite similar, but that the design of the experiment has a large impact on the generalization of the models (how do the models behave outside the training domain). In this example, the GPM models seem to be less sensitive to this problem. Overall, these observations emphasizes again the importance of a good design of the experiment.

\section{Examples IV:  Slate-Grey Model (Block Oriented Model): Hydraulic Crane}
\subsection{The Process}
\begin{figure}
  \centering
   \includegraphics[scale=0.3,angle=90]{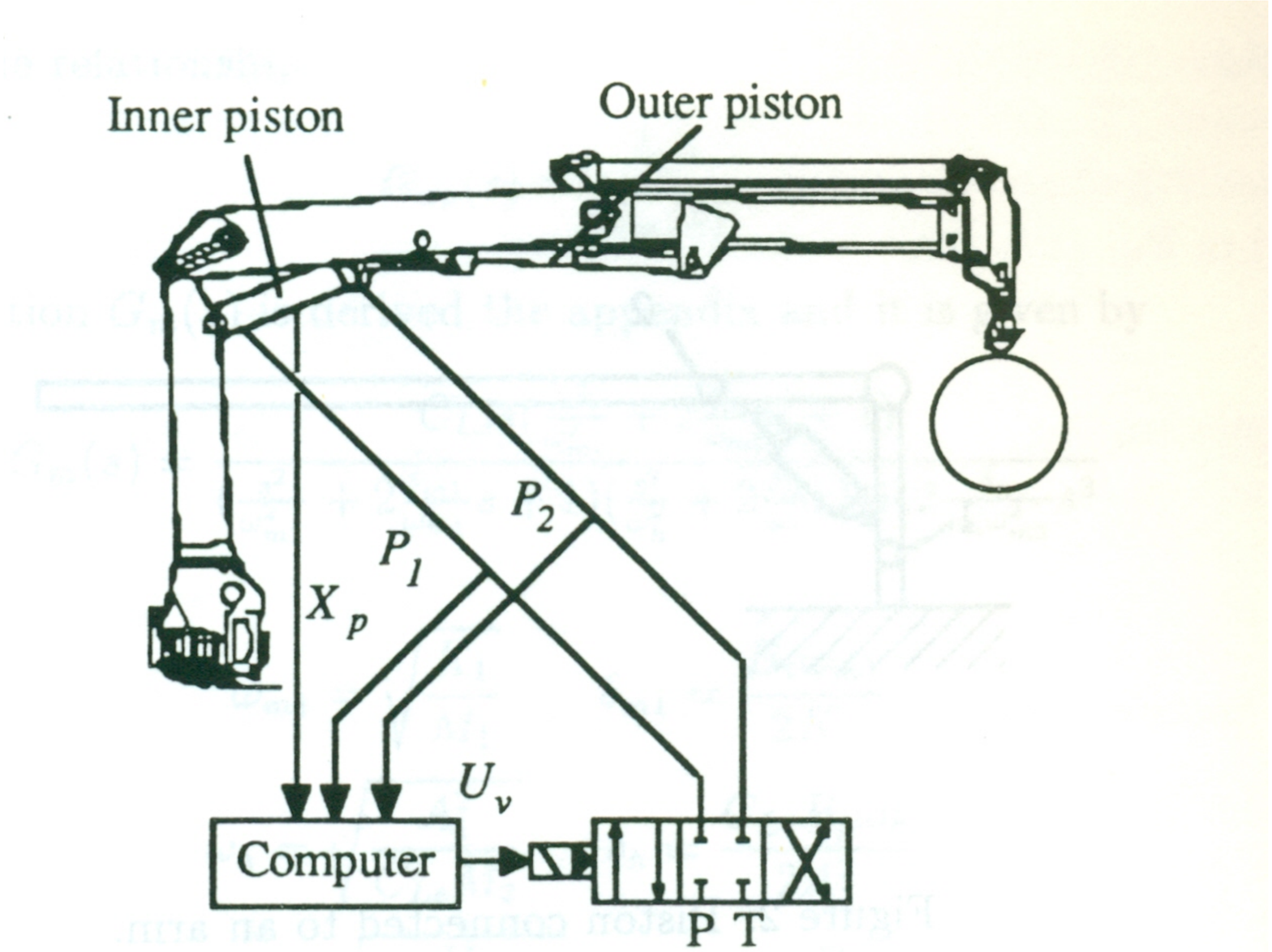}
  \caption{The forest harvest machine: A hydraulic crane.}
  \label{fig:hydcrane}
\end{figure}

When handling tree-logs in forest harvesting, huge hydraulic cranes do
all the lifing and moving. The  cranes can be thought of as industrial robots
that are contolled by hydraulic pressure and the conrolled output is
the position of the gripper at the top of the crane. See Figure \ref{fig:hydcrane}.
The crane shows oscillitory behaviour at the gripper and in order to design a good regulator, a  model has to be developed
\subsection{The Data}
\begin{figure}
  \centering
  \includegraphics[scale=0.5]{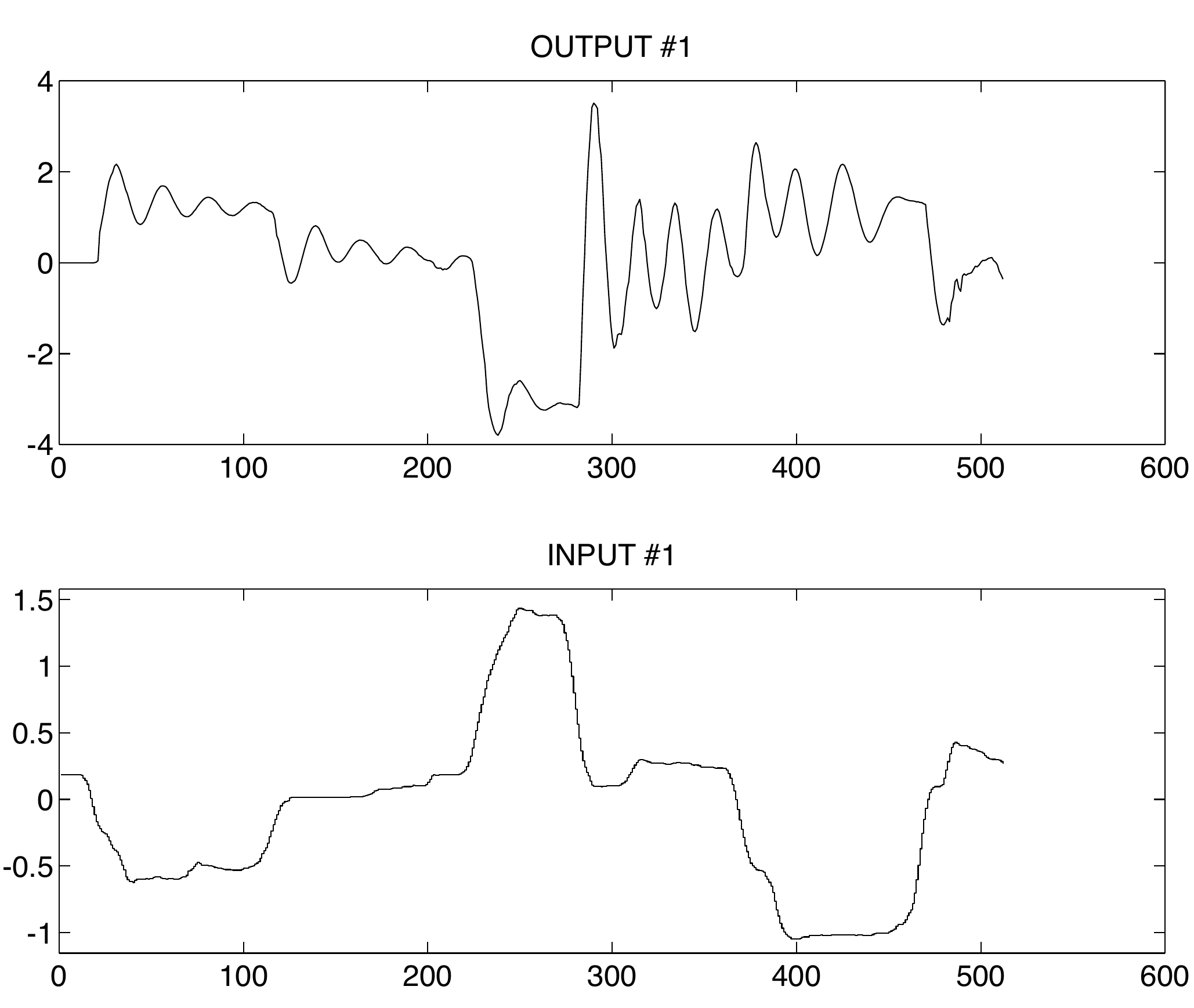}
  \caption{Data from the hydraulic crane. Input: Hydraulic
    Pressure. Output: Position of the tip.}
  \label{fig:hyddata}
\end{figure}

Some collected data from a particular crane are shown in Figure \ref{fig:hyddata}.
Clearly the dynamics is quite resonant.

\subsection{A Linear Model}
\begin{figure}
  \centering
   \includegraphics[scale=0.4]{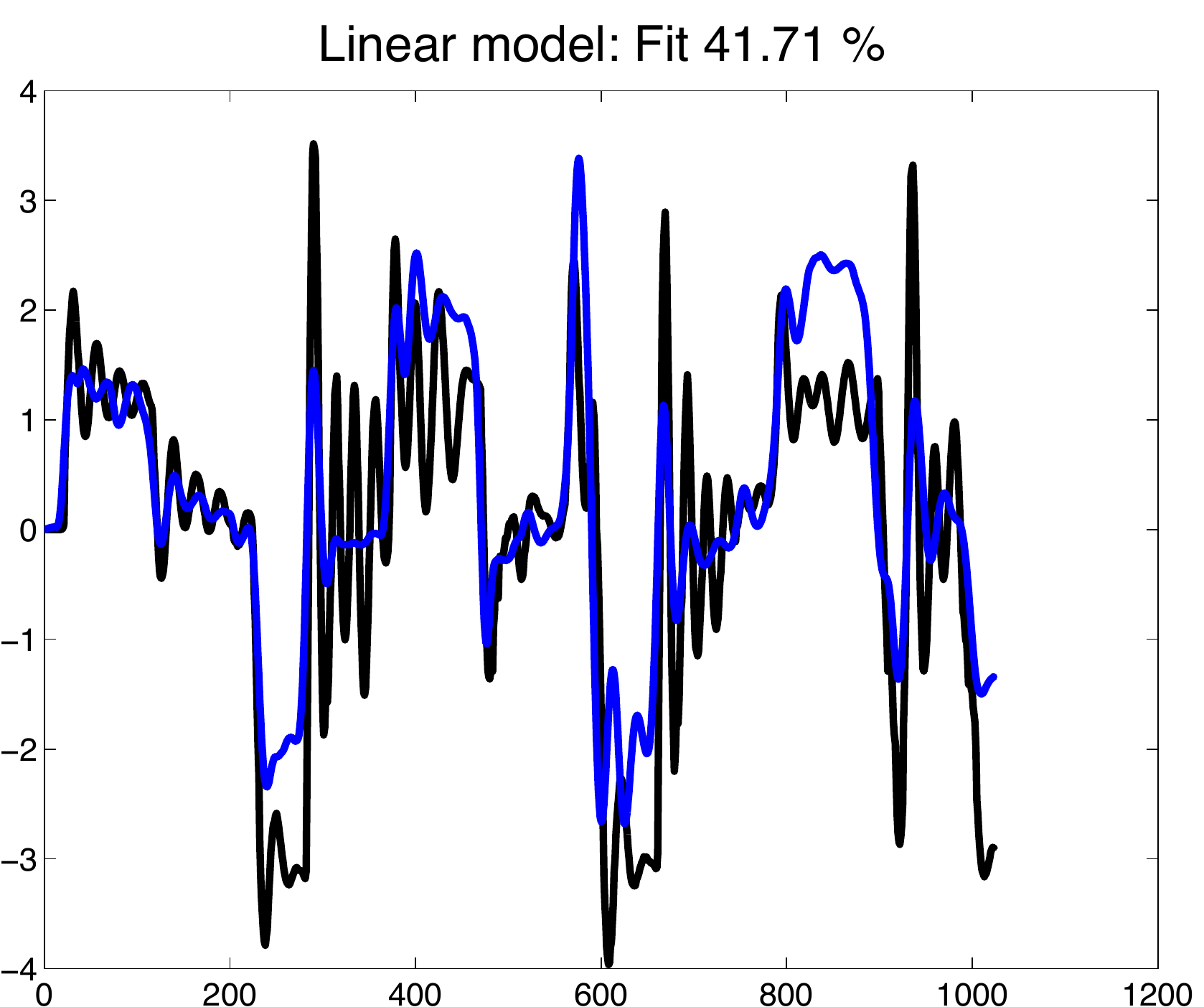}
  \caption{Comparison between the best linear model's simulated output (blue)
    and the measured output (black). The first half of the data was used for
    estimation.}
  \label{fig:hydrolin}
\end{figure}

To find  a model, 
first build a linear model of the crane using the data.  The first
half of the data sequence in Figure \ref{fig:hyddata} was used to
estimate the model and the second half was used to evaluate the fit
between the model's simulated output and the measured output. After some
experimentation the best model was obtained as a 5th order linear
state space model. The comparison between model and measured outputs
is shown in Figure \ref{fig:hydrolin}
 
This best fit for a linear model (42 \%) is not very  impressive, and not good enough for control design. 
\subsection{Nonlinear models: A hammerstein model}
\begin{figure}
  \centering
   \includegraphics[scale=0.4]{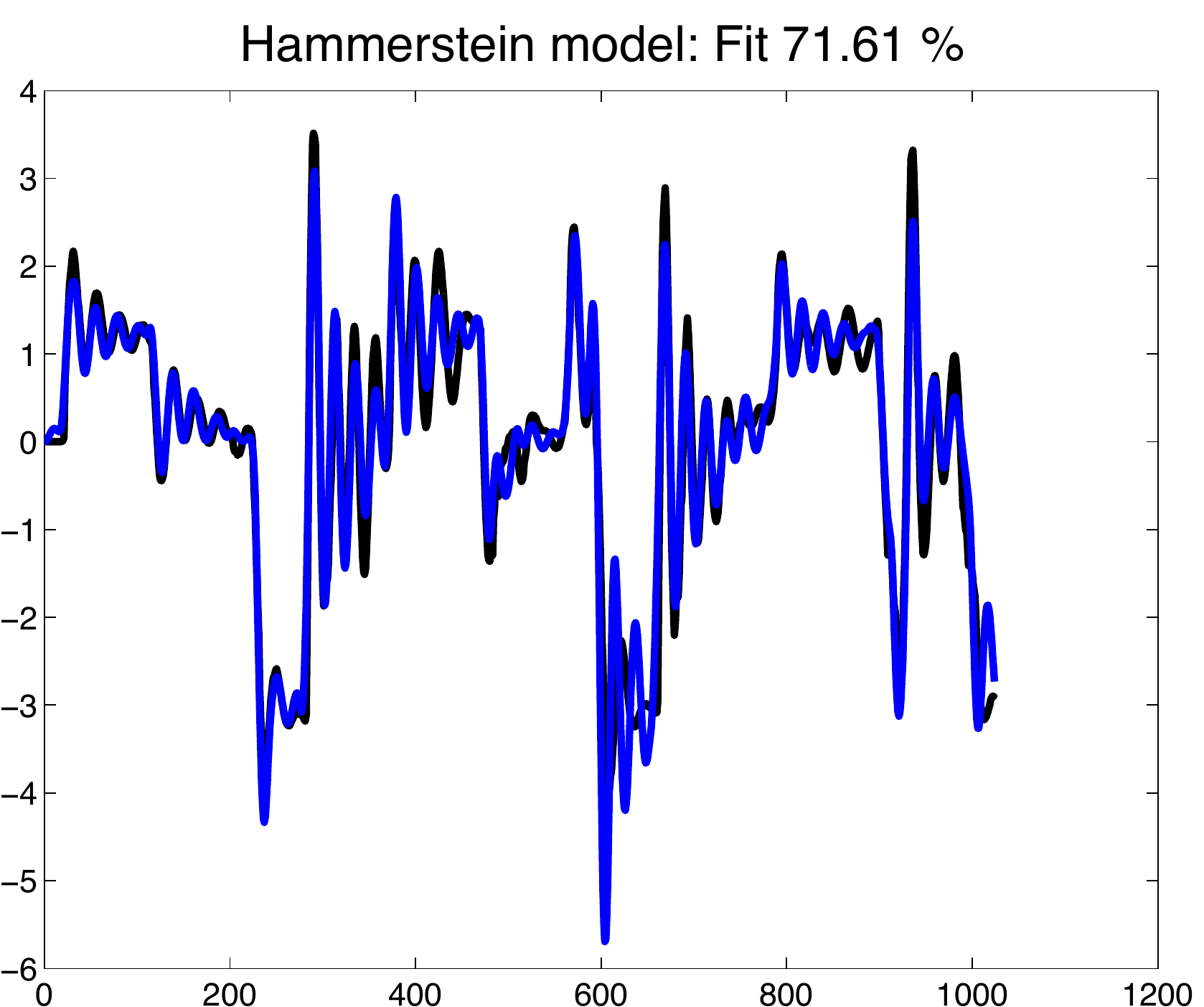}
  \caption{Same comparison as in Figure \ref{fig:hydrolin} but for a
    Hammerstein model.} 
  \label{fig:hydroham}
\end{figure}

\begin{figure}
  \centering
     \includegraphics[scale=0.4]{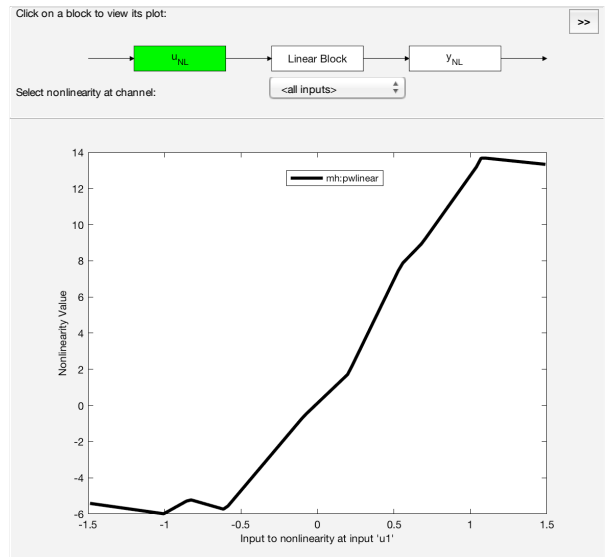}
  \caption{The estimated piecewise linear (10 breakpoints allowed) input nonlinearity.}
\label{fig:Mhnl}
\end{figure}

The lack of succss with linear models may indicate that there are nonlinear effects in the system. A simple test to see if nonlinearities can improve the fit is to try a Hammestein model, cf. Figure \ref{fig:EID1}. A Hammerstein model with a 5th order linear system, preceeded by a non-linear (piece-wise linear) static nonlinearity estimated from the first half of the data record  gave a model fit depicted in Figure \ref{fig:hydroham}. The estimated nonlineariry at the input is shown in Figure \ref{fig:Mhnl}


The impovement from 42 to 72\% fit with the input nonlinearity is
quite impressive! In retrospect a ``semiphysical'' explanation can be
given: Most of the resonance dynamics is due to the mechanical
contruction of the crane. The transformation from the hydraulic input
pressure to actual forces on the mechanical parts of the crane is more
complicated, and in that way a model with an unknown nonlinear static
transformation of the input acting on a linear system becomes
physically feasible. Note that the estimated input nonlinearity essentially is a saturation.
 
\section{Examples V(a): Black Box Volterra Model of the Brain}
\subsection{Volterra model of the brain}
In collaboration with the Department of Biomechanical Engineering of the Delft University of Technology, the Netherlands, a regularized Volterra model has been identified for a part of the human sensorimotor system, as explained in Figure \ref{fig:BrainSystem}. Detailed information on this experiment is reported in \cite{Vlaar2016, Vlaar2018}.

\subsection{The system}
\begin{figure}[h] 
\centering
\includegraphics[scale=0.45]{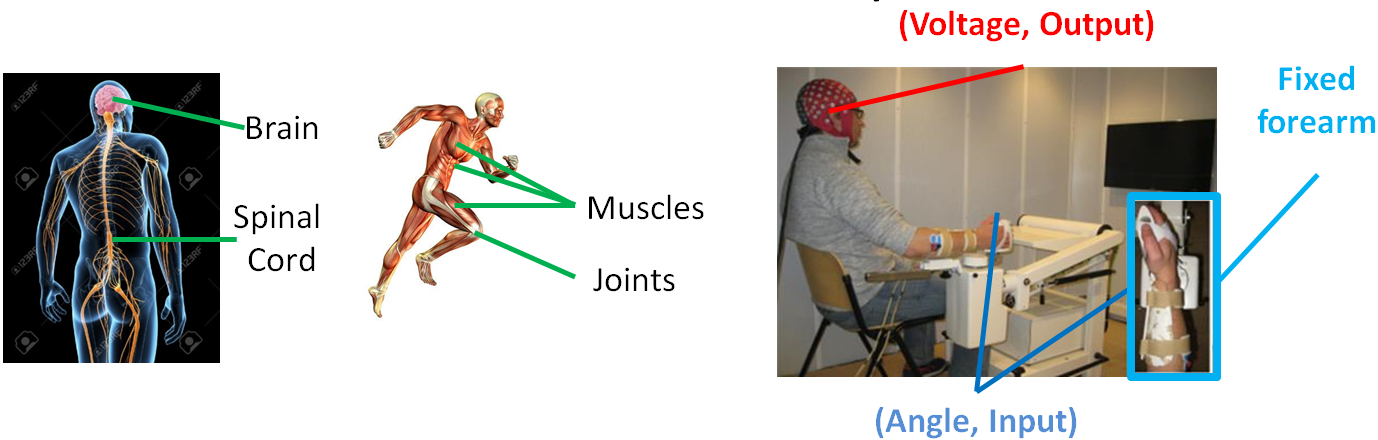}
\caption{The sensorimotor system refers to all the mechanisms in the human body that contribute to what we call human motion. The brain, the spinal cord as well as the muscles and the joints between the bones constitute the basic parts of the sensorimotor system. In this experiment, the relation between the wrist joint motion and the electroencephalography (EEG) signals measured at a selected position on the skull is modeled. The right forearm of the subject is strapped into an armrest and the right hand is strapped to the handle, requiring no hand force to hold the handle. The imposed wrist motion includes circular motion of the wrist around a fixed reference (zero angle). The angle of the wrist constitutes the input to the system while the EEG signal in the brain, as a reaction to this motion, is the system output. The measured output is contaminated with measurement noise while the input signal is assumed to be exactly known.}
\label{fig:BrainSystem}
\end{figure}

In the experiment, the relation between the wrist joint motion and the electroencephalography (EEG) signals is modeled. The experiment setup used to obtain the EEG data evoked as a reaction to the wrist perturbation is depicted in Figure \ref{fig:BrainSystem}. 

\subsection{Volterra models}
\begin{figure}[h] 
\centering
\includegraphics[scale=0.4]{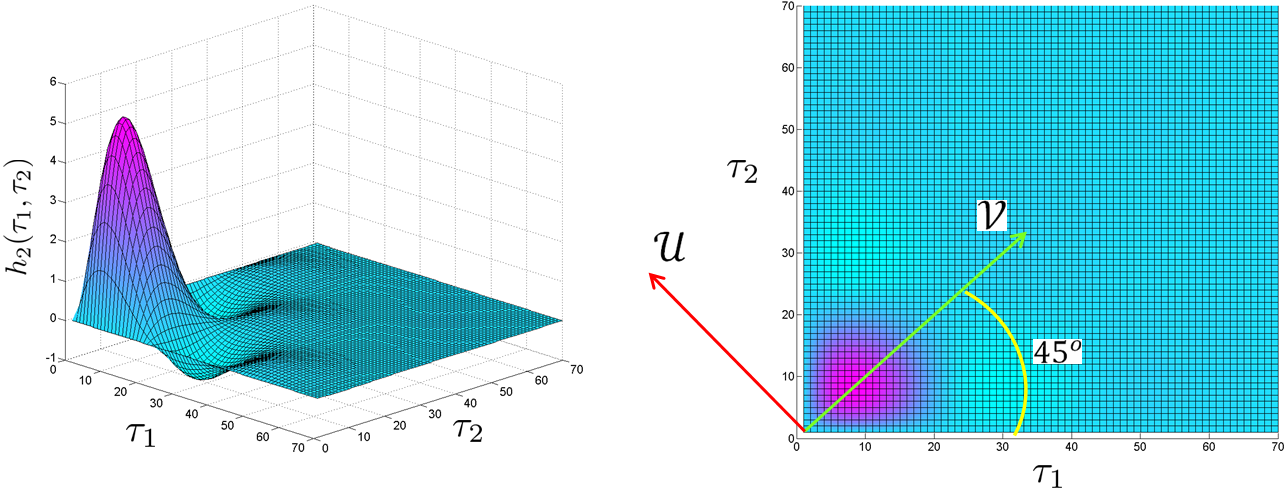}
\caption{Example of a second degree Volterra kernel with imposed smoothness and exponential decay in the $u$ and $v$ direction. The smoothness and decay are set by hyper parameters that are tuned during the optimization step by adding a regularization term to the cost function. This approach is generalized to $n$ dimensional kernels, using $n$ regularization directions \cite{Birpoutsoukis2017}.}
\label{fig:BrainVolterraRegularization}
\end{figure}

\emph{Regularized Volterra models} are a nonparametric representation of the system, using multidimensional impulse responses \eqref{eq:Volterra_u}, called Volterra kernels. Because the number of parameters grows very fast with the degree of the kernel, an additional constraint will be imposed on it to express that the kernel should be smooth in some directions, and decays exponentially to zero along these directions. This is illustrated in Figure \ref{fig:BrainVolterraRegularization} for a second degree kernel. By using constrained models, it is still possible to identify the kernels from relatively short data sets. This turned nonparametric Volterra modeling into a handy tool \cite{Birpoutsoukis2017}.

\emph{The choice of the maximum degree} in \eqref{eq:Volterra_u} is critical. Choosing it too high results in a very fast growing number of parameters to be estimated, choosing it too low creates large structural model errors. A prior nonparametric analysis \cite{SchoukensJ2016} showed that a linear model can capture only 10\% (in terms of Variance Accounted For, VAF) of the characteristics of the wrist joint - brain system while more than 70\% is attributed to even nonlinear behavior \cite{Vlaar2018}. On the basis of these results, a second degree Volterra model will be estimated, 
\begin{equation}
\label{eq:2ndDegreeModel}
\begin{aligned}
\hat y(t|\theta)  &= \sum_{\alpha=0}^2 y_0^\alpha(t)
\end{aligned}
\end{equation}
including a DC-offset ($\alpha=0$), a linear kernel ($\alpha=1$), and a quadratic kernel ($\alpha=2$).  The output $\hat y(t)$ is calculated by  direct evaluation of \eqref{eq:Volterra_u}

\emph{Memory length}: The average memory length of the selected models was 33 samples corresponding to approximately 130 ms at a sampling rate of 256 Hz. This choice resulted in the smallest structural model errors.

\subsection{The data}
 \emph{Experiment design}: The perturbation signals used were random phase multisines with a period of 1 s, resulting in a fundamental frequency $f_{0}=1$ Hz. Only odd harmonics of the fundamental frequency were excited, namely 1, 3, 5, 7, 9, 11, 13, 15, 19, and 23 Hz (odd random phase multisine). Exciting the nonlinear system using different phase realizations of a multisine signal (i.e. same amplitude per frequency, yet new random phases) allows for using different data for estimation and validation when modeling. Seven different multisine realizations were generated After transient removal, 210 periods are available for each of the seven realizations. The aforementioned choices for the duration of the excitation resulted in a total duration of clear perturbation equal to 24.5 minutes per subject. Including the extra time intervals corresponding to the preparation of the equipment, preparation of the subject as well as the pauses necessary for the safety and convenience of the subject, the total experiment time results to more than 2 hours!

\emph{Preprocessing of the data}: The EEG measured signals were high-pass filtered with a cut-off frequency of 1 Hz, and also the 50 Hz disturbance from the mains was removed. In a second step, the data were averaged over the periods, resulting in an SNR of about 20 dB.  For all participants the recorded output signals were shifted in time to impose a time delay of 20ms.

\subsection{Cost function}
The  model  parameters are  estimated  by minimizing a regularized least squares cost function \cite{Birpoutsoukis2017}, using a Baysian perspective \cite{Pillonetto2014} consisting of the sum of the squared differences between the averaged measured output and the modeled output $\hat y(t|\theta)$ \eqref{eq:2ndDegreeModel}, plus a regularization term (see also \eqref{eq:CostNLTDReg} in ``Black Box Models Complexity: Keeping the Exploding Number of Parameters under Control; Increased Structural Insight; Model Reduction''). 
\begin{equation} \label{eq:VolterraCost}
V=\frac{1}{N}\sum_{t=1}^{N}(y(t)-\hat y(t|\theta))^{2}+[\theta_{1}^{T} \theta_{2}^{T}]
\left[\begin{array}{cc}
P_{1}^{-1}&0\\
0&P_{2}^{-1}
\end{array}\right]
\left[\begin{array}{c}
\theta_{1}\\
\theta_{2}
\end{array}\right]
\end{equation}

The regularization matrix is a block diagonal matrix, where each block $P_{i}$ accounts for the regularization of the Volterra kernel of degree $i$. Due to this choice, and the selection of an odd excitation (only odd frequency components are present in the multisine), the identification of the linear part is completely decoupled of the even nonlinear part (the zero and second degree kernels). The hyperparameters in the regularization matrices $P_{i}$ are tuned using a marginalized maximum likelihood estimator \cite{Pillonetto2014}  which leads to a nonlinear numerical optimization in the hyperparameters (the model parameters $[\theta_{0} \theta_{1}^{T} \theta_{2}^{T}]]$ are eliminated in the marginalizing step). Once the hyperparameters are fixed, the remaining problem is linear in the model parameters $[\theta_{0} \theta_{1}^{T} \theta_{2}^{T}]]$ and are directly obtained by solving the linear least squares problem (\ref{eq:VolterraCost}).

\subsection{Results}
\begin{figure}[h] 
\centering
\includegraphics[scale=0.4]{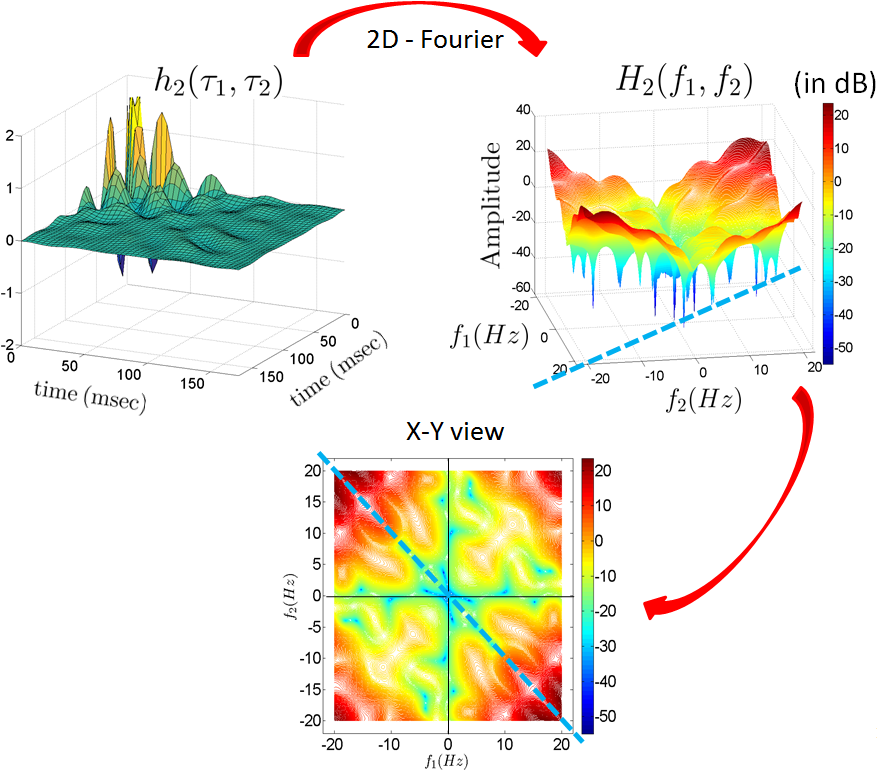}
\caption{The estimated second order Volterra kernel for the wrist angle – EEG system is shown up left. Up left: 2D Frequency Response Function (FRF) obtained by applying the 2D-Fourier transform to the 2D impulse response (Volterra kernel) is shown. Down, an X-Y view of the 2D FRF is shown. From these results it can be concluded that the system behaves as a high pass system (the lower frequencies are strongly attenuated) that transfers the intermediate high frequency power down to the low frequencies at the output (the kernel is mainly concentrated around the anti-diagonal).}
\label{fig:BrainResult}
\end{figure}

The results are shown in Figure \ref{fig:BrainResult}. Only the second degree kernel is shown. The linear part of the model captured about 10\% of the output (in terms of Variance Accounted For, VAF) while more than 70\% was attributed to even nonlinear behavior in the nonparametric analysis \cite{Vlaar2018}. The Volterra series, combined with the regularization technique described above was used to model the nonlinear system behavior. 

The obtained models were able to achieve on average 46\% of the VAF on validation datasets across different participants. Keeping in mind that 10\% of the output is due to noise, it indicates that about 44\% of the output variance remains unmodeled. In this case, a fourth order kernel would be needed to improve the results further on, but even with regularization this remains still an unattractive problem. Richer and longer experiments would be needed, and that is not feasible.  Although the second degree Volterra model still suffers from large structural model errors, it provides already very useful insights in the wrist-brain system.

In \cite{Birpoutsoukis2018} a new experiment was made using a richer excitation so that the risk for overfitting is further reduced. Instead of exciting only the odd frequencies, also the even frequencies were excited. At that moment, the estimation of the linear part can no longer be decoupled from the nonlinear part of the model. With this rich data set, the VAF for increases to almost 60\%, but the qualitative conclusions remained the same. 

\newpage 
\section{Examples V(b):  Nonlinear State Space Black Box Model\\  a Li-Ion Battery}

Lithium-ion batteries are attracting significant and growing interest, because their high energy and high power
density render them an excellent option for energy storage, particularly in hybrid and electric vehicles. In this section, a nonlinear state-space model is proposed for the operating points at the cusp of linear and nonlinear
regimes of the battery's short-term electrical operation. This point is selected on the basis of a nonparametric distortion analysis as a function of state-of-charge (SOC) and temperature. More detailed information is available in  \cite{Relan2017} for a fixed SOC (10\%) and temperature model (25\textdegree{}C). In \cite{Relan2017b}, a nonlinear state space model is developed that covers a varying temperature (from 5\textdegree{}C to 40\textdegree{}C) and SOC (2\% to 10\%). For higher SOC values, it follows from the nonparametric distortion analysis that a linear model can be used.

\subsection{The system and the experimental setup}
A high energy density Li-ion polymer battery [EIG-ePLB-C020, Li(NiCoMn)] with the following electrical characteristics: the nominal voltage of 3.65 V, the nominal capacity of 20 Ah, and ac impedance (1 kHz) $< 3$  m$\Omega$ along with the PEC battery tester SBT0550 with 24 channels is used for the data acquisition. The tests are performed on a preconditioned battery inside a temperature controlled chamber at 25\textdegree{}C.

\subsection{Nonlinear state space model}
A discrete time nonlinear state space model (\ref{eq:DTNLSS}) is selected to approximate the continuous time system (see "Approximating A Continuous Time NLSS With A Discrete Time NLSS Model"). The NLSS in \eqref{eq:bbgen} is rewritten as
\begin{equation}\label{eq:BatteryNLSS}
\begin{aligned}
x(k+1) & =  \tilde{F}(x(k),u(k))=Ax(k)+Bu(k)+F(x(k),u(k))\\
y(k) & =  \tilde{G}(x(k),u(t))=Cx(k)+Du(k)+G(x(k),u(t)).
\end{aligned}
\end{equation}
Although the split between the linear part and the nonlinear terms $F(x(k),u(k))$, and $G(x(k),u(t))$ in (\ref{eq:BatteryNLSS}) is not unique, it is convenient to write the equations like that because the initialization procedure that will be presented below starts from the best linear approximation of the nonlinear system. It results in initial estimates for $A,B,C,D$.

 The nonlinear terms $F(x(k),u(k))$, and $G(x(k),u(t))$ are multivariate nonlinear functions. These will be written as a linear combination of nonlinear basis functions. The whole palette of possibilities can be used here, ranging from polynomials, hinge functions, to Gaussian processes. In this example, a polynomial representation will be used.

\subsection{The data}
An odd-random phase multisine signal is used as an input excitation signal. The band of excitation is kept between 1 and 5 Hz, because the dynamic range of interest of the battery for hybrid and  electric vehicles applications is covered well within this band of excitation. It also takes into consideration the limitations of the battery tester in terms of the sampling frequency. The excitation signal has a period of 5000 samples, and the sample frequency $f_{s}$ is set to 50 Hz resulting in a frequency resolution of $f_{0}=50/5000=0.01$ Hz. The input is zero mean with an rms value of 20A. A detailed description of the whole measurement procedure (charging, discharging, $\ldots$) is given in \cite{Relan2017}.

\subsection{Cost function}
The cost function is formulated in the frequency domain on the difference between  $Y(k)$ and $Y_{mod}(k,\theta)$ that are the discrete Fourier coefficients of the measured and modeled output.
\begin{equation}
\label{eq:BatteryCost}
V=\frac{1}{F}\sum_{k\in B}^{N}\frac{|Y(k)-\hat Y(k| \theta)|^{2}}{W(k)},
\end{equation}
where $B$ is the set of frequencies of interest, and $W(k)$ a user selected weighting. Because structural model errors dominate, the weighting $W(k)$ is put equal to 1 (see "Impact Of Structural Model Errors").
\subsection{Results}

\subsubsection{Nonparametric distortion analysis}
\begin{figure}[h] 
\centering
\includegraphics[scale=0.8]{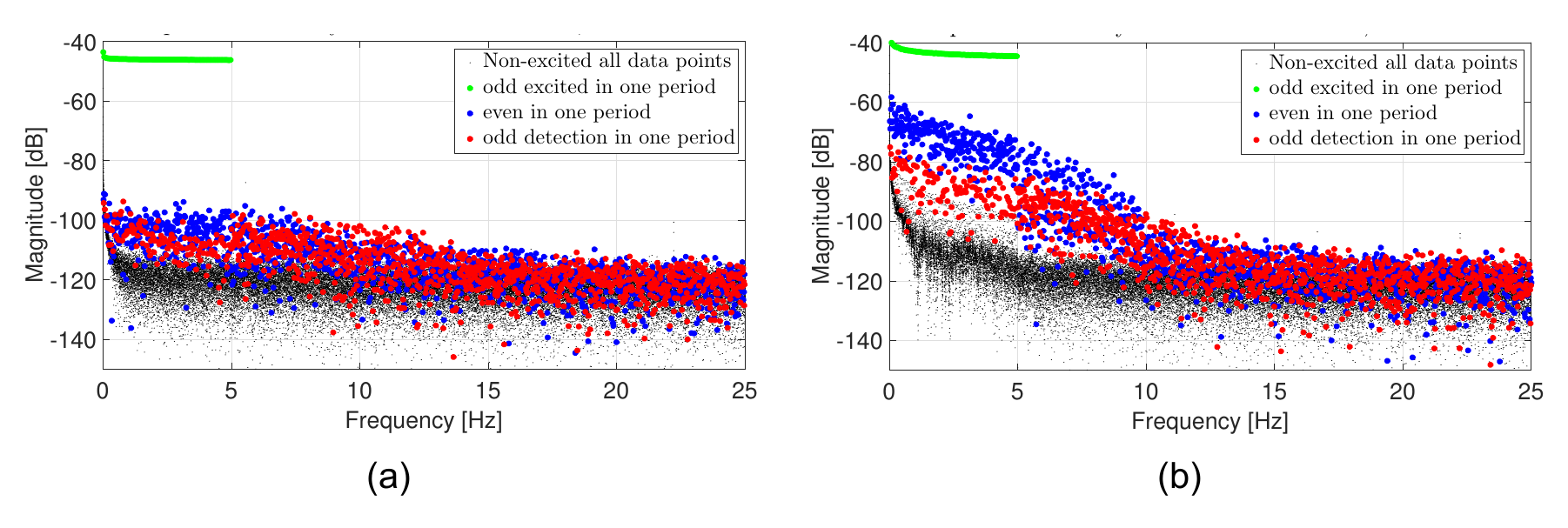}
\caption{Nonlinear distortion analys for a SOC of 90\% (a) and 10\% (b) at a temperature of 25\textdegree{}C. The SNR of the measurements is very good. The output at the excited frequencies (green) is more than 60dB about the noise floor (black). At 90\% SOC, the nonlinear distortions are at -50 dB (less than 1\%). At 10\% SOC, the even nonlinear distortions (blue) are well above the odd distortions (red), at 10\% (20 dB) of the output.}
\label{fig:BatteryNLdist}
\end{figure}

At the start of the identification process, a nonlinear distortion analysis (see ``Nonparametric Noise and Distortion Analysis Using Periodic Excitations'') provides a lot of insight about the quality of the data, and the possible gain that can be made with a nonlinear model. In this case, it is seen in Figure \ref{fig:BatteryNLdist} that the SNR of the data is very high (above 50 dB). The nonlinear distortions are very low for a high SOC, while they increase above 10\% for low SOC (10\%). More results for other temperatures, SOC, and amplitude ranges are available in \cite{Relan2017b}. On the basis of these results, it was decided to focus the nonlinear modeling effort on the low SOC range.

\subsubsection{Parametric nonlinear state space model}
The nonlinear state space model is identified in two steps. In the initialization step, the best nonlinear approximation is identified (using one of the classical linear identification methods). The noise weighting $W(k)$ in \eqref{eq:BatteryCost} during this step is set by the nonlinear distortions. These results are used to find initial values for the $A,B,C,D$ matrices. In the second step the nonlinear terms are added to the model. These are linear-in-the-parameters and initialized at zero. Eventually a nonlinear numerical optimization is used to minimize the cost function with respect to all the parameters (see \cite{Relan2017} for more details).  In this and the next steps, $W(k)=1$ because model erros dominate.

In this case, a polynomial nonlinear state space model is used with three internal states, and multivariate polynomials up to degree 3.

A major problem that often shows up in many applications is the appearance of instabilities. Even if the identified model remained stable on the test data, it happened frequently that the model still becomes unstable on the validation data. This is mainly due to the polynomial basis functions that were used in this example. Outside the state space domain covered by the test data, the polynomials have the tendency to grow very fast, which results in instabilities of the model. 

\begin{figure}[h] 
\centering
\includegraphics[scale=0.8]{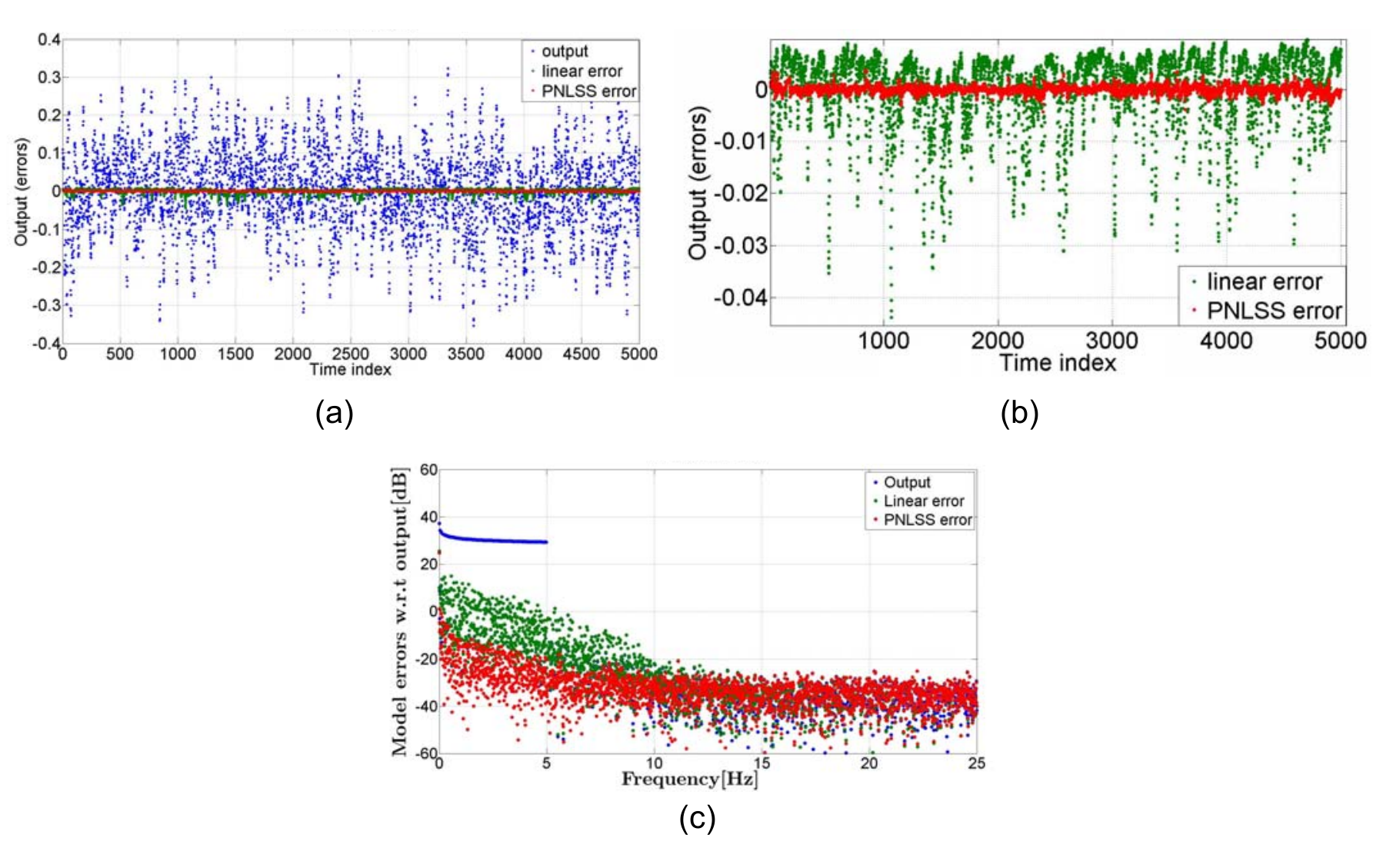}
\caption{Simulation error of the best linear (green) and the NLSS (red) model on a validation set. Figure (b) shows a zoom of the results in (a). Observe that the error is strongly asymmetric distributed, which is well in agreement with the presence of dominating even nonlinearities. This results in large spikes for the linear model that are completely removed by the nonlinear model. In (c) the same results are shown in the frequency domain. Although the errors of the NLSS are 20 dB below those of the linear model, they are still far above the noise floor that was measured in the nonparametric noise analysis (see Figure \ref{fig:BatteryNLdist}), so that in this case the model errors dominate.}
\label{fig:BatteryErrorTDFD}
\end{figure}

The results are discussed in Figure \ref{fig:BatteryErrorTDFD}. The nonlinear model outperforms the linear model with a factor 10 on the validation data, but the errors are still far above the noise floor.

\emph{Remark: Nonlinear state space models that include output feedback}: It is also possible to replace the states $x(k)$ by a well selected set of outputs $y(k)$ as the input to the multivariate nonlinear functions, leading back to the full expression \eqref{eq:NLSSuy}. Although such a representation can always be reduced to that in (\ref{eq:BatteryNLSS}), it can be much more compact. A typical application are vibrating mechanical systems where the nonlinearities depend often on the displacement of the structure close to the nonlinearities which can be selected to be the states of the nonlinear equations \cite{Noel2015}. This choice also simplifies the initialization of the identification problem, because in these applications the states are directly measurable outputs.
  
 \section{Conclusions}

In this article, a full overview of the nonlinear system identification process is given. Nonlinear SI is a very rich topic with many different aspects. The selection of the topics, and the organization of the discussion is strongly influenced by the authors' experiences. The discussion is aligned along the following topics: 1) Is a nonlinear modeling approach needed, or is it still possible to address the users questions using a linear design; 2) The lead actors of the SI process: experiment design, model structure selection, choice of the cost function, validation; 3) Illustration of the wide variety of methods on a series of experiments. The sidebars provide more detailed technical background information on some issues. 

The main conclusions are formulated as a set of guidelines and summaries. More detailed guidelines are provided throughout the article.

\emph{Guidelines - Summary}
 \begin{itemize}
 \item \emph{Linear or nonlinear SI?} The need for nonlinear modeling starts where linear modeling fails to solve the problems.  Nonlinear SI is more involved than linear SI because models require higher flexibility, and the presence of structural model errors is difficult to avoid. Tools are available to check the level of nonlinearity in a nonparametric preprocessing step, allowing the reader to make a well informed decision.
 \item\emph{ The main actors of the nonlinear SI process}
  \begin{itemize}
 \item \emph{Experiment design}
  \begin{itemize}
  \item Use periodic excitations whenever it is possible.
  \item Cover the domain of interest to keep structural model errors under control. Covering the amplitude range and the frequency band of interest are necessary but not sufficient conditions to guarantee that no internal extrapolation will show up in later use of the model.
  \item The experiment should be informative to keep the noise induced uncertainty low.
   \end{itemize}
    
    \item \emph{Selection of a model structure}: The choice of the model class is  driven by i) the user preference (white box - black box models), ii) the system behavior  (open loop or closed loop NL system), iii) Models for simulation or prediction.
   \begin{itemize}
  \item User choice: Decide how much physical insight will be injected. The Palette provides an overview of models ranging from white to black box modeling
  \item System behavior: Fading memory (nonlinear open loop models: the nonlinearity is not captured in a dynamic loop) are much easier to identify than nonlinear closed loop models where the nonlinearity is part of a dynamic closed loop. However dealing with complex behaviors like shifting resonances, changing damping, or even chaotic behavior requires nonlinear closed loop model structures.
  \item The model complexity and effort is strongly affected  by the later use of the model: prediction or simulation. The development of a good prediction model is  less demanding than obtaining a good simulation model.
   \end{itemize}
  
   \item \emph{Choice of the cost function} 
   \begin{itemize}
   \item Do structural model errors dominate?
   \begin{itemize}
     \item Yes: the choice of the cost function is set by user criteria to shape the model errors.
     \item No: the choice of the cost function is set by the noise properties.
     \end{itemize}
  \item Regularization  is a very powerful tool that can help to keep the model complexity under control. It can be used in black box models to create either sparse models, or to impose smooth solutions.
   \end{itemize}
   
  \item Validation: does the model meet the user needs? Are the errors small enough in the domain of interest? Is there information left in the data?
   \begin{itemize}
   \item Test the model on new data and check the internal domain
  \item Check the linear validation criteria: Whiteness test of the output residuals, and cross-correlation test between the input and output residuals?
  \item Nonlinear validation criteria: for example higher order moment tests.
   \end{itemize}
   \end{itemize}
   \item \emph{Uncertainty bounds}: If structural model errors are present (the model does not pass the validation test, but it is good enough for the intended application), no reliable theoretical uncertainty bounds are available. The variability of the model should be obtained from repeated experiments with varying excitations.
 
 \end{itemize}
  
\setcounter{equation}{0}
\renewcommand{\theequation}{S\arabic{equation}}
\setcounter{figure}{0}
\renewcommand{\thefigure}{S\arabic{figure}}

\section{Sidebar\\Static Nonlinearities}
A static nonlinearity is the basic building block in nonlinear
models. Consider the mapping
\begin{align}
  \label{eq:snl}
  z =H(x);\quad z\in R^{p} ,    \; x \in R^m
\end{align}
If $z$ and $x$ are time varying signals in the model, the mapping is
applied for each $t$ : $x(t)=H(x(t))$, so the mapping is
\emph{static}. Such  a mapping can occur in many contexts in a
nonlinear model, and is indeed what constitutes the nonlinear
behavior. See also (\ref{eq:genstr}). In a state-space model, $H$ can
be mapping from the state at time $t$, to the state at the next time
instant, \eqref{eq:bb1}. In a regression model like (\ref{eq:NLARX}),
$H$ can be the mapping from the regressors to the model output. In
block oriented models, Figure \ref{fig:EID1}, the static nonlinearity
block is a fundamental component. In this sidebar, the
parameterization of static nonlinearities, $H(x,\theta)$, will be discussed.
\subsection{SISO nonlinearities}
 Consider the simple SISO case  ($p,m=1$) which  brings
out all the essential ideas. This is the case of describing a one-dimensional curve.
\subsubsection{Breakpoint-based: Piecewise Constant and Piecewise
  Linear}
A very simple idea to parameterize a curve is to define its values at
a number of \emph{breakpoints}, $\{x_1,\ldots, M\}$, so 
\begin{align}
  \label{eq:brkp}
  \theta = \{x_k, z_k=H(x_k), k=1,\ldots M\}
\end{align}
would be the parameterization of the curve. An interpolation rule must
then be applied to define $H(x)$ at the intermediate points:
\begin{itemize} 
\item \emph{Piecewise constant} : the value $H(x)$ equals $H(x_k)$ where
$x_k$ is the breakpoint immediately to the left of $x$.
\item \emph{Piecewise linear}: The value $H(x)$ is linearly
  interpolated from its two neighbouring breakpoints.
\end{itemize}
Clearly, more sophisticated interpolation rules could also be used.

\subsubsection{Basis function Expansions}
A common way to parameterize functions is to choose a system of
\emph{basis functions}, $\{\rho_k(x), k = 1,\ldots, M\}$ and parameterize
the curve as
\begin{align}
  \label{eq:basisf}
  H(x,\theta) = \sum_{k=1}^M \theta_k \rho_k(x,\theta)
\end{align}
Where the basis functions $\rho$ may or may not depend on $\theta$.
Very many basis function expansions are possible, and a few will be
reviewed here:
\paragraph{\textbf{Custom Regressors}}
With some physical insights, -- or ``semiphysical modelling'' --, the user can come up with specially chosen basis functions $\rho_k^c(x)$ that reflect typical nonlinearities for the application in question, e.g. using $\rho(x) = \sqrt{x}$ if $z=H(x)$ describes the output in a free level flow system on $x$ is the level in the flw system. This gives the expansion
\begin{align}
  \label{eq:cust}
   H(x,\theta) = \sum_{k=1}^M \theta_k \rho_k^c(x)
\end{align}
\paragraph{Polynomial Expansion}
The most common  ``black box'' expansion is the polynomial expansion ($\rho_k(x)=x^k$), the Taylor
series
\begin{align}
  \label{eq:poly1}
  H(x,\theta)  = \sum_{k=1}^M \theta_k x^k
\end{align}
Such polynomial expansions are in modelling contexts where $x$ consists delayed inputs, also known as a Volterra model, see (\ref{eq:Volterra_u}). (See also ``External or internal nonlinear dynamics: Volterra Models'').
\paragraph{Linear Regressions}
Note that if the basis functions $\rho_k$ are known (do not depend on the parameter $\theta$, as in the previous two subsection)  then (\ref{eq:basisf}) is a linear regression, so it is easy to estimate $\theta$ from measurement of $z$ and $x$
\paragraph{Local Basis Functions}
It is also possible to construct local basis functions, ''pulses'' that
are nonzero only over certain intervals:
\begin{align}
  \label{eq:local}
  \rho_k(x,\gamma_k,\beta_k) =
  \begin{cases}
    1 \; \text{if} \;  \gamma_k \le x < \gamma_k+1/\beta_k\\
0 \; \text{else}
  \end{cases}
\end{align}
Then, if $\gamma_{k+1}=\gamma_k+1/\beta_k$ the expansion
\begin{align}
  \label{eq:loc1}
  H(x,\theta) = \sum_{k=1}^M \alpha_k \rho_k(x,\gamma_k,\beta_k)
\end{align}
will be a piecewise constant function, like (\ref{eq:brkp}) with
breakpoints $\{\gamma_k\}$. It can approximate any reasonable function
arbitrarily well with sufficiently large $M$. Note that
\begin{align}
  \label{eq:loc2}
   \rho_k(x,\gamma_k,\beta_k) = \kappa(\beta_k(x-\gamma_k))
\end{align}
where $\kappa(\cdot)$ is the unit pulse indicator, 1 for $0\le x < 1$
and zero elsewhere.
\paragraph {Neural Networks}
Inspired by the approximation capability of (\ref{eq:loc1},\ref{eq:loc2}) a
  ``mother function'' $\kappa(x)$ (also known as \emph{activation
  function}) can be selected,  translated it by $\gamma_k$ and dilated it by $\beta_k$ and
  the expansion
\begin{align}
  \label{eq:nn1}
  H(x,\theta) = \sum_{k=1}^M \alpha_k \kappa(\beta_k(x-\gamma_k))
\end{align} 
can be be defined.
Common activation functions are the Gaussian Bell (a ``soft pulse'')
\begin{align}
  \label{eq:gauss}
  \kappa(x) = e^{-x^2}
\end{align}
and the sigmoid function (a ``soft step'')
\begin{align}
  \label{eq:sigm}
  \kappa(x) = \frac{1}{1 + e^{-x}}
\end{align}
Another common choice of the activation function is $\kappa(x)$ = ReLU(x) 
(Rectified Linear Unit) 
\begin{align*}
  \kappa(x) =
  \begin{cases}
    0 & \text{if} \: x<0\\
x &  \text{if} \: x\ge  0
  \end{cases}
\end{align*}
this makes $H$ piecewise linear  and continuous in $x$.

The expansion (\ref{eq:nn1}) represents the simplest \emph{neural networks}.
\paragraph{Trees}
A useful mapping $z=H(x)$ can be defined via (binary) trees. Such
trees are popular  mappings and used in e.g. decision trees and
classification trees. Here a regression tree will be defined. Loosely
speaking, the value $x$ will be subjected to a number of binary
questions at different nodes. All the binary answers will be collected
and based on those a value $z=H(x)$ will be assigned. 

More specifically, at the root node, for a scalar $x$ the
basic question is $x\ge c_0?$ for some number $c_0$.
Depending on the answer a new question will be asked at the next level
 $x\ge c_1^r?, r=1,2$. And so it continues, so at level $k$ there are
$2^k$ nodes with similar questions $x\ge c_k^r?, r=1,...,2^k$. For a
tree with depth $n$ the final level is $k=n$ with $2^n$ nodes, called
leaves. They correspond to a partition of the $x$-axis into $2^n$
parts.
 Each leaf has a value $d_k$ for $H(x)$ at the $x$-value that leads to that
 point.
[Some branches may be cut before level $n$, then the leaves are
defined at a lower level.] So the tree defines a function  $H(x)$
which is piecewise constant - all values of $x$ that belong to the
same partition give the same value $d_k$. So a tree is an alternative
to piecewise constant functions (\ref{eq:brkp}) where the breakpoints
 are defined in a somewhat contrived way be the tree partitioning. For
 a regression tree it is customary to add an interpolation step to
 compute the function value:
 \begin{align}
   \label{eq:treeint}
   H(x) = d_k + xL_k
 \end{align}
where $d_k$ is the value produced by the leaf of the tree and $L_k$ is a scalar
also provided by the leaf in question. In that way the tree $H(x)$ becomes a
\emph{piecewise linear function} of $x$.
\paragraph{Gaussian Processes}
Gaussian processes (GPs) offer a nonparametric approach to model  $z=H(x)$   in \eqref{eq:snl} that can be put in a Bayesian framework \cite{Liu2018}. This static model can then be used to form dynamic  models as in  $f(x(t),u(t),\theta)$ and $h(x(t),\theta)$ in \eqref{eq:NLSSu} or NARX models and other examples in this sidebar.

The idea is
that function to be estimated is embedded in a stochastic framework so
  that $H(x)$ is seen as a \emph{realization of a (Gaussian) stochastic
  process.}
 
So there is a  \emph{prior}  (before any measurements)
distribution with a (zero) \emph{mean}, (large) \emph{variance} and \emph{covariance
function} that describes the smoothness of $H$. 

 When $z_k=H(x_k),\; k=1,\ldots N$ have been measured,
the \emph{posterior} distribution $H^p(x|z)$ can be formed for any
$x$. The mean of that function will  be the estimate of the function
$H$ and is formed by \emph{interpolation and extrapolation $[z_k,x_k]$ using
the probabilistic relationships}.  The reliability of the estimate can
also be assessed from the \emph{posterior variance}. [But is of course a
reflection of the prior assigned distributions.]

If all variables are jointly Gaussian all this can be  computed by
\emph{simple and efficient linear algebraic expressions.}

\begin{figure}
  \centering
\includegraphics[scale=0.3]{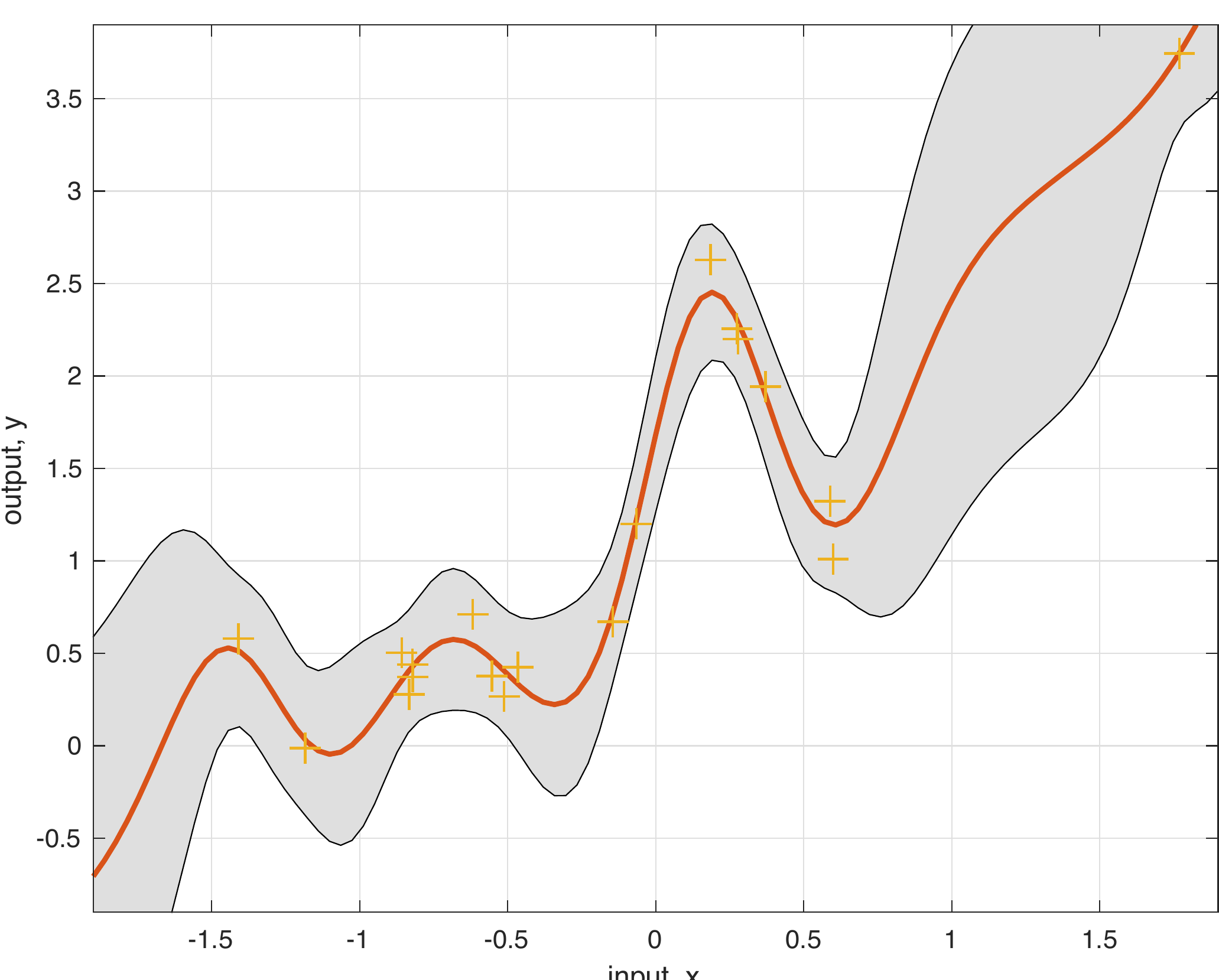}
  \caption{Yellow crosses: the observed $x$ and $y$ values\\
Red line: The estimate of the $f$ function\\
Shaded region: Reliability: the standard deviation around the mean $f$.}
\label{fig:GPex}
\end{figure}

The idea is depicted in figure \ref{fig:GPex}.
The framework can be seen as a generalization of the regularization approach \cite{Pillonetto2014}, see the example \eqref{eq:VolterraCost} and the discussion in ``Black Box Models Complexity''. 
The reader is referred to \cite{RasmussenW:2006} for a full introduction to the topic. A recent tutorial to apply GPs to modeling of dynamic systems is given in \cite{Liu2018}.

\subsection{MISO nonlinearities}
When the  nonlinearity is multi input ($m>1$) $x$ will be an
$m$-vector. Most of the ideas for SISO models carry over to the MISO
case. Some specific aspects will be noted here.
\subsubsection{Polynomial Expansion}
In the polynomial expansion (\ref{eq:poly1}) the term $\theta_k x^k$
should be interpreted as the sum of all those terms that can be
created from exponentials of the components $x_i^{r_i}$ whose
exponents sum up to $k$. So each term $x^k$ expands into several new terms. Each of these terms require its own
parameter, so the number of parameters in (\ref{eq:poly1}) rapidly increases with $m$ and $M$. 
A common special case of polynomial expansion is the Volterra model.

\subsubsection{{Volterra models}}
A Volterra system \cite{Schetzen1980} is a NFIR system with a multivariate polynomial nonlinearity. Let the regressors be $\varphi(t)=[u(t),u(t-1),u(t-m+1)]$. Then the model is
\begin{equation}
\label{eq:Volterra_u}
\begin{aligned}
\hat y(t|\theta)  &= \sum_{\alpha=1}^M y_0^\alpha(t)\\
y_0^\alpha(t) &= \sum_{r_1=0}^{m-1}\ldots \sum_{\tau_M=0}^{m-1}g_\alpha(\tau_1,\ldots,\tau_M)u(t-\tau_1)u(t-\tau_2)\cdots u(t-\tau_M)
\end{aligned}
\end{equation}
The kernel $g_{\alpha}(\tau_{1},\ldots,\tau_{\alpha})$ is the multidimensional
impulse response of degree $\alpha$ and it correspond to the model parameters $\theta_k$.

The number of terms in the Volterra models is $O(m^M)$, and it becomes very large when the memory length grows. For that reason, Volterra models were only applied to problems with short memory length $m$ and moderate polynomial orders $M$. Only recently, more complex problems could be handled \cite{Birpoutsoukis2017, Vlaar2018} by using the regularization framework \cite{Pillonetto2014} to reduce the impact of the exploding number of parameters (see ``Black box models complexity: Keeping the exploding number of parameters under control; Increased structural insight; Model reduction'').
\subsubsection{Neural Networks}
For the neural network model (\ref{eq:nn1}) it is customary to stick
to the activation function $\kappa(x)$ with a scalar argument and
rather reinterpret the argument with vector $x$ in either of two ways
\begin{itemize}
\item \textbf{Ridge construction} Let $\beta_k$ be an $m$-vector and
  interpret
$\kappa(\beta_k(x-\gamma_k))$ as $\kappa(\beta_k^Tx-\gamma_k)$. That
means that the basis function assumes the same value for all $x$ in
the hyperplane $\beta_k^Tx = constant$, thus creating a ridge
structure for the function values. With a sigmoid activation function
this leads to the celebrated \emph{one hidden layer feedforward
  sigmoid neural net}.
\item \textbf{Radial construction}. Let $\gamma_k$ be an $m$-vector of
  translation and let $\beta_k$ by a positive definite
  matrix of dilation coefficients (or a scaled version of the identity
  matrix) and interpret $\kappa(\beta_k(x-\gamma_k))$ as
  $\kappa(\|x-\gamma_k\|_{\beta_k})$, where
  $\|x\|^2_{\beta_k}=x^T\beta_kx$ is a quadratic norm defined by
  $\beta_k$. This means that the contribution  of the basis function
  only depends on the distance between $x$ and a given centerpoint,
  thus giving a certain radial symmetry of the function. With the
  Gaussian activation function, this gives the well known \emph{radial
    basis neural network}.
\end{itemize}
\subsubsection{Trees}
With an $m$-vector $x$, the questions asked at the nodes will be $[1
x^T]^TC_k^r >0?$ for $m+1$ vectors $C_k^r$. The linear interpolation
term $L_k$ in (\ref{eq:treeint}) will also be a $m$-vector provided by
the leaf in question.
\subsection{MIMO nonlinearities}
Turning to the Multioutput case, $p>1$  there is a common and simple
solution: \emph{Treat it as $p$ independent MISO cases!} If models are
built from data, the actual models for the different outputs may come
up in different shapes - and different model orders. Grey box thinking
and/or experimental evidence may suggest that the same parameters
may be used in different output channels, so as to get a more
efficient representation. Some thinking of this kind is treated in the
Sidebar ``Decoupling of Multivariate Polynomials.'' But this does not
change the picture that the essential features of MIMO representation
of static nonlinearities are captured when MISO models are
constructed.


\section{Sidebar\\ Nonparametric Noise and Distortion Analysis Using Periodic Excitations}

In this section, tools will be presented that allow the user to detect and analyze the presence of nonlinear distortions during the initial
tests. Using a well designed periodic excitation \eqref{eq:Multisine}, the frequency response function of the BLA (see ``Linear Models of Nonlinear Systems''), the power spectrum of the disturbing noise, and the level of the nonlinear distortions will be obtained from a nonparametric analysis without any user interaction. The user can set the desired
frequency resolution and the desired power spectrum of the excitation signal. The phase will be chosen randomly on $[0,2\pi)$. In this article, only a brief introduction is given, the reader is referred to \cite{Pintelon2012, Schoukens2016} for a more extensive discussion, and to \cite{Pintelon2013} for measurements under nonlinear closed loop conditions.

\subsubsection{The response of a nonlinear system to a periodic excitation\label{sub:Response NL system to periodic signal}}
A linear time-invariant system cannot transfer power from one frequency to another while a nonlinear system does. Consider a nonlinear system $y=u^{\alpha}$, excited at the frequencies $\pm\omega_{k},k=1,\ldots,F.$ The frequencies
at the output of such a system are given by making all possible combinations of $\alpha$ frequencies, including repeated frequencies, selected
from the set of $2F$ excited frequencies (see also Figure \ref{fig:NLdetectSine} for an illustration on a sine excitation with frequencies \{-1,1\}) \cite{Pintelon2012}:
\begin{equation}
\sum_{i=1}^{\alpha}\omega_{k_{i}},\textrm{ with}\;\omega_{k_{i}}\in\{-\omega_{F},\ldots,-\omega_{1},\omega_{1},\ldots,\omega_{F}\}.\label{eq:NLS output frequencies}
\end{equation}
Using Volterra series models \eqref{eq:Volterra_u} \cite{Schetzen2006}, this result can be generalized to dynamic fading-memory systems that include also discontinuous nonlinear systems \cite{Boyd1985, Pintelon2012, Schoukens2012}. Chaotic systems are excluded from this study, because these have no periodic output for a periodic input.

In the next section, the nonlinear distortions will be detected using a multisine excitation where some amplitudes $U_{k}$ in \eqref{eq:Multisine}
are put equal to zero for a well-selected set of frequencies.

\subsubsection{Detection and characterization of the nonlinear distortions}
\begin{figure}
\includegraphics[scale=0.6]{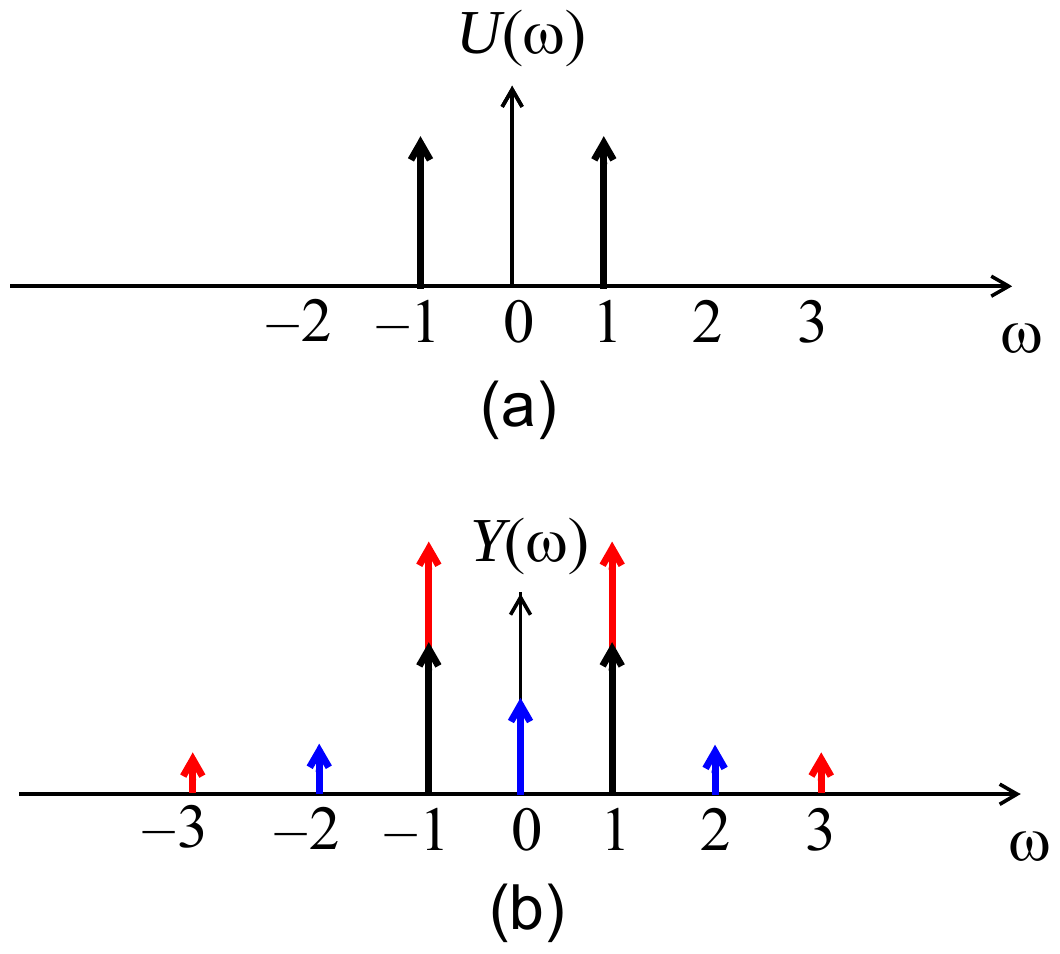}
\centering
\caption{Detection of a nonlinear behavior using a sine excitation. The input spectrum (a) shows two spikes at the postive and negative excitation frequency. The output spectrum (b) shows the linear (black), quadratic (blue), and cubic (red) contributions of the output. At frequency 2 and 3, the presence of even and odd nonlinearities is detected.}
\label{fig:NLdetectSine}
\end{figure}

\emph{Sine test}: The most simple nonlinearity test is a sine test. As shown in Figure \ref{fig:NLdetectSine}, nonlinear operations, like $x^2$ or $x^3$ create harmonic components \eqref{eq:NLS output frequencies}  that can be detected at the output of the system and reveal the nonlinear behavior of the system.  This result can be generalized to dynamic nonlinear systems. The sine test method is very popular, for example, in mechanical engineering. To speedup the measurement, the sine is replaced by a sweeping sine \cite{Noel2014, SchoukensJ2016}. A sine test is not very robust because the higher harmonics that indicate the presence of nonlinear behavior can be amplified or attenuated by the linear dynamics of the system. In highly resonating or bandpass systems, the presence of nonlinearities can be strongly underestimated. For that reason, more robust tests were developed using multisine excitations \eqref{eq:Multisine}.

\begin{figure}

\includegraphics[scale=0.6]{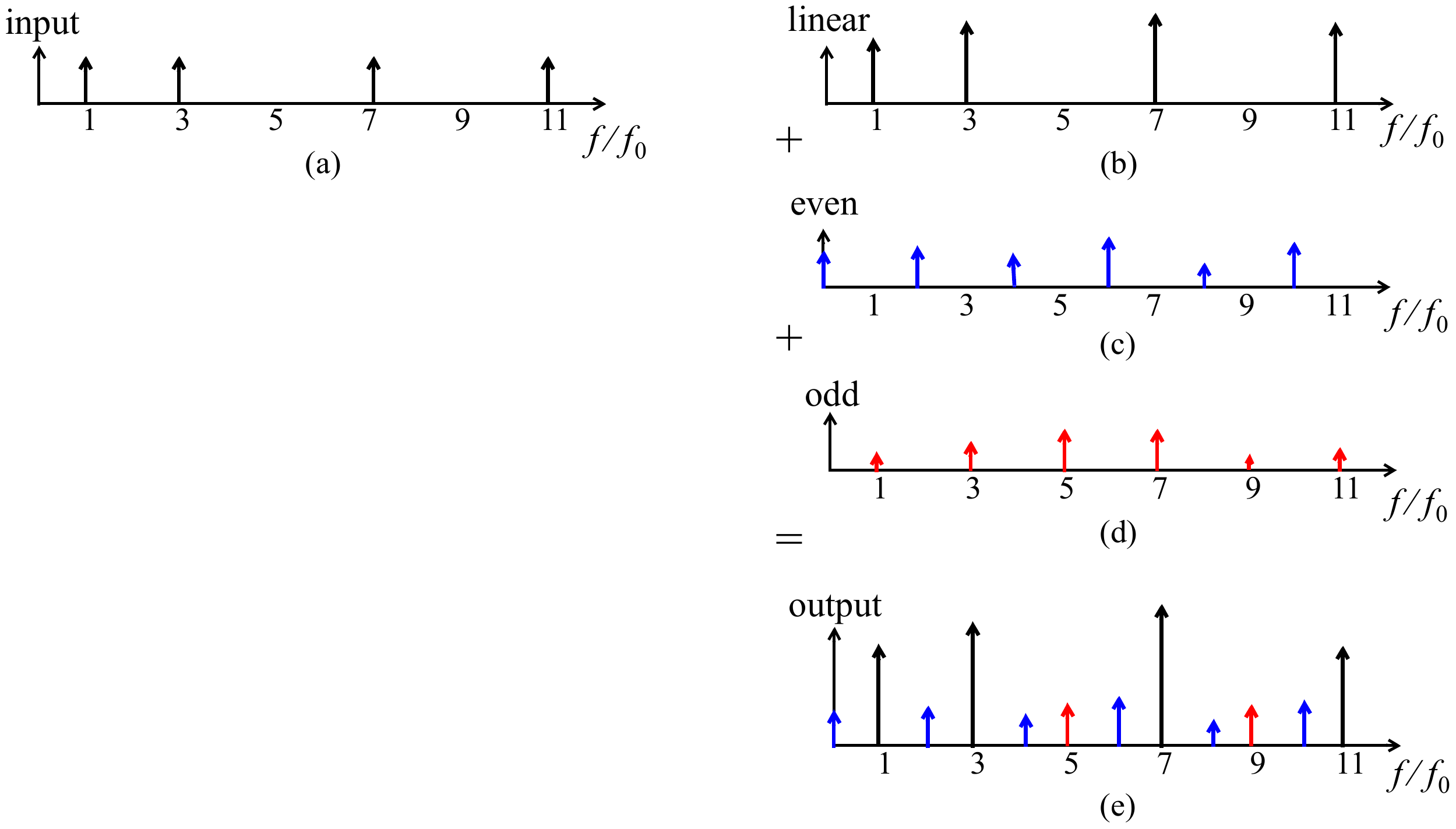}

\caption{Design of a multisine excitation for a nonlinear analysis. (a): Selection of the excited frequencies at the input (left side); At the output (right side), from top to bottom: linear (b), even (c), odd (d) contributions, and total output (e). For simplicity, only the positive frequencies are shown.}
\label{fig:Multisine NonLin detection}
\end{figure}

\emph{Multisine test}: Using well designed multisines, the nonlinear sine test is robustified (more reliable estimate of the nonlinear level) and speeded up \cite{SchoukensJ2016}. The basic idea, illustrated in Figure \ref{fig:Multisine NonLin detection}, is very simple and starts from a multisine \eqref{eq:Multisine} that excites a well-selected
set of odd frequencies (odd frequencies correspond to odd values of $k$ in \eqref{eq:Multisine}). This excitation signal
is applied to the nonlinear system under test. Even nonlinearities show up at the even frequencies because an even number of odd frequencies
is added together. Odd nonlinearities are present only at the odd frequencies because an odd number of odd frequencies is added together.
At the odd frequencies that are not excited at the input, the odd nonlinear distortions become visible at the output because the linear
part of the model does not contribute to the output at these frequencies (for example, frequencies 5 and 9 in Figure \ref{fig:Multisine NonLin detection}). By using a different color for each of these contributions, it becomes easy to recognize these in an amplitude spectrum plot of the output
signal.

\subsubsection{Disturbing noise characterization}
\begin{figure}
\centering
\includegraphics[scale=0.55]{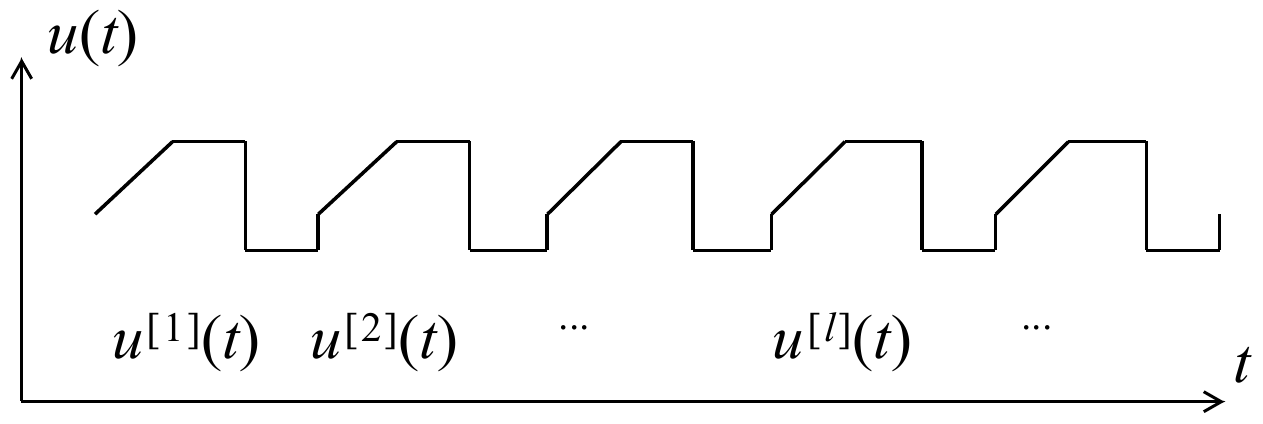}
\caption{Calculation of the sample mean and variance of a periodic signal.}
\label{fig:Periodic Signal Variance Mean}
\end{figure}

In the next step, the disturbing noise analysis is made. By analyzing the variations of the periodic input and output signals over the measurements
of the repeated periods, the sample mean and the sample (co-)variance of the input and the output disturbing noise is calculated, as
a function of the frequency. Although the disturbing noise varies from one period to the other, the nonlinear distortions do not, so
they remain exactly the same. This results eventually in the following simple procedure: consider the periodic signal $u(t)$ in Figure
\ref{fig:Periodic Signal Variance Mean}. The periodic signal is measured
over $P$ periods. For each subrecord, corresponding to a period, the discrete Fourier transform is calculated using the fast Fourier
transform (FFT) algorithm, resulting in the FFT spectra of each period $U^{[l]}(k),Y^{[l]}(k),$ for $l=1,\ldots,P$. 
The sample means $\hat{U}(k),\hat{Y}(k)$ and noise (co)variances
$\hat{\sigma}_{U}^{2}(k),\hat{\sigma}_{Y}^{2}(k),\hat{\sigma}_{YU}^{2}(k)$
at frequency $k$ are then given by
\[
\hat{U}(k)=\frac{1}{P}\sum_{l=1}^{P}U^{[l]}(k)\quad\hat{Y}(k)=\frac{1}{P}\sum_{l=1}^{P}Y^{[l]}(k),
\]

and 

\begin{equation}
\begin{array}{c}
\hat{\sigma}_{U}^{2}(k)=\frac{1}{P-1}\sum_{l=1}^{P}|U^{[l]}(k)-\hat{U}(k)|^{2},\\
\hat{\sigma}_{Y}^{2}(k)=\frac{1}{P-1}\sum_{l=1}^{P}|Y^{[l]}(k)-\hat{Y}(k)(k)|^{2},\\
\hat{\sigma}_{YU}^{2}(k)=\frac{1}{P-1}\sum_{l=1}^{P}(Y(k)-\hat{Y}(k))(U(k)-\hat{U}(k))^{H}.
\end{array}\label{eq:Sample variances}
\end{equation}
In \eqref{eq:Sample variances}, $(.)^{H}$ denotes the complex conjugate. The variance of the estimated mean values $\hat{U}(k)$ and $\hat{Y}(k)$ is $\hat{\sigma}_{U}^{2}(k)/P$ and $\hat{\sigma}_{Y}^{2}(k)/P,$ respectively. Adding together all this information in one figure
results in a full nonparametric analysis of the system with information about the system (the FRF), the even and odd nonlinear distortions,
and the power spectrum of the disturbing noise, as shown in Figure \ref{fig:SilverboxNLdistortions} for the forced Duffing oscillator.

\section{Sidebar\\ Simulation Errors and Prediction Errors}
There are two basic uses of a model: \emph{Simulation} and \emph{Prediction}.

Simulation means that the model is subject to an input sequence $U^t=[u(1), u(2), \ldots, u(t)]$ and its response to that input is computed. Such applications are important to evaluate the system's behaviour under new situtations without having to do actual experiments.

Prediction means that an observed input-ouput sequence $Z^{t}=[u(1),y(1),u(2),y(2),\ldots, u(t),y(t)]$ is given and a prediction of the next output $\hat y(t+1|t)$ is sought. (Or outputs $k$ steps ahead $\hat y(t+k|t)$, in which case also tentative future inputs $u(t+1),\ldots,u(t+k)$ should be supplied.) Such applications are important for system prediction, but primarily for control design - control of a system subject to disturbances can be seen as control of the predicted output. 

In (\ref{eq:pred}) it was stated that a model is essentially a predictor for the output. The model is a \emph{simulation model} if the prediction only depends on past inputs, so it will be focused on the simulation task. It is a \emph{prediction model} if the prediction also depends on past outputs.

\begin{figure}
\centering
\includegraphics[scale=0.55]{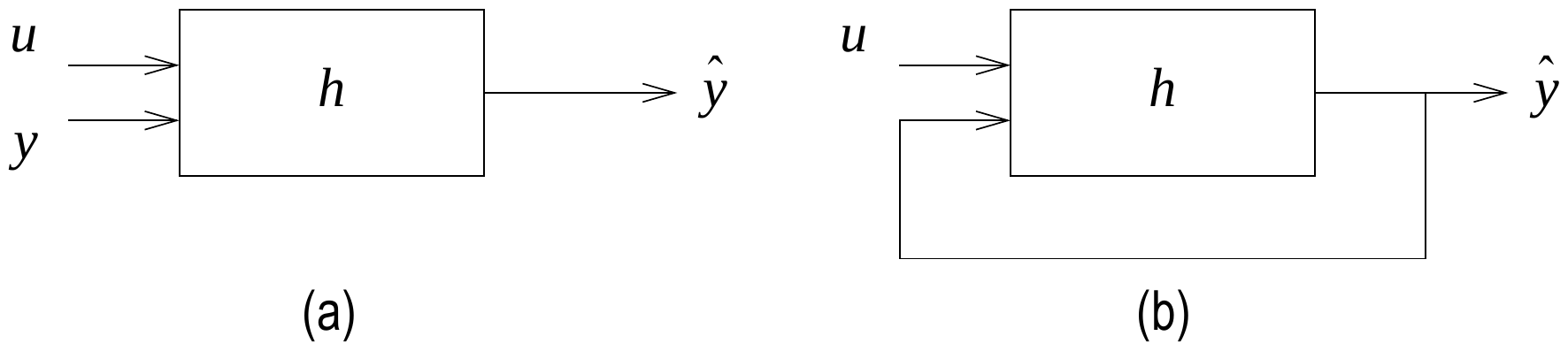}
\caption{A prediction model (a) starts from the measured inputs $u^{t-1}$ and outputs $y^{t-1}$ to estimate the output $y(t)$, while in a simulation model (b), the measured output $y^{t-1}$ is replaced by the estimated output $\hat{y}^{t-1}$.}
\label{fig:SimulationPrediction}
\end{figure}

Note that a prediction model $h(Z^t)$ (omit the parameter vector $\theta$ for simplicity) can also be used as a simuation model (simulating future outputs without access to past outputs), simply y replacing, recursively, the outputs in the pedictor by predicted outputs:
Let
\begin{equation}
\begin{aligned}
  \label{eq:predsim}
Z_0^{t}=[u(1),y_0(1),u(2),y_0(2),\ldots, u(t),y_0(t)]\\
\text{where}\\
 y_0(s) = h(Z_0^{s-1}),\; s=1,2...
\end{aligned}
\end{equation}
This is depicted in Figure \ref{fig:SimulationPrediction}.

What can be said about the error created by simulation and prediction models, respectively?
Let $h_0^p$ be the true prediction description of a system (based on the correct statisical properties. Let $h_0^s(U^t)$ be the simulation model created from this using (\ref{eq:predsim}). Then $\hat y_0(t)=h_0^s(U^t)$ will be the noise-free output response to the input $U^t$, and
\begin{align}
  \label{eq:truesim}
  v(t)=y(t)-\hat y_0(t)
\end{align}
will be the true output error disturbance.

If $v$ happens to be white noise, the true model structure is  a simulation model and $\hat y_0(t)$ are indeed the optimal predictions.


Otherwise, the disturbance $v(t)$ has a smooth or correlated behavior  so that its future values can be partly predicted from its past. This property is used to improve the quality of the prediction.  The past values of $v(t)$ can be estimated as the difference between the past model and measured output values $\hat{v}(t)=y(t)-\hat{y}_0(t)$. This is the intrinsic idea that is used in the development of optimal prediction models. 


Whether it is worth while to further develop an improved predictor or not depends on the nature of the disturbance $v(t)$.
\begin{itemize}
\item $v(t)$ \emph{is dominated by measurement or sensor noise}: sensor or measurement noise are not related at all with the process of interest. In that case the main goal is the elimination of this disturbance so that $y_{0}(t)$ is the signal of interest, and hence a simulation model is the natural choice.
\item $v(t)$ \emph{is dominated by process noise}: process noise is an intrinsic part of the system output. It models that part of the system output that is due to inputs that are not known to the user. In control applications good predictions are important
 so that the noise disturbance needs also to be included in the model. Moreover, the past outputs are available to decide on the next control action which turns the prediction model into a natural tool for these applications.
\item $v(t)$ \emph{is dominated by structural model errors}: If the model set is not rich enough to capture the true system, structural model errors appear (See ``Impact of Structural Model Errors'' which can also be represented as a 'disturbance'.
 
A predictor can use past outputs to be informed about structural errors and find a better predictor than a simulation model. To take a trivial example: Consider a simple process $y(t)=y(t-1)+u(t)$ which ignores a structural error in the true system: $y(t)=y(t-1)+u(t)+ C$ ($C$ ignored in the model). The predictor $y(t)=y(t-1)+u(t)$ will have an error $C$, while the simulated output cannot avoid  an error $Ct$.

 This is one of the reasons why prediction is an `easier' task than simulation,  and why it is more demanding to get a small simulation error than a small prediction error.

\end{itemize}

\section{Sidebar\\ Extensive Case Study: The Forced Duffing Oscilator}

Throughout this article many results are illustrated on the forced Duffin oscillator, sometimes called the Silverbox in nonlinear benchmark studies.  This section describes the setup and the experiments that are used throughout this article. A detailed description of the experiments is given in \cite{SchoukensJ2016}.

\subsection{System, experimental setup, experiments}
\begin{figure}[h] 
\centering
\includegraphics[scale=0.70]{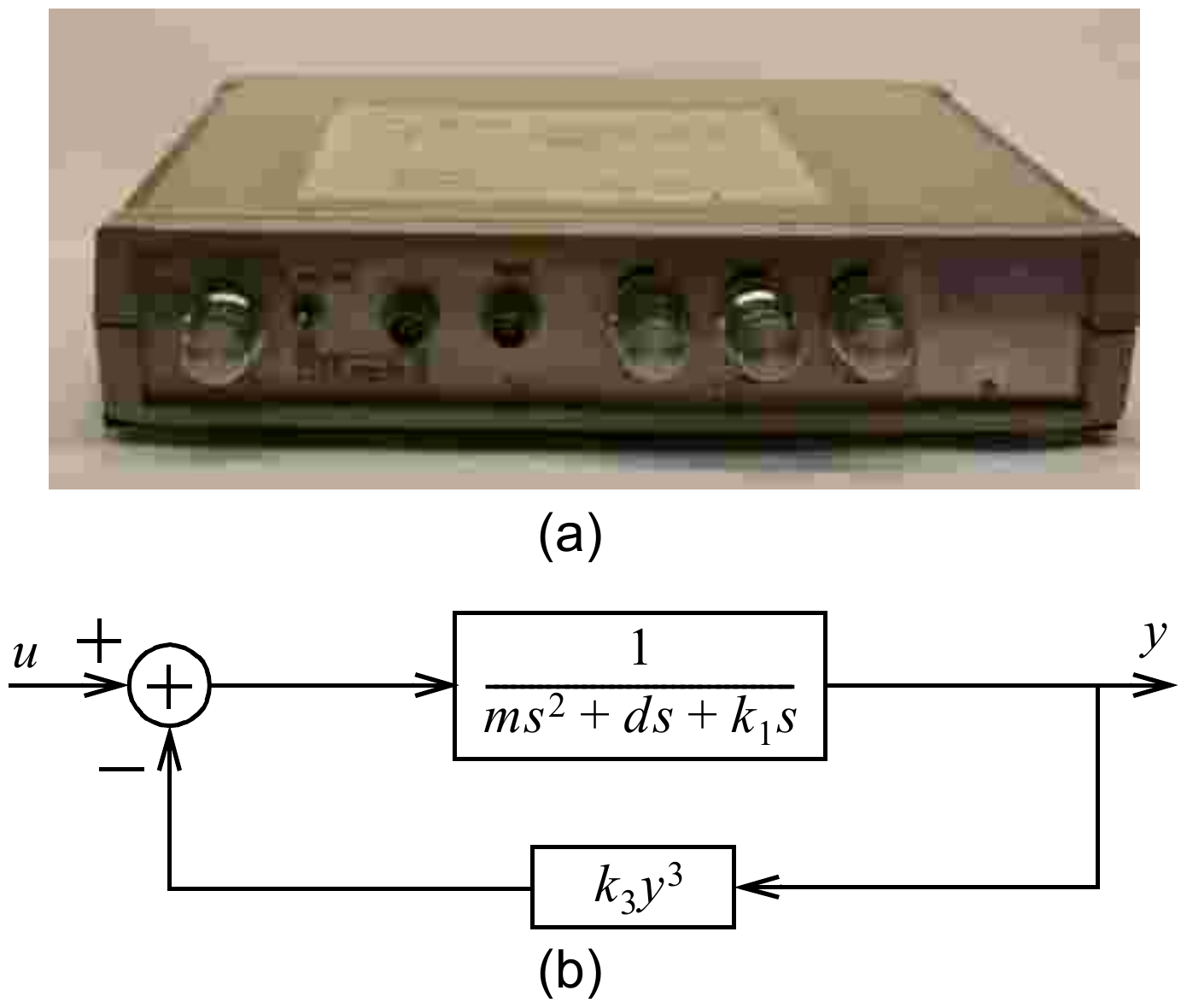}
\caption{Forced Duffing oscillator (a): The electronic circuit mimics a nonlinear mechanical system with a hardening spring. Such a system is sometimes called a forced Duffing oscillator. The system is excited with an input $u(t)$  (the applied force to the mechanical system). The output of the system corresponds to the displacement $y(t)$. The schematic representation of the system is given in (b) as a second-order system with a nonlinear feedback.}
\label{fig:SB1}
\end{figure}

\emph{The system} is an electronic circuit that mimics a nonlinear mechanical system with a cubic hardening spring as shown in Figure \ref{fig:SB1}. This class of nonlinear systems has a very rich behavior, including regular and chaotic motions, and the generation of sub-harmonics \cite{Thompson1986,Ueda1991}. 

\emph{The experimental setup} consists of a generator and two data acquisition cards that are  synchronized to avoid leakage errors in the spectral analysis. The generator starts from the ZOH (zero-order-hold) reconstruction \cite{SchoukensJ1994} $u_{\text{ZOH}}$ of a discrete time sequence $u_{d}(k)$ that is passed through a lowpass generator filter $G_{\text{gen}}$ to eliminate the higher harmonics of the ZOH reconstruction, $u(t)=G_{\text{gen}}u_{\text{ZOH}}$. The sampling frequency is $f_{s}=10 \text{MHz}/2^{14}\approx610\text{Hz}$.  The data acquisition cards are alias protected (the signals are passed through a lowpass filter before sampling) and sample the input and output at a rate $f_{s}$. High impedant buffers are used to eliminate the interaction between the plant and the measurement setup.

\begin{figure}
\centering
\includegraphics[scale=0.70]{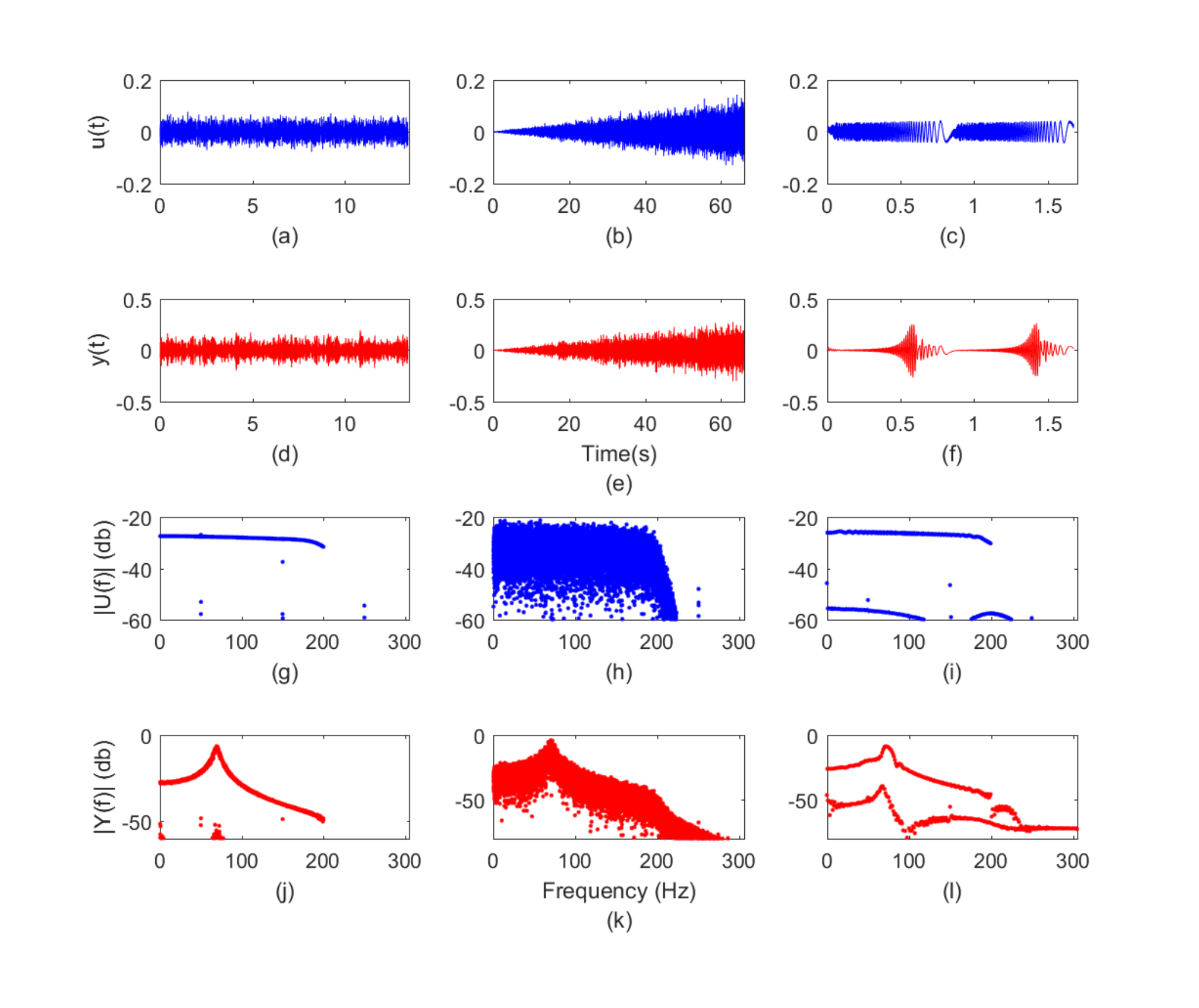}
\caption{Experiments on the Duffing oscillator for a random phase multisine \eqref{eq:Multisine} (a,d, g, j), growing noise (b,e,h,k), and a 'sweeping sine' (c,f,i,l) experiment. The  input signals are shown in blue, the output signals in red. The two top lines show the time domain signals, and the two lines at the bottom show the amplitude spectra.}
\label{fig:SBsignals}
\end{figure}

\emph{The experiments} are shown in Figure \ref{fig:SBsignals} and consist of three parts using a 'tail' (a), 'arrow' (b), and 'sweeping sine' (c) input $u(t)$. The output (d),(e),(f) of the circuit corresponds to the displacement $y(t)$. The following observations can be made
\begin{itemize}
\item The tail (a) and swept sine excitation (c)  have about the same RMS value. The arrow signal (b), that has  a maximum amplitude that is twice that of the other signals, will be used to verify the extrapolation capabilities of the models that are identified on the other excitations. The swept sine excitation is a Schroeder multisine \cite{Pintelon2012} that has a dominant odd behavior because the amplitude of the excited odd frequencies is about 30 dB dB above the level of the excited even frequencies (see (f),(l)).
\item All input signals have the same bandwidth (see Figure \ref{fig:SBsignals} (g),(h),(i)).
\item  Observe that the output (f) for the sweeping sine (c) becomes very large around the resonance frequency, even if the input level remains constant. This results in an internal extrapolation for models that are identified on the tail (a).
\item In the input spectrum (g), it can be seen that there are spurious components at the odd multiples of the mains frequency (50 Hz). The signal is picked up by the circuit and acts as process noise.
\item The nonparametric distortion analysis in Figure \ref{fig:SilverboxNLdistortions} shows that he SNR of these measurements is 40 to 60 dB (measurement and process noise at the 1\% to 0.1\% level).
\end{itemize}

\subsection{Linear simulation and prediction of the forced Duffing oscillator}
The behavior of simulation and prediction errors (see ``Simulation Errors and Prediction Errors'') is illustrated on  the forced Duffing oscillator.   Because the SNR of these measurements is very high (noise disturbances well below 1\%),  the simulation and prediction errors in the study below are completely dominated by structural model errors. The simulation and prediction errors will be shown for linear Box-Jenkins and ARX models \cite{Ljung1987, Soderstrom1989}, and a nonlinear NARX model. 

\begin{figure}[h] 
\centering
\includegraphics[scale=0.70]{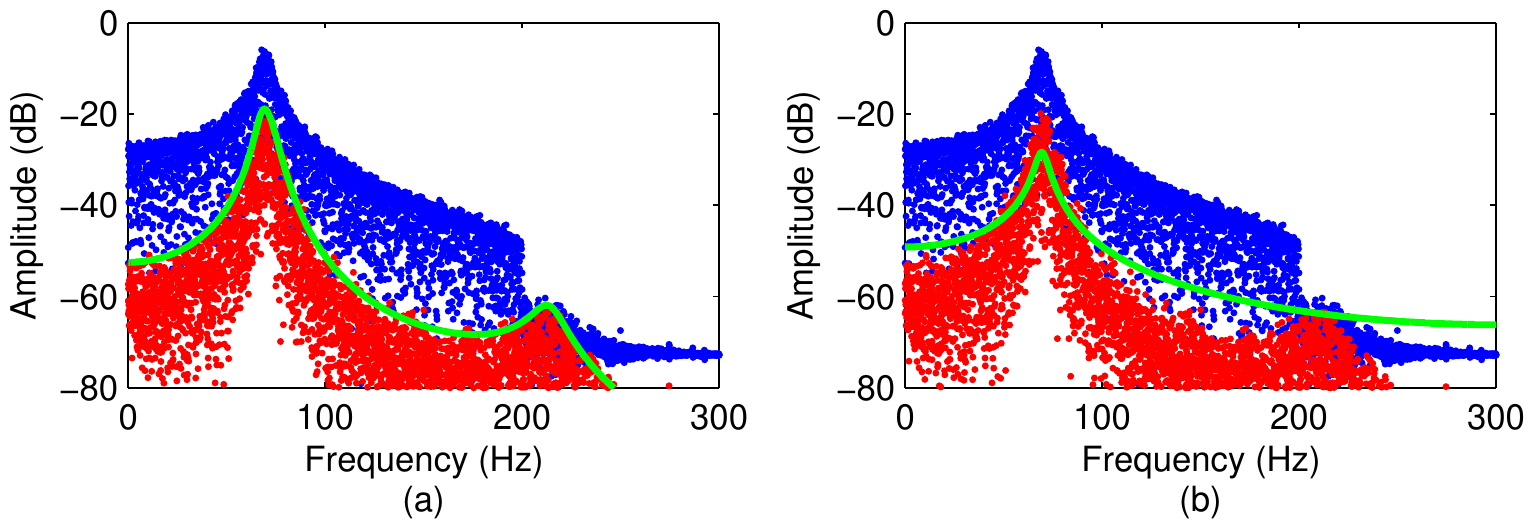}
\caption{The output of the forced Duffing oscillator is simulated using an estimated Box-Jenkins model (plant model order 2 poles and 2 zeros, noise model order 6 poles and 6 zeros) (a), or an ARX model (plant model order 2 poles and 2 zeros) (b). The amplitude of the discrete Fourier transform of the measured output and the simulation error are shown. The blue dots are the measured output, the red dots are the simulation error. The green line is the 95\% error level that is calculated from the estimated noise model. Observe that the simulation error for both models is very similar. The BJ-noise model describes the disturbances very well, while the ARX model under estimates the errors around the resonance frequency.}
\label{fig:SB2}
\end{figure}

\subsubsection{Linear simulation and prediction of the Duffing oscillator}
In a first step a linear model
\begin{equation}
\begin{aligned}
y(t)&=y_{0}(t)+v(t)\\&=G(q,\theta)u(t)+H(q,\theta)e(t),
\end{aligned}
\end{equation}
with
\begin{equation}
\begin{aligned}
G(q,\theta)&=\frac{B(q,\theta)}{A(q,\theta)}=\frac{b_{0}+b_{1}q^{-1}+\ldots+b_{n_{b}}q^{-n_{b}}}{1+a_{1}q^{-1}+\ldots+a_{n_{a}}q^{-n_{a}}},\\
H(q,\theta)&=\frac{C(q,\theta)}{D(q,\theta)}=\frac{1+c_{1}q^{-1}+\ldots+c_{n_{c}}q^{-n_{c}}}{1+d_{1}q^{-1}+\ldots+d_{n_{d}}q^{-n_{d}}},
\end{aligned}
\end{equation}
is estimated on the 'tail' experiment using the prediction error framework \cite{Ljung1987, Soderstrom1989}. The order of the BJ-model is $n_{a}=n_{b}=2$ and $n_{c}=n_{d}=6$, the plant and noise model $G,H$ are independently parametrized.  For the ARX-model the order of the plant model is also $n_{a}=n_{b}=2$, while in this case the noise model is $C=1, D(q,\theta)=A(q,\theta)$. The latter fits with the assumption that disturbances origin mainly at the input of the system so that it shares its dominant dynamics with the input signal. The BJ and ARX simulation errors are shown in Figure \ref{fig:SB2} and compared to the 95\% amplitude levels calculated from the estimated noise models. The BJ-noise model describes the disturbances very well, while the ARX model under estimates the errors around the resonance frequency.

\begin{figure}
\centering
\includegraphics[scale=0.70]{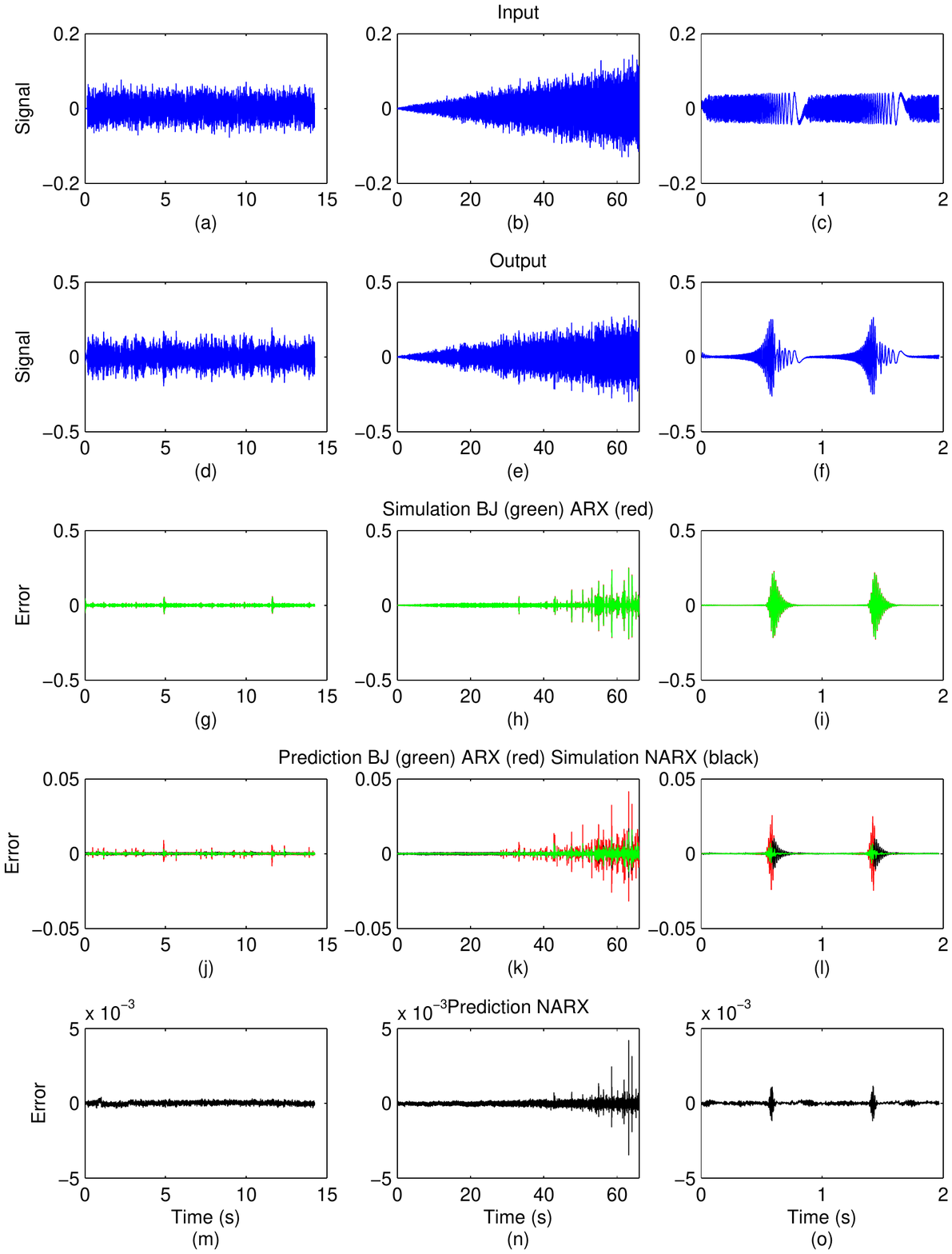}
\caption{Study of the Duffing oscillator for the 'tail' (a), (d) ,'arrow' (b), (e), and 'sweeping sine' (c), (e) experiment. The  BJ (green) and ARX (red) simulation errors, shown in (g), (h), (i), are almost equal to each other. However, the prediction error for the BJ model are well below those of the ARX model, especially for larger amplitudes of the output (j), (k), (l). This is mainly due to the better BJ noise model as shown in Figure \ref{fig:SB2}. The nonlinear NARX simulation (black) has the same quality as the linear predictions in (j), (k) (lines on top of each other), while it is in between the ARX and BJ linear predictions in (l).  The NARX prediction errors are a factor 10 smaller. The simulation error is shown in Figure \ref{fig:NARXresults1}.} 
\label{fig:SB3}
\end{figure}

These results are used to simulate $\hat{y}_{0}(t)=G(q,\theta)u(t)$ (see Figure \ref{fig:SB3} (g),(h),(i)), and predict $\hat{y}(t)=H^{-1}(q,\theta)G(q,\theta)u(t)+(1-H^{-1}(q,\theta))y(t)$  (see Figure \ref{fig:SB3} (j),(k),(l)) the output for the three experiments \cite{Ljung1987,Soderstrom1989}. The simulation errors of both the BJ and ARX model are on top of each other in these figures. However, the better BJ noise model results in significantly smaller prediction error for the BJ models compared to those of the ARX model. By increasing the model orders $n_{a}, n_{b}$ for the ARX model, these results could be improved. However, that would increase the number of terms in the NARX model. Therefore, and also for didactic reasons, the discussion is continued with the too simple model. In practice, the user has to balance the model complexity against the required model quality.

\subsubsection{Summary}
\begin {itemize}
\item Linear prediction can provide good results in the presence of nonlinear distortions. This is much harder for linear simulation.
\item The quality of the `noise' model has a strong impact on the quality of the predictions.
\item A changing nature of the excitatition signal results in a changing behavior of the residuals (due to structural model errors). This can turn a good prediction model in a poor one.
\end {itemize}

\subsection{Nonlinear simulation and prediction of the forced Duffing oscillator using a NARX and a Nonlinear State Space Model}
A polynomial NARX  \eqref{eq:NLARX} and a polynomial nonlinear state space model \eqref{eq:NLSSu} is estimated on one of the realizations of the random phase multisines in the tail.

\subsubsection{The NARX and PNLSS model}
\emph{The NARX model} is of order $n_{a}=2, n_{b}=2,$ and polynomial expansion of degree 3 with arguments $R=[u(t),u(t-1),u(t-2),y(t-1),y(t-2))]^T$.  The selection of the nonlinear degree and arguments was done following the results in \cite{Westwick2018a} that were obtained on a trial and error basis. 
\begin{equation} \label{eq:NARX1}
\begin{aligned}
y_{0}(t)&=h(u(t),\ldots,u(t-n),y(t-1),\ldots,y(t-n_{a}))\\
&=\sum_{\textnormal{all combinations}}c_{p,m-p}(k_{1},k_{2}),\ldots,k_{m})\prod_{i=1}^{p}y(t-k_{i})\prod_{i=p+1}^{p}u(t-k_{i}).
\end{aligned}
\end{equation}
 It might also be possible to get better results by replacing the polynomials by other basis functions.  Most important is to observe that his model is linear in the parameters $c_{p,m-p}$, which will reduce the identification to a problem that is linear-in-the-parameters for a quadratic cost function.

\emph{The polynomial NLSS} is of order 2 (two states) and polynomial degree 3 with arguments $R=[x_{1}(t) x_{2}(t) u(t)]^T$:
\begin{equation} \label{eq:NLSSSilverbox}
		\begin{aligned}
		\begin{bmatrix}x_{1}(k+1) \\ x_{2}(k+1)\end{bmatrix}&=A\begin{bmatrix}x_{1}(k) \\ x_{2}(k)\end{bmatrix}+Bu(k)+\textcolor{blue}{\begin{bmatrix}f_{1}(x_{1}(k), x_{2}(k), u(k)) \\ f_{2}(x_{1}(k), x_{2}(k), u(k))\end{bmatrix}} \\
		y(k)&=C\begin{bmatrix}x_{1}(k) \\ x_{2}(k)\end{bmatrix}+Du(k)
		\end{aligned}
\end{equation}

Observe that only the state transition equation has a nonlinear term, it turned out that adding a nonlinear term to the output equation did not improve the results.

\subsubsection{Cost function}
NARX models with an expansion in known basis functions are linear-in-the-parameters, and hence the cost function, given by the squared equation errors,  is minimized by solving a linear set of equations, so that no initialization problem needs to be solved. This is the major advantage of this class of NARX models, compared to many of the other models presented in this article. Because the noise enters nonlinearly into the model, a bias will appear, but as long a the SNR is decent (for example better than 20 dB), this will not be the major issue. For that reason, NARX models became one of the most popular methods, and they are very widely applied in many different fields.

For the NLSS, the cost function is nonlinear in the parameters (non convex), and a numerical method is used to minimize the costfunction.

\begin{equation} \label{eq:NARX1}
\begin{aligned}
e(t)&=y(t)-h(u(t),\ldots,u(t-n),y(t-1),\ldots,y(t-n_{a})),\\
V&=\frac{1}{N}\sum_{t=1}^{N}e(t)^2,
\end{aligned}
\end{equation}
where $u(t),y(t)$ are the measured values.

\subsubsection{Results}
\begin{figure}[h] 
\centering
\includegraphics[scale=0.7]{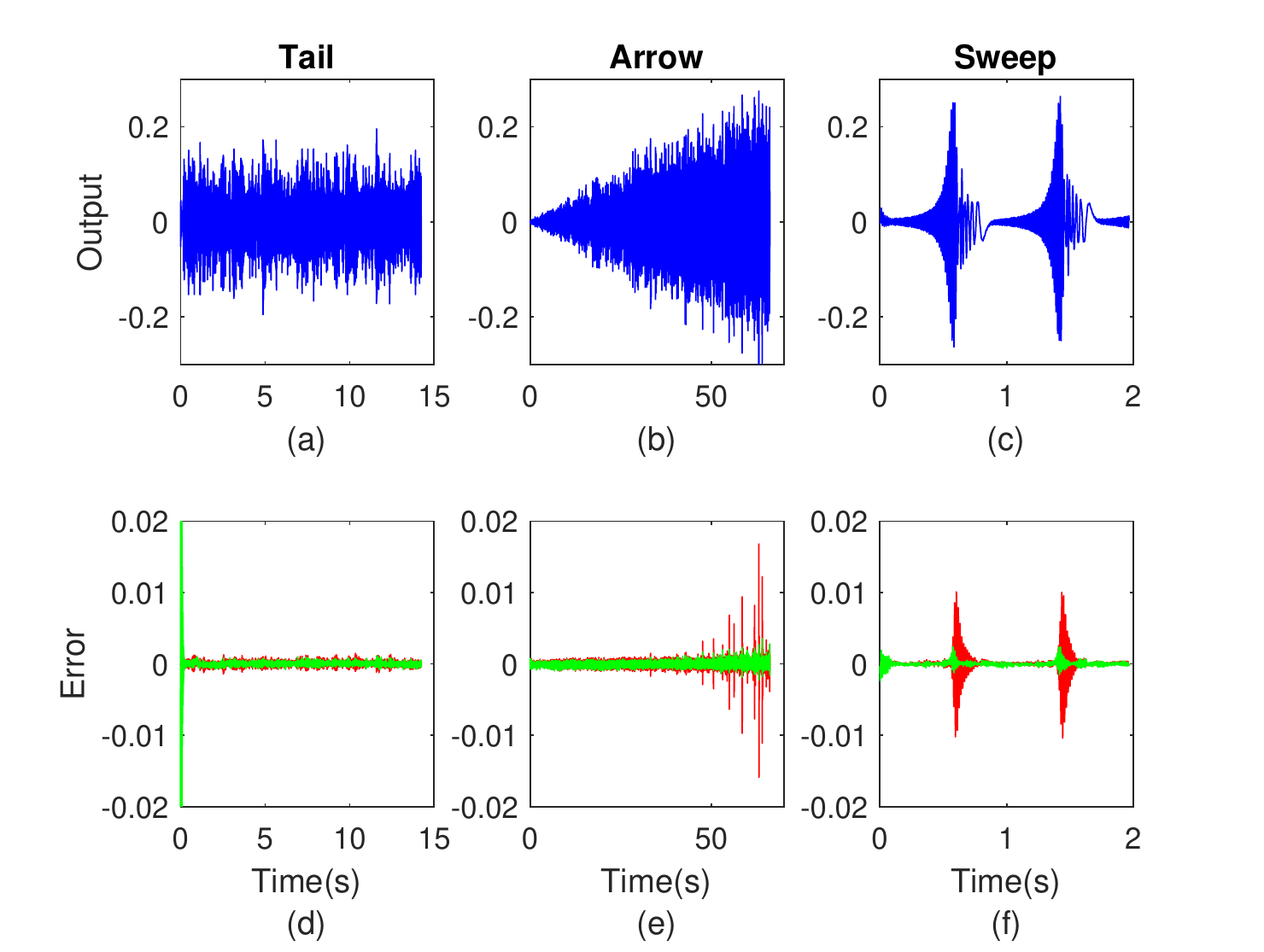}
\caption{Modeling the Silverbox using a discrete time NARX and polynomial NLSS model. Both models are identified on a section of the Tail data. The simulation error is shown for the NARX (red) and the NLSS (green) on validation data in the Tail (d), Arrow (e), and Sweep (f). For smaller output levels, both models are quite comparable, but for larger outputs, the errors of the NARX model are two to three times larger. The large initial green spike in (d) is due to transient effects because the initial states were not estimated.}
\label{fig:NARXresults1}
\end{figure}

The results for the NARX and NLSS models are shown and discussed in Figure \ref{fig:NARXresults1}. For smaller output levels, the two models are quite comparable, but for larger outputs, the errors of the NARX model are two to three times larger than those of the NLSS model.

\subsubsection{Summary}
\begin{itemize}
\item NARX models  with an expansion in known basis functions are very attractive because these are very simple to use: no initialization, and no problems with local minima show up. The PNLSS results in (slightly) better results because it was better tuned.
\item The nonlinear prediction models do significantly better than the linear ones, even if they are not optimally tuned. 
\item The polynomial regression functions can be replaced with more convenient alternatives from machine learning (Gaussian bells, hinge functions, etc.) as discussed also in ``Static Nonlinearities''.
\item The NARX method can be combined with pruning and structure revealing methods to simplify the models.
\end{itemize}

\subsection{Pruning, data driven structure retrieval, and model reduction}

\subsubsection{NARX model pruning}
The major drawback of NARX models is the combinatorial grow of the number of parameters with the number of regressors (in this example 5) and the degree (in this example 3). No further pruning of the model was made in the results presented here. It is well known pruning can improve the model quality significantly, especially if the number of regressors grows \cite{Billings2013}.  

In \cite{Billings2013}, different strategies are presented to include the dominating terms gradually, for a growing complexity of the NARX model. 

Recently, a top down approach is proposed using the decoupling strategy presented in "Decoupling Of Multivariate Polynomials" \cite{Westwick2018a}. First a full model (including all regressor combinations) is identified, resulting in a single output multivariate polynomial which is next decoupled as a sum of univariate polynomials $P_k$  that act on linear combinations of the regressors. In \cite{Westwick2018a}, the NARX Silverbox model is decoupled with 4 polynomials of degree 3: $y(t)=\sum_{k=1}^{4}P_{k}(v_{k}^{T}R)$. Especially for a large number of regressors, this method offers a systematic approach to model pruning.

\subsubsection{NLSS model reduction and model data driven structure retrieval}
\begin{figure}[h] 
\centering
\includegraphics[scale=0.6]{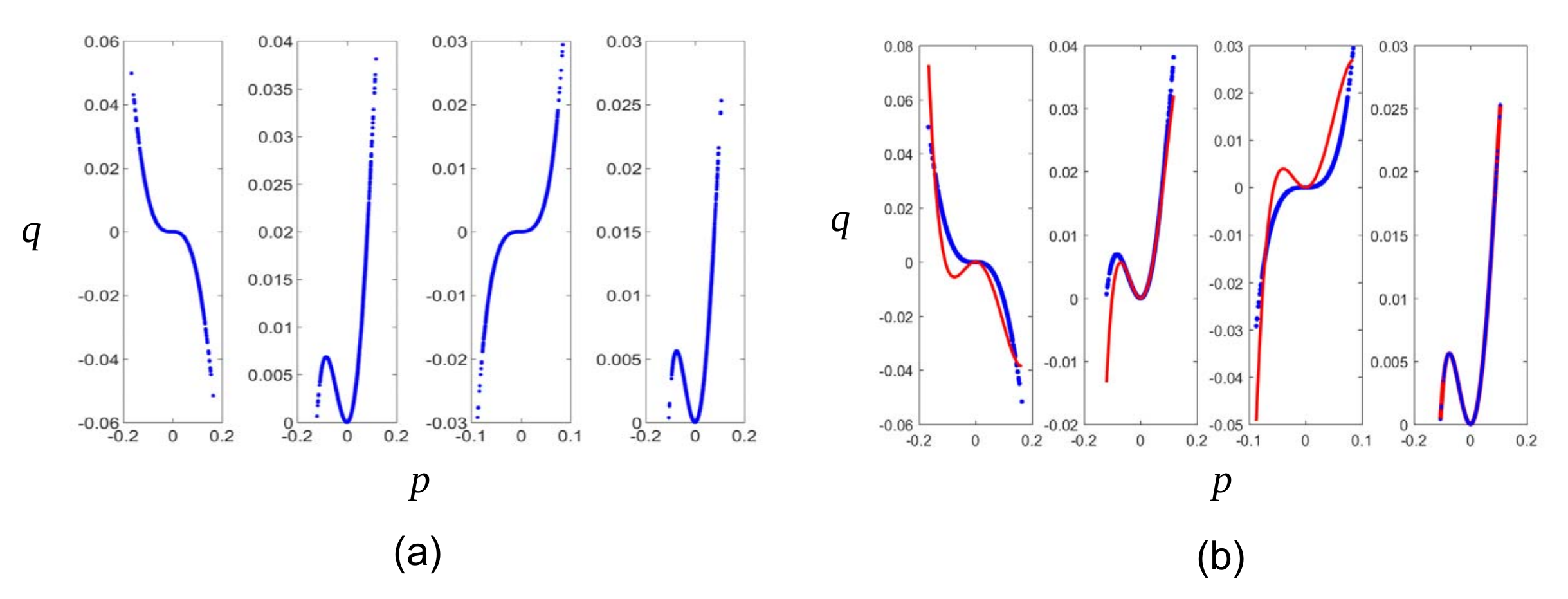}
\caption{The multivariate nonlinear polynomial vector function (blue term in \eqref{eq:NLSSSilverbox}) is decoupled using 4 SISO branches, shown in (a). In (b), these functions are approximated by a single function after proper scaling of the input $p$ and the output $q$ in each of the branches.}
\label{fig:SilberboxDecoupled}
\end{figure}

\emph{Decoupling}: Similar to the NARX model, it is also possible to decouple the multivariate nonlinear vector function $[f_1, f_2]^t$ in \eqref{eq:NLSSSilverbox} using 
the decoupling strategy presented in "Decoupling Of Multivariate Polynomials". It turned out that 4 internal SISO branches were needed in the decoupled representation (see Figure \ref{fig:decoupling}) that are shown in Figure \ref{fig:SilberboxDecoupled}. In a second step the degree of the polynomials was increased from 3 to 5, and the cost function was minimized for this decoupled model. This reduced the RMS error of the decoupled model to 0.40\%, while the original full $3^{th}$ degree model had an RMS error of 0.49\% (see Table \ref{table:Model reduction}).

\begin{figure}[h] 
\centering
\includegraphics[scale=0.3]{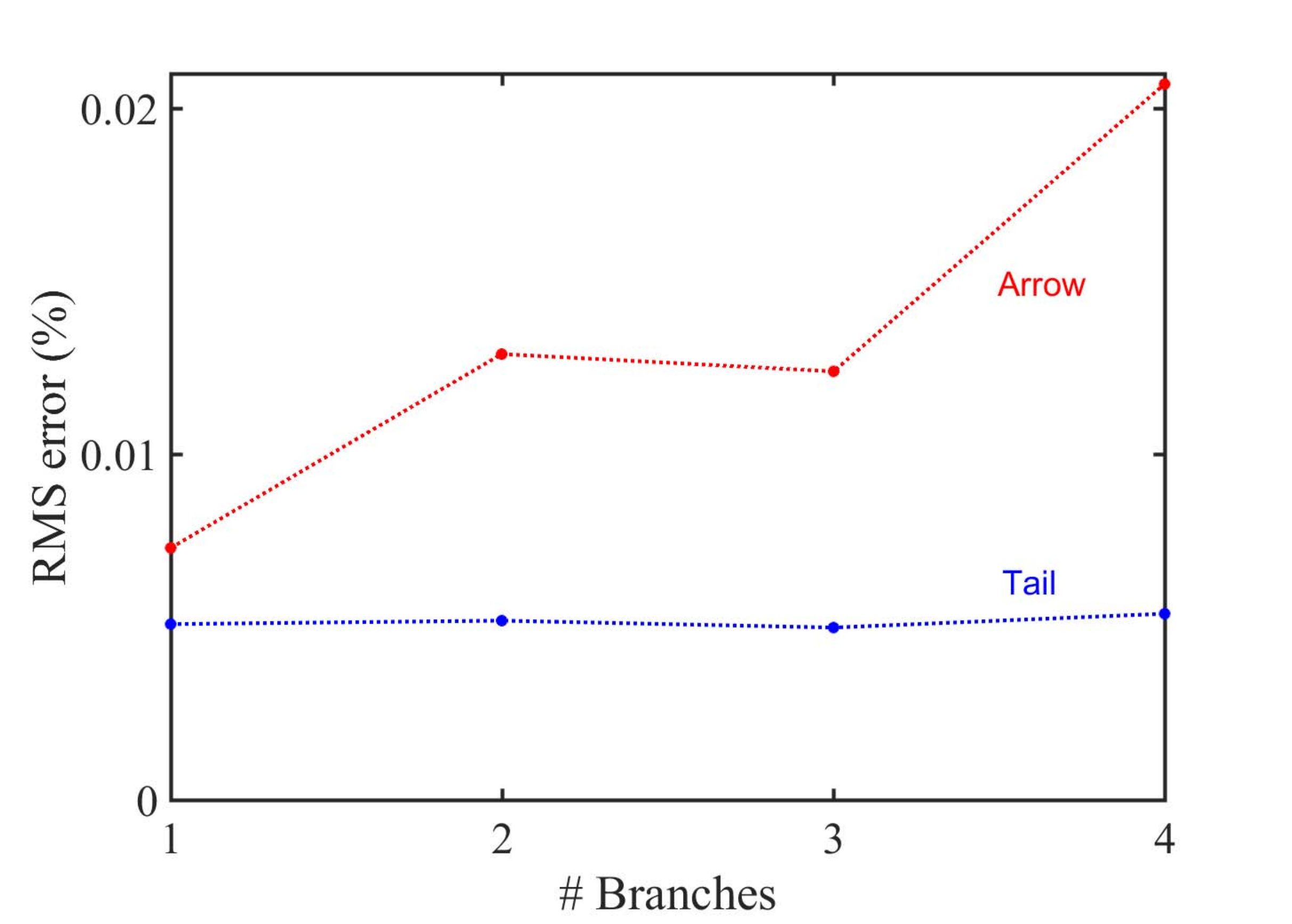}
\caption{Evolution of the RMS error on the Tail and Arrow data as a function of the number of branches.}
\label{fig:SilberboxDecoupledReduced}
\end{figure}

\emph{Model reduction:}
In the following step, the number of branches of the decoupled model was reduced. In each reduction step, a new cost function minimization was done. The RMS errors of the reduced models on the Tail and Arrow validation data is shown in Figure \ref{fig:SilberboxDecoupledReduced}. Although it could be expected that the error would start to grow when the model is simplified, it turns out that the errors remain constant on the Tail data and even drops significantly on the Arrow data. This indicates that the true system can be described with only one nonlinear branch in the decoupled model (see Figure \ref{fig:decoupling}). The reduced error on the Tail data indicates a better generalization (extrapolation) capability of the reduced model. The models are estimated on the Tail data that cover a smaller domain than the Arrow and Sweep data as is shown in Figure \ref{fig:SilverboxPhasePlane}.

The evolution of the RMS error on the Tail validation data, the number of linear parameters $n_{\theta_{L}}$, and the number of nonlinear parameters $n_{\theta_{NL}}$ as a function of the model complexity during the model reduction process is given in Table \ref{table:Model reduction}.

\emph{Data driven structure retrieval:}
The final NLSS model is
\begin{equation}
\label{eq:NLSS Duffin model}
		\begin{aligned}
		\begin{bmatrix}x_{1}(k+1) \\ x_{2}(k+1)\end{bmatrix}&=A\begin{bmatrix}x_{1}(k) \\ x_{2}(k)\end{bmatrix}+Bu(k)+\textcolor{blue}{\begin{bmatrix}w_{1} \\ w_{2}\end{bmatrix}g(p),} \\ 
		y(k)&=C\begin{bmatrix}x_{1}(k) \\ x_{2}(k)\end{bmatrix}+Du(k), \\\
		\textcolor{blue}{p}&\textcolor{blue}{=v_{1}x_{1}(k)+v_{2}x_{2}(k)+v_{3}u(k).}
		\end{aligned}
\end{equation}

In these equations, $g(p)$ is a SISO polynomial of $p$ that is a linear combination of the states $x_1, x_2$ and the input $u$ (with coefficients $v_1,v_2,v_3$). The coefficients $w_1, w_2$ scale the polynomial output $g(p)$.

\begin{figure}[h] 
\centering
\includegraphics[scale=0.3]{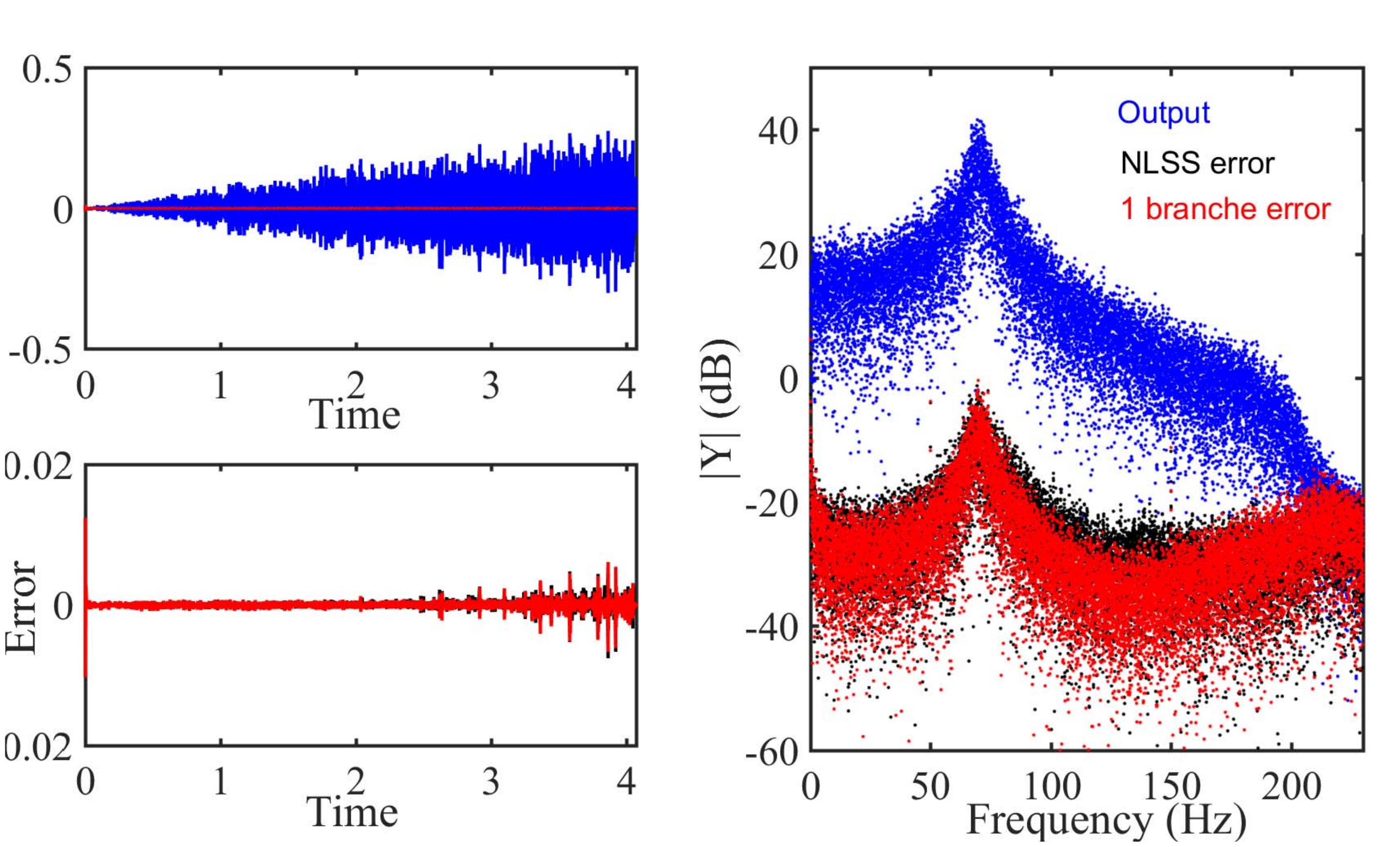}
\caption{Validation of the final model \eqref{eq:NLSS Duffin model} on the Arrow data in the time and the frequency domain. The measured output is shown in blue, the validation of the final model in red, and the full model in black. }
\label{fig:SilberboxDecoupledReducedValidation}
\end{figure}

The validation results on the Arrow data are given in Figure \ref{fig:SilberboxDecoupledReducedValidation}. Observe that the full and the final reduced model have almost the same quality.

\begin{table}
\caption{Evolution of the RMS error on the Tail validation data, the number of linear parameters $n_{\theta_{L}}$, and the number of nonlinear parameters $n_{\theta_{NL}}$ as a function of the model complexity during the model reduction process.} \label{table:Model reduction}
\begin{center}
\begin{tabular}{|c|c|c|c|c|c|}
\hline 
 & BLA & NLSS & Decoupled & Equal Branches & Single Branch \tabularnewline
\hline 
RMS error &12\% & 0.49\% & 0.40\% & 0.40\% & 0.40\% \tabularnewline
\hline 
$n_{\theta_{L}}$ & 5 & 5 & 5 & 5 & 5\tabularnewline
\hline 
$n_{\theta_{NL}}$ & 0 & 30 & 12 & 6 & 3 \tabularnewline
\hline
\end{tabular}
\end{center}
\end{table}
Because the forced Duffing oscillator is a lab setup, this model can be compared to the physical model. It turns out that in this case the data driven retrieved model structure and the physical model structure are identical. 	This is a very motivating result for the black box identification framework. However, this conclusion cannot be generalized because it is not obvious that the most simple model coincides with the physical model. Moreover, the structure of many (non)linear systems is unidentifiable from input output data due to indistinguishability problems: the same data can be represented by multiple models that can not be distinguished from each other without having additional information (for example the measurement of an internal signal). In the case of the forced Duffing oscillator, it is also possible to represent the same input-output behavior by a system with a nonlinearity in the forwards path instead of the feedback path \cite{Schoukens2008}. 

\subsubsection{Remark}
An alternative approach to obtain a highly structured model for the forced Duffing oscillator in a black box modeling framework is presented in \cite{Young2018}. It is based on a Data-Based Mechanistical Modeling approach as explained in \cite{Young2011}, where the objective is to obtain a model that can be interpreted in the mechanistic terms that are most appropriate to the nature of the dynamic system. The nonlinear nature of the system is captured by using state dependent parameters. The basic idea is to start with a linear model and check for data dependent variations of (some of) the parameters in the model. For the Duffing oscillator, this resulted in a transfer function model that is very similar to the final model \eqref{eq:NLSS Duffin model}.

\subsubsection{Summary}
\begin{itemize}
\item The decoupling method offers a data driven systematic approach to model pruning.
\item Tuning the number of branches offers a data driven model reduction approach.
\item The simplified models can significatnly improve the intuitive insight in the nonlinear model behavior.
\item The underlying physical model structure can sometimes be retrieved, although no guarantee can be given that a simple model coincides with the physical model.
\end{itemize}

\subsection{Conclusions}
In this case study structural model errors dominate the noise disturbances. The following general observations can be made:
\begin{itemize}
\item The prediction errors for a given plant model are smaller than the simulation errors. This was expected because prediction methods use explicitly the output measurements to reduce the simulation error. 
\item The prediction method decreases the impact of the structural model errors compared to the simulation method. This holds true as well for the linear simulation/prediction models as for the nonlinear models.
\item The quality of the prediction depends strongly on the quality of the noise model. The better BJ noise model results also in better predictions than those of the NARX model.
\item A nonlinear model captures better the system behavior than a linear model does. This results in smaller simulation and prediction errors.
\item The decoupling methods are a powerful tool to pruning, data driven structure retrieval, and model reduction
\end{itemize}

\section{Sidebar\\   Approximating a Continuous Time NLSS with a Discrete Time NLSS Model}
Although real-world systems evolve in continuous time, discrete time models are preferred in many (control) applications because most simulations and many controllers are implemented digitally \cite{Goodwin2013}. However, without special care, the discrete time approximation can have a completely different behavior than the original continuous time system \cite{Pearson2004, Pearson2006}. To provide a better understanding of these problems, the discretization of linear systems is first considered, and next the discrete time approximation of continuous time nonlinear systems is discussed.

\subsection{Discretization of linear systems}

\emph{ZOH-discretization}: In linear system identification, it is well known that a continuous time system that is excited by a zero-order-hold (ZOH) excitation (also called piecewise constant excitation \cite{Ljung1987}) can be replaced by a discrete time model that gives an exact description of the discrete time input-output relations \cite{Ljung1987, Soderstrom1989,SchoukensJ1994}. Generalizing this result to nonlinear systems is not always possible because the ZOH nature can be lost inside the system. For example, the output of the feedback in a nonlinear closed loop systems will be no longer ZOH. For that reason, solutions that do not rely explicitly on the ZOH-nature are needed.

\begin{figure}[h] 
\centering
\includegraphics[scale=1]{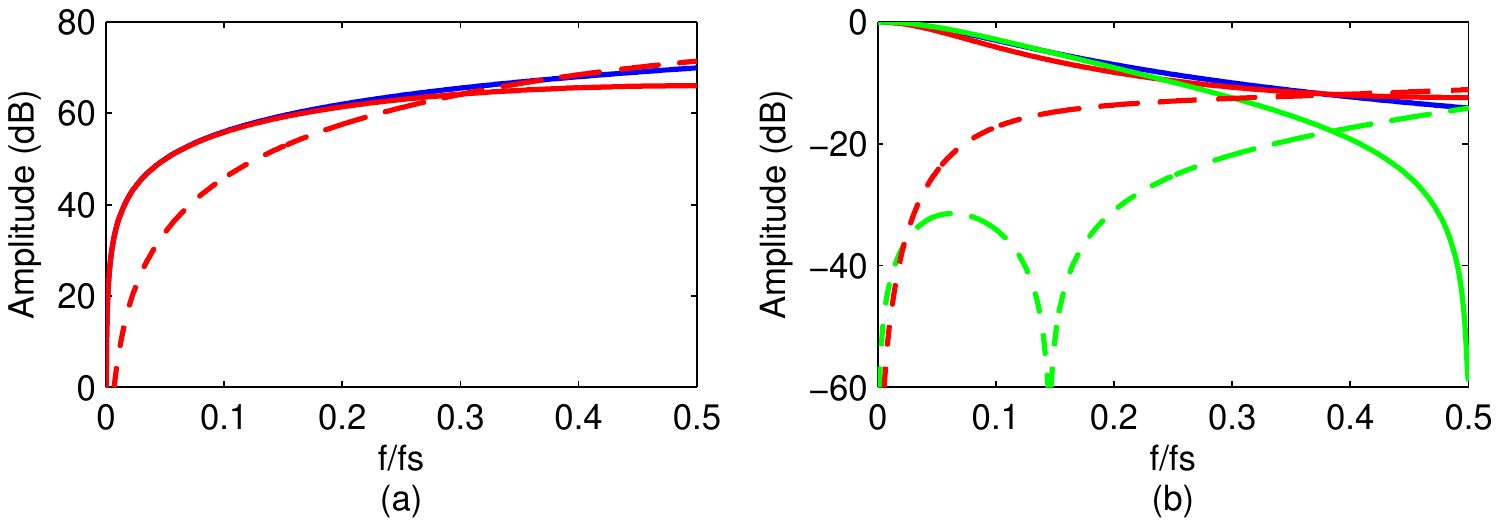}
\caption{Approximation of a continuous time system by a discrete time system. (a) Discrete time approximation of a differentiator: $G(s)=s$ (blue) and $G(z)=(1-z^{-1})f_{s}$ (red). Observe that the error (broken line) grows fast with the frequency to 100\% at $f/f_{s}\approx0.3$; (b) Approximation of a continuous time first order system $G(s)$ (blue) by a discrete time system $G_{1}(z)$ obtained by a finite difference transformation $s\rightarrow (1-z^{-1})f_{s}$ (red), and a first order discrete time system $G_{2}(z)$ (green) that is fitted in least square sense in the bandwidth $[0  0.25]f_{s}$.  Observe that error of the finite difference solution is very small around $f=0$, but grows very fast.}
\label{fig:CTDT1}
\end{figure}

\emph{Alternative discretizations}: Replacing the continuous time differentiation by a finite difference is intuitively very appealing:
\begin{equation}
\frac{d x(t)}{dt}\approx\frac{x(t)-x(t-T_{s})}{T_{s}}=(x(t)-x(t-T_{s}))f_{s},
\end{equation}
or in the Laplace- and Z-domain
\begin{equation}
s\rightarrow (1-z^{-1})f_{s}.
\label{eq:DTdiff1}
\end{equation}
In the frequency domain, $s=j2\pi f$ and  $z^{-1}=e^{-j2\pi f/f_{s}}\approx1-j2\pi f/f_{s}+O((f/f_{s})^{2}$, and (\ref{eq:DTdiff1}) becomes
\begin{equation}
s\rightarrow (1-z^{-1})f_{s}\approx j2\pi f + O((f/f_{s})^{2}.
\label{eq:DTdiff2}
\end{equation}
This simple solution works only well if the sample frequency $f_s$ is much larger than the frequency band of interest. In Figure \ref{fig:CTDT1} (a) it can be seen that the relative error grows to 100\% for $f> 0.3 f_{s}$.  In Figure  \ref{fig:CTDT1} (b) the transformation (\ref{eq:DTdiff1}) is applied to a continuous time first order system
\begin{equation}
G(s)=\frac{1}{1+\tau s}\rightarrow  G_{1}(z)=\frac{1}{1+\tau f_{s} -\tau f_{s}z^{-1}}.
\end{equation}
Observe that also here the error is very small around the origin, but grows very fast to 100\% for  $f> 0.3 f_{s}$.

A very popular approach in the system identification community is to identify directly, in the frequency band of interest, a discrete model $G(z)$ for he continuous time system $G(s)$. This is also illustrated in  Figure  \ref{fig:CTDT1} (b). A first order discrete time model $G_{2}(z)$ is  obtained by a least squares fit in the frequency band $[0 0.25]f_{s}$. Observe that the error of $G_{2}(z)$ is much smaller than that of  $G_{1}(z)$. 

The reader is referred to the signal processing literature for more information on the classical solutions to this problem (impulse invariant transformation, bilinear transformation, ...) \cite{Oppenheim1997}.

\subsection{Discretization of nonlinear systems}

Finding good discrete time approximations for continuous time nonlinear systems has been studied for a long time \cite{Rao2006, Billings2013,Garnier2015,Zhang2015}. As was done for linear systems, the discussion starts with dedicated methods for ZOH excitions, followed by an approach to deal with the more general class of lowpass excitations. 

\emph{ZOH framework}: An exhaustive overview of the discretization problem is made in \cite{Goodwin2013}, and references therein. The authors look for an approximate discrete time representation, operating in a ZOH-framework. Some of their major conclusions are: 
\begin{itemize}
\item The popular 'forward Euler method' should be used with extreme care because it results in relative errors that grow fast with $f/f_{s}$ leading to the need for a very high oversampling. This confirms the observation made in Figure \ref{fig:CTDT1}. 
\item The dynamics of sampled data models can be different from those of the continuous time system, numerical 'sampling zeros' are created. 
\item A 'Truncated Taylor Series Approximate Model (TTS)' is proposed to approximate a continuous time nonlinear state space equation by a discrete time equivalent. The global fixed-time truncation error (the error when integrating over a fixed time interval) drops to zero, proportional with the inverse of the sampling frequency $f_{s}$ as an $O(\Delta=1/f_{s})$ 
\end{itemize}

\begin{figure*}[h]
    \centering
    \begin{subfigure}[t]{0.5\textwidth}
        \centering
        \includegraphics[scale=0.55]{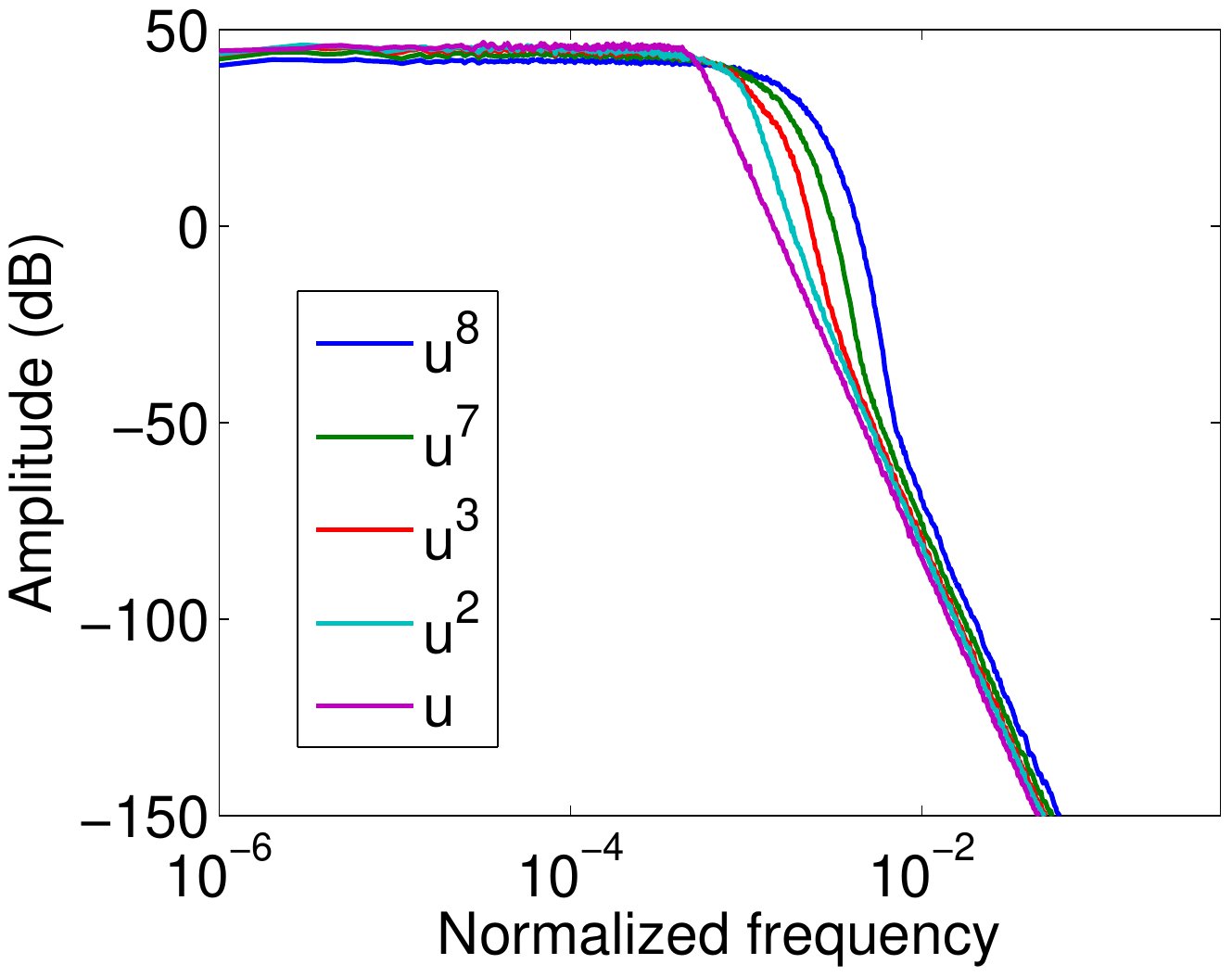}
        \caption{}
    \end{subfigure}%
    ~ 
    \begin{subfigure}[t]{0.5\textwidth}
        \centering
        \includegraphics[scale=0.55]{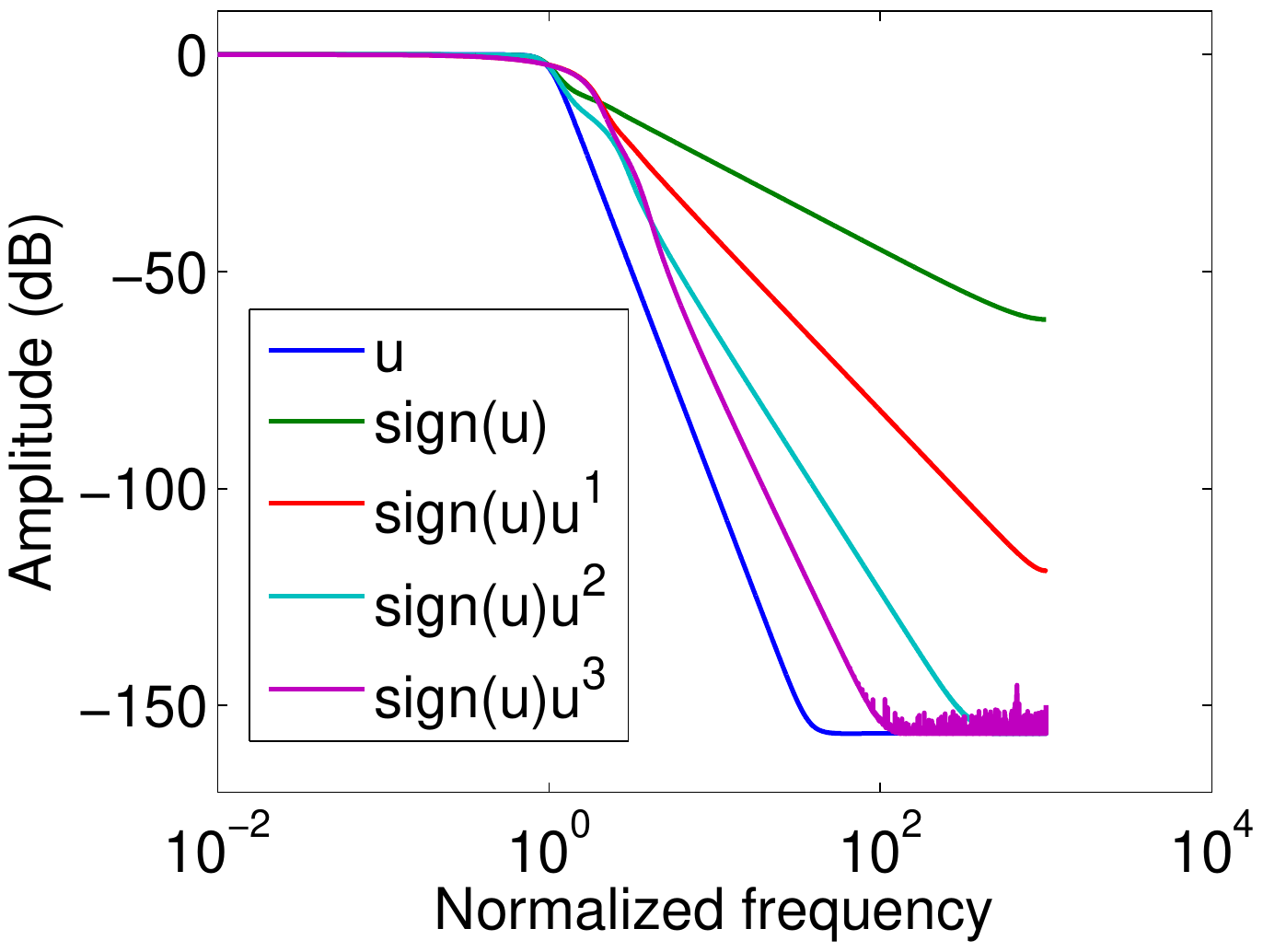}
        \caption{}
    \end{subfigure}
    \caption{Study of the amplitude spectrum of a lowpass signal of relative degree  $d_{u}=5$ that passed through a nonlinear system. The amplitude spectrum of the output normalized to an RMS value of 1 is shown, the DC-value is not shown. (a) Results for a static nonlinear system $y=p^{n},$ $n=1,2,3,7,8$. Observe that all the signals have the same relative degree, independent of $n$. In this case the relative degree of the output is set by the relative degree of the input. (b) Results for a static nonlinear system $y=\text{sign}(u)u^{n}$ with $n=1,2,3$. In this case the relative degree is set by the nonlinear system: $d_{F}=n+1$ for $d_{F}\leq d_{u}$.}
\label{fig:NLScharacteristics}
\end{figure*}

\emph{Lowpass framework}: An alternative approach for the approximation of a continuous time nonlinear statespace models is presented in \cite{SchoukensJ2017}. Consider
\begin{equation}
\begin{aligned}
\frac{dx(t)}{dt} & =  f(x(t),u(t))\\
y(t) & =  h(x(t),u(t)),
\label{eq:CTNLSS}
\end{aligned}
\end{equation}
that is excited with an input signal $u(t)$ that is a lowpass signal of degree $d_{u}$ (the square root of the power spectrum is an $O(1/f^{d_{u}})$  for $f>f_{c}$). Observe that a ZOH-excitation is a lowpass signal with $d_{u}=1$. Similar to the identification approach for the linear first order example in the previous section, a discrete time nonlinear state space 
\begin{eqnarray}
x_{d}(k+1) & = & F_{d}(x_{d}(k),u_{d}(k))\label{eq:DTNLSS}\\
y_{d}(k) & = & G_{d}(x_{d}(k),u_{d}(t))\nonumber 
\end{eqnarray}
is identified directly to the data. The discrete time signals $x_{d}(k), y_{d}(k)$  equal the sampled continuous time signals $x(t),y(t)$ within an error:
\begin{equation}
\begin{aligned}
x_{d}(k)&=x(kT_{s})+O(1/f_{s}^{d_{x}-1.5})+\varepsilon_{I},\\
y_{d}(k)&=y(kT_{s})+O(1/f_{s}^{d_{x}-1.5})+\varepsilon_{I}.
\end{aligned}
\end{equation}
$d_{x}=\textrm{min}(d_{u},d_{F})+1$, with $d_{F}$ a characteristic of the nonlinear system as explained in Figure \ref{fig:NLScharacteristics} \cite{SchoukensJ2017}. For a a static discontinuous nonlinear function, $d_{F}=1$, and $d_{F}=n+1$ if the $n^{th}$ derivative exists on the domain of interest \cite{Baum1982, SchoukensJ2017}. 
The error $\varepsilon_{I}$ is a user choice that is set by the choice of the complexity of the discrete time nonlinear state space. The discretization error is dominated by the aliasing effect that is due to the sampling of the input, output, and internal signals. Observe that the error in the direct identification approach is smaller than that of the TTS  ZOH-approach. This is again due to the fact that the discrete time model is tailored to the continuous time data in the least squares minimization step.

This result provides a strong theoretical foundation for the popular and successful practice to identify directly explicit discrete time models of continuous time nonlinear systems. 

\subsection{Illustration: discretization of the Duffing oscillator}

The results of the previous section are illustrated on the Duffing oscillator setup (Figure \ref{fig:SB1}) described in `` Experimental setup: The forced Duffin oscillator''.   In this case, the ZOH-output $x_{\text{ZOH}}(t)$  of the generator filter is filtered by a $4^{th}$-order  lowpass filter with a cut-off frequency of 200 Hz. A wide band zero mean excitation with a bandwidth of $f_{s}/2$ is used as the excitation signal in the frequency band [0 - 39062 Hz]. With these settings, the signal $u_{c}(t)$  has a flat amplitude spectrum up to 200 Hz, to be compared to the bandwidth of the second order system that is below 100 Hz.

\begin{figure}[h] 
\centering
\includegraphics[scale=0.25]{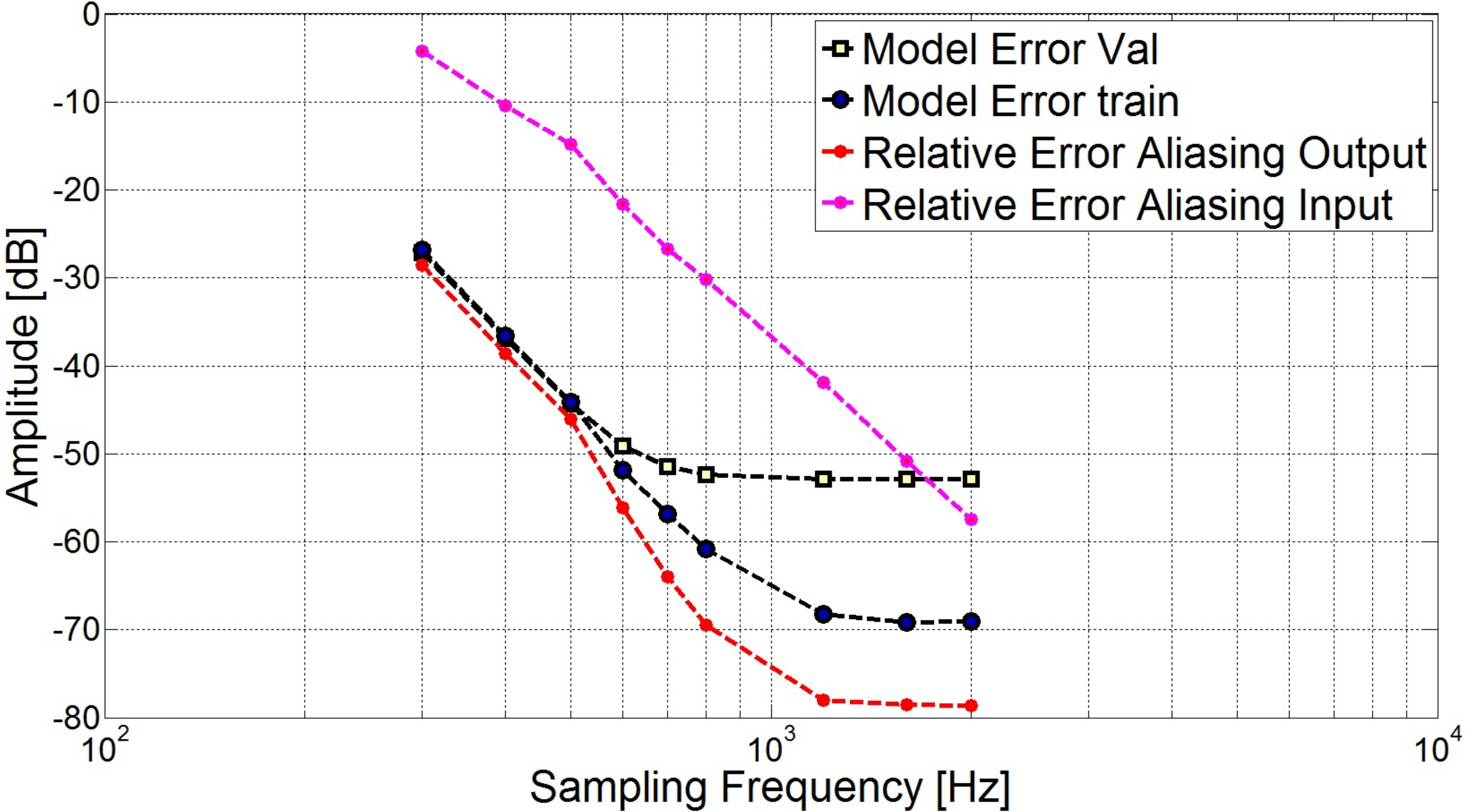}
\caption{Experimental verification on the Duffing oscillator shown in Figure \ref{fig:SB1}. A discrete time model nonlinear state space model is identified, and the RMS value of the simulation error on the estimation set and a validation set is plotted as a function of the sample frequency. It is also compared to the relative alias error of the input and output. The alias error drops with about 70 dB/decade, which is in perfect agreement with the presence of the 4th order filter in the generator path. It is also seen that the error on the modeled output follows the slope of the aliased power of the input and output, as was expected from the theory. At lower error levels, structural model errors dominate. These could not be further reduced by increasing the complexity of the model.}
\label{fig:SilverboxDTerrors}
\end{figure}

The continuous time input and output are sampled at a high sampling frequency $f_{s}=78125\text{ Hz}$, and next the data are subsampled at different rates \cite{SchoukensJ2017}. For each subsampled data set a discrete time nonlinear state space model is identified on the measurements using a DT nonlinear polynomial state space model (PNLSS), using the methods described in \cite{Paduart2010} and ``Extensive case study: The forced Duffing oscilator''. The model has 2 states, and the degree of the internal multivariate polynomial is 3 (it depends on both the states and the input). Higher orders and degrees were tested, but this did not significantly improve the results. The results are shown in Figure \ref{fig:SilverboxDTerrors}. It shows that the errors drop as an $O(1/f_{s}^{4})$ till the noise floor or the structural model error floor is reached. The drop rate is proportional to the drop of the alias errors, and much faster than $1/f_{s}$.

\subsection{Summary}

Different options are available to approximate a continuous time system/model by a discrete time model.

\begin{itemize}
\item \emph{Data driven approach}: A nonlinear discrete time model can be identified directly from the experimental data. The approximation error is dominated by the alias error that is $O(1/f_{s}^{d_{x}-1.5})$. The model complexity of the discrete time model can be higher than that of the continuous time counterpart in order to keep the discretization error below a user defined error level.
\item \emph{Model based approach}: If a continuous time model is available, it is possible to turn it into a discrete time model. Two options are open for the user:
\begin{itemize}
\item \emph{Truncated Taylor Series Approximate Model}: The continuous time equations are transformed into a set of discrete time equations using a Taylor series approximation. The major advantage is that there is a close connection with the original (physical) equations. The main drawbacks are i) the restriction to ZOH excitations, ii) the approximation error is $O(f_{s}^{-1})$ which drops slowly with the sample frequency. 
\item \emph{Simulation-Identification approach}: The conitnuous time model is used to create a rich data set that is used as the input for a direct fit of a discrete time model. The major advantages are that i) a compact model is obtained with a user controllablle balance quality/complexity. ii) By a proper design of the data set, the user can focus the model on the intended application. The major disadvantage is the loss of physical interpretablility of the model.
\end{itemize}
\end{itemize}

\section{Sidebar\\External or Internal Nonlinear Dynamics}
\subsection{Dual representation of linear systems}
\emph{Infinite impulse response (IIR)  models}: A linear system can be modeled by the recurrent representation \eqref{eq:EIDyn2}
\begin{equation}
\begin{aligned}
y(t)=&b_{0}u(t)+b_{1}u(t-1)+\ldots+b_{n_{b}}u(t-n_{b})\\
&-a{1}y(t-1)-\ldots-a_{n_{a}}y(t-n_{a})\\
=&h_{IIR}(u(t),\ldots,u(t-n_{b}),y(t-1),\ldots,y(t-n_{a})).
\end{aligned}
\label{eq:EIDyn2}
\end{equation}
 
 The system \eqref{eq:EIDyn2} can have an infinite memory (its impulse response has an infinite length) and is called for that reason an infinite impulse response (IIR) model. 

\emph{Finite impulse response (FIR) models}: 
The equivalent impulse response representation of  \eqref{eq:EIDyn2} is $y(t)=g(t)*u(t)=\sum_{k=0}^{\infty}g(k)u(t-k).$
Truncating this inifinte long impulse response to a finite length $n$ leads to the finite impulse response (FIR) model
\begin{equation}
\begin{aligned}
y(t)&=g(t)*u(t)\\
&=\sum_{k=0}^{n}g(k)u(t-k)\\
&=h_{FIR}(u(t),\ldots,u(t-n)).
\end{aligned}
\label{eq:EIDyn1}
\end{equation}

\emph{Dual representation of linear systems}: From the previous discussion, it follows that linear systems can be either represented by an IIR model $y(t)=h_{IIR}(u(t),\ldots,u(t-n_{b}),y(t-1),\ldots,y(t-n_{a}))$ or an FIR model $y(t)=h_{FIR}(u(t),\ldots,u(t-n))$ (where $n$ can grow to infinity). 

From behaviour point of view there is no difference between both representations. However, a structural difference is how the memory (dynamics) is created in both representations. The FIR model $h_{FIR}$ makes  no use of an  'internal' memory, the dynamic behavior is obtained by using delayed inputs. For that reason it will be called a model with ``external'' memory. The IIR model $h_{IIR}$ includes also delayed outputs to create an 'internal' memory. 

\subsection{NFIR and NIIR models}
\emph{Internal or external dynamics}: By choosing $h$ in (\ref{eq:EIDyn1}) and (\ref{eq:EIDyn2}) to be nonlinear, the linear FIR-IIR classification can be generalized towards nonlinear NIIR systems $y(t)=h_{NIIR}(u(t),\ldots,u(t-n_{b}),y(t-1),\ldots,y(t-n_{a}))$ having internal dynamics (\ref{eq:EIDyn1}) or NFIR systems $y(t)=h_{NFIR}(u(t),\ldots,u(t-n_{b}))$ having external dynamics (\ref{eq:EIDyn2}) \cite{Sjoberg1995, Nelles2001}. For notational simplicity, the $h_{NFIR}$ and $h_{NIIR}$ will be both written as $h$, the difference between both models follows from the arguments of the function.

For linear systems, it is a free users choice to select the FIR or IIR representations, there is a full equivalence between both representation. This is no longer the case for nonlinear systems. Not all NIIR systems can be modeled with a NFIR model as explained below. For that reason, the choice between NFIR and NIIR systems/models affects the structure, the behavior, and the stability properties. Also the numerical methods how to deal with these systems are strongly dependent upon it.

\emph{Structural aspects}: The NFIR/NIIR nature of a system is uniquely linked to its topology: for NFIR systems there can be no dynamic closed loop around the nonlinearity while for NIIR systems the nonlinearity is captured in a dynamic closed loop. 

This property is used in the structural detection framework  \cite{Schoukens2015}. It starts from the best linear approximation (BLA) that is identified for a varying input offset or amplitude \cite{Esfahani2016}. The poles of the BLA remain fixed for NFIR systems while they move for NIIR systems. 

Remark: The class of systems that is included in the NIIR representation can be further generalized by including also the nonlinear state space models that have a nonlinear feedback over some of the states, such that internal dynamic nonlinear closed loops are created.

\begin{figure}[h] 
\centering
\includegraphics[scale=0.25]{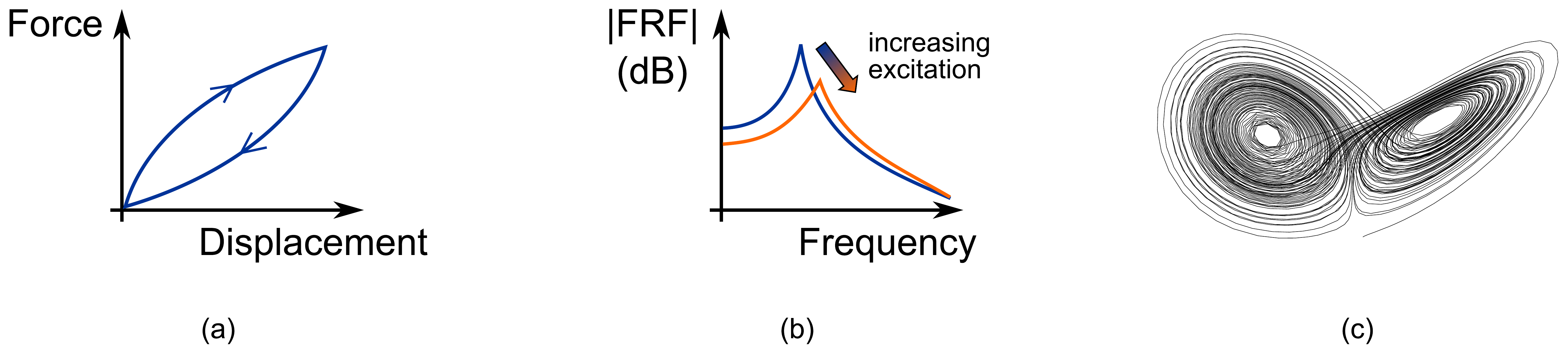}
\caption{Hysteresys (a), moving resonance frequency (b), and chaotic behavior (c) are typical nonlinear phenomena that can only be modeled by systems with a nonlinear feedback loop.}
\label{fig:EID0}
\end{figure}

\emph{Behavior aspects}: Some typical nonlinear behaviors like chaos, bifurcations, jumps, moving resonances, autonomous oscillations (see Figure \ref{fig:EID0}) can only  be generated by NIIR systems. NFIR models cover fading memory systems whose behavior is 'closer' related to that of a linear system. Fading memory systems \cite{Boyd1985}, forget the past inputs asymptotically over time. Loosely spoken, it can be stated that a nonlinear system that has no fading memory requires a NIIR model.

NFIR systems are a subset of fading memory systems. All stable NFIR systems have a fading memory, unless the nonlinearity would be discontinuous. The most simple example of such an exception is a static discontinuous system.  An NIIR system can have a fading memory behavior on a restricted input domain (for example the Duffing oscillator in "Linear simulation and prediction of the forced Duffing oscillator" in ``Extensive case study: the forced Duffing oscilator''), and on that input domain NFIR models can be used to approximate the NIIR system. This is illustrated on the Duffing oscillator in \cite{Schoukens2003}.

However, the output of nonlinear NIIR systems can also show bifurcations and jump phenomena (small variations at the input can result in an a sudden 'qualitative' or topological changes in the output), even for systems with smooth nonlinearities \cite{Worden2001, Kerschen2006, Thompson1986, Ueda1991}. The output can become even chaotic, or autonomous oscillations can appear. An NFIR model cannot model these phenomena, illustrated in Figure \ref{fig:EID0}.

\emph{Stability}: While the stability of NFIR systems can be guaranteed under very general conditions, it is much harder to analyse the stability of NIIR systems. Although general theories exist to analyse and impose stability on NIIR systems \cite{Suykens1997}, often the user is left with extensive simulations in order to check stability \cite{Nelles2001}.

\emph{Numerical aspects}: The numerical optimization aspects of nonlinear systems/models are strongly affected by moving from NFIR to NIIR. Calculating the derivative of the output of a NIIR system with respect to the model parameters boils down to computing the output of another nonlinear model \cite{Narendra1990}. 

Table \ref{table:NLmemory} gives an overview of NFIR and NIIR systems that are considered in this article. The reader is referred to \cite{Pearson2003} for a more extensive table of different models and behaviors.

\begin{table}
\begin{center}
\caption{Systems/models with external or internal nonlinear dynamics} \label{table:NLmemory}
\begin{tabular}{|c|c|}
\hline 
External nonlinear memory & Internal nonlinear memory \tabularnewline
no nonlinear output feedback & nonlinear output feedback \tabularnewline
\hline 
\hline 
NFIR&NARX\tabularnewline
\hline 
Volterra  & $$\tabularnewline
\hline 
Open loop block oriented & Closed loop block-oriented\tabularnewline
\hline 
NLSS lower triangular & NLSS full\tabularnewline
\hline
\end{tabular}

\end{center}
\end{table}

\subsection{Summary - User guidelines}
\begin{itemize}
\item \emph{Internal or External dynamics models}: The choice between an external dynamics NFIR model (no nonlinear closed loop) or an internal dynamics NIIR model (nonlinear closed loop present) strongly influences the complexity of the identification problem. 
\item \emph{Initial tests or insight}: Initial tests or insight can help to choose between the two model classes. This avoids that a lot of time and effort is wasted by trying to fit the wrong model structure to the data. 
\item \emph{External dynamics model class}: Only systems with a fading memory  behavior can be modeled, with a small error, with an external dynamics model. Typical models are: NFIR, Volterra, open loop block-oriented models (e.g. Wiener, Hammerstein, Wiener-Hammerstein and Hammerstein-Wiener). These models can only be used if the poles of the BLA do not move for varying experimental conditions.
\item \emph{Internal dynamics model class}: Strong nonlinear behaviors like moving resonances, jump phenomena (bifurcations), hysteresis, autonomous oscillations  can only be modeled with internal dynamics (NIIR) models. External dynamics  (NFIR) models are not able to represent such phenomena. Typical models are: NARX, block-oriented models with feedback, nonlinear state space models. Whenever the poles of the BLA move for varying experimental conditions, this class of models is to be preferred.
\end{itemize}

\section{Sidebar\\   Impact of Structural Model Errors}

\subsection{Introduction}
Any estimated model has some error. It is important to distinguish between two error sources
\begin{itemize}
\item \emph{Structural model errors}: These are errors that come from deficiencies in the model structure. The model is simply not capable of producing correct model outputs. Even with an infinite amount of perfect estimation data, the model output will have errors.
\item \emph{Random model errors:} The disturbances present in the estimation data will affect the model, so even when there are no structural model errors, the model output will have errors. 
\end{itemize}
Here these concepts will be defined more precisely. For simplicity only the case of \emph{output error models} (or \emph{simulation models}), will be treated, see ``Simulation Errors and Prediction Errors'' \eqref{eq:truesim}: $y(t) = y_{0}(t)+v(t)$.
The term $v(t)$ models the disturbances which are assumed, for simplicity in this section, to be additive and independent of the input. 

 For a simulation model, the predictions of $y(t)$ will only depend on past inputs:
\begin{align}
  \label{eq:1}
  \hat y(t|\mathcal S) &=y_0(t)=h_0(u^{t-1})\quad \text{prediction with the true system}\\
\hat y(t|\mathcal M(\theta))&=h(u^{t-1},\theta) \quad \text{prediction with a model for parameter value $\theta$}
\end{align}
Clearly, the output $y(t)$ contains an \emph{innovation} $\nu(t) =
y(t)-\hat y(t|\mathcal S)$ which cannot be predicted by any model, even not with the true system.  
``Model errrors'' $y_{\varepsilon}$ will always be in addition to $\nu$.
Consider two cases
\begin{itemize}
\item \emph{The true system is in the model class}, $\mathcal S \in \mathcal M$
  
  There exists a $\theta_0$ such that $h_0(u^{t-1})=h(u^{t-1},\theta_0)$. Then normally the estimate $\hat \theta_N \to
  \theta_0$ as $N\to\infty$ \cite{Ljung1987}, and the model parameter error $\theta_0-\hat \theta_N$ will be a random error caused by data
  disturbances $v(t)$. So the model error $h(u^{t-1},\theta_0)-h(u^{t-1},\hat\theta_N)$
  will be defined by the random error in $\theta_0-\hat\theta_N$, and
  \begin{equation}
  y(t)-h(u^{t-1},\hat\theta_N)=\nu(t)+[h(u^{t-1},\theta_0)-h(u^{t-1},\hat\theta_N)].
  \end{equation}

\item \emph{The true system is not in the model class}, $\mathcal S \notin \mathcal M$. 

Define 
  \begin{align}
    \label{eq:2}
    \theta^*_u=\lim_{N \to \infty} \hat \theta_N
  \end{align}
 The subscipt $u$ is to emphasize that $ \theta^*_u$ will
depend on the (statistical properties of the) input $u$.
The output model error $y(t)-\hat y(t|\mathcal M (\hat \theta_N))$ can
now be decomposed as

\begin{equation}
\begin{aligned}
  \label{eq:4}
  y(t)- h(u^{t-1},\hat\theta_N))= \hphantom{+}&\nu(t)  \enspace \text{\emph{innovation}}\\
 +& [h_0(u^{t-1}) - h(u^{t-1},\theta^*_u)] \enspace \text{\emph{structural model error}}\\
 +& [h(u^{t-1},\theta^*_u)-h(u^{t-1},\hat\theta_N))] \enspace \text{\emph{random model error}}
\end{aligned}
\end{equation}
\end{itemize}

Denote the structural model  error in \eqref{eq:4} as $y_{\varepsilon}(t)=h_{\varepsilon}(u^{t-1})$. Observe that this error depends upon the input.

\subsection*{Remarks}
\begin{enumerate}
\item Under mild ergodicity properties of estimation data disturbances
  the parameter $\theta^*_u$ in (\ref{eq:2}) obeys
\begin{align}
  \label{eq:3}
   \theta^*_u = {\rm{arg min}}_\theta E\|y(t)-h(u^{t-1},\theta)\|^2
\end{align}
which stresses that $\theta^*_u$ is the ``best model available in the
set $\mathcal M$ (for the chosen input)''.
\item For some applications it  may be of interest to consider a
  family of different input properties and configurations, $u \in
  \mathcal X$. If these configurations are equipped with a probability
  measure, the ``best model for the family $\mathcal X$'', $\theta_{BAM}$ can be defined
  as
  \begin{align}
    \label{eq:5}
    \theta_{BAM} = E_{u\in\mathcal X} \theta^*_u
  \end{align}
\end{enumerate}

\subsection{Experiment design}
A system describes that part of reality that is of importance for the user. The main idea of system theory is to make this description independent of the actual inputs that are applied, creating a clear split between the system characteristics and the signals on which the system acts. For the system identification theory, it implies that the excitation signal does not affect the system, and hence the model should not depend upon it. In the presence of structural model errors and approximate modeling, this paradigm does no longer hold. The approximate model is only valid around a given working point in a restricted input domain where the structural model error $h_{\varepsilon}(u(t))$ is acceptably small. The approximate model depends on the working point and on the class of inputs, so that a major advantage of the system theoretic framework is lost. Nevertheless, this is the best that can be done if it turns out that a complete model class that includes the true system would be too complex.

\subsubsection{Illustration on a static nonlinearity $y_{s}(t)=au(t)$} 
\begin{figure}[h] 
\centering
\includegraphics[scale=1]{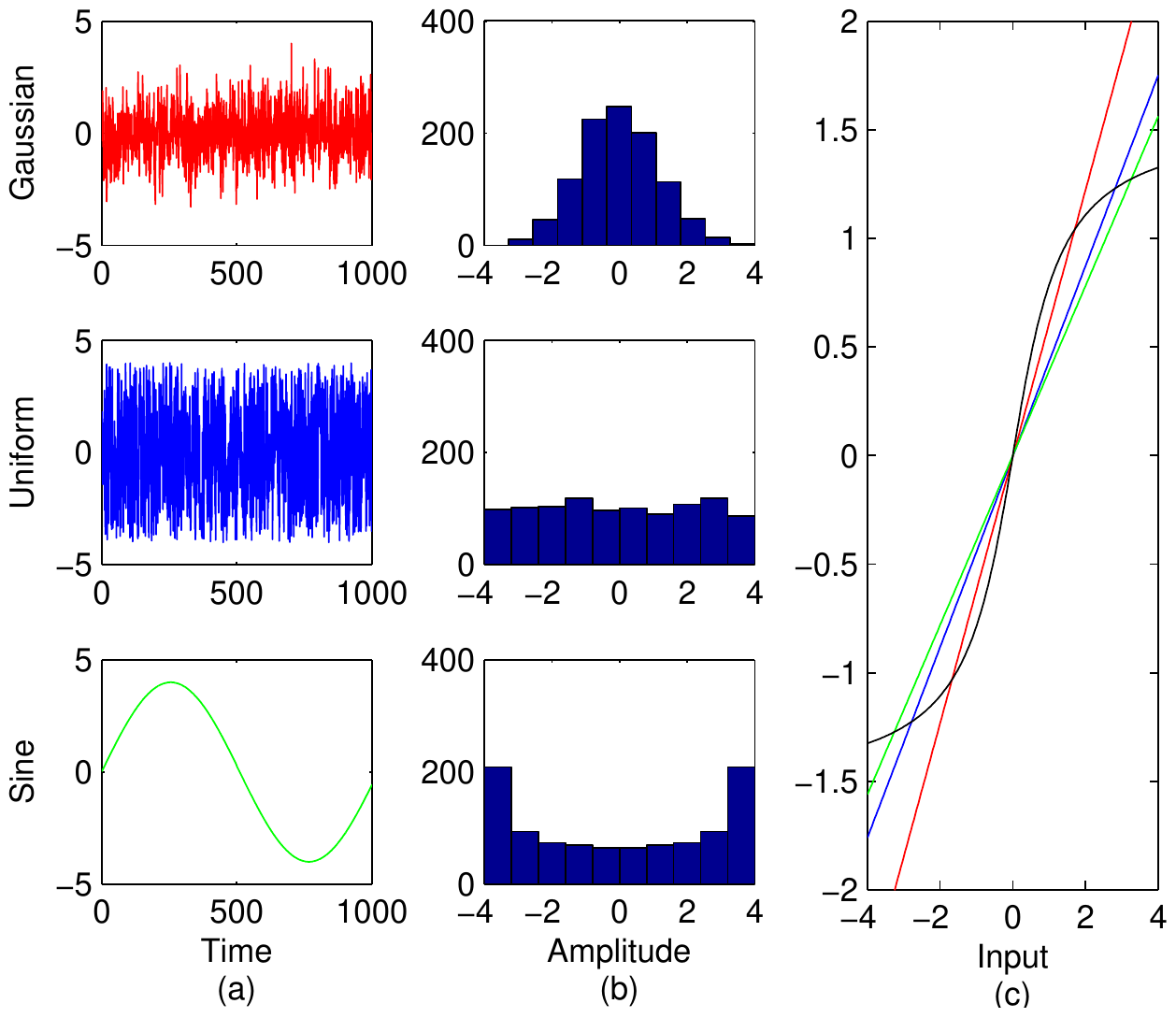}
\caption{The nonlinear system $y=\arctan (u)$ is approximated with a linear model $y=au$ for three excitation signals with a different amplitude distribution  (a): Gaussian (red), uniform (blue), and sine (green) excitations  are applied. All signals are scaled to have the same peak value. The histogram (for 1024 samples) for each of the excitation signals is shown in (b). The approximate linear models, plotted in (c), depend strongly on the distribution of the excitation signal. Since most of the probability mass of a Gaussian distribution is around the origin (see Gaussian histogram), the Gaussian excitation results in the best fit in that domain (red). A sine excites mostly the extreme values (see the histogram of the sine excitation), and it results in a fit that better approximates the nonlinear function for these extreme values (green). This comes at a cost of larger approximation errors around the origin. The behavior of the uniform distribution is in between these two extreme distributions, and this is also true for the corresponding fit (blue).}
\label{fig:20}
\end{figure}

The dependency of the model on the excitation class is illustrated in  Figure \ref{fig:20} where the true system is given by $h_{0}(u(t))=\arctan(u(t))$, and the simplified model is $y_{s}(t)=au(t)$. The disturbing noise $v(t)$  is put to zero to focus completely on the impact of the excitation class on the approximate model. A clear dependency of $\theta^*_u$ on the input distribution can be observed. 

The results of this simple illustration are generally valid. Whenever a complex system is approximated with a simplified model, the results will depend on the actual applied inputs. For that reason the experiment should be designed such that it covers the input domain of interest. It should be tuned to the problem to be solved so that the structural model errors  $y_{\varepsilon}(t)$ remain acceptable small for the user. This will set the actual subdomain that needs to be covered. This is illustrated on the Duffing oscillator example in Figure \ref{fig:SilverboxPhasePlane} and Figure \ref{fig:NARXresults1}. Within this restircted domain, it still remains important to select signals that are sufficiently rich to collect as much information as possible within the tolerable cost of the experiment.

\subsubsection{Example: Design of an excitation for the Wet-Clutch setup}
\begin{figure}[h] 
\centering
\includegraphics[scale=0.5]{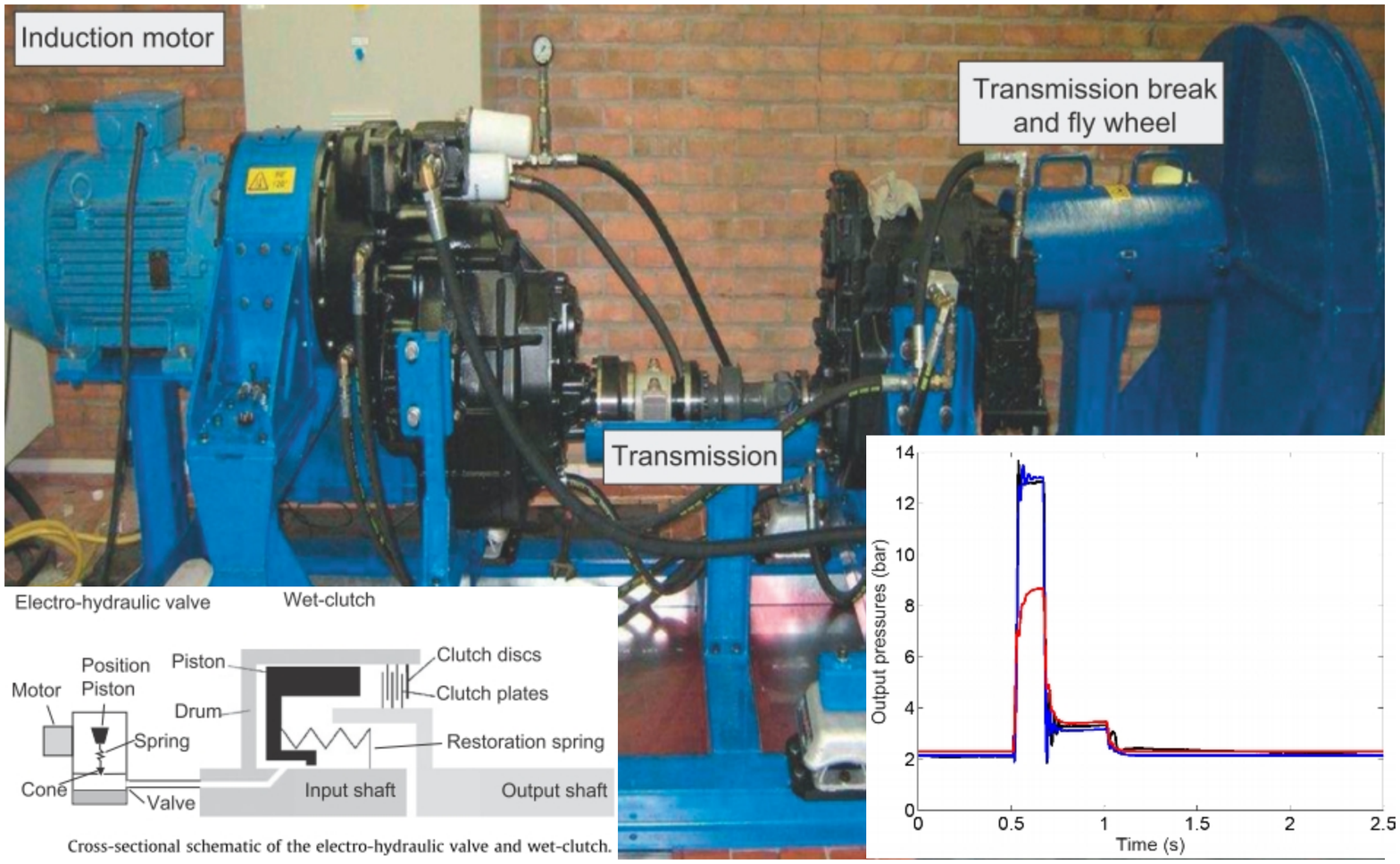}
\caption{A wet-clutch device transmits torque from an input axis to an output axis via fluid friction prior to the engagement of the friction plates. Such devices are commonly used in automatic transmissions for off highway vehicles and agricultural machines to transfer torque from the engine to the load. An electro-hydraulic proportional valve regulates the pressure inside the clutch which causes the engagement of the piston with the friction plates. A model describing the relation between the current applied to the motor of the elctro-hydraulic valve and the resulting pressure during the filling stage of the clutch is built, to bring about a smooth engagement. A linear model (red line) fails to model the true oil pressure (black line) for the spiky input, while a nonlinear state space model (blue line) matches very well with the real data \cite{Widanage2011}. The example is further discussed in Figure \ref{fig:Wet-ClutchInput}}
\label{fig:WetClutch}
\end{figure}

\begin{figure}[h] 
\centering
\includegraphics[scale=0.7]{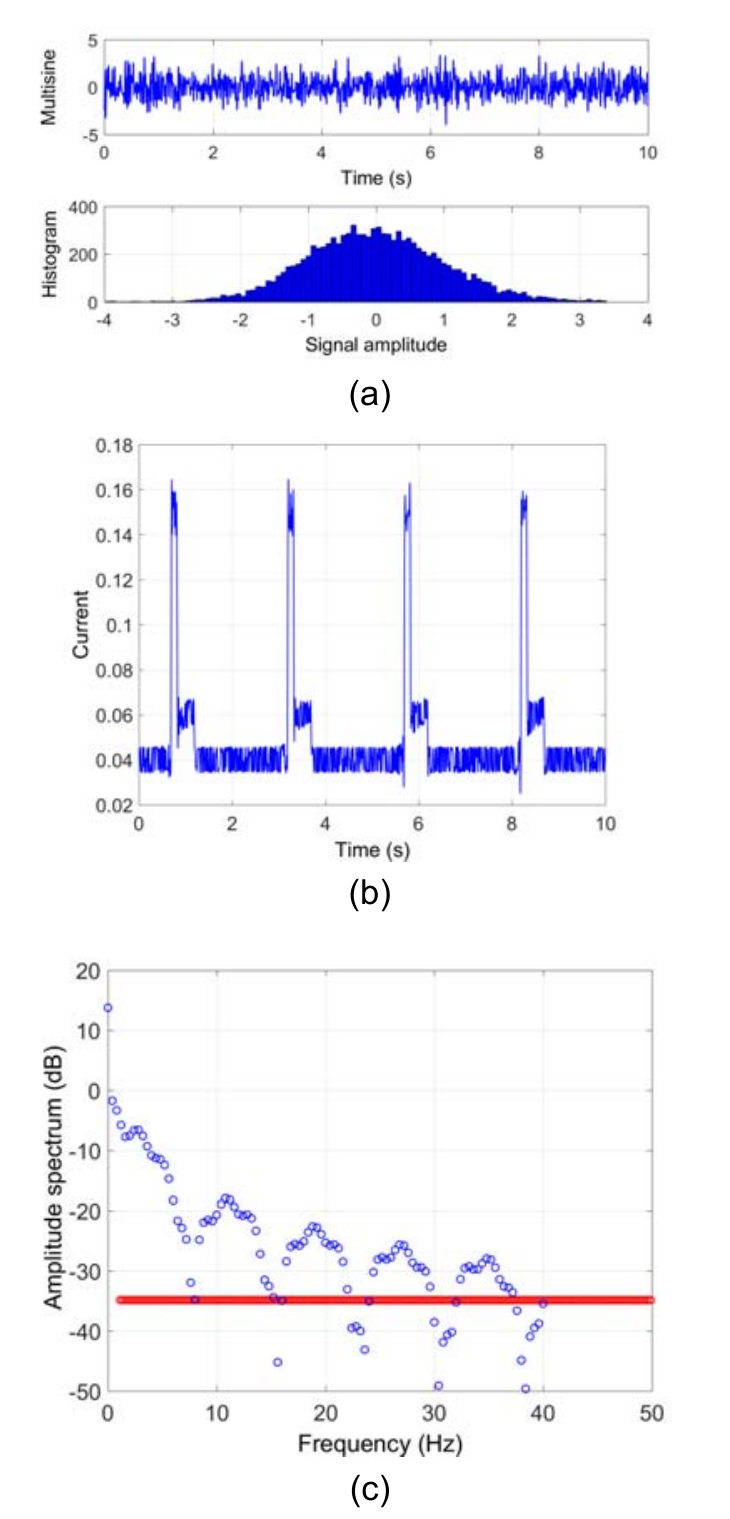}
\caption{Design of excitation signals for the Wet-Clutch setup in Figure \ref{fig:WetClutch}. A random phase multisine (a) has a Gaussian amplitude distribution. This does not fit with the spiky signals that are typically applied to the Wet-Clutch as shown in Figure \ref{fig:WetClutch}. The nonlinear state space models that were identified with these signals failed completely to simulate the system for the spiky signals that were applied during the ILC (iterative learning control) design. The models became even unstable. For that reason, a band limited periodic signal was designed that consists of the sum of two signals (b). The first signal is a band limited approximation of the spiky signal, see blue spectrum in (c). A multisine with a flat and small amplitude spectrum, red in (c), is added to it. The red components excite the dynamics around the spiky profile. This results in a rich excitation that mimics very well the future use of the model. This guarantees that the structural model errors will be small for the intended application.}
\label{fig:Wet-ClutchInput}
\end{figure}

A simulation model of the Wet-Clutch setup in Figure \ref{fig:WetClutch} was used during an iterative control design (ILC) for the system. The goal was to obtain a fast but smooth engagement of the clutch. To reach that goal, spiky signals with a large amplitude are used to get a short filling phase, followed by a small excitation to get a smooth engagement. The design of rich excitation signals that mimic this behavior is discussed in Figure \ref{fig:Wet-ClutchInput}.

\subsection{Choice of the cost function}
\emph{No structural model errors present}

In linear system identification, the classical choice for the cost function is \cite{Ljung1987,Soderstrom1989}
\begin{equation}\label{eq:CostClassicTD}
\begin{aligned}
V_{N}(\theta)&=\frac{1}{N}\sum_{t=1}^{N}\frac{1}{2}\epsilon^{2}(t,\theta),\\
\epsilon(t,\theta)&=H^{-1}(q,\theta)[y(t)-G(q,\theta)u(t),
\end{aligned}
\end{equation}
with $G$ the plant model, and $H$ the noise model. This can also be written in the frequency domain\cite{ Ljung1987} (neglecting the begin and end effects that create leakage errors \cite{SchoukensJ2018a}) as
\begin{equation}\label{eq:CostClassicFD}
V_{N}(\theta)=\frac{1}{N}\sum_{k=1}^{N}\frac{1}{2}\frac{|Y(k)-G(k,\theta)U(k)|^{2}}{|H(k,\theta)|^{2}}.
\end{equation}
$U(k),Y(k)$ are the DFT of $u,y$ evaluated at the frequency $f_{k}=\frac{kf_{s}}{N}$, and $G(k,\theta),H(k,\theta)$ are the plant and noise transfer function evaluated at $f_{k}$. 
In the prediction error framework, the parametric plant and noise model are simultaneously estimated by minimizing the cost fucntion $V_{N}(\theta)$ with respect to $\theta$. It is shown \cite{Ljung1987} that this is the maximum likelihood formulation of the identification problem if the disturbing noise $v$ is Gaussian distributed.

An alternative is to replace the parametric noise model by a nonparametric measurement of the noise variance $\sigma_{v}^{2}(k)=|H(k,\theta)|^{2}$. The variance $\sigma_{v}^{2}(k)$ is directly estimated from the data in a nonparametric preprocessing step, using periodic excitations \cite{Pintelon2012,SchoukensJ2018a} (see ``Nonparametric Noise and Distortion Analysis Using Periodic Excitations''), or starting from arbitrary excitations using the recently developed nonparametric estimation methods \cite{Schoukens2009}. Observe that in this approach, the estimation of the noise model does not depend upon the plant model, so that a too simple plant model will not affect the noise model. The cost function becomes
\begin{equation}\label{eq:CostNonParFD}
V_{N}(\theta)=\frac{1}{N}\sum_{k=1}^{N}\frac{1}{2}\frac{|Y(k)-G(k,\theta)U(k)|^{2}}{\sigma_{v}^{2}(k)}.
\end{equation}

In the absence of structural model errors there is a full equivalence between both approaches \cite{SchoukensJ1999a,Pintelon2012}, the differences are mostly on the implementation side \cite{SchoukensJ2011}. This picture changes completely if structural model errors are present.

The discussion in the section can be directly generalized to nonlinear systems by replacing the linear model $G(q,\theta)u(t)$ by the nonlinear model $h(u,\theta)$. A further generalization would be to include also a nonlinear noise model to deal for example with process noise (see "Process Noise in Nonlinear System Identification").

\emph{Structural Model errors present}

In the presence of structural model errors, the parametric noise model $|H(k,\theta)|^{2}$ will account for the power in both the disturbing noise $v(t)$ and the  structural model errors $y_{\varepsilon}(t)$. Moreover, the maximum likelihood motivation is no longer valid because the structural model errors are not independent of the input $u$. 

The nonparametric noise analysis based on periodic excitations will still estimate $\sigma_{v}^{2}(k)=|H(k,\theta)|^{2}$, while for the advanced nonparametric methods \cite{SchoukensJ2009, Pintelon2010a,Pintelon2010b} a combination of the disturbing noise variance and the mean squared structural model errors will be retrieved.

This raises the question what is the best choice for the weighting function in  (\ref{eq:CostClassicTD}), (\ref{eq:CostClassicFD}), (\ref{eq:CostNonParFD})  in the presence of structural model errors.  It makes no sense to weight the structural model errors with the noise variance if the structural model errors dominate the noise. The weighting function should rather reflect in what frequency band larger structural model errors can be tolerated and where small structural model errors are needed.  This is done by replacing the noise variance based weighting $H^{-1}(q,\theta)$ in \eqref{eq:CostClassicTD} or  $1/\sigma_{v}^{2}(k)$ in \eqref{eq:CostClassicFD} by a predefined frequency weighting $L^{-1}(q)$ or $1/w^{2}(k)$, chosen by the user, that reflects the most acceptable behavior of the structural model errors for the application in mind:

\begin{equation}
     \begin{aligned} \label{eq:CostNLTD}
V_{N}(\theta)&=\frac{1}{N}\sum_{t=1}^{N}\frac{1}{2}\epsilon^{2}(t,\theta),\\
\epsilon(t,\theta)&=L^{-1}(q)[y(t)-G(q,\theta)u(t).
\end{aligned}
\end{equation}
 Observe that, in contrast to  \eqref{eq:CostClassicTD}, the weighting filter $L^{-1}(q)$ does no longer depend on $\theta$. In the frequency domain, the cost function  \eqref{eq:CostNLTD} is   
\begin{equation}
       \hat \theta_N = \text{argmin}_\theta\sum_{f\in F} \frac{|Y(f)-\hat Y(f|\theta)\|^2}{|L(f)|^{2}}.
 \end{equation}
By taking the sum only over the frequencies of interest $f\in F$, the fit is focused on the frequency band of interest. 

\subsection{Covariance matrix expressions}
The covariance matrix of the parameter estimates $C_{\theta}$ for the linear-least-squares problem $y=K(u)\theta+w$, where $y,w \in \mathbb{R}^{N\times 1}$, $\theta \in \mathbb{R}^{n_{\theta}\times 1}$, and $K(u)\in \mathbb{R}^{N\times n_{\theta}}$ is given by \cite{Ljung1987,Soderstrom1989,Pintelon2012}
\begin{equation}
\label{eq:CovExpression}
C_{\theta}=E_{u,w}[(\frac{1}{N}K(u)^{T}K(u))^{-1}\frac{1}{N}K(u)^{T}ww^{T}K(u)\frac{1}{N}(K(u)^{T}K(u))^{-1}].
\end{equation}

\subsubsection{No structural model errors present: $w(t)=v(t)$ \emph and $y_{\varepsilon}(t)=0$}
In this case, $w(t)$ is independent of $u$, and \eqref{eq:CovExpression} converges for $N\rightarrow\infty$ to
\begin{equation} \label{eq:CovGen}
C_{\theta}=E_{u}[K(u)^{T}K(u)]^{-1}]E_{u,w}[K(u)^{T}ww^{T}K(u)]E_{u}[K(u)^{T}K(u)]^{-1}].
\end{equation}
Making use of the independency of $w$ and $u$, it follows that $w$ is independent of $K(u)$, so that  
\begin{equation}
E_{u,w}[K(u)^{T}ww^{T}K(u)]=E_{u}[K(u)^{T}E_{w}[ww^{T}]K(u)]=E_{u}[K(u)^{T}C_{w}K(u)],
\end{equation}
and the covariance matrix becomes
\begin{equation}
C_{\theta}=E_{u}[K(u)^{T}K(u)]^{-1}]E_{u}[K(u)^{T}C_{w}K(u)]E_{u}[K(u)^{T}K(u)]^{-1}].
\end{equation}
For $w$ being white noise, $C_{w}=\sigma^{2}_{w}I$, and
\begin{equation} \label{eq:VocSimplified}
C_{\theta}=\sigma^{2}_{w}E_{u}[K(u)^{T}K(u)]^{-1}.
\end{equation}

\subsubsection{Structural model errors present: $w(t)=v(t)+y_{\varepsilon}(t)$}
In the structural model error case, $K(u)$ and $w$ depend both upon the input $u$, so that the independency is lost, and $E_{u,w}[K(u)^{T}w^{T}wK(u)]\neq E_{u}[K(u)^{T}E_{w}[w^{T}w]K(u)]$. In that case, higher order moments of $u$ show up in the calculation of $E_{u,w}[K(u)^{T}w^{T}wK(u)]$, and the general expression (\ref{eq:CovGen}) should be used. Since the dependency of $w$ on $u$ is not known (the structural model error is unknown), it becomes in general impossible to get closed form expressions for the covariance matrix $C_{\theta}$. This creates a huge problem because the classical but simplified expression (\ref{eq:VocSimplified}) underestimates the variability, providing the user with a far too optimistic estimate of the uncertainty on the estimates \cite{Hjalmarsson1992}. Estimating the variability directly from a set of repeated experiments with varying inputs is a pragmatic solution to this problem, but provides of course no closed form expressions \cite{Hjalmarsson1992}.

\emph{Example 1: linear approximation of $y=u^{n}$}
It is shown in \cite{SchoukensJ2010a} that the underestimation of the variance is maximal for static nonlinear systems. These can be approximated arbitrary well in mean square sense using a polynomial representation. For that reason, the study of static nonlinear systems $y_{0}(t)=u(t)^{n}$ is very informative. For such a system, excited with Gaussian noise, the best linear approximation $G_{BLA}$ is constant (see "Linear Models of Nonlinear Systems: More on the best linear approximation $G_{BLA}$") \cite{Bussgang1952, Gelb1968, Enqvist2005a, Enqvist2011, Pintelon2012}, and hence, $G_{BLA}(q, \theta) = a_{BLA},$ which is only different from zero for $n$ odd. It is also possible to explicitly calculate the ratio between the full nonlinear-induced variance (structural model errors present, input dependent residuals) and the classical variance (no structural model errors, input independendent residuals)  of $\hat{a}_{BLA}$ \cite{SchoukensJ2010a}
\begin{equation}
\frac{\sigma^{2}_{Ass_(dependant_errors)}}{\sigma^{2}_{Ass_(independent_errors)}}=2n+1.
\end{equation}
This shows that the underestimation of the variance by (\ref{eq:VocSimplified}) grows with the nonlinear
degree $n$.

\begin{figure}[h] 
\centering
\includegraphics[scale=1]{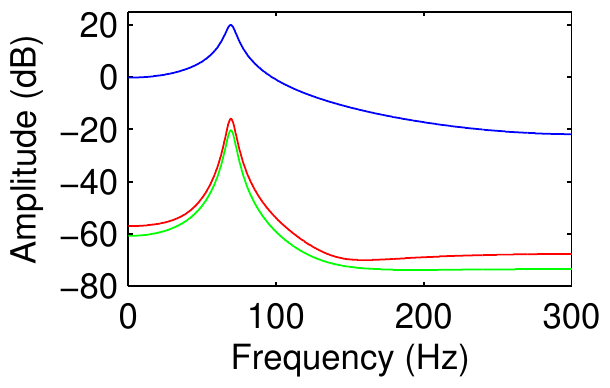}
\caption{The amplitude of the estimated transfer function model is shown (blue). Green line: the theoretic standard deviation of the estimated plant model, calculated under the assumption that there are no structural model errors ($w$ is independent of $u$). Red line: the actual observed standard deviation of the estimated plant model, estimated from the variations of the estimated plant model over the 10 subrecords. It can be seen that the actually observed standard deviation is underestimated with 4 dB by the simplified theoretical analysis. This leads to too small error-bounds.}
\label{fig:SilverboxVariability}
\end{figure}

\emph{Example 2: the forced Duffing oscillator}
A linear approximating model is estimated to the silverbox from the experimental data (tail part) discussed in "Simulation errors and prediction errors. The tail is split in 10 sub-records with a length of 8692 points, and each of these is used to identify a second-order discrete-time plant model and a sixth-order noise model using the Box-Jenkins model structure of the prediction-error method \cite{Ljung1987, Soderstrom1989}. The estimated second-order plant transfer function, is shown in Figure \ref{fig:SilverboxVariability}. The estimation procedure resulted in the plant and noise model. From this information it is possible in the structural model error free system identification approach to obtain also an estimate of the variance on the results. In Figure \ref{fig:SilverboxVariability}, the estimated standard deviation of the transfer function is compared with the sample standard deviation that is calculated from the repeated estimates on the 10 subrecords. Both curves look very similar, but the model-based estimated value (green) underestimates the actual observed standard deviation (red) by 50\% or more. This is due to the fact that the system identification framework fails to estimate precisely the uncertainty in the presence of nonlinear structural model errors. The user should keep in mind that, whenever structural model errors are present, the confidence bounds are wrong.

\begin{figure}[h] 
\centering
\includegraphics[scale=0.5]{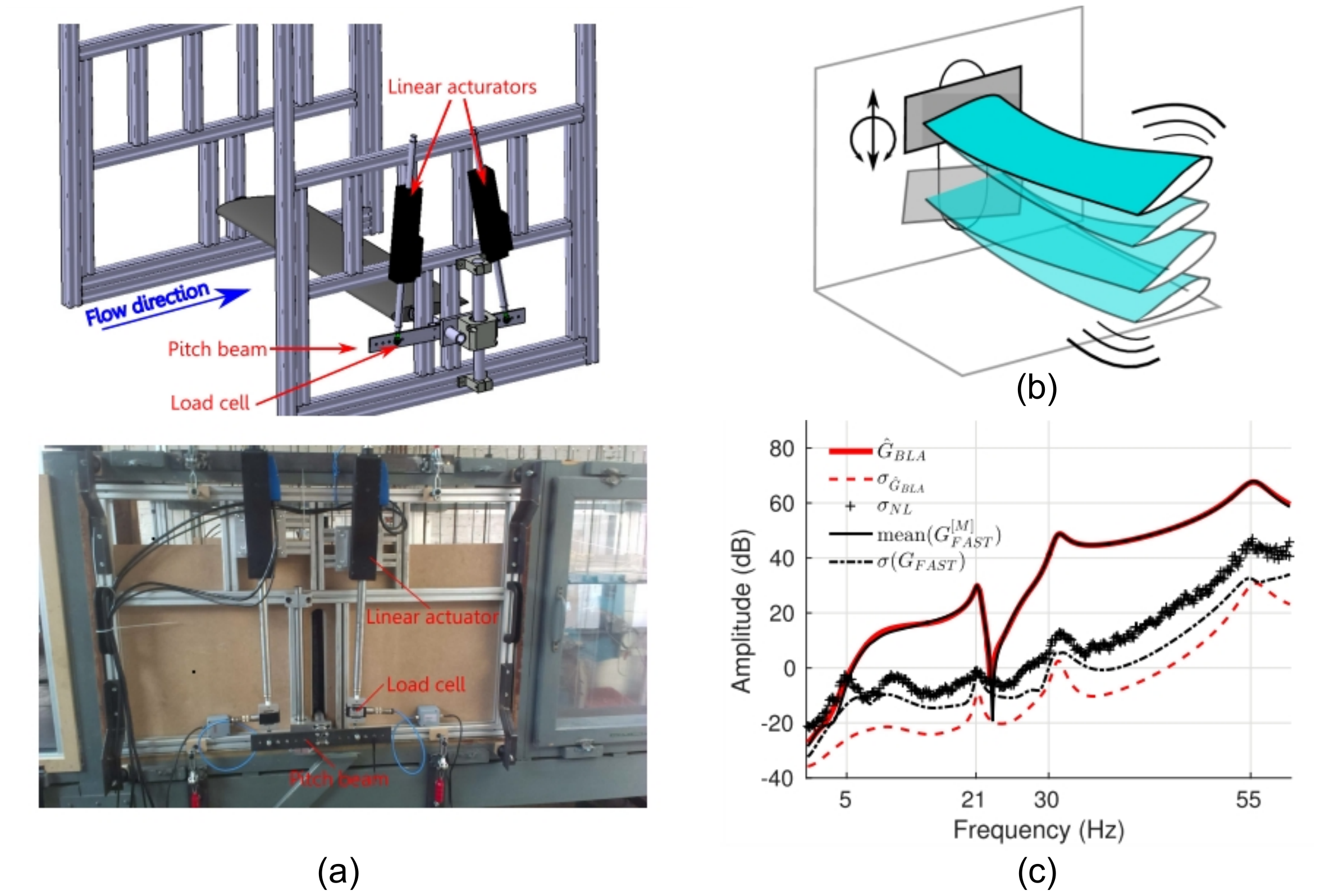}
\caption{Illustration on a wind tunnel experiment \cite{Ertveldt2015} of the under estimated standard deviation on parametric models in the presence of nonlinear model errors. In this experiment \cite{Ertveldt2017}, the transfer function is measured from the forced displacement (a random phase multisine) \cite{Pintelon2012} at the root of a wing mounted in a wind tunnel (a,b), to the acceleration of the tip of the wing (b). The windspeed is 8 m/s, and the angle of attack  is $17.5^{\circ}$. The nonlinear distortion level (+), obtained from a nonparametric analysis, is 20 dB above the noise floor (not shown on this figure). The measurements are repeated for 8 realizations of the random phase multisine input.  (bottom). The best linear approximation (BLA) for this nonlinear system is estimated. The solid red line in (c) shows the parametric transfer function model of the BLA obtained by processing all realizations together. The BLA is also estimated for each individual realization. The black solid line that coincides with the red line shows the mean value of these individual estimates. The theoretical standard deviation (broken red line) that is obtained under the assumption that the errors are independent of the input, under estimates the actual observed standard deviation with 11 dB (about a factor 3). This is a typical result when structural model errors dominate the noise.}
\label{fig:WindTunnel}
\end{figure}

\emph{Example 3: Wind tunnel experiment}
In Figure \ref{fig:WindTunnel}, the under estimation of the variability in the presence of nonlinear model errors is illustrated on a wind tunnel experiment.  In this experiment \cite{Ertveldt2017}, the transfer function of the best linear approximation (see ``Linear Models of Nonlinear Systems'') is measured from the forced displacement (a random phase multisine) \cite{Pintelon2012} at the root of a wing mounted in a wind tunnel (a,b), to the acceleration of the tip of the wing (b). The simplified variance analysis (neglecting the dependency of $w$ on the input) and the actual observed variance obtained from different realizations of the experiment are shown. The simplified analysis under estimates the actual observed standard deviation with 11 dB (about a factor 3).

\subsection{Optimal strategy to generate simplified models}
A key issue in system identification is how to cope with high system complexity. Sometimes structural model errors are unavoidable because too complex models are needed to include the system in the model class. In other situations structural model errors are deliberately created because simple models are needed in the next phase of the (control) design process.  In that case the experiment can be designed such that some complex system behavior is concealed and the simple model still performs well on the domain of interest \cite{Hjalmarsson2009}. 

Identifying simplified models leads to structural model errors and brings all the questions discussed before into the picture. This raises also the question for the best strategy: 1) Identify first the most complex model that is affordable (reduce the structural model errors as much as possible), followed by a model reduction step, or 2) Identify directly a too  simple model, dealing directly with the structural model errors. 

In Appendix VI: A Separation Principle of \cite{Hjalmarsson2009} it is shown, using the Maximimum Likelihood Invariance principle, that the first strategy is the best choice to follow, if it is affordable, because it results in an asymptotically efficient estimate (smallest variance) that is consistent (retrieve the 'true' value if the number of data goes to infinity). Moreover, it allows the user to separate the identification and model reduction step: use the maximum likelihood framework in the first step (resulting in efficient estimates with the lowest uncertainty), followed by a  model reduction step using an application oriented cost function. This two step approach allows the user also to make a proper characterization of the reliability of the simplified model. For that reason this is the best strategy whenever it is affordable.

However, often this first option is not affordable, and then the only solution is to identify directly a simple model in a setting with structural model errors. In that case, the following user guidelines help to face that situation.

\subsection{User guidelines: how to deal with structural model errors?} 
\begin{itemize}
\item \emph{Experiment design}: The experiment should come as close as possible to the future applications so that the structural model errors are guaranteed to be small under these conditions. For example, the best simplified model can be completely different for a Gaussian or a sine excitation. The same holds true for varying power spectra, amplitude ranges, and operating points.
\item \emph{Choice of the cost function}: When structural model errors dominate, the maximum likelihood paradigm that proposes a cost function based on the disturbing noise properties is no longer the natural choice. The cost function should reflect the users needs and express where smaller structural model errors are needed and larger structural model  errors can be tolerated. This can be done by using for example a relative error in the cost, or by proposing a user defined weighting function. 

\item \emph{Frequency weighting}: The frequency weighting of the errors should no longer reflect the disturbing noise variance when the structural model errors $y_{\epsilon}$ dominate the disturbing noise $v$. Instead, the weighting should be chosen to make sure that the structural model errors remain small in the frequency band of interest.

\item \emph{Did we loose 50 years our time?} From the discussions in this section, the reader could get the impression that the past efforts on the development of a system identification framework are in vane when structural model errors show up. This is certainly not true! The lessons learned from the classical system identification approach should not be forgotten, the clearly structured picture that is provided in the classical text books like \cite{Ljung1987,Soderstrom1989,Pintelon2012} is still valid at full power and provides still a road-map how to organize the system identification process. 

Improper data handling can over-emphasize small noise disturbances, making the identification process again vulnerable to noise disturbances that are far below the structural model error level. For that reason the user is still strongly advised to make a clear split between the experiment design, the model class selection, the choice of the cost function, and the choice of the numerical procedures used to minimize the cost function. The consistency analysis developed in the structural model error free framework still provides insight about the convergence of the algorithms in the presence of structural model errors. The major open issue that is still unsolved today is how to generate reliable uncertainty bounds in the presence of structural model errors.

\item \emph{Uncertainty bounds}: The covariance matrix that is generated by the structural model error free system identification framework is wrong in the presence of structural model errors. This is still like that if the noise model is tuned to include also the structural model errors. The covariance matrix underestimates the true value because it does not account for the dependency between the structural model errors and the input. This is well in line with the rule of thumb: do not pay any attention to the model's uncertainty bounds if the validation test fails. At this moment there is no theoretical framework available that can provide more reliable bounds. For that reason the user is advised to repeat the experiment for different excitation signals (for example different realizations of a random excitation) and to look directly to the variability of the results under these conditions (see Figure \ref{fig:SilverboxVariability}) \cite{Hjalmarsson1992}.

\end{itemize}

\section{Sidebar\\ -- Linear Models of Nonlinear Systems}
Nonlinear models are clearly  much more versitile and complicated than linear models. A general linear model can be represented and fully characterized by a \emph{Transfer function} $G(s)$ which is the Laplace transform of the model's impulse response $g(\tau)$:
\begin{align}
  \label{eq:tfct}
  G(s) = \int_{\tau=0}^\infty g(\tau)e^{-s\tau}d\tau
\end{align}
(In discrete time the transfer function $G(z)$ is the $Z$-transform of the impulse response.)

Due to the simplicity of linear models, it is tempting and common to work linear approximations of nonlinear systems. Possibly one can work with several linear models to capture different aspects of the nonlinear system. Two ways of defining linear approximations will be used:
\subsection{Linearization around an Equilibrium}
Consider a general nonlinear state space model:
\begin{subequations}
 \label{eq:dss_l}
  \begin{align}
  \dot x(t) &= f(x(t),u(t))\\
y(t) &= h(x(t)
\end{align}
\end{subequations}

Suppose the input is constant: $u(t)=u^*$, and that there is a corresponding state equilibrium $x^*; \quad f(x^*,u*)=0$. Define the deviations 
\begin{subequations}
\label{eq:sdelta}
\begin{align}\Delta u(t) &= u(t)-u^*\\ \Delta x(t) &= x(t)-x^*\\
\Delta y(t) &= y(t)-y^*=y(t)-h(x^*)
\end{align}
\end{subequations}

By expanding the nonlinear functions $f,h$ in Taylor series around $x^*,u^*,y^*$ and neglecting terms of higher order, we obtain a linear state space equation
\begin{subequations}
\label{eq:sslinear}
\begin{align}
\dot \Delta x(t) &= A\Delta x(t) + B \Delta u(t) \\
\Delta y(t) &= C\Delta x(t)\\
A &= \frac{\partial}{\partial x}f(x,u)|_{x^*.u^*}\quad B = \frac{\partial}{\partial u}f(x,u)|_{x^*,u^*} \\
 C&=\frac{\partial}{\partial x}h(x,u)|_{x^*,u^*}
\end{align}
\end{subequations}

So in the vicinity of the equilibrium, the nonlinear system (\ref{eq:dss_l}) can be approximated by the linear transfer function $G(s)=C(sI-A)^{-1}B$. This is a well known and commonly used linearization. Corresponding expressions apply in discrete time.
\subsection{Stochastic Linearization}
Suppose the nonlinear system (\ref{eq:dss_l}) is excited by an input $u$ with a spectrum $\Phi_u(\omega)$, and the cross-spectrum between output and input $\Phi_{yu}(\omega)$ is well defined.  The \emph{second order properties} of the input and output signals (i.e the covariance functions and spectra) are thus well defined. 
For this input,  define the  linear model \cite{Ljung2001}:
\begin{align}
\label{eq:SL}
G_{BLA}(\omega)=\Phi_{yu}(\omega)[\Phi_u(\omega)]^{-1}
\end{align}
that has the same second order properties as the nonlinear system. By considering second order signal properties (for this input) the nonlinear system  (\ref{eq:dss_l}) \emph{thus cannot be distinguished from the linear model} $G_{BLA}$, They are \emph{second order equivalents}.

This also means that   a linear model is estimated with standard linear identification methods (that only use the second order propertires of the signals) the estimate will converge to $G_{BLA}$. For that reason, the linear second order equivalent is also called \emph{BLA: Best Linear Approximation}. Note that the BLA of a nonlinear system will depend on the input signal spectrum!

Note also that the definition of spectra does not require a stochastic setting. It is sufficient  that the following limita exist  (in discrete time)
\begin{subequations}
\begin{align}
  \label{eq:spec}
 R_{yu}(\tau)&=\lim_{N\to\infty}\frac{1}{N}\sum_{t=1}^Ny(t)u(t-\tau)\\
\Phi_{yu}(\omega)&= \sum_{\tau=-\infty}^\infty R_{yu}(\tau)e^{-i\omega \tau}
\end{align}
\end{subequations}
for the input -output signals \cite{Ljung2001}, and correspondingly for $\Phi_u$.

\subsection{More on The best linear approximation $G_{BLA}$}
In this section, the best linear approximation model is further analyzed. The reader is referred to \cite{SchoukensJ2016} for a more extensive introduction.

\emph{The best linear approximation $G_{BLA}$} in \eqref{eq:SL} represented either by its impulse response
$g_{BLA}(t)$, or its FRF $G_{BLA}(\omega)$ is the solution of \cite{Pintelon2012, Enqvist2005, Schoukens2009}
\begin{equation}
G_{BLA}=\arg\min_{G}E\left\{ \left|y_{0}\left(t\right)-G\left(q\right)u\left(t\right)\right|^{2}\right\},
\label{eq: 2 def Gbla}
\end{equation}
with $q$ the shift operator for a discrete time model. Similar expressions can be given for continuous time models. All expected values $E\left\{ \right\} $ are taken with respect to the random input $u(t)$.
In most applications, the DC-value of the input and output signal should be removed in order to obtain a model that is valid around a given setpoint. 

The transfer function $G_{BLA}(k)$ depends on the characteristics of the input signal. Changing the power spectrum or the amplitude distribution (for example, replacing a Gaussian  by a uniform distribution) of the excitation will change the BLA.

\emph{The nonlinear 'noise' source $y_{s}(t)$}: The difference between the output of the nonlinear system and that of the BLA $y_{s}(t)=y(t)-G_{BLA}(q)u(t)$ is called the stochastic nonlinear contribution or nonlinear noise. Although this name might be misleading (the error is deterministic for a given input signal), it is still preferred to call it a stochastic contribution because it looks very similar to a noise disturbance for a random excitation \cite{Schoukens1998,Pintelon2012}. Because $y_{s}(t)$ is the residual of a least squares fit, it is uncorrelated with the input. However, in general, it is still dependent on the input. The properties of $G_{BLA}$ and $Y_{s}$ are well known for Gaussian and Rieman-equivalent \cite{Schoukens2009} excitations.

\emph {A new paradigm}: Combining both results, the output of the nonlinear system can be written as

\begin{equation}
\begin{aligned}
y(t) & =  y_{0}(t)+v(t) \\
y_{0}(t) & =  G_{BLA}(q)u(t)+y_{S}(t),
\end{aligned}
\end{equation}
where $y_{s}(t)$ is uncorrelated put dependent on the input $u(t)$.

\begin{figure}[h] 
\centering
\includegraphics[scale=0.75]{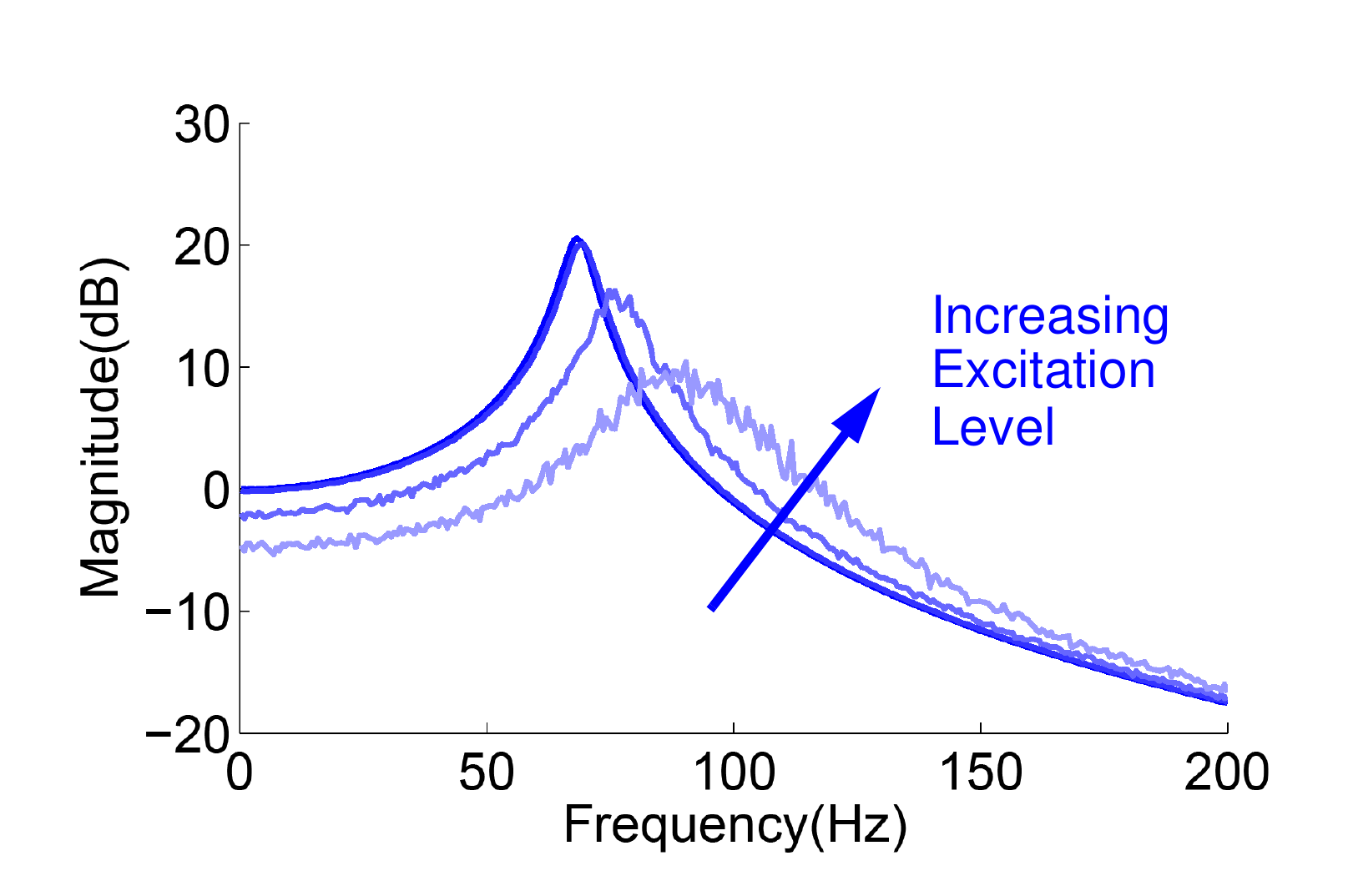}
\caption{Measured frequency response function of the BLA of the forced Duffing oscillator. Observe that there is a systematic shift, and the disturbances grow with the excitation level. The shift is due to the systematic nonlinear contributions that create a shift in the dynamics of $G_{BLA}$. The increased apparent noisy behavior is due to the stochastic nonlinearities $y_s$  that grows with the excitation level}
\label{fig:SilverboxBLA}
\end{figure}

\emph{Experimental illustration on the Duffing oscillator}:
In this example, measurements on the forced Duffing oscillator are shown (see \cite{SchoukensJ2016} for more details). The FRF is measured for 4 different excitation levels and shown in Figure \ref{fig:SilverboxBLA}. For each excitation level, the FRF is averaged over 50 realizations of the input signal to obtain a smoother result. Two observations can be made: i) The resonance frequency shifts to the right for increasing excitation levels, and ii) the measurements become more noisy. Both effects are completely due to the nonlinear distortions. 

\emph{The BLA of a static nonlinearity: a simple example}: The best linear approximation of a static nonlinearity excited with a Gaussian excitation is a constant \cite{Bussgang1952}. For example, the BLA of $y=u^3$ is
\begin{equation}
\begin{aligned}
y(t)&=u(t)^3 \text{  with  } u\sim N(0,\sigma^{2}_{u}),\\
G_{BLA}&=E\{y(t)u(t)\}/E\{u(t)^{2}\}=3\sigma^2_{u}.
\end{aligned}
\label{eq:BLAStatic1}
\end{equation}

Observe that 
\begin{equation}
\begin{aligned}
y_{s}(t)&=y(t)-G_{BLA}u(t)\\
&=u(t)^{3}-3\sigma^2_{u}u(t) 
\end{aligned}
\label{eq:BLAStatic2}
\end{equation}
is uncorrelated but dependent upon $u$. This is a general valid observation.

\emph{Impact of process noise on the BLA}:
Process noise $w(t)$ coming into the system before the nonlinearity creates mixing terms with the input signal so that at the output of the system the process noise contributions are no longer independent of input.  This will also affect the BLA, and a generalization of the framework is needed. The reader is referred to  \cite{SchoukensM2018} for a full discussion, here the simple example (\ref{eq:BLAStatic1}) is extended as an illustration. Assuming that the process noise $w(t)$ is independent of the input $u(t)$, the BLA becomes

\begin{equation}
\begin{aligned}
y(t)&=(u+w)^{3}\text{  with  } u\sim N(0,\sigma^{2}_{u}),w\sim N(0,\sigma^{2}_{w}),\\
G_{BLA}&=E\{y(t)u(t)\}/E\{u(t)^{2}\}=3\sigma^2_{u}+3\sigma^2_{w}.
\end{aligned}
\label{eq:BLAStatic3}
\end{equation}
This shows that the 'averaged' behavior of a nonlinear system is strongly affected by the presence of process noise entering the system before the nonlinearity.

\subsection{Summary}
\begin{itemize}
\item \emph{Best linear approximation $G_{BLA}$}: It is possible to identify a simplified representation, for example the linear model $G_{BLA}$, for a nonlinear system.
 \item \emph{Stochastic nonlinearities $y_{s}$}: For a random excitation, the structural nonlinear model errors, called stochastic nonlinearities  $y_{s}$, look like noise. $y_{s}(t)$ is uncorrelated but not independent of the input $u(t)$. For an untrained user, it is very hard to distinguish process or measurement noise from $y_{s}$ (the nonlinear model errors). Different actions are needed to deal properly with both effects (see \cite{SchoukensJ2016}). 
\item \emph{$G_{BLA}$ depends on the nature of the input}: The best simplified approximation ($G_{BLA}$) depends on the nature of the excitation. Changing the power spectrum or input distribution can change $G_{BLA}$.
\item \emph{Structure detection}: The variations of $G_{BLA}$ for varying experimental conditions provide insight on the structure of the nonlinear system. For example, moving poles (resonances) are only possible for nonlinear closed loop systems. The reader is referred to \cite{Schoukens2015} for a full overview.
\item \emph{Process noise}: Unmeasured random inputs (process noise) coming into the system before the nonlinearity can affect the output in a systematic way. The BLA with respect to the known input depends on the process noise properties.
\end{itemize}
 
\section{Sidebar\\   Process Noise in Nonlinear System Identification}

Process noise $w(t)$, as defined in \eqref{eq:ProcNoiseMeasNoise}, affects also the system's internal signals and not only the measurements. In this section, the impact of process noise is studied in more detail. The process noise $w(t)=H_{w}(q)e_{w}(t)$ is considered to be a zero mean, white or colored, noise variable.

\emph{Process noise in linear systems}: It is well known that for linear systems, the effects of process noise can be collected at the output as an additive noise term $H_{w}(q)w(t)$, where it is combined with the measurement noise $v(t)=H_{v}(q)e(t)$ into one noise term:
\begin{equation}
\begin{aligned}
 y(t)&=y_{0}(t)+H_{w}(q)e_{w}(t)+H_{v}(q)e_{v}(t)\\
&=y_{0}(t)+H(q)\tilde{e}(t)\\
&= y_{0}(t)+\tilde{v}(t).
\end{aligned}
\label{eq:ProcNoiseLin}
\end{equation}
The disturbances $e_{w}(t),e_{v}(t)$ are mutually independent distributed, and also to be independent of the input $u(t)$. Under these conditions, the combined effect of the process and the measurement noise result in a zero mean distributed output disturbance that is independent of the input.

\emph{Process noise in nonlinear systems}: In nonlinear systems, process noise can have a structural impact on the identified models as was illustrated in \eqref{eq:PN1} and \eqref{eq:BLAStatic3}. Zero mean process noise can change for example the linear gain of a nonlinear system, and the apparent disturbances at the output can become nonstationary and depend upon the input as discussed in more detail later in this section.

\subsection{Process noise: curse or blessing?} 
\emph{Curse}: In most applications, process noise is very disturbing. It is more difficult to control a system in the presence of process noise. Also the system identification methods become much more involved as explained below. For that reason, process noise is mostly considered as a very annoying effect.   

\emph{Blessing}: However, in some applications process noise is used to reduce or linearize the averaged effect of abrupt nonlinearities, this is called dithering. Desired effects of dithering are augmenting the linearity of the open or closed loop system, increasing the robustness and asymptotic stability \cite{Gelb1968, Arfei2013,Atherton1982}. Dither can be used to reduce the effect of Coulomb friction, dead zones in hydraulic valves valves and hysteresis effects, but this can come with an increased wear due to the rapid motions \cite{Atherton1982}. Dithering is also employed in digital instrumentation and data acquisition systems in order to improve their measurement-related characteristics. A typical example is the use of dithering in an analog-to-digital converter (ADC) to improve the resolution, dynamic range, and spectral purity below the quantization level \cite{Carbone2000}.  

\begin{figure}[h] 
\centering
\includegraphics[scale=0.5]{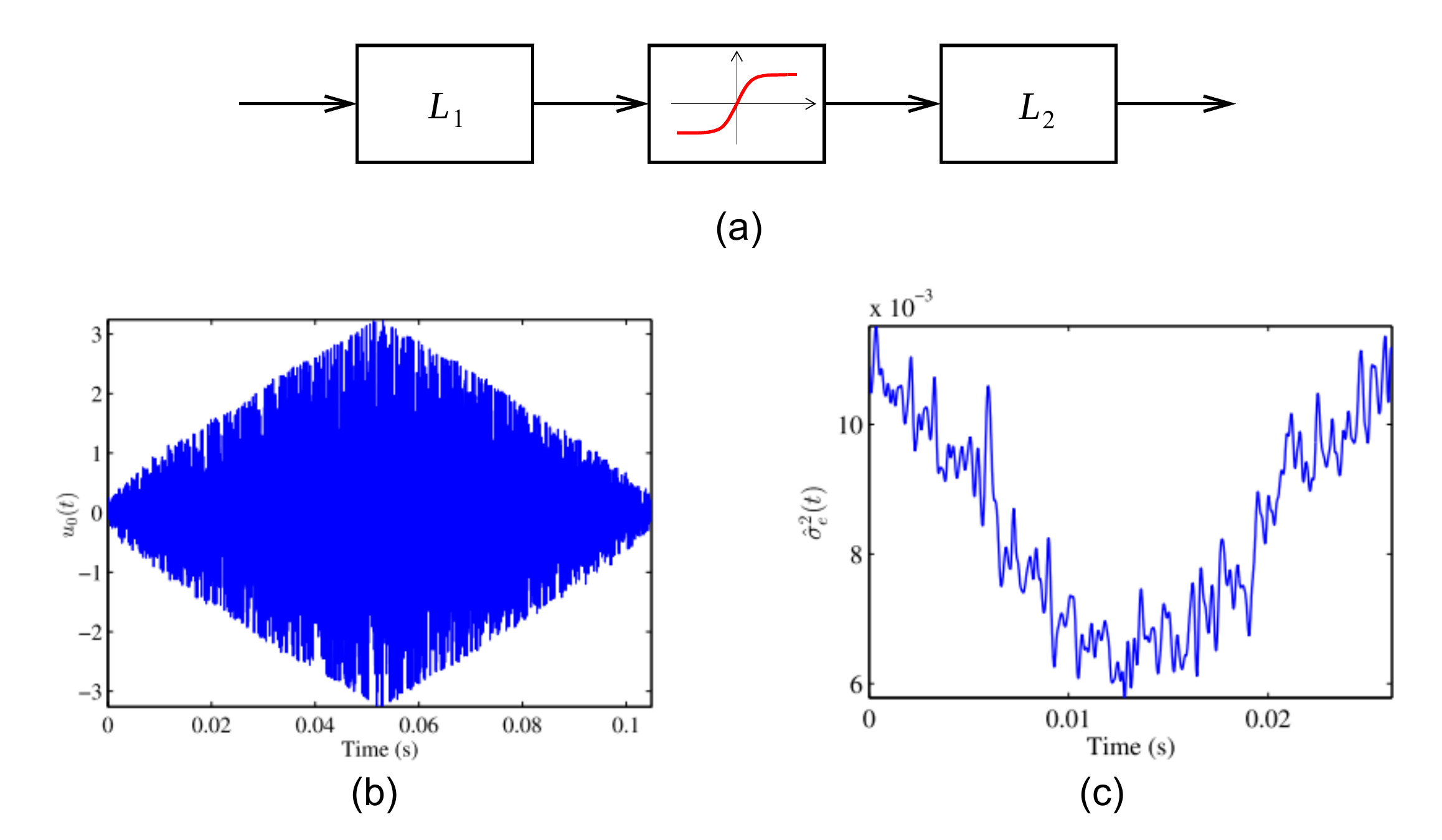}
\caption{Detection of process noise: A nonstationary input (b) is periodically repeated  and applied to a Wiener-Hammerstein system (a) with a saturating nonlinearity  that is sandwiched between two lowpass linear filters $L1,L2$. The process noise is coming in before the nonlinearity \cite{Zhang2017}. This system was studied and discussed in detail at the nonlinear benchmark workshop \cite{SchoukensM2016c}. The presence of the process noise is revealed through its nonstationary behavior at the output. The smoothed variance of the output noise (c) varies over time. Observe that it becomes small where the excitation is large, and large where the excitation is small, pointing to process noise that is injected before a saturating nonlinearity.}
\label{fig:ProcessNoise}
\end{figure}

\emph{Detection of the presence of process noise}: The dependence of the process noise output $y_{p}(t)$ on the input $u(t)$ can be used to detect it, even in the presence of measurement noise. In \cite{Zhang2017}, periodic non-stationary input signals are used. Assuming that a periodic input results in a periodic output, the process and measurement noise $\tilde{v}(t)$ can be estimated as the non-periodic part of the output. Next the variance $\sigma_{\tilde{v}}(t)$ is estimated. Assuming that the measurement noise is stationary, the presence of the process noise is revealed by a time-varying variance as illustrated in Figure \ref{fig:ProcessNoise}. 

\begin{figure}[h] 
\centering
\includegraphics[scale=0.7]{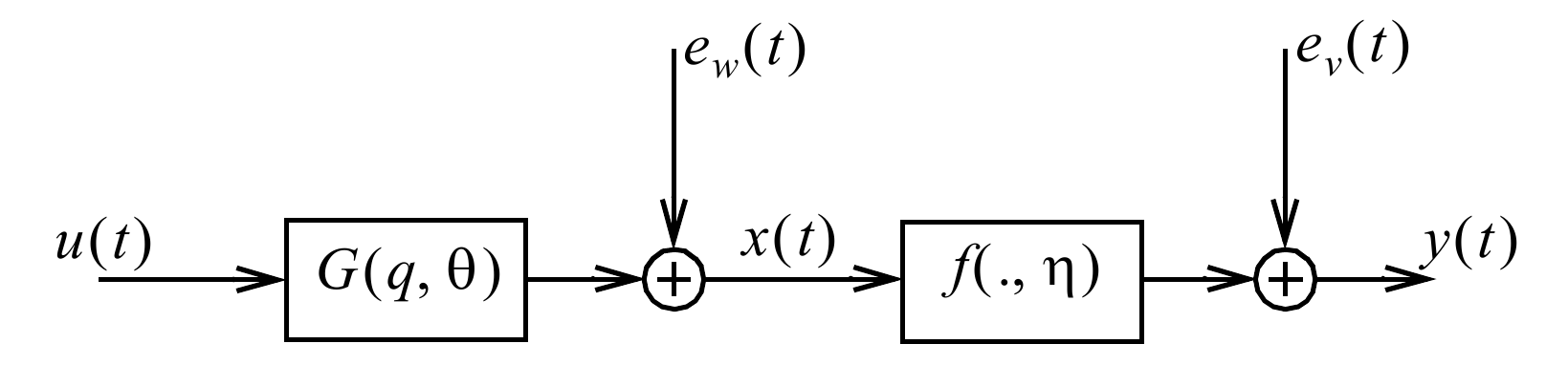}
\caption{Wiener system with white process noise $e_{w}(t)$ and measurement noise $e_{v}(t)$ . The signals $u(t),y(t)$ are available to identify the system, the nuissance signal is not measured.}
\label{fig:22}
\end{figure}

\emph{Impact of process noise on the system identification problem}:
The presence of process noise increases the complexity of the system identification problem significantly. The nonlinear operations change the distribution of the process noise $w(t)$, and can turn a least squares problem formulation at the output of the system to be very inefficient and even strongly biased. For that reason, the output error framework, that assumes that all noise enters at the output of the system has to be abandoned. Instead, the procedure should start from the joint (Gaussian) distribution of $w(t),v(t)$. The discussion in the following is focused on the identification of a Wiener system  to set the ideas. Following the ideas in \cite{Hagenblad2008}, the likelihood function can be easily generated using the intermediate nuissance variable $x(t)$ in Figure \ref{fig:22} from the following probability density function
\begin{equation}
p_{y}(\theta,\eta)=\int _{x\in\mathbb{R}}p_{v}(y-f(x,\eta))p_{w}(x-G(q,\theta))dx.
\label{eq:PN2}
\end{equation}
The calculation of this integral becomes very difficult because it is not possible to eliminate $x(t), t=1,\ldots,N$ analytically, with $N$ the number of data points.  Today, dedicated numerical methods are developed to deal with these high dimensional integrals. The reader is referred to ``Identifying nonlinear dynamical systems in the presence of process noise'' for a first introduction.


\subsection{Summary}
\begin{itemize}
\item \emph{Process noise}: The process noise contributions to the output of a nonlinear system may no longer Gaussian distributed and independent of the input.
\item \emph{Detection}: Applying nonstationary periodic excitations, it is possible to detect the presence of process noise.
\item \emph{Identification}: If no process noise is detected, the simpler output error formulation can be used to identify the nonlinear model. If the process noise turns out to be large, more advanced identification approaches are needed to guarantee consistent and efficient estimates.
\end{itemize}

\section{Sidebar\\ Identifying Nonlinear Dynamical Systems in the Presence of Process Noise}
This section was written by Thomas B. Sch{\"o}n, Department of Information Technology, Uppsala University, Sweden.

A rather general formulation to be used when identifying nolinear dynamical systems is arguably provided by the nonlinear state-space model which represent a system with input signal $u_t$ and output signal $y_t$ in terms of a latent Markovian state $x_t$,
\begin{subequations}
	\label{eq:SSM}
	\begin{align}
		x_{t+1} &= f(x_t, u_t; \theta) + w_t(\theta),\\
		y_t &= h(x_t, u_t; \theta) + v_t(\theta).
	\end{align}
\end{subequations}
Here the nonlinear functions $f(\cdot)$ and $h(\cdot)$ represent the dynamics and the measurements, respectively. The variables $w_t$ and $v_t$ describe the process noise and the measurement noise, respectively. Finally, the unknown static model parameters are denoted by~$\theta$ and the initial state is given by $x_0\sim p(x_0)$ for some distribution $p(\cdot)$.

The problem to be solved is the identification of the unknown parameters $\theta$ in~\eqref{eq:SSM} based on observed inputs $u^T=\{u_t\}_{t=1}^{T}$ and the corresponding outputs $y^T=\{y_t\}_{t=1}^{T}$.

%
The maximum likelihood formulation amounts to

\begin{align}
	\label{eq:MaximumLikelihood}
	\widehat{\theta} = \arg\max_{\theta}{p(y^T;\theta)},
\end{align}

where the nature of the intractability of the likelihood is revealed by
\begin{align}
	\label{eq:Likelihood}
	p(y^T; \theta) = \prod_{t=1}^{T}p(y_t\mid y^{t-1}, \theta)
	= \prod_{t=1}^{T}\int p(y_t\mid x_t, \theta) p(x_t\mid y^{t-1}, \theta)d x_t.
\end{align}
To stress the point that the process noise enters this integral, the following alternative integral can be considered:
\begin{align}
	p(y^T; \theta) = \prod_{t=1}^{T}\int p(y_t\mid x_t, \theta)p(x_t\mid x_{t-1}) p(x_{t-1}\mid y^{t-1}, \theta)d x_{t-1:t},
\end{align}
where it is clear that the process noise enters the integral via the term $p(x_t\mid x_{t-1}) = p_{w_t}(x_t - f(x_{t-1}, u_{t-1};\theta))$.

More specifically, the challenge lies in that the predictive state distribution $p(x_t\mid y^{t-1}, \theta)$ cannot be explicitly computed, and  approximations have to be made. The sequential Monte Carlo (SMC) methods  \cite{DoucetJ:2011} like particle filters and smoothers can be used to compute this distribution arbitrarily well. These methods were introduced in the beginning of the 1990s \cite{Gordon:1993}, but it is still a research area which is very much alive with new results emerging all the time. The idea behind these methods is to maintain an empirical distribution
\begin{align}
	\label{eq:SMCest}
	\widehat{p}(x_t\mid y_{1:t-1}) = \sum_{i=1}^{N}W_t^{i}\delta_{x_t^{i}}(x_t),
\end{align}
made up of samples $\{x_t^{i}\}_{i=1}^N$ (sometimes referred to as  particles) and their corresponding weights $\{W_t^{i}\}_{i=1}^N$. Here $\delta_{x_t^{i}}(x_t)$ denotes the Dirac delta. The SMC methods describe how to update these weights over time in such a way that the estimate~\eqref{eq:SMCest} converge to the true underlying
as the number of samples $N\rightarrow\infty$. The theoretical basis underpinning these methods is by now quite extensive, an entry-points into this literature is \cite{Delmoral:2004,DoucetJ:2011}. 
What is perhaps most relevant for our present discussion is that the SMC method is capable of producing \emph{unbiased} estimators of the likelihood~\cite{Delmoral:2004,PittSGK:2012}. The likelihood estimate is obtained by inserting~\eqref{eq:SMCest} into~\eqref{eq:Likelihood}. 

Based on the above, noisy unbiased estimates of the cost function in the maximum likelihood problem can be computed. As with any optimization problem---especially when faced with a stochastic optimization problem as it is here---it is easier to solve if there are also gradients available. The SMC method can be used for this as well, see e.g. \cite{PoyiadjisDS:2011,DoucetJR:2015}. 
Driven by deep learning, the state of the art when it comes to solving stochastic optimization problems is evolving quite rapidly at the moment, see e.g. \cite{BottouCN:2018}.

As an alternative to the above solution, the expectation maximization (EM) method \cite{DempsterLR:1977} can be used to  solve~\eqref{eq:MaximumLikelihood} in the presence of process noise, see e.g. \cite{SchonWN:2011,OlssonDCM:2008}. EM is an iterative algorithm that based on a current iterate $\theta_k$ computes an approximation of the so-called intermediate quantity
\begin{align}
	\label{eq:Qfunction}
	\mathcal{Q}(\theta, \theta_k) = \int \log p(x^T, y^{T}; \theta)p(x^{T}\mid y^{T}, \theta_k)dx_{0:T},
\end{align}
which is then maximized to find the next iterate $\theta_{k+1}$. This procedure is then repeated until convergence and it is guaranteed to stop at a stationary point of the likelihood surface. The SMC method is key here as well, in that it allows us to approximate the smoothing distribution $p(x^T\mid y^T, \theta_k)$ in~\eqref{eq:Qfunction} arbitrarily well. 

%
%
The Bayesian approach is of course also highly interesting for nonlinear system identification in general \cite{Peterka:1981} and for the problem when there is process noise present in particular. The slight variation to the above is that now the unknown parameters are assumed to be random variables instead and rather than computing a point estimator (like~\eqref{eq:MaximumLikelihood})  the posterior distribution $p(\theta\mid y^T)$ is computed. The breakthrough came in 2010 with the introduction of the so-called particle Markov chain Monte Carlo (PMCMC) methods \cite{AndrieuDH:2010}. 

Tutorial overviews of how to identify nonlinear state-space models using SMC are provided by \cite{SchonLDWNSD:2015,Kantas:2015}. Concrete system identification examples were there is significant process noise available are provided in for example \cite{SchonWN:2011,WillsSLN:2013,LindstenJS:2013}. 

\section{Sidebar\\  Black Box Models Complexity: \\ Keeping the Exploding Number of Parameters under Control; Increased Structural Insight; Model Reduction}

Black box modeling is a very flexible method that requires little or no physical insight of the user. This comes with the cost of an exploding number of model parameters for a growing complexity, leading to an increased risk of overfitting \cite{Ljung1987, Soderstrom1989, Pintelon2012}. Regularization and data driven structure retrieval tools are developed to keep this number of  growing parameters or their effect on the modeled output under control. Both approaches are discussed below.

\subsection{Regularization}
The most simple approach to obtain a simplified model is to set the 'least significant' parameters equal to zero in a model pruning step using manual trial and error methods. Regularization methods replace manual tuning by automatic procedures \cite{PillonettoDCNL:2014,Sliwinski2017}. The basic idea is to add an additional term $R(\theta)$ to the cost function \eqref{eq:CostNLTD}, imposing an extra constraint on the parameters
\begin{equation} \label{eq:CostNLTDReg}
V_{N}(\theta)=\frac{1}{N}\sum_{t=1}^{N}\frac{1}{2}\epsilon^{2}(t,\theta)+R(\theta).
\end{equation}
The choice of $R$ sets the behavior of the regularized solution, leading to sparse or smooth solutions.

\emph{Sparse models}: Putting $R(\theta)=\lambda |\theta|$  leads to LASSO \cite{Tibsharani:96}, a method that puts as many parameters as possible equal to zero which leads to sparse models. 

\begin{figure}[h] 
\centering
\includegraphics[scale=0.8]{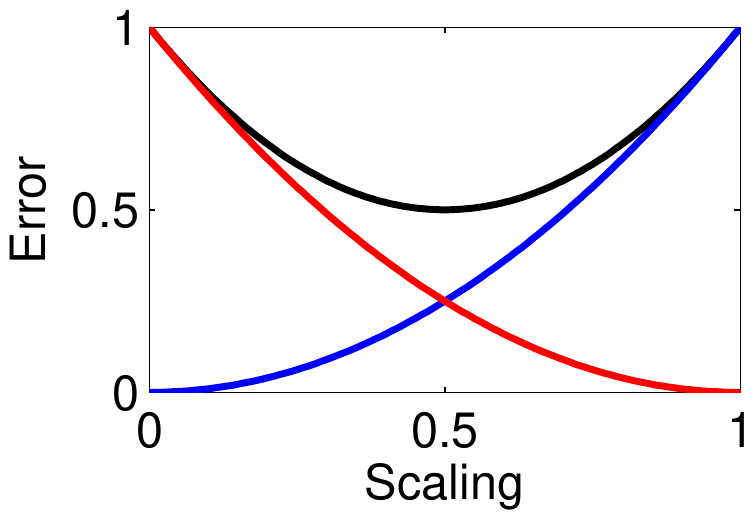}
\caption{Bias and variance trade-off of a scaled stimator: evolution of the total mean square error (black), squared bias (red), and variance error (blue) as a function of the scaling factor. An error $e$ can always be written as the sum $e=b+v$ of its mean value $b=E\{e\}$, called the bias, and the remaining part $v=e-b$ with variance $\sigma^{2}$ The total mean square error is $e_{MS}=b^{2}+\sigma^{2}.$ Depending on the preference, either the bias $b$ or the mean square error $e_{MS}$ should be as small as possible. It is always possible to scale an unbiased estimator (no bias present) towards zero such that $e_{MS}$ drops. This is illustrated on a simple scalar example.  Assume that $\hat{\theta}$ is an unbiased estimate  of the true parameter $\theta_{0}=1$,  with variance $\sigma^{2}=1$. Consider next the scaled estimator $\tilde{\theta}=\lambda \hat{\theta}.$
The bias of  $\tilde{\theta}$ is $b=(1-\lambda)$, and the variance $\tilde{\theta}$ is $\sigma^{2}_{\tilde{\theta}}=\lambda^{2}.$
The MSE becomes $ e_{MS}=(1-\lambda)^{2}+\lambda^{2}.$ Black: mean square error, Red: bias error, Blue: variance error.}
\label{fig:BiasVarianceTradeOff}
\end{figure}

\emph{Smooth models}: A quadratic regularization $R(\theta)=\theta^{T}P^{-1}_{\theta}\theta$ leads to a milder regularization than LASSO. The  regularization  term  pulls  the  estimates  toward  zero,  resulting in a reduced output variance at a cost of an increased bias. An optimal bias/variance balance is made that minimizes the mean square error of the output as illustrated in Figure \ref{fig:BiasVarianceTradeOff}.  

The  success  of  this  approach  strongly  depends  on  a proper choice of the regularization matrix $P$  that reflects the additional user knowledge or user desires, also called prior information, that is added on top of the data. It can be based on physical insight, or impose for example the user desire for a smooth solution. The reader is referred to \cite{PillonettoDCNL:2014}.  Observe that in this approach the number of parameters is not reduced, but instead their freedom to vary independently is restricted, leading to a smaller ``effective number of parameters''. 

Imposing smooth and exponentially decaying solutions is illustrated for nonparametric Volterra models in the examples section on Black Box Volterra Model of the Brain \cite{Birpoutsoukis2017}. 

\subsection{Data driven structure retrieval}
An alternative approach to reduce the number of model parameters is to impose more structure on the model. In black box modeling, the structural information should be retrieved from the data rather than using physical information.  The explosive growth of the number of parameters in nonlinear black box modeling is due to the parameterization of the multivariate nonlinear function \eqref{eq:genstr} that is present in every nonlinear model. Retrieving more efficient presentations of the function $F\in   R^{n_{q}\times n_{p}}$ is a key to reduce the number of parameters, and keep the model flexibility under control. Recently, decoupling methods are developed that allow the multivariate function $F$ to be written as a combination of linear transformations and a well selected set of single-input single-output (SISO) nonlinear functions as shown in Figure \ref{fig:decoupling}.  In ``Decoupling of Multivariate Polynomials'', a tensor based decoupling approach is explained in more detail. 

A decoupled representation offers major advantages:

	\begin{itemize}
    \item The combinatorial grow of the number of parameters is reduced to a linear grow as a function of the complexity. For example, if $F$ is a multivariate polynomial of degree $d$, the number of parameters drops from $n_qO(n_p^d)$ to $n_qO(rd)$, with $r$ the number of internal SISO branches.
    \item The decoupled representation of $F$ is easier to interprete, giving more intuitive access to the nonlinear behavior of the system. For example, it is much simpler to plot a set of SISO nonlinear functions than to make a graphical representation of a high dimensional multivariate nonlinear function because only slices of the multivariate function can be shown.
    \item Tuning the number of branches in the decoupled representation is not only a tool to reduce the flexibility of the model, it can also be used to make a user selected balance between structural model errors and model complexity which opens the road for model complexity reduction.
    \end{itemize}

\emph{Applications}: The decoupling approach can be applied on a variety of problems. In the ``Extensive case study: The forced Duffing oscilator'', a detailed illustration of the decoupling approach is given.  
An early application of the decoupling strategy \cite{Favier2009} proposes a tensor based decoupling method to decouple on a separate basis the Volterra kernels of different degree. Next, this idea was further refined and applied to the identification of parallel Wiener and parallel Wiener-Hammerstein block-oriented models \cite{SchoukensM2012a, Tiels2013, SchoukensM2014a}. The full decoupling method \cite{Dreesen2015} that is described in this section is used in \cite{Esfahani2018} to decouple a nonlinear state space model for the Bouc-Wen hysteresis problem. Recently, the ideas were further generalized to apply the decoupling idea to the pruning of NARX models \cite{Westwick2018a}. Eventually, the decoupling ideas are also directly applied to the design of decoupled controllers \cite{Stoev2017}.

\subsubsection{Decoupling of Multivariate Polynomials}
\begin{figure}[h] 
\centering
\includegraphics[scale=1]{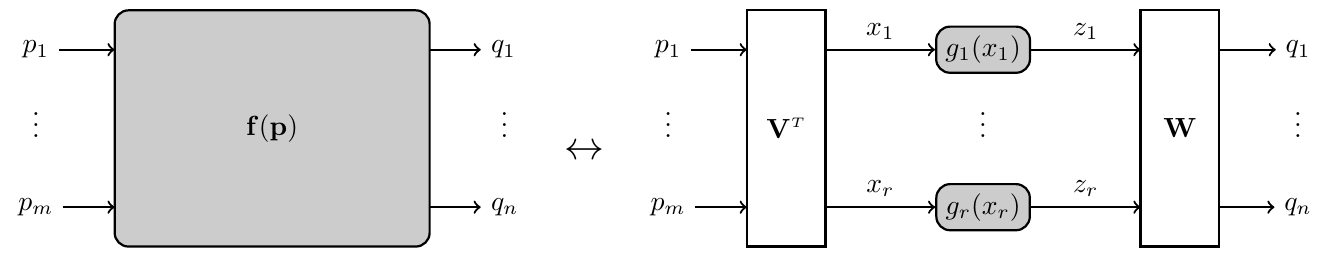}
\caption{The multivariate nonlinear function $q=f(p)$ is replaced by a decoupled representation \cite{Dreesen2015}. }
\label{fig:decoupling}
\end{figure}

Representations that increase the structural insight, and reduce the number of model parameters are most welcome. The approach that is briefly described in this section is discussed in full detail in \cite{Dreesen2015}. The starting point is a set of multivariate basis functions, for example polynomials, that needs to be unraveled into a simplified structure. The cross-links among input variables are `decoupled' into single-variable functions by considering linear transformations at the input and output of the static non-linearity.  These transformations reveal internal variables between which univariate relations hold. Consider the variables $\textbf{p},\textbf{q}$ and the multivariate vector function $\textbf{f}$ with proper dimensions.
\begin{equation}
\begin{aligned}
\textbf{q} &= \textbf{f}(\textbf{p})  \\
&= W \textbf{g}(V^\top \textbf{p}).
\end{aligned}
\end{equation}
The entry $i$ of $\textbf{g}$ is a univariate function $\textbf{g}_{i}=g_{i}(x_{i})$, $i=1,\ldots,r$, with $\textbf{x}=V^\top \textbf{p}$. In this way, the model can be given again a physical/intuitive interpretation while at the same time the number of parameters decreases. A graphical representation is given in Figure \ref{fig:decoupling}, each univariate function is called a branch, and $r$ is the number of branches that is used in the decoupled presentation. In \cite{Dreesen2015b} an exact decomposition method is proposed to obtain a decoupled representation. 

\emph{Example: Exact decomposition of a multivarite polynomial}
Consider the polynomials $f_1(p_1,p_2)$ and $f_2(p_1,p_2)$ of total degree $d=3$, given as
\begin{alignat}{3}
     q_1 & =  f_1(p_1,p_2) \nonumber\\
        & = 54 p_1^3 - 54 p_1^2 p_2 + 8 p_1^2 + 18 p_1 p_2^2 + 16 p_1 p_2 - 2 p_2^3 + 8 p_2^2 + 8 p_2 + 1,\label{eq:exampleeq1}\\ \nonumber
   q_2 &= f_2(p_1,p_2) \nonumber\\ 
        &= - 27 p_1^3 + 27 p_1^2 p_2 - 24 p_1^2 - 9 p_1 p_2^2 - 48 p_1 p_2 - 15 p_1 + p_2^3 - 24 p_2^2 - 19 p_2 - 3. \label{eq:exampleeq2} 
\end{alignat}

The equations~(\ref{eq:exampleeq1})--(\ref{eq:exampleeq2}) can be represented under the following decoupled structure:
\[
\left[ \begin{array}{c} q_1 \\ q_2 \end{array} \right] =
\left[ \begin{array}{rr} 1 & 2 \\ -3 & -1 \end{array}\right]
\left[ \begin{array}{c} 2 x_1^2 - 3 x_1 + 1 \\ x_2^3 - x_2 \end{array}\right]
,
\quad \textrm{with} \quad
\left[ \begin{array}{c} {x}_1 \\ {x}_2 \end{array} \right] =
\left[ \begin{array}{rr} -2 & -2 \\ 3 & -1 \end{array}\right]
\left[ \begin{array}{c} p_1 \\ p_2 \end{array}\right],
\]
revealing the internal univariate polynomials and the linear transformations at the input and output of the structure.

 \emph{Uniqueness}: It is shown that the decoupled models are not (always) unique \cite{Dreesen2015}. Hence decoupling will not lead to 'the' physical underlying representation, but nevertheless it will still provide an increased intuitive insight in the behavior of the system.

\emph{Exact and approximate decomposition}

\begin{figure}[h] 
\centering
 \centering
    \begin{subfigure}[t]{0.5\textwidth}
        \centering
        \includegraphics[scale=0.3]{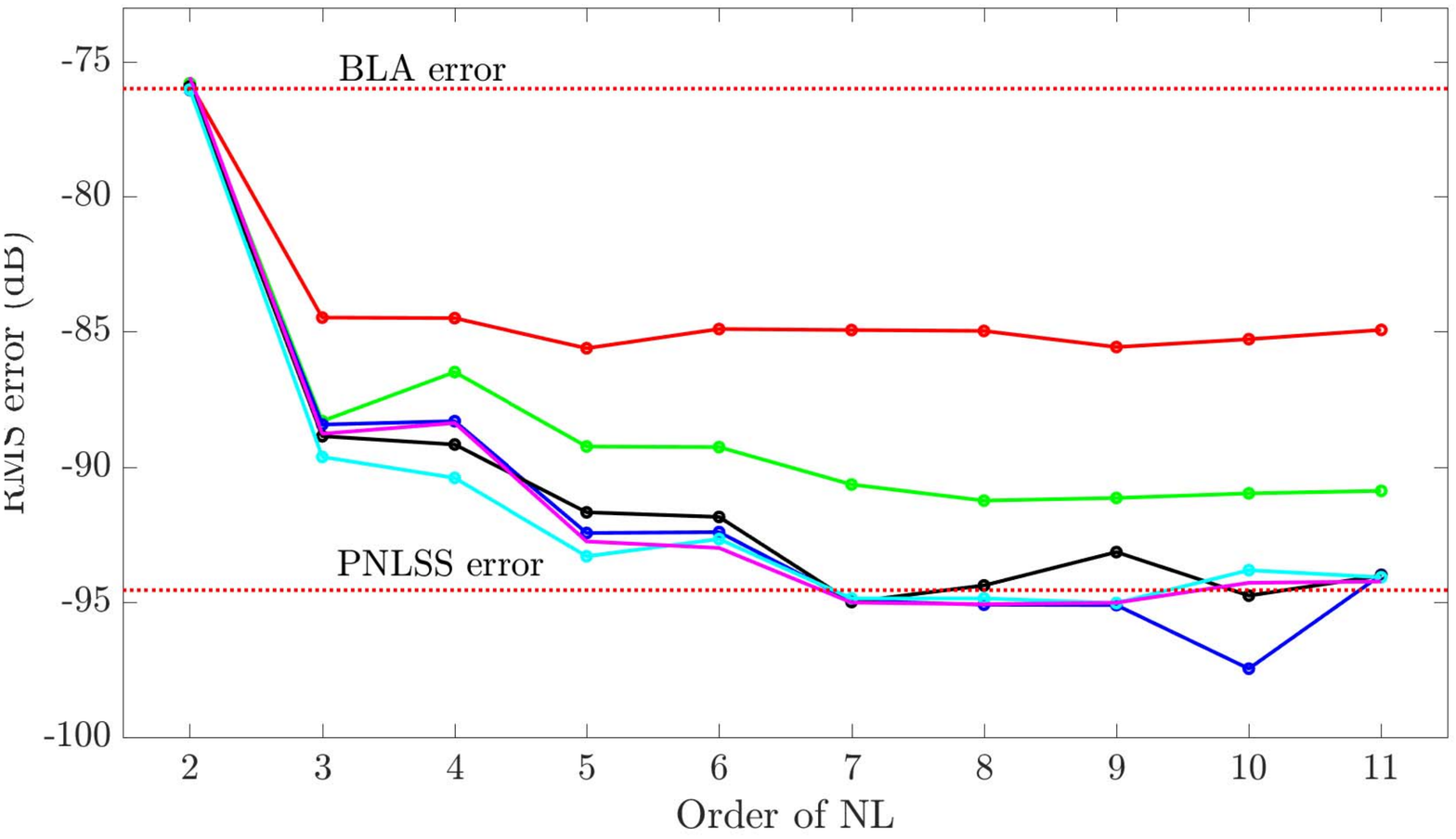}
        \caption{}
    \end{subfigure}%
    ~ 
    \begin{subfigure}[t]{0.5\textwidth}
        \centering
        \includegraphics[scale=0.3]{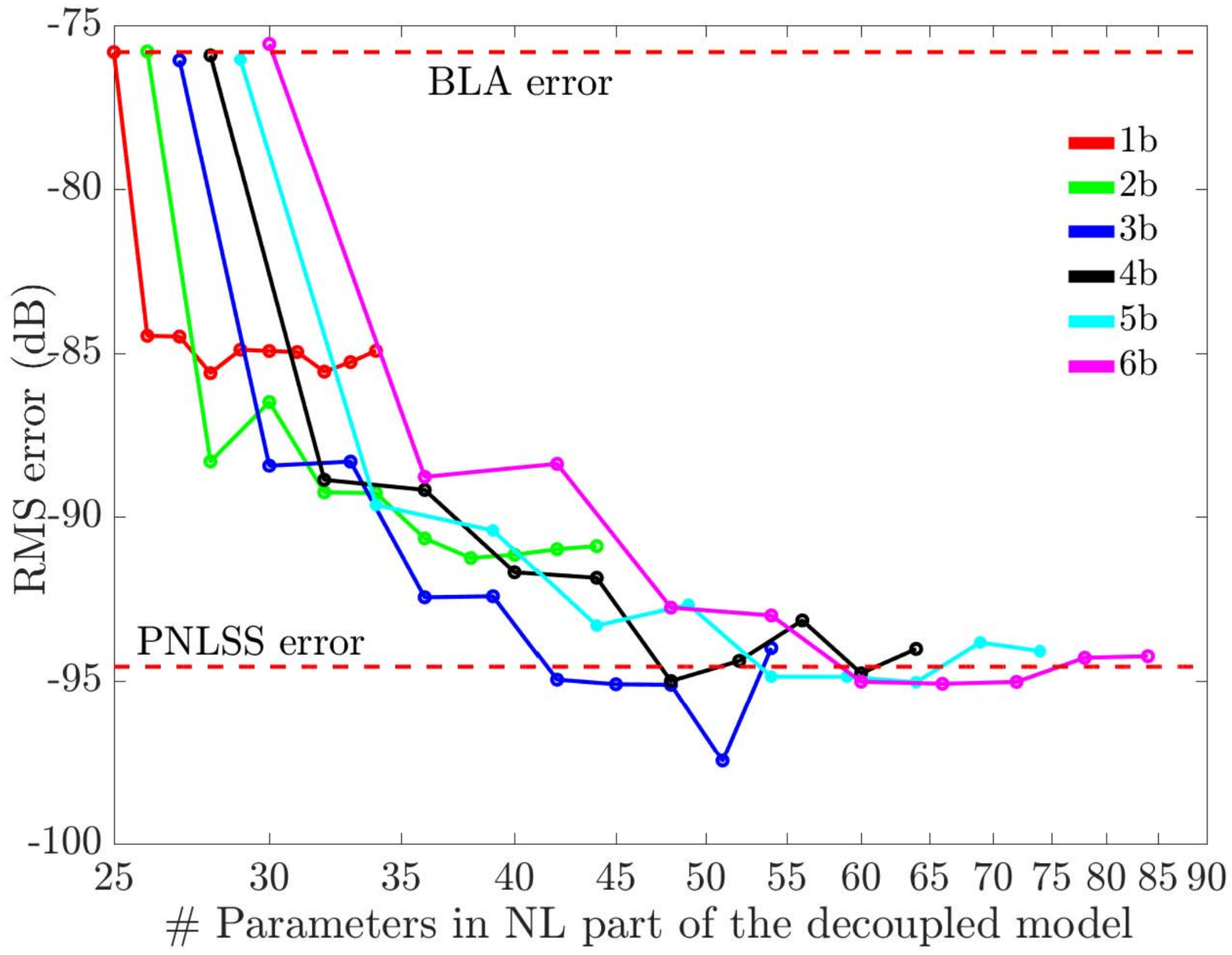}
        \caption{}
    \end{subfigure}

\caption {Illustration of the approximate decoupling on the Bouc-Wen hysteresis problem \cite{Esfahani2018}. The multivariate nonlinear function of the full PNLSS model (3 states, multivariate polynomial of degree 7 on the states and the input, 364 parameters) is replaced by a decoupled representation with an increasing number of branches (1 to 6). Next, the degree of the univariate polynomials is varied between 3 and 11. The decoupled model is tuned to the data using a least squares cost function. In the figure, the RMS error on a validation set is plotted for a varying number of branches, ranging from 1 to 6. In (a) the error is plotted as a function of the degree of the univariate polynomial, in (b) as a function of the number of parameters (depends on the degree of the univariate polynomials). The best results are obtained for a model with 4 branches and polynomial degree 11. The MSE of the BLA is -76 dB, the error of the best nonlinear models is 10 times smaller. The RMS error of the coupled PNLSS model is indicated with a *.}
\label{fig:22Bis}
\end{figure}

In an exact decoupling of an arbitrary multivariate function, the number of branches $r$ might become very large. A truncated decoupling, with a reduced number of branches can be used to approximate the multivariate function.  This problem is studied in \cite{Hollander2018}. The approximation error is tuned using a weighted least squares criterion. The number of branches $r$ can be used as a handle to balance the complexity of the model against the level of the tolerated structural model errors. This is illustrated on the forced Duffing oscilator in Figure \ref{fig:SilberboxDecoupledReduced}, and on the Bouc-Wen model in Figure \ref{fig:22Bis} that are reported in detail in \cite{Esfahani2018}.

\begin{figure}[h] 
\centering
\includegraphics[scale=0.5]{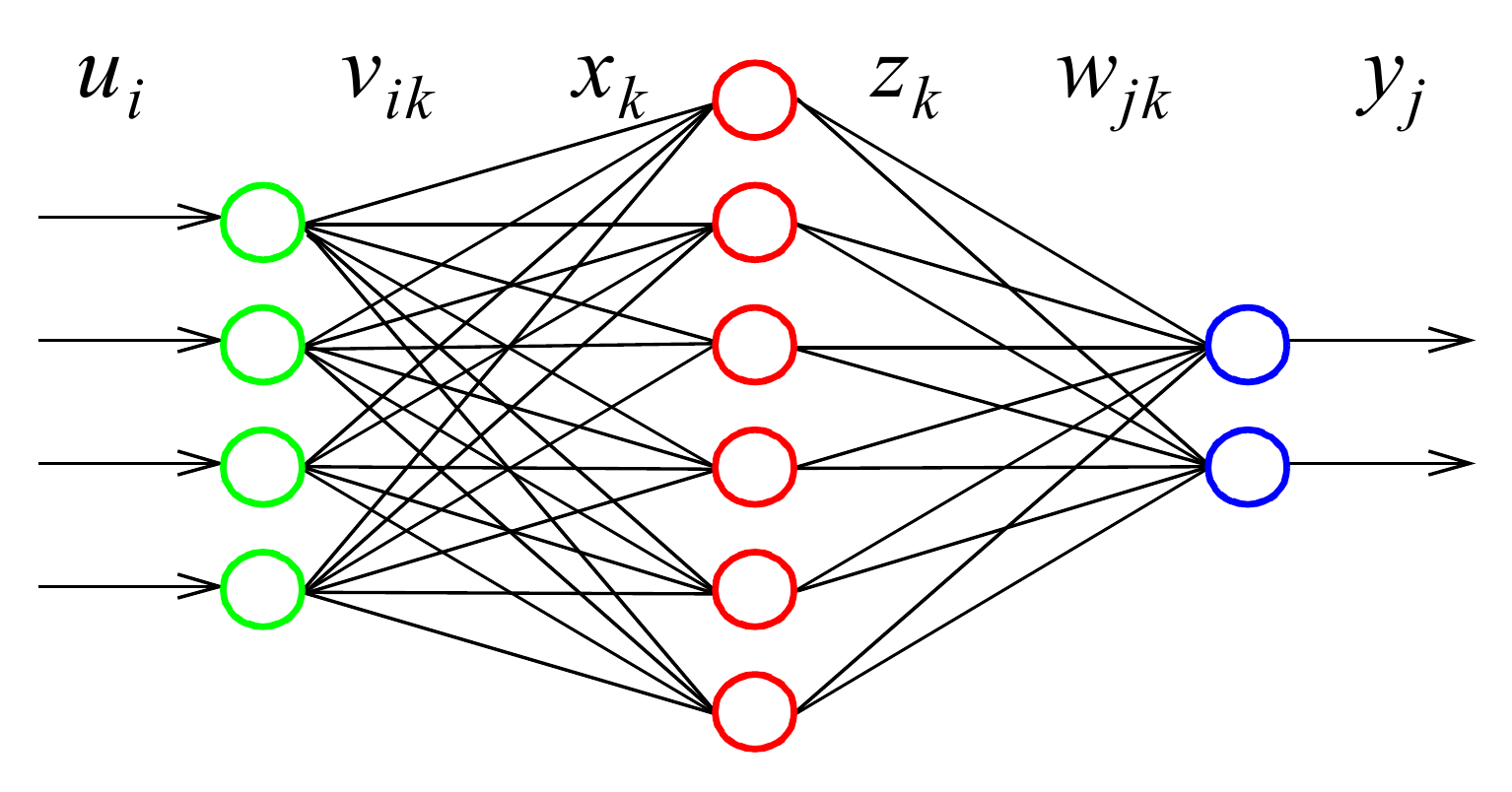}
\caption{Neural network model: the inputs $u_{i}$ (green) are linearly combined with the weights $v_{ik}$ to generate the inputs $x_{k}$ for the neurons (red). The output $z_{k}$ of each neuron is calculated by a uni-variate non-linear function on its input $x_{k}$, and is linearly combined with the wieghts $w_{kj}$ to generate the outputs $y_j$ (blue). This is exactly the same structure as was used in the decoupling approach. The major difference between the neural network and the decoupling method is that in the latter the nonlinear functions vary from one node to the other, while in the neural network they are all equal to each other. This additional flexibility results in a faster convergence so that less branches are needed.}
\label{fig:23}
\end{figure}

\emph{Relation with neural networks}
The structure of the decoupled representation in Figure \ref{fig:decoupling} is very similar to that of a neural network as shown in Figure \ref{fig:23}. In a neural network, the univariate functions belong all to the same family (for example sigmoids or hinge functions). It is known that neural nets are universal approximators \cite{Hornik1989, Sjoberg1995, Suykens1996} that converges as $O(1/\sqrt{r})$. In the  decoupling method, the univariate functions follow from the decoupling and are tuned to the specific problem. This additional degree of freedom will reduce the number of branches that is needed in the decoupled representation of the multivariate function.

\section{Sidebar\\ Software Support}
The algorithms needed for nonlinear model definition, estimation and analys are typically quite complex, and need sophisticated computer software support. Several program packages for such support is publically available, some for free download and some for license agreements. Here some such packages will be briefly described.
\subsection{The System Identification Toolbox for use with \textsc{Matlab}}
The System Identification Toolbox, \cite{Ljungtb:18} is a commercially available program package, distributed by MathWorks Inc. It contains implementations of several of the methods/models discussed in this article
\begin{itemize}
\item \texttt{idnlarx} is a model object that handles the NARX model (\ref{eq:NLARX}) with several possible nonlinearties.
\item \texttt{idhw} handles the Hammerstein-Wiener block-oriented model in Fig \ref{fig:EID1},
\item \texttt{idnlgrey} treats nonlinear state-space grey-box models (\ref{eq:NLSSu}) where $ƒ$ and $h$ have to be programmed by the user either in MATLAB or C-code.
\end{itemize}
Static nonlinearities (\ref{eq:snl}) can be chosen for example as
\begin{itemize}
\item \texttt{pwlinear}: (\ref{eq:brkp})
\item \texttt{poly1d}: (\ref{eq:poly1})
\item \texttt{sigmoidnet}: (\ref{eq:nn1})
\item \texttt{treepartition}: (\ref{eq:treeint})
\item \texttt{customnet}: (\ref{eq:cust})
\end{itemize}
for use in \texttt{idnlarx} and \texttt{idnlhw}.
All model validation and evaluation commands like, \texttt{compare} (for cross validation), \texttt{resid} (for (linear) residual analysis), \texttt{sim, predict, forecast} (for simulation and forecasting) are available analogous to linear models, and can be used also to compare linear and nonlinear models in the same command.
 
\subsection{Nonparametric Noise Analysis}
 The freely available frequency-domain identification toolbox FDIDENT can be used to make the nonparametric noise and distortion analysis (``Nonparametric Noise and Distortion Analysis Using Periodic Excitations'')  (http://home.mit.bme.hu/\char`\~kollar/fdident/). This toolbox also includes the tools to design the random-phase multisines  \eqref{eq:Multisine} and to perform the  nonparametric nonlinear analysis. In \cite{Schoukens2012}, all the procedures for the noise and distortion analysis that are presented in this article are discussed in full detail, and the related Matlab software can be freely downloaded from booksupport.wiley.com.
 
 \subsection{Nonlinear State Space Model}
 A package to identify a polynomial nonlinear state space model (see ``Extensive Case Study: The Forced Duffing Oscilator'') is freely available at http://homepages.vub.ac.be/~jschouk/.
 
 \subsection{Nonlinear Benchmark Website}
 
 The website http://www.nonlinearbenchmark.org/ hosts many experimental data sets that are well documented. For each benchmark, a detailed description of the setup and the experiments is given. In addition, a list of references is provided to publications that process these data. The data for ``Extensive Case Study: The Forced Duffing Oscilator'' are available on the this website. 
 
\section{Acknowledgement}

This work was supported in part by the Fund for Scientific Research (FWO-Vlaanderen), the Vrije Universiteit Brussel (VUB), and by the ERC advanced grant SNLSID, under contract 320378.
The authors thank Julien Ertveldt (Vrije Universiteit Brussel) for the results on the wind tunnel experiments, including Figure \ref{fig:WindTunnel}; Dhammika Widanalage and Julian Stouv (Vrije Universiteit Brussels) for the wet-clutch results shown in Figure \ref{fig:WetClutch}; Ga\"{e}tan Kerschen and Jean-Philippe No\"{e}l (University of Li\`{e}ge, and Bart Peeters (Siemens Product Life-cycle Management Belgium for the results on the ground vibration test shown in Figure \ref{fig:F16}; Torbjr\"{o}n Wigren (University Uppsala), Maarten Schoukens (Vrije Universiteit Brussel), and Keith Worden and his team (University of Sheffield) for the results on the cascaded water tanks, including Figure \ref{fig:CascadedTanksWhiteBoxWorden}; Oliver Nelles (University of Siegen) and his team for the High Pressure Fuel Supply System example, including Figures \ref{fig:NellesHILOMOT}, \ref{fig:NellesResults}, and \ref{fig:NellesResultsGPM}; Martijn Vlaar and his team (Technical University Delft), and Georgios Birpoutsoukis (Vrije Universiteit Brussel) for the results on the brain modeling (Figure \ref{fig:BrainSystem}); Rishi Relan (Vrije Universiteit Brussel) for the results on the battery model in Figures \ref{fig:BatteryNLdist} and \ref{fig:BatteryErrorTDFD}; Erliang Zhang (Zhengzhou University) and Maarten Schoukens (Vrije Universiteit Brussel) for the results on process noise detection in Figure \ref{fig:ProcessNoise}; Alireza Fakhrizadeh Esfahani (Vrije Universiteit Brussel) and Jean-Philippe No\"{e}l (University of Li\`{e}ge) for the results on the Bouc-Wen hysteresis problem (Figure \ref{fig:22Bis}).

\section{Author Information}

\noindent Johan Schoukens received both the Master's degree in electrical engineering in 1980, and the PhD degree in engineering sciences in 1985 from the Vrije Universiteit Brussel (VUB), Brussels, Belgium. In 1991 he received the degree of Geaggregeerde voor het Hoger Onderwijs from the VUB, and in 2014 the degree of Doctor of Science from The University of Warwick. From 1981 to 2000, he was a researcher of the Belgian National Fund for Scientific Research (FWO-Vlaanderen) at the Electrical Engineering Department of the VUB, From 2000 to 2018 he was a full-time professor in electrical engineering, since 2018 he is emiritus professor at the department INDI of the VUB and at the Department of Electrical Engineering of the TU/e (The Netherlands). From 2009 to 2016, he was visiting professor at the Katholieke Universiteit Leuven (Belgium). His main research interests include system identification, signal processing, and measurement techniques. He has been a Fellow of IEEE since 1997. He was the recipient of the 2002 Andrew R. Chi Best Paper Award of the IEEE Transactions on Instrumentation and Measurement, the 2002 Society Distinguished Service Award from the IEEE Instrumentation and Measurement Society, and the 2007 Belgian Francqui Chair at the Universit\'e Libre de Bruxelles (Belgium). Since 2010, he is a member of Royal Flemish Academy of Belgium for Sciences and the Arts. In 2011 he received a Doctor Honoris Causa degree from the Budapest University of Technology and Economics (Hungary). Since 2013, he has been an honorary professor of the University of Warwick. 

\noindent Lennart Ljung  received his Ph.D. in Automatic Control from Lund Institute of Technology in 1974. Since 1976
he is Professor of the chair of Automatic Control In Link\"{o}ping, Sweden. He has held visiting positions at
Stanford and MIT and has written several books on System Identification and Estimation. He is an IEEE Fellow, an
IFAC Fellow and an IFAC Advisor. He is a member of the Royal Swedish Academy of Sciences (KVA), a member of the Royal Swedish Academy of Engineering Sciences (IVA), a member of Academia Europaea,  an Honorary Member of the Hungarian Academy of Engineering, an Honorary Professor of the Chinese Academy of Mathematics and Systems Science, and a Foreign Member of the US National Academy of Engineering (NAE). He has received honorary doctorates from the Baltic State Technical University in St. Petersburg, from Uppsala University, Sweden, from the Technical University of Troyes, France, from the Catholic University of Leuven, Belgium and from Helsinki University of Technology, Finland. In 2002 he received the Giorgio  Quazza Medal from IFAC and in 2017 he received the Nathaniel B Nichols medal also from IFAC. In 2003 he received the Hendrik W. Bode Lecture Prize from the IEEE Control Systems Society, and he was the 2007 recipient of the IEEE Control Systems Award.
He was honored with the Great Gold Medal from the Royal Swedish Academy of Engineering Sciences in 2019.

\bibliographystyle{unsrt}      													 
\bibliography{ReferencesLibraryV4,llref,bibThomas}

\begin{thebibliography}{100}

\bibitem{Zadeh1962}
L.A. Sadeh.
\newblock From circuit theory to system theory.
\newblock {\em Proceedings of the IRE}, pages 856--865, 1962.

\bibitem{Astrom1971}
K.J. Åström and P.~Eykhoff.
\newblock System identification-a survey.
\newblock {\em Automatica}, 7:123--167, 1971.

\bibitem{BoxandJenkins1970}
G.E.P. Box and G.M. Jenkins.
\newblock {\em Time Series Analysis, Forecasting and Control}.
\newblock Holden-Day, 1970.

\bibitem{Astrom1970}
K.J. {\AA}str{\"o}m.
\newblock {\em Introduction to Stochastic Control Theory}.
\newblock Academic Press, 1970.

\bibitem{Eykhoff1974}
P.~Eykhoff.
\newblock {\em System Identification and State Estimation}.
\newblock Wiley, 1974.

\bibitem{Ljung1987}
L.~Ljung.
\newblock {\em {System Identification: Theory for the User}}.
\newblock Prentice Hall, Upper Saddle River, New Jersey, 1987 (2nd edition in
  1999).

\bibitem{Soderstrom1989}
T.~S{\"o}derstr{\"o}m and P.~Stoica.
\newblock {\em {System Identification}}.
\newblock Prentice Hall International (UK) Ltd, Hemel Hempstead, Hertfordshire,
  1989.

\bibitem{Pintelon2012}
R.~Pintelon and J.~Schoukens.
\newblock {\em {System Identification: A Frequency Domain Approach}}.
\newblock Wiley-IEEE Press, Hoboken, New Jersey, 2nd edition, 2012.

\bibitem{Pintelon2015}
R.~Pintelon, E.~Louarroudi, and J.~Lataire.
\newblock Nonparametric estimation of time-variant dynamics.
\newblock In {\em 17th IFAC Symposium on System Identification, Beijing,
  China.}, 2015.

\bibitem{Vaes2015}
M.~Vaes, J.~Schoukens, B.~Peeters, J.~Debille, T.~Dossogne, J.P. No{\"e}l, and
  G.~Kerschen.
\newblock {Nonlinear ground vibration identification of an F-16 aircraft. Part
  I: fast nonparametric analysis of distortions in FRF measurements}.
\newblock In {\em 16th International Forum on Aeroelasticity and Structural
  Dynamics (IFASD), Sint Petersburg, Russia}, Saint Petersburg, Russia, 2015.

\bibitem{Dossogne2015}
T.~Dossogne, J.P. Noël, G.~Grappasonni, Kerschen G., B.~Peeters, and
  J.~Schoukens.
\newblock Nonlinear ground vibration identification of an f-16 aricraft - part
  2: Understanding nonlinear behaviour in aerospace structures using sine-sweep
  testing.
\newblock In {\em 16th International Forum on Aeroelasticity and Structural
  Dynamics (IFASD), Sint Petersburg, Russia}, 2015.

\bibitem{Hjalmarsson2009}
H~Hjalmarsson.
\newblock System identification of complex and structured systems.
\newblock {\em European Journal of Control}, pages 275--310, 2009.

\bibitem{Criens2016}
C.~Criens, T.~van Keulen, F.~Willems, and M.~Steinbuch.
\newblock A control oriented multivariable identification procedure for
  turbocharged diesel engines.
\newblock {\em International Journal of Powertrains}, 5:95--119, 2016.

\bibitem{Wernholt2008}
E.~Wernholt and S.~Gunnarsson.
\newblock Estimation of nonlinear effects in frequency domain identification of
  industrial robots.
\newblock {\em IEEE Transactions on Instrumentation and Measurement},
  57:856--863, 2008.

\bibitem{SchoukensJ2016}
J.~Schoukens, M.~Vaes, and R.~Pintelon.
\newblock {Linear system identification in a nonlinear setting: Nonparametric
  analysis of the nonlinear distortions and their impact on the best linear
  approximation}.
\newblock {\em {IEEE Control Systems Magazine}}, {36}:{38--69}, {2016}.

\bibitem{IEEEstandard1057}
1057-2007 - ieee standard for digitizing waveform recorders. revision of ieee
  std 1057-1994.
\newblock Technical report.

\bibitem{Hjalmarsson1992}
H.~Hjalmarsson and L.~Ljung.
\newblock Estimating model variance in the case of undermodeling.
\newblock {\em IEEE Transaction on Automatic Control}, 37(7):1004--1008, 1992.

\bibitem{Ljung2010}
L.~Ljung.
\newblock {Perspectives on system identification}.
\newblock {\em Annual Reviews in Control}, 34(1):1--12, 2010.

\bibitem{Pearson2003}
R.K. Pearson.
\newblock Selecting nonlinear model structures for computer control.
\newblock {\em Journal of Process Control}, 13(1):1--26, 2003.

\bibitem{SchoukensJ1994}
J.~Schoukens, R.~Pintelon, and H.~Van~hamme.
\newblock {Identification of linear dynamic systems using piecewise-constant
  excitations - use, misuse and alternatives}.
\newblock {\em {Automatica}}, {30}({7}):{1153--1169}, {1994}.

\bibitem{Tiller:01}
Michael Tiller.
\newblock {\em Introduction to physical modeling with Modelica}.
\newblock Kluwer Academic Publishers, Boston, 2001.

\bibitem{brenan:89}
K.~E. Brenan, S.~L. Campbell, and L.~R. Petzold.
\newblock {\em Numerical solution of initial-value problems in
  differential-algebraic equations}.
\newblock North-Holland, 1989.

\bibitem{LjungG:16}
L.~Ljung and T.~Glad.
\newblock {\em Modeling and Identification of Dynamic Systems}.
\newblock Studentliteratur, Lund, Sweden, 2016.

\bibitem{AndersonM:79}
B.~D.~O. Anderson and J.~B. Moore.
\newblock {\em Optimal Filtering}.
\newblock Prentice-Hall, New Jersey, 1979.

\bibitem{Schon2011}
T.B. Sch\"on, A.~Wills, and B.~Ninness.
\newblock {System identification of nonlinear state-space models}.
\newblock {\em Automatica}, 47(1):39--49, 2011.

\bibitem{DoucetT:03}
A.~Doucet and V.~B. Tadi\'c.
\newblock Parameter estimation in general state-space models using particle
  methods.
\newblock {\em Ann. Inst. Statist. Math.}, 55(2):409--422, 2003.

\bibitem{fritzson14}
Peter Fritzson.
\newblock {\em Principles of Object-Oriented Modeling and Simulation with
  Modelica 3.3: A Cyber-Physical Approach}.
\newblock Wiley-IEEE Press, 2nd edition, 2014.

\bibitem{simscape}
Mathworks.
\newblock wwww.mathworks.com/products/simscape.

\bibitem{Bohlin:06}
T.~Bohlin.
\newblock {\em Practical Grey-box Process Identification}.
\newblock Springer Verlag, London, 2006.

\bibitem{Nelles2001}
O~Nelles.
\newblock {\em Nonlinear System Identification}.
\newblock Springer, 2001.

\bibitem{MurrayJ:97}
R.~Murray-Smith and T.~A. Johansen, editors.
\newblock {\em Multiple Model Approaches to Modeling and Control}.
\newblock Taylor and Francis, London, 1997.

\bibitem{TakagiS:85}
T.~Takagi and M.~Sugeno.
\newblock Fuzzy identification of systems and its application to modeling and
  control.
\newblock {\em IEEE Trans. Syst. Man and Cyber.}, 15(1):116--132, 1985.

\bibitem{LeeP:99}
L.~Lee and K.~Poolla.
\newblock Identification of linear parameter-varying systems using non-linear
  programming.
\newblock {\em ASME Journal of Dynamic Systems, Measurement and Control},
  121:71--78, 1999.

\bibitem{Toth10}
R.~Toth.
\newblock {\em Modeling and Identification of Linear Paramter-Varying Systems}.
\newblock Springer, 2010.

\bibitem{Tothetal:12}
R.~Toth, P.~C.~S. Heuberger, and P.~M. J~Van den Hof.
\newblock Prediction-error identification of {LPV} systems: present and beyond.
\newblock In J.~J. Mohammadpour and C.~W. Scherer, editors, {\em Control of
  Linear Parameter Varying Systems with Applocations}, chapter~2, pages 27--58.
  Springer, 2012.

\bibitem{Boyd1985}
S.~Boyd and L.O. Chua.
\newblock {Fading Memory and the Problem of Approximating Nonlinear Operators
  with Volterra Series}.
\newblock {\em IEEE Transactions on Circuits and Systems},
  {32}({11}):{1150--1161}, {1985}.

\bibitem{Palm1979}
G.~Palm.
\newblock {On representation and approximation of nonlinear systems Part II:
  Discrete Time}.
\newblock {\em Biological Cybernetics}, 34:49--52, 1979.

\bibitem{Wills2013}
A.~Wills, T.B. Sch\"{o}n, L.~Ljung, and B.~Ninness.
\newblock {Identification of Hammerstein-Wiener models}.
\newblock {\em Automatica}, 49(1):70--81, 2013.

\bibitem{SchoukensTiels2017}
M.~Schoukens and K.~Tiels.
\newblock Identification of block-oriented nonlinear systems starting from
  linear approximations: A survey.
\newblock {\em Automatica}, 85:272--292, 2017.

\bibitem{Giri2010}
F.~Giri and E.W. Bai, editors.
\newblock {\em {Block-oriented Nonlinear System Identification}}, volume 404 of
  {\em Lecture Notes in Control and Information Sciences}.
\newblock Springer, Berlin Heidelberg, 2010.

\bibitem{Schoukensm2015a}
M.~Schoukens, A.~Marconato, R.~Pintelon, G.~Vandersteen, and Y.~Rolain.
\newblock {Parametric identification of parallel Wiener-Hammerstein systems}.
\newblock {\em Automatica}, 51(1):111 -- 122, 2015.

\bibitem{Schoukens2008}
J.~Schoukens, L.~Gomm\'{e}, W.~Van~Moer, and Y.~Rolain.
\newblock {Identification of a Block-Structured Nonlinear Feedback System,
  Applied to a Microwave Crystal Detector}.
\newblock {\em IEEE Transactions on Instrumentation and Measurement},
  57(8):1734--1740, 2008.

\bibitem{RasmussenW:06}
C.~E. Rasmussen and C.~K.~I. Williams.
\newblock {\em Gaussian Processes for Machine Learning}.
\newblock MIT Press, Cambridge, MA, 2006.

\bibitem{Billings2013}
S.A. Billings.
\newblock {\em {Nonlinear System Identification: NARMAX Methods in the Time,
  Frequency, and Spatio-Temporal Domains}}.
\newblock John Wiley \& Sons, Ltd., West Sussex, United Kingdom, 2013.

\bibitem{Leontaritis1985}
I.J. Leontaritis and S.A. Billings.
\newblock {Input output parametric models for non-linear systems. Part II:
  stochastic non-linear systems}.
\newblock {\em International Journal of Control}, 41(2):329--344, 1985.

\bibitem{Liu2018}
Miao Liu, Girish Chowdhary, Bruno~Castra da~Silva, Shih-Yuan Liu, and
  Jonathan~P. How.
\newblock Gaussian processes for learning and control: A tutorial with
  examples.
\newblock {\em {IEEE} Control Systems}, 38(5):53--86, oct 2018.

\bibitem{RasmussenW:2006}
C.~E. Rasmussen and C.~K.~I. Williams.
\newblock {\em Gaussian processes for machine learning}.
\newblock MIT Press, 2006.

\bibitem{Sjoberg1995}
J.~Sj\"oberg, Q.H. Zhang, L.~Ljung, A.~Benveniste, B.~Delyon, P.Y. Glorennec,
  H.~Hjalmarsson, and A.~Juditsky.
\newblock Nonlinear black-box modeling in system identification: A unified
  overview.
\newblock {\em Automatica}, 31:1691--1724, 1995.

\bibitem{Juditsky1995}
A.~Juditsky, H.~Hjalmarsson, A.~Benveniste, B.~Delyon, L.~Ljung, J.~Sjoberg,
  and Q.H. Zhang.
\newblock Nonlinear black-box models in system identification: Mathematical
  foundations.
\newblock {\em Automatica}, 31:1725--1750, 1995.

\bibitem{FanG:96}
J.~Fan and I.~Gijbels.
\newblock {\em Local Polynomial Modelling and Its Applications}.
\newblock Number~66 in Monographs on Statistics and Applied Probability.
  Chapman \& Hall, 1996.

\bibitem{Nad:64}
E.~Nadaraya.
\newblock On estimating regression.
\newblock {\em Theory of Prob. and Applic.}, 9:141--142, 1964.

\bibitem{Epanechnikov:69}
V.~A. Epanechnikov.
\newblock Non-parametric estimation of a multivariate probability density.
\newblock {\em Theory of Probability and its Applications}, 14:4--20, 1969.

\bibitem{RollNL:05a}
Jacob Roll, A.~Nazin, and Lennart Ljung.
\newblock Non-linear system identification via direct weight optimization.
\newblock {\em Automatica}, 41(3):475--490, March 2005.

\bibitem{Tenenbaumetal:00}
J.~B. Tenenbaum, V.~de~Silva, and J.~C. Langford.
\newblock A global geometric framework for nonlinear dimensionality reduction.
\newblock {\em Science}, 290(5500):2319--2323, 2000.

\bibitem{RoweisS:00}
S.~T. Roweis and L.~K. Saul.
\newblock Nonlinear dimension reduction by locally linear embedding.
\newblock {\em Science}, 290:2323--2326, December 2000.

\bibitem{SchoukensJ1988}
J.~Schoukens, R.~Pintelon, E.~Vanderouderaa, and J.~Renneboog.
\newblock Survey of excitation signals for fft based signal analyzers.
\newblock {\em {IEEE Transactions on Instrumentation and Measurement}},
  {37}({3}):{342--352}, {1988}.

\bibitem{Schoukens2016}
J.~Schoukens, M.~Vaes, and R.~Pintelon.
\newblock {Linear system identification in a nonlinear setting: Nonparametric
  analysis of the nonlinear distortions and their impact on the best linear
  approximation}.
\newblock {\em IEEE Control Systems}, 36:38--69, 2016.

\bibitem{Schoukens2012}
J.~Schoukens, R.~Pintelon, and Y.~Rolain.
\newblock {\em {Mastering System Identification in 100 Exercises}}.
\newblock John Wiley \& Sons, Hoboken, New Jersey, 2012.

\bibitem{Geerardyn2013}
E.~Geerardyn, Y.~Rolain, R.~Pintelon, and J.~Schoukens.
\newblock Design of quasi-logarithmic multisine excitations for robust broad
  frequency band measurements.
\newblock {\em IEEE Transaction on Instrumentation and Measurement},
  62:1364--1372, 2013.

\bibitem{Schoukens1988}
J.~Schoukens and T.~Dobrowiecki.
\newblock Design of broadband excitation signals with a user imposed power
  spectrum and amplitude distribution.
\newblock In {\em IEEE Instrumentation and Measurement Technology Conference
  St. Paul, USA,}, 1998.

\bibitem{Mehra1974}
R.~K. Mehra.
\newblock Optimal input signals for parameter estimation in dynamic
  systemssurvey and new results.
\newblock {\em IEEE Transactions on Automatic Control}, 19:753--768, 1974.

\bibitem{Goodwin1977}
G.~C. Goodwin and R.~L. Payne.
\newblock {\em Dynamic System Identification: Experiment Design and Data
  Analysis}.
\newblock Academic, 1977.

\bibitem{Pronzato2008}
L.~Pronzato.
\newblock Optimal experimental design and some related control problems.
\newblock {\em Automatica}, 44:303--325.

\bibitem{Rivera2003}
D.E. Rivera, H.e Le, M.W. Braun, and H.D. Mittelman.
\newblock "plant-friendly" system identification: a challenge for the process
  industries.
\newblock {\em {IFAC} Proceedings Volumes of the 13th IFAC Symposium on System
  Identification (SYSID 2003) )}, 36(16):891--896, 2003.

\bibitem{Gevers2005}
M.~Gevers.
\newblock Identification for control: From the early achievements to the
  revival of experiment design.
\newblock {\em European Journal of Control}, 11:335--352, 2005.

\bibitem{Bombois2012}
X.~Bombois and G.~Scorletti.
\newblock Design of least costly identification experiments: The main
  philosophy accompanied by illustrative examples.
\newblock {\em J. Eur. des Syst. Automat.}, 46:587--610, 2012.

\bibitem{Forgione2015}
M.~Forgione, X.~Bombois, and P.~M.~J. Van~den Hof.
\newblock Data-driven model improvement for model-based control.
\newblock {\em Automatica}, 52:118--124, 2015.

\bibitem{annegren2017}
M.~Annergren, C.A. A.~Larsson, H.~Hjalmarsson, X.~Bombois, and B.~Wahlberg.
\newblock Application-oriented lnput design in system identification: Optimal
  input design for control.
\newblock {\em IEEE Control Systems Magazine}, 37:31--56, 2017.

\bibitem{Mahata2016}
K.~Mahata, J.~Schoukens, and A.~De~Cock.
\newblock Information matrix and d-optimal design with gaussian inputs for
  wiener model identification.
\newblock {\em Automatica}, 69:65--77, 2016.

\bibitem{Gevers2012}
M.~Gevers, M.~Caenepeel, and J.~Schoukens.
\newblock Experiment design for the idenification of a simple wiener system.
\newblock In {\em 51st IEEE Conference on Decision and Control, Maui, Hawaii,
  USA 2012, pp. 7333-7338}.

\bibitem{Birpoutsoukis2018}
G.~Birpoutsoukis.
\newblock {\em Volterra series estimation in the presence of prior knowledge}.
\newblock PhD thesis, Vrije Universiteit Brussel, 2018.

\bibitem{Forgione2014}
M.~Forgione, X.~Bombois, P.~M.~J. Van~den Hof, and H.~Hjalmarsson.
\newblock Experiment design for parameter estimation in nonlinear systems based
  on multilevel excitation.
\newblock pages 25--30, 2014.

\bibitem{DeCock2016}
A.~De~Cock, M.~Gevers, and J.~Schoukens.
\newblock D-optimal input design for nonlinear fir-type systems: A
  dispersion-based approach.
\newblock {\em Automatica}, pages 88 -- 100, 2016.

\bibitem{DeCock2017PhD}
A.~De~Cock.
\newblock {\em D-Optimal Input Design for the Identification of Structured
  Nonlinear Systems}.
\newblock PhD thesis, Vrije Universiteit Brussel, 2017.

\bibitem{Chianeh2011}
H.A. Chianeh, J.D. Stigter, and K.J. Keesman.
\newblock Optimal input design for parameter estimation in a single and double
  tank system through direct control of parametric output sensitivities.
\newblock {\em Journal of Process Control}, pages 111 -- 118, 2011.

\bibitem{Ljungtb:18}
L.~Ljung.
\newblock {\em System Identification Toolbox for use with \textsc{Matlab}.
  Version 9.}
\newblock The MathWorks, Inc, Natick, MA, 9th edition, 2018.

\bibitem{modval}
N.~Draper and H.~Smith.
\newblock {\em Applied Regression Analysis}.
\newblock Wiley, 1981.

\bibitem{Enqvist2007}
M.~Enqvist, J.~Schoukens, and R.~Pintelon.
\newblock Detection of unmodeled nonlinearities using correlation methods.
\newblock In {\em IMTC 2007 - IEEE Instrumentation and Measurement Technology
  Conference Warsaw, Poland, May 1-3, 2007}, 2007.

\bibitem{SchoukensM2016b}
M.~Schoukens, P.~Mattson, T.~Wigren, and J.P. No\"el.
\newblock {Cascaded tanks benchmark combining soft and hard nonlinearities}.
\newblock In {\em Workshop on Nonlinear System Identification Benchmarks},
  pages 20--23, Brussels, Belgium, April 2016.

\bibitem{Rogers2017}
T.J. Rogers, G.R. Holmes, E.J. Cross, and K.~Worden.
\newblock On a grey box modelling framework for nonlinear system
  identification.
\newblock In {\em Dervilis N. (eds) Special Topics in Structural Dynamics,
  Volume 6. Conference Proceedings of the Society for Experimental Mechanics
  Series. Springer, Cham}, 2017.

\bibitem{Andersson&Pucar:95}
Torbj{\"{o}}rn Andersson and Predrag Pucar.
\newblock Estimation of residence time in continuous flow systems with
  dynamics.
\newblock {\em Journal of Process Control}, 5:9--17, February 1995.

\bibitem{Heinz2017}
T.O. Heinz, M.~Schillinger, B.~Hartmann, and O.~Nelles.
\newblock Excitation signal design for nonlinear dynamic systems.
\newblock In Karsten R{\"o}pke and Clemens G{\"u}hmann, editors, {\em
  International Calibration Conference - Automotive Data Analytics, Methods,
  DoE}, pages 191--208, 2017.

\bibitem{Heinz2018}
T.O. Heinz and O.~Nelles.
\newblock Excitation signal design for nonlinear dynamic systems with multiple
  inputs - a data distribution approach.
\newblock {\em at - Automatisierungstechnik}, 66(9):714--724, 2018.

\bibitem{Tietze2014}
N.~Tietze, U.~Konigorsk, C.~Fleck, and D.~Nguyen-Tuong.
\newblock Model-based calibration of engine controller using automated
  transient design of experiment.
\newblock In {\em Proceedings}, pages 1587--1605. Springer Fachmedien
  Wiesbaden, 2014.

\bibitem{Nelles2006}
O.~Nelles.
\newblock Axes-oblique partitioning strategies for local model networks.
\newblock In {\em 2006 {IEEE} Conference on Computer Aided Control System
  Design, 2006 {IEEE} International Conference on Control Applications, 2006
  {IEEE} International Symposium on Intelligent Control}. {IEEE}, oct 2006.

\bibitem{Tietze2015}
Tietze N.
\newblock {\em Model-based Calibration of Engine Control Units Using Gaussian
  Process Regression}.
\newblock PhD thesis, 2015.

\bibitem{Vlaar2016}
M.~Vlaar, T.~Solis-Escalante, A.~Vardy, F.~{Van der Helm}, and A.~Schouten.
\newblock {Quantifying Nonlinear Contributions to Cortical Responses Evoked by
  Continuous Wrist Manipulation}.
\newblock {\em IEEE Transactions on Neural Systems and Rehabilitation
  Engineering}, 2016.

\bibitem{Vlaar2018}
M.P. Vlaar, G.~Birpoutsoukis, J.~Lataire, M.~Schoukens, A.C. Schouten,
  J.~Schoukens, and van~der Helm~F.C.T.
\newblock Modeling the nonlinear cortical response in {EEG} evoked by wrist
  joint manipulation.
\newblock {\em {IEEE} Transactions on Neural Systems and Rehabilitation
  Engineering}, 26(1):205--215, 2018.

\bibitem{Birpoutsoukis2017}
G.~Birpoutsoukis, A.~Marconato, J.~Lataire, and J.~Schoukens.
\newblock Regularized nonparametric volterra kernel estimation.
\newblock {\em Automatica}, 82:72--82, 2017.

\bibitem{Pillonetto2014}
G.~Pillonetto, F.~Dinuzzo, T.~Chen, G.~De~Nicolao, and L.~Ljung.
\newblock {Kernel methods in system identification, machine learning and
  function estimation: A survey}.
\newblock {\em {Automatica}}, {50}({3}):{657--682}, {2014}.

\bibitem{Relan2017}
R.~Relan, Y.~Firouz, J.M. Timmermans, and J.~Schoukens.
\newblock Data-driven nonlinear identification of li-ion battery based on a
  frequency domain nonparametric analysis.
\newblock {\em IEEE Transactions on Control Systems Technology}, 25:1825--1832,
  2017.

\bibitem{Relan2017b}
R.~Relan.
\newblock {\em Data-Driven Discrete-Time Identification of Continuous Time
  Nonlinear Systems and Nonlinear Modelling of Li-ion Batteries}.
\newblock PhD thesis, Vrije Universiteit Brussel, 2017.

\bibitem{Noel2015}
J.P No\"el, J.~Schoukens, and G.~Kerschen.
\newblock Grey-box nonlinear state-space modelling for mechanical vibrations
  identification.
\newblock In {\em Preprints of the 17th IFAC Symposium on System
  Identification, October 19-21, 2015. Beijing, China}, 2015.

\bibitem{Schetzen1980}
M.~Schetzen.
\newblock {\em {The Volterra and Wiener Theories of Nonlinear Systems}}.
\newblock Wiley, New York, 1980.

\bibitem{Pintelon2013}
R.~Pintelon and J.~Schoukens.
\newblock {FRF Measurement of Nonlinear Systems Operating in Closed Loop}.
\newblock {\em IEEE Transactions on Instrumentation and Measurement},
  62(5):1334--1345, May 2013.

\bibitem{Schetzen2006}
Martin Schetzen.
\newblock {\em The {Volterra} \& {Wiener} Theories of Nonlinear Systems}.
\newblock Krieger Publishing Company, Malabar, Florida, 2006.

\bibitem{Noel2014}
J.P. No{\"e}l, L.~Renson, and G.~Kerschen.
\newblock {Complex dynamics of a nonlinear aerospace structure: Experimental
  identification and modal interactions}.
\newblock {\em Journal of Sound and Vibration}, 333(12):2588--2607, 2014.

\bibitem{Thompson1986}
J.M.T. Thompson and H.B. Stewart.
\newblock {\em Nonlinear Dynamics and Chaos}.
\newblock Wiley, New York, 1986.

\bibitem{Ueda1991}
Y.~Ueda.
\newblock Survey of regular and chaotic phenomena in the forced duffing
  oscillator.
\newblock {\em Chaos, Solutions \& Fractals}, 1:199--231, 1991.

\bibitem{Westwick2018a}
D.T. Westwick, G.~Hollander, K.~Karami, and J.~Schoukens.
\newblock Using decoupling methods to reduce polynomial {NARX} models.
\newblock {\em {IFAC}-{PapersOnLine} 18th IFAC Symposium on System
  Identification, SYSID 2018}, 51(15):796--801, 2018.

\bibitem{Young2018}
Peter~C. Young and A.~Janot.
\newblock Efficient parameterisation of nonlinear system models: a comment on
  nöel and~schoukens~(2018).
\newblock {\em International Journal of Control}, pages 1--5, sep 2018.

\bibitem{Young2011}
Peter~C. Young.
\newblock {\em Recursive Estimation and Time-Series Analysis: An Introduction
  for the Student and Practitioner}.
\newblock Springer, 2011.

\bibitem{Goodwin2013}
G.C. Goodwin, J.C. Agüero, M.E.C. Garrido, M.E. Salgado, and J.I. Yuz.
\newblock Sampling and sampled-data models.
\newblock {\em IEEE Control Systems Magazine}, pages 34--53, 2013.

\bibitem{Pearson2004}
R.K. Pearson and U.~Kotta.
\newblock Nonlinear discrete-time models: state-space vs. i/o representations.
\newblock {\em Journal of Process Control}, 14:533--538, 2004.

\bibitem{Pearson2006}
R.K. Pearson.
\newblock Nonlinear empirical modeling techniques.
\newblock {\em Computers and Chemical Engineering}, 30:15141528, 2006.

\bibitem{Oppenheim1997}
A.V. Oppenheim and A.S. Willsky.
\newblock {\em Signals and Systems}.
\newblock Prentice-Hall, London, 1997.

\bibitem{Rao2006}
G.P. Rao and H.~Unbehauen.
\newblock Identification of continuous-time systems.
\newblock {\em IEE Proc.-Control Theory Appl.}, pages 185--220, 2006.

\bibitem{Garnier2015}
H.~Garnier.
\newblock Direct continuous-time approaches to system identification. overview
  and benefits for practical applications.
\newblock {\em European Journal of Control}, 24:50--62, 2015.

\bibitem{Zhang2015}
B.~Zhang and S.A. Billings.
\newblock Identification of continuous-time nonlinear systems: The nonlinear
  difference equation with moving average noise (ndema) framework.
\newblock {\em Mechanical Systems and Signal Processing}, pages 810--835, 2015.

\bibitem{SchoukensJ2017}
J.~Schoukens, R.~Relan, and M.~Schoukens.
\newblock Discrete time approximation of continuous time nonlinear state space
  models.
\newblock In {\em Proceedings of the 20th World Congress of the International
  Federation of Automatic Control, 9-14 July 2017 in IFAC-PapersOnLine, Volume
  50, Issue 1, July 2017, Pages 8339-8346}, 2017.

\bibitem{Baum1982}
R.F. Baum.
\newblock The correlation function of gaussian noise passe through nonlinear
  devices.
\newblock {\em IEEE Trans. on Information Theory}, 15:448--456, 1982.

\bibitem{Paduart2010}
J.~Paduart, L.~Lauwers, J.~Swevers, K.~Smolders, J.~Schoukens, and R.~Pintelon.
\newblock {Identification of nonlinear systems using polynomial nonlinear state
  space models}.
\newblock {\em Automatica}, 46(4):647--656, 2010.

\bibitem{Schoukens2015}
J.~Schoukens, R.~Pintelon, Y.~Rolain, M.~Schoukens, K.~Tiels, L.~Vanbeylen,
  A.~{Van Mulders}, and G.~Vandersteen.
\newblock {Structure discrimination in block-oriented models using linear
  approximations: A theoretic framework}.
\newblock {\em Automatica}, 53:225--234, 2015.

\bibitem{Esfahani2016}
A.~Fakhrizadeh~Esfahani, F.~Schoukens, and L.~Vanbeylen.
\newblock {Using the Best Linear Approximation With Varying Excitation Signals
  for Nonlinear System Characterization}.
\newblock {\em IEEE Transactions on Instrumentation and Measurement},
  65(5):1271--1280, May 2016.

\bibitem{Schoukens2003}
J.~Schoukens, J.G. Nemeth, P.~Crama, Y.~Rolain, and R.~Pintelon.
\newblock {Fast approximate identification of nonlinear systems}.
\newblock {\em Automatica}, {39}({7}):{1267--1274}, {2003}.

\bibitem{Worden2001}
G.R. Worden, K.~andTomlinson.
\newblock {\em Nonlinearity in Structural Dynamics: Detection, Identification
  and Modelling}.
\newblock Institute of Physics Publishing IOP, 2001.

\bibitem{Kerschen2006}
G.~Kerschen, K.~Worden, A.F. Vakakis, and J.~Golinval.
\newblock Past, present and future of nonlinear systems identification in
  structural dynamics.
\newblock {\em Mechanical Systems and Signal Processing}, 20:505--592, 2006.

\bibitem{Suykens1997}
J.A.K. Suykens, J.~Vandewalle, and B.L.R. De~Moor.
\newblock Nl theory: Checking and imposing stability of recurrent neural
  networks for nonlinear modeling.
\newblock {\em IEEE Trans. Signal Processing}, 45:2682 -- 2691, 1997.

\bibitem{Narendra1990}
K.~Narendra and K.~Parthasarty.
\newblock Identification and control of dynamical systems using neural
  networks.
\newblock {\em IEEE Trans. Neural Networks}, 1:4--27, 1990.

\bibitem{Widanage2011}
W.D. Widanage, J.~Stoev, A.~{Van Mulders}, J.~Schoukens, and G.~Pinte.
\newblock Nonlinear system-identification of the filling phase of a wet-clutch
  system.
\newblock {\em Control Engineering Practice}, 19(12):1506 -- 1516, 2011.

\bibitem{SchoukensJ2018a}
J.~Schoukens, K.~Godfrey, and M.~Schoukens.
\newblock Nonparametric data driven modeling of linear systems. estimating the
  frequency response and impulse response function.
\newblock {\em IEEE Control Systems Magazine}, page Accepted for publication,
  2018.

\bibitem{Schoukens2009}
J.~Schoukens, J.~Lataire, R.~Pintelon, G.~Vandersteen, and T.~Dobrowiecki.
\newblock {Robustness issues of the best linear approximation of a nonlinear
  system}.
\newblock {\em IEEE Transactions on Instrumentation and Measurement},
  58(5):1737--1745, 2009.

\bibitem{SchoukensJ1999a}
J.~Schoukens, R.~Pintelon, and Y.~Rolain.
\newblock Study of conditional ml estimators in time and frequency-domain
  system identification.
\newblock {\em Automatica}, 35:91--100, 1999.

\bibitem{SchoukensJ2011}
J.~Schoukens, Y.~Rolain, G.~Vandersteen, and R.~Pintelon.
\newblock User friendly box-jenkins identification using nonparametric noise
  models.
\newblock In {\em 50th IEEE Conference on Decision and Control and European
  Control Conference (CDC-ECC), Orlando, FL, USA, December 12-15}, 2011.

\bibitem{SchoukensJ2009}
J.~Schoukens, G.~Vandersteen, K.~Barbe, and R.~Pintelon.
\newblock {Nonparametric Preprocessing in system identification: a powerful
  tool}.
\newblock {\em {European Journal of Control}}, {15}:{260--274}, {2009}.

\bibitem{Pintelon2010a}
R.~Pintelon, J.~Schoukens, G.~Vandersteen, and K.~Barbe.
\newblock {Estimation of nonparametric noise and FRF models for multivariable
  systems-Part I: Theory}.
\newblock {\em Mechanical Systems and Signal Processing}, {24}({3}):{573--595},
  {2010}.

\bibitem{Pintelon2010b}
R.~Pintelon, J.~Schoukens, G.~Vandersteen, and K.~Barbe.
\newblock {Estimation of nonparametric noise and FRF models for multivariable
  systems-Part II: Extensions, applications}.
\newblock {\em Mechanical Systems and Signal Processing}, {24}({3}):{596--616},
  {2010}.

\bibitem{SchoukensJ2010a}
J.~Schoukens and Pintelon.
\newblock Study of the variance of parametric estimates of the bestllinear
  approximation of nonlinear systems.
\newblock {\em IEEE Transactions on Instrumentation and Measurement},
  59:3159--3167, 2010.

\bibitem{Bussgang1952}
J.J. Bussgang.
\newblock {Cross-correlation functions of amplitude-distorted Gaussian
  signals}.
\newblock Technical Report 216, MIT Laboratory of Electronics, 1952.

\bibitem{Gelb1968}
A.~Gelb and W.E. Vander~Velde.
\newblock {\em Multiple-Input Describing Functions and Nonlinear System
  Design}.
\newblock McGraw-Hill, 1968.

\bibitem{Enqvist2005a}
M.~Enqvist and L.~Ljung.
\newblock {Linear approximations of nonlinear FIR systems for separable input
  processes}.
\newblock {\em Automatica}, 41(3):459--473, 2005.

\bibitem{Enqvist2011}
M.~Enqvist.
\newblock {Separability of scalar random multisine signals}.
\newblock {\em Automatica}, 47(9):1860--1867, 2011.

\bibitem{Ertveldt2017}
J.~Ertveldt.
\newblock {\em Application of frequency-domain system identification to wind
  tunnel experiments on the Active Aeroelastic Test Bench}.
\newblock PhD thesis, Vrije Universiteit Brussel, 2017.

\bibitem{Ertveldt2015}
J.~Ertveldt, J.~Schoukens, R.~Pintelon, and S.~Vanlanduit.
\newblock Experiments on the active aeroelastic test bench (aatb) for the
  identification of unsteady aerodynamics.
\newblock In {\em Proceedings of the International Forum on Aeroelasticity and
  Structural Dynamics (IFASD), Saint-Petersburg, Russia}, 2015.

\bibitem{Ljung2001}
L.~Ljung.
\newblock {Estimating Linear Time-invariant Models of Nonlinear Time-varying
  Systems}.
\newblock {\em European Journal of Control}, 7(2-3):203--219, 2001.

\bibitem{Enqvist2005}
M.~Enqvist.
\newblock {\em {Linear Models of Nonlinear systems}}.
\newblock PhD thesis, Institute of technology, Link\"{o}ping University,
  Sweden, 2005.

\bibitem{Schoukens1998}
J.~Schoukens, T.~Dobrowiecki, and R.~Pintelon.
\newblock {Parametric and non-parametric identification of linear systems in
  the presence of nonlinear distortions. A frequency domain approach}.
\newblock {\em IEEE Transactions on Automatic Control}, 43(2):176--190, 1998.

\bibitem{SchoukensM2018}
M.~Schoukens, R.~Pintelon, T.~Dobrowiecki, and J.~Schoukens.
\newblock Extending the best linear approximation framework to the process
  noise case.
\newblock {\em arXiv:1804.07510}, 2018.

\bibitem{Arfei2013}
F.~Arfaei~Malekzadeh, R.~Mahmoudi, and A.~van Roermund.
\newblock {\em Analog Dithering Techniques for Wireless Transmitters}.
\newblock Springer, 2013.

\bibitem{Atherton1982}
D.P. Atherton.
\newblock {\em Nonlinear Control Engineering}.
\newblock Van Nostrand Reinhold, 1982.

\bibitem{Carbone2000}
P.~Carbone and D.~Petri.
\newblock Performance of stochastic and deterministic dithered quantizers.
\newblock {\em IEEE Transactions on Instrumentation and Measurement}, pages
  337--340, 2000.

\bibitem{Zhang2017}
E.~Zhang, M.~Schoukens, and J.~Schoukens.
\newblock Structure detection of wienerhammerstein systems with process noise.
\newblock {\em IEEE Transactions on Instrumentation and Measurement},
  66:569--576, 2017.

\bibitem{SchoukensM2016c}
M.~Schoukens and J.P. No\"el.
\newblock Wiener-hammerstein benchmark with process noise.
\newblock In {\em 2nd Workshop on Nonlinear System Identification Benchmarks.
  Brussels, 2016. http://www.nonlinearbenchmark.org}, 2016.

\bibitem{Hagenblad2008}
A.~Hagenblad, L.~Ljung, and A.~Wills.
\newblock {Maximum likelihood identification of Wiener models}.
\newblock {\em Automatica}, 44(11):2697--2705, 2008.

\bibitem{DoucetJ:2011}
A.~Doucet and A.~M. Johansen.
\newblock A tutorial on particle filtering and smoothing: Fifteen years later.
\newblock In D.~Crisan and B.~Rozovsky, editors, {\em Nonlinear Filtering
  Handbook}. Oxford University Press, 2011.

\bibitem{Gordon:1993}
N.~J. Gordon, D.~J. Salmond, and A.~F.~M. Smith.
\newblock Novel approach to nonlinear/non-{G}aussian {B}ayesian state
  estimation.
\newblock In {\em IEE Proceedings on Radar and Signal Processing}, volume 140,
  pages 107--113, 1993.

\bibitem{Delmoral:2004}
P.~Del~Moral.
\newblock {\em Feynman-Kac formulae: Genealogical and Interacting Particle
  Systems with Applications}.
\newblock Springer, New York, USA, 2004.

\bibitem{PittSGK:2012}
M.~K. Pitt, R.~dos Santos~Silva, R.~Giordani, and R.~Kohn.
\newblock On some properties of {M}arkov chain {M}onte {C}arlo simulation
  methods based on the particle filter.
\newblock {\em Journal of Econometrics}, 171(2):134--151, 2012.

\bibitem{PoyiadjisDS:2011}
G.~Poyiadjis, A.~Doucet, and S.S. Singh.
\newblock Particle approximations of the score and observed information matrix
  in state space models with application to parameter estimation.
\newblock {\em Biometrika}, 98(1):65--80, 2011.

\bibitem{DoucetJR:2015}
A.~Doucet, P.~E. Jacob, and S.~Rubenthaler.
\newblock Derivative-free estimation of the score vector and observed
  information matrix with application to state-space models.
\newblock Technical report, arXiv:1304.5768, July 2015.

\bibitem{BottouCN:2018}
L.~Bottou, F.~E. Curtis, and J.~Nocedal.
\newblock Optimization methods for large-scale machine learning.
\newblock {\em SIAM Review}, 60(2):223--311, 2018.

\bibitem{DempsterLR:1977}
A.~Dempster, N.~Laird, and D.~Rubin.
\newblock Maximum likelihood from incomplete data via the {EM} algorithm.
\newblock {\em Journal of the {R}oyal {S}tatistical {S}ociety, {S}eries {B}},
  39(1):1--38, 1977.

\bibitem{SchonWN:2011}
T.~B. Sch\"{o}n, A.~Wills, and B.~Ninness.
\newblock System identification of nonlinear state-space models.
\newblock {\em Automatica}, 47(1):39--49, January 2011.

\bibitem{OlssonDCM:2008}
J.~Olsson, R.~Douc, O.~Capp{\'{e}}, and E.~Moulines.
\newblock Sequential {M}onte {C}arlo smoothing with application to parameter
  estimation in nonlinear state-space models.
\newblock {\em Bernoulli}, 14(1):155--179, 2008.

\bibitem{Peterka:1981}
V.~Peterka.
\newblock Bayesian system identification.
\newblock {\em Automatica}, 17(1):41--53, 1981.

\bibitem{AndrieuDH:2010}
C.~Andrieu, A.~Doucet, and R.~Holenstein.
\newblock Particle {M}arkov chain {M}onte {C}arlo methods.
\newblock {\em Journal of the Royal Statistical Society. Series B
  (Methodological)}, 72(2):1--33, 2010.

\bibitem{SchonLDWNSD:2015}
T.~B. Sch\"{o}n, F.~Lindsten, J.~Dahlin, J.~W{\aa}gberg, A.~C. Naesseth,
  A.~Svensson, and L.~Dai.
\newblock Sequential {M}onte {C}arlo methods for system identification.
\newblock In {\em Proceedings of the 17th IFAC Symposium on System
  Identification (SYSID)}, Beijing, China, October 2015.

\bibitem{Kantas:2015}
N.~Kantas, A.~Doucet, S.~S. Singh, J.~M. Maciejowski, and N.~Chopin.
\newblock On particle methods for parameter estimation in state-space models.
\newblock {\em Statistical Science}, 30(3):328--351, 2015.

\bibitem{WillsSLN:2013}
A.~Wills, T.~B. Sch\"{o}n, L.~Ljung, and B.~Ninness.
\newblock Identification of {H}ammerstein-{W}iener models.
\newblock {\em Automatica}, 49(1):70--81, 2013.

\bibitem{LindstenJS:2013}
F.~Lindsten, T.~B. Sch\"{o}n, and M.~I. Jordan.
\newblock Bayesian semiparametric wiener system identification.
\newblock {\em Automatica}, 49:2053--2063, 2013.

\bibitem{PillonettoDCNL:2014}
G.~Pillonetto, F.~Dinuzzo, T.~Chen, G.~De~Nicolao, and L.~Ljung.
\newblock Kernel methods in system identification, machine learning and
  function estimation: a survey.
\newblock {\em Automatica}, 50(3):657--682, 2014.

\bibitem{Sliwinski2017}
P.~Sliwinski, A.~Marconato, P.~Wachel, and G.~Birpoutsoukis.
\newblock Non-linear system modelling based on constrained volterra series
  estimates.
\newblock {\em IET Control Theory and Applications}, 11:2623--2629.

\bibitem{Tibsharani:96}
R.~Tibsharani.
\newblock Regression shrinkage and selection via the lasso.
\newblock {\em Journal of Royal Statistical Society B (Methodological)},
  58(1):267--288, 1996.

\bibitem{Favier2009}
G.~Favier and T.~Bouilloc.
\newblock {Parametric complexity reduction of Volterra models using tensor
  decompositions}.
\newblock In {\em 17th European Signal Processing Conference (EUSIPCO)}, pages
  2288--2292, Glasgow, Scotland, Aug. 2009.

\bibitem{SchoukensM2012a}
M.~Schoukens and Y.~Rolain.
\newblock {Crossterm elimination in parallel Wiener systems using a linear
  input transformation}.
\newblock {\em IEEE Transactions on Instrumentation and Measurement},
  61(3):845--847, 2012.

\bibitem{Tiels2013}
K.~Tiels and J.~Schoukens.
\newblock {From coupled to decoupled polynomial representations in parallel
  Wiener-Hammerstein models}.
\newblock In {\em 52nd IEEE Conference on Decision and Control (CDC)}, pages
  4937--4942, Florence, Italy, Dec. 2013.

\bibitem{SchoukensM2014a}
M.~Schoukens, K.~Tiels, L.~Ishteva, and J.~Schoukens.
\newblock {Identification of parallel Wiener-Hammerstein systems with a
  decoupled static nonlinearity}.
\newblock In {\em 19th World Congress of the International Federation of
  Automatic Control}, pages 505--510, Cape Town, South Africa, Aug. 2014.

\bibitem{Dreesen2015}
P.~Dreesen, M.~Schoukens, K.~Tiels, and J.~Schoukens.
\newblock {Decoupling Static Nonlinearities in a Parallel Wiener-Hammerstein
  System: A First-order Approach}.
\newblock In {\em IEEE International Instrumentation and Measurement Technology
  Conference (I2MTC)}, Pisa, Italy, May 2015.

\bibitem{Esfahani2018}
A.~Fakhrizadeh~Esfahani, P.~Dreesen, K.~Tiels, J.~No\"el, and Schoukens J.
\newblock Parameter reduction in nonlinear state-space identification of
  hysteresis.
\newblock {\em Mechanical Systems and Signal Processing}, page Accepted for
  publication, 2018.

\bibitem{Stoev2017}
J.~Stoev, J.~Ertveldt, T.~Oomen, and J.~Schoukens.
\newblock Tensor methods for mimo decoupling and control design using frequency
  response functions.
\newblock {\em Mechatronics}, 45:71--81, 2017.

\bibitem{Dreesen2015b}
P.~Dreesen, M.~Ishteva, and J.~Schoukens.
\newblock Decoupling multivariate polynomials using first-order information and
  tensor decompositions.
\newblock {\em Siam Journal on Matrix Analysis and Applications}, pages
  864--879, 2015.

\bibitem{Hollander2018}
G.~Hollander, P.~Dreesen, M.~Ishteva, and Schoukens J.
\newblock Approximate decoupling of multivariate polynomials using weighted
  tensor decomposition.
\newblock {\em Numerical Linear Algebra with Applications - accepted for
  publication}, 2018.

\bibitem{Hornik1989}
K.~Hornik, M.~Stinchcombe, and H.~White.
\newblock Multilayer feedforward networks are universal approximators.
\newblock {\em Neural Networks}, 2(5):359--366, 1989.

\bibitem{Suykens1996}
J.A.K. Suykens, J.P.L. Vandewalle, and B.L.R. De~Moor.
\newblock {\em Artificial neural networks for modeling and control of
  non-linear systems}.
\newblock Kluwer Academic Publishers, 1996.

\end{thebibliography}

\end{document}